%% file: main.tex
\documentclass[  conference]{IEEEtran}
\usepackage{cite}
\usepackage{amsmath,amssymb,amsfonts}
\usepackage{graphicx}
\usepackage{textcomp}
\usepackage{xcolor}
\usepackage[hyphens]{url}
\usepackage[caption=false]{subfig}
\usepackage{multirow}
\usepackage{arydshln}
\usepackage[linesnumbered,ruled,vlined]{algorithm2e}
\usepackage{pgfplots}
\usepackage{svg}
\usepackage{import}

\def\BibTeX{{\rm B\kern-.05em{\sc i\kern-.025em b}\kern-.08em
    T\kern-.1667em\lower.7ex\hbox{E}\kern-.125emX}}

\newcommand*\circled[1]{\tikz[baseline=(char.base)]{
    \node[shape=circle,draw,inner sep=1pt] (char) {#1};}}

\usepackage[noend]{algpseudocode}
\algnewcommand{\LeftComment}[1]{\(\triangleright\)#1} 
\input{comments}

\title{
L2C2: Last-Level Compressed-Cache NVM and a Procedure to Forecast Performance and Lifetime}

\author{\IEEEauthorblockN{Carlos Escuin\IEEEauthorrefmark{1}, Pablo Ibañez\IEEEauthorrefmark{1}, Teresa Monreal\IEEEauthorrefmark{2}, Jose M. Llaberia\IEEEauthorrefmark{2}, Victor Viñals\IEEEauthorrefmark{1}}
\IEEEauthorblockA{\IEEEauthorrefmark{1}University of Zaragoza}
\IEEEauthorblockA{\IEEEauthorrefmark{2}Universitat Politècnica de Catalunya}}

\begin{document}
\maketitle
\thispagestyle{plain}
\pagestyle{plain}


\begin{abstract}
Several emerging non-volatile (NV) memory technologies are rising as interesting alternatives to build the Last-Level Cache (LLC). Their advantages, compared to SRAM memory, are higher density and lower static power, but write operations wear out the bitcells to the point of eventually losing their storage capacity. In this context, this paper presents a novel LLC organization designed to extend the lifetime of the NV data array and a procedure to forecast in detail the capacity and performance of such an NV-LLC over its lifetime.

From a methodological point of view, although different approaches are used in the literature to analyze the degradation of an NV-LLC, none of them allows to study in detail its temporal evolution. In this sense, this work proposes a forecast procedure that combines detailed simulation and prediction, allowing an accurate analysis of the impact of different cache control policies and mechanisms (replacement, wear-leveling, compression, etc.) on the temporal evolution of the indices of interest, such as the effective capacity of the NV-LLC or the system IPC.

We also introduce L2C2, a LLC design intended for implementation in NV memory technology that combines fault tolerance, compression, and internal write wear leveling for the first time. Compression is not used to store more blocks and increase the hit rate, but to reduce the write rate and increase the lifetime during which the cache supports near-peak performance. 
In addition, to support byte loss without performance drop, L2C2 inherently allows N redundant bytes to be added to each cache entry. Thus, L2C2+N, the endurance-scaled version of L2C2, allows balancing the cost of redundant capacity with the benefit of longer lifetime.

For instance, as a use case, we have implemented the L2C2 cache with STT-RAM technology. It has affordable hardware overheads compared to that of a baseline NV-LLC without compression in terms of area, latency and energy consumption, and increases up to 6-37 times the time in which 50\% of the effective capacity is degraded, depending on the variability in the manufacturing process. Compared to L2C2, L2C2+6 which adds 6 bytes of redundant capacity per entry, that means 9.1\% of storage overhead, can increase up to 1.4-4.3 times the time in which the system gets its initial peak performance degraded.

\end{abstract}


\input{I-Introduction}

\input{II-Background}
\input{III-L2C2}
\input{IV-Forecasting}

\input{V-Methodology}

\input{VI-Validation}
\input{VII-Evaluation}

\input{VIII-Conclusions}

\input{Appendix}

\section*{Acknowledgements}
All authors acknowledge support from grants (1) PID2019-105660RB-C21 and PID2019-107255GB-C22 / AEI / 10.13039/501100011033 from Agencia Estatal de Investigación (AEI) and European Regional Development Fund (ERDF), (2) gaZ: T58\_20R research group from Aragón Government and European Social Fund (ESF), and (3) 2014-2020 "Construyendo Europa desde Aragón" from European Regional Development Fund (ERDF). The funders had no role in study design, data collection and analysis, decision to publish, or preparation of the manuscript.


\bibliographystyle{IEEEtranS}
\bibliography{refs}


\end{document}

%% file: comments.tex
\usepackage{ifthen}
\usepackage{amssymb}
\usepackage{xcolor}

\newboolean{showcomments}
\setboolean{showcomments}{true}

\makeatletter
\newcommand{\mynote}[3]{%
  \ifthenelse{\boolean{showcomments}}{%
   \fbox{\bfseries\sffamily\scriptsize#1}%
   {\small$\blacktriangleright$\textsf{\emph{\color{#3}{#2}}}$\blacktriangleleft$}}%
  {%
   \@bsphack
   \@esphack
  }%
}
\makeatother

\definecolor{asparagus}{rgb}{0.53, 0.66, 0.42}
\definecolor{amber}{rgb}{1.0, 0.75, 0.0}
\definecolor{fuchsia}{rgb}{1.0, 0.10, 1.0}
\definecolor{customblue}{rgb}{0.4, 0.4, 1.0}

\usepackage{verbatim}

\usepackage{caption}
\usepackage{graphicx}
\usepackage{multirow}
\usepackage{arydshln}
\usepackage{booktabs}

%% file: I-Introduction.tex
\section{Introduction, Motivation and Contributions}

The goal of the cache subsystem in a shared memory multiprocessor is to reduce the number of main memory accesses. Specifically, the shared last-level cache (LLC) filters requests from the lower-level caches turning slow main memory accesses into fast LLC hits, saving main memory bandwidth, power, and increasing system performance. However, the number of cores/threads integrated on a chip grows faster than the bandwidth with main memory. Therefore, it is necessary to improve the hit ratio of the LLC by increasing not only  total size but also  size per core/thread.
Most LLCs are implemented with 6T-SRAM cells, a technology that does not scale well in terms of density and static power~\cite{Sakhare-18-LLC-STT-RAM-5nm}. 

In the short to medium term, non-volatile memory (NVM) technologies rise as an alternative to SRAMs due to their higher density and lower static power.
Among these technologies we can mention phase change (PCM)~\cite{Lee-09-PCM, Qureshi-11-PCM, Joo-10-PCMLLC}, magnetic tunnel junction (STT-RAM)~\cite{Apalkov-13-STTRAM, Korgaonkar-18-DensityTradeoffs-WCAB-VHC, Salehi-17-STT-RAM-Survey, Sakhare-18-LLC-STT-RAM-5nm}, or resistive (ReRAM)~\cite{Xu-15-OvercomingCrossbarResistive, Zhang-16-ReRAM}.
However, write operations on most NVMs cause noticeable wear on their bitcells, making their lifetime much shorter than that of SRAMs. The simplest way to deal with an uncorrectable fault in a bitcell is to disable the memory region to which it belongs, with a size that depends on the context: a whole memory page, a cache frame\footnote{From now on we will use the term cache \textit{frame} to designate the set of physical bitcells of the data array holding a cache \textit{block}, compressed or not.}, or a byte.

In this paper we present two contributions to the design and evaluation of NV-LLCs made up with memory bitcells that wear out with writes. First, L2C2, a new fault-tolerant last-level cache organization intended for NV technologies that relies on byte disabling and data compression to increase lifetime while keeping performance. The design called L2C2+N is the endurance-scaled version of L2C2 with no more than adding N spare bytes. Second, we introduce a procedure to forecast the time evolution of effective capacity and performance, suitable for modeling either frame disabling or byte disabling with compression.

\subsection{L2C2: Last-Level Compressed-Contents NV cache}

Proposals on NV-LLC organizations usually focus on mechanisms to decrease and/or balance the number of writes, seeking to increase the lifetime and at the same time, if possible, counteract the high cost in energy and latency of writes. 

To reduce the number of writes, it has been proposed, for example, to reduce the number of inserted cache blocks, using some kind of filtering~\cite{Rodriguez-18-Reuse,Cheng-16-LAP, Ahn-2014-DASCA}, or collaborating with the private levels~\cite{Korgaonkar-18-DensityTradeoffs-WCAB-VHC}. Other techniques to reduce writes are closely tied to particular bitcell designs, supporting e.g., read-before-write~\cite{Joo-10-PCMLLC}, or early-write-termination ~\cite{zhou2009durable, Yazdanshenas-13-CodingLLC}.
Finally, it is worth mentioning the proposals for hybrid
SRAM/NVM LLCs, which stand out for their great potential to reduce writes, in exchange for a more complex design that seeks to send as many write requests as possible to the SRAM part without losing performance or increasing power consumption~\cite{Wang-14-AdaptivePlacementHybrid, Cheng-16-LAP}.


Wear-leveling mechanisms focus on evenly distributing write operations throughout all the NV-LLC dimensions: cache sets, ways within sets, and bytes within frames~\cite{Wang-13-I2WAP, Farbeh-16-FloatingECC, Agarwal-20-LinovoThesis, Joo-10-PCMLLC}. These works seek to slow down write wear by avoiding the formation of hot spots, but none of them considers how to prolong service in the presence of defective bitcells, possibly allowing for a gradual loss of performance.

The possibility of storing compressed blocks in an NV-LLC has hardly been explored. For example, Choi et al.~\cite{Choi-14-AdaptiveCacheCompression} explores an adaptation of the DCC compression scheme proposed for SRAM caches~\cite{Sardashti-2013-DCC}. Just as in DCC caches their goal is to increase the effective capacity by allowing the total number of compressed blocks stored in a cache set to exceed the nominal associativity. Using a set dueling mechanism they dynamically adjust the activation/deactivation of compression to balance the miss rate vs. write rate tradeoff, concluding that their proposal increases energy efficiency, but decreases lifetime by 8\% with respect to a cache without compression.

Thus, in relation to our goal of designing an NVM cache that continues to provide service as bitcells fail, it is appropriate to broaden the scope to look for related solutions that could serve our purpose.  

Any memory structure is subject to experience a bitcell fault during its operation, transient or permanent. To avoid a system crash, the memory controller must detect the error and correct it. To do this, fault-tolerant cache memories may protect every cache frame with an error correction code (ECC) mechanism, capable of at least detecting up to two errors and correcting one (SECDED). If the error comes from a permanent bitcell fault, the SECDED mechanism no longer ensures correct operation, since a new error in another bitcell of the same frame would be unrecoverable.
The simplest solution is \textit{frame disabling}, already present in commercial processors long time ago~\cite{Chang-07-BD1, Wuu-05-Asynchronous}. It consists of disabling the entire cache frame when a permanent fault is detected in one of its bitcells. Other approaches propose to add more redundancy, seeking to correct more than one error in each frame. Redundancy can be included in the error correction code itself, allowing to correct N errors instead of just one~\cite{Kim-07-MultiBit}. However, the overhead required by such ECCs increases rapidly with N, to the point of making it impractical.

Alternatively, redundancy can be added outside the ECC mechanism by noting permanently failed bitcells and correcting their value~\cite{Schechter-10-UseECPnotECC, Seong-10-SAFER, Yoon-11-FREE-p}. For example, Schechter et al., in the context of main memory proposes the Error-Correcting Pointers (ECP) mechanism that stores for each faulty bitcell its frame position and the value it should store, e.g. a 9-bit pointer for a 64-byte memory frame and a one bit data, respectively~\cite{Schechter-10-UseECPnotECC}.
The extra storage cost limits this approach to a moderate number of faulty cells. In fact, the authors evaluate the mechanism for up to N=6 defective bitcells (ECP-6).
 
Other work proposes to take advantage of memory frames with defects without having to disable them entirely. For example, Ipek et al. proposes the Dynamically Replicated Memory (DRM) technique to store a memory page in two partially faulty page frames~\cite{Ipek-10-DynamicallyReplicatedMemory}. Or, with a higher complexity, Jadidi et al. advocate the use of compression to harden main memory~\cite{Jadidi-17-CollaborativeCompression}. They assume a PCM memory with ECP-6 protection for each 64-byte frame. Their mechanism allows storing a compressed block in a degraded frame, as long as there is a contiguous chunk within the frame, called compression window, of size greater than or equal to the compressed block, and with no more than 6 bitcell faults. This allows a memory frame to be used even if it has more than 6 faults, as long as they are outside the compression window. In summary, this proposal increases memory lifetime by three aggregate effects: it has a repair mechanism, it decreases the write rate by the same amount as the compression rate achieved, and it does not create write hot spots because it has an intra-frame write leveling mechanism. However, although its ideas are inspiring, this proposal has been developed to collaborate with OS paging system and its direct transfer to cache memory hardware is not straightforward at all.

Data compression has also been proposed in the context of caches operating at near-threshold voltage. Ferrerón et al. propose the Concertina cache, which  provides each frame with a bit vector or a few pointers identifying the bytes that fault when the supply voltage drops~\cite{Ferreron-15-Concertina}. These metadata are calculated once, by scanning the cache when entering in low-voltage mode, and do not change as long as the supply voltage remains constant. 
Before inserting a new block, a simple null subblock compression mechanism searches in LRU order for the existence of a frame with enough live bytes. Concertina does not need or seek to level write wear, nor requires a high-coverage compression mechanism, but part of its design will be useful for our proposal.

\textbf{Contributions.}
L2C2 is the first NV-LLC capable of tolerating byte faults in NVM bitcells. It uses data block compression and intra-frame write leveling to extend the lifetime of degraded frames. In contrast to current alternatives, it is able to maintain high performance for a longer time, or in other words, for a given time of use it achieves higher performance, and it does so at a reasonable hardware cost. Moreover, its design is inherently scalable in terms of lifetime: simply adding N additional spare bytes to each frame, without modifying the design ideas, results in L2C2+N, the endurance-scaled version of L2C2, which is able to support the nominal capacity for longer.

On the one hand, the design of L2C2 carefully considers previous concepts of non-volatile main memory management and SRAM caches, namely:

\begin{itemize}
\item Support for byte disabling~\cite{Ferreron-15-Concertina}, by incorporating the necessary metadata to identify non-operational bytes. Besides, a SECDED mechanism is incorporated with the ability to trigger an Operating System routine that disables a byte by modifying such metadata.
\item BDI compression~\cite{Pekhimenko-12-BaseDeltaImmediate}. This data compression mechanism is selected because it provides high coverage and a good compression ratio. These two characteristics allow, simultaneously, to reduce the number of bitcells written (more duration) and to increase the possibilities of saving the block in frames of reduced size (more performance). In addition, its hardware implementation has low decompression latency. 
\item LRU-Fit replacement algorithm~\cite{Ferreron-15-Concertina}. After appropriate experimentation, this option is selected. LRU-Fit is a locality-aware replacement algorithm, which selects the LRU victim cache frame among all those that are large enough to allocate the incoming compressed block (Fit). 
\end{itemize}

On the other hand, L2C2 incorporates two original enhancements, which are crucial to maintain high performance for a longer time, namely:

\begin{itemize}
\item Intra-frame wear-leveling and compressed block rearrangement within the frame. We propose a new mechanism that achieves three key objectives: (a) wear out the live bytes of each frame evenly as the rest is failing, (b) upon inserting a compressed block into L2C2, rearrange the byte layout of the compressed block to write the appropriate subset of live bytes of the frame, and (c) the same but in the reverse direction, i.e., in case of a L2C2 hit, reconstruct the original layout of a compressed block, which is scattered in a partially broken frame, to supply it to the decompressor. Using VLSI synthesis, that circuitry has been shown to be feasible in terms of area, latency and power consumption.
\item Because the above mechanism is scalable, it is possible to add an arbitrary number of N redundant bytes to each frame, privately and without any change in the design. L2C2+N, the version of L2C2 with redundancy, thus has frames with 64+N data bytes that cooperate in storing compressed blocks from the beginning, extending the cache lifetime in proportion to the built-in N degree of redundancy.
\end{itemize}

\subsection{Forecasting capacity and performance evolution of NV-LLCs}

Previous work often highlights the difficulty of accurately modeling aging in NV memory and its effects on performance. In the absence of a standard procedure, practical solutions have been proposed, designed to assess specific aspects of one or another mechanism.

A first group of papers related to the evaluation of reliability improvement in NV main memory or cache, focuses exclusively on measuring either the reduction in the number of writes or their variability~\cite{Dgien-14-CompressionArchitecture, Palangappa-18-CASTLE, Wang-13-I2WAP}. For instance, Wang et al. compare wear-leveling mechanisms in NV-LLCs by calculating the elapsed time from startup to the first bitcell fault~\cite{Wang-13-I2WAP}. Such cache lifetime is computed by dividing the maximum number of writes supported by a bitcell by the number of writes per unit time (write rate) in the cache line that accumulates the most writes 
The procedure consists of a single cycle-accurate simulation to record write variability, followed by an aging prediction that assumes such variability to be constant throughout the life of the cache. This procedure is simple, fast and allows the production of performance metrics such as the number of instructions per cycle (IPC), but does not take into account the manufacturing variability in bitcell endurance. More importantly, it also does not allow to calculate the time evolution of the capacity or performance in degraded mode of operation, in which cache frames are progressively lost.

A second group of works, focused on extending the main memory lifetime, already incorporate process variability, modeling bitcell endurance by means of a normal probability distribution~\cite{Ipek-10-DynamicallyReplicatedMemory, Schechter-10-UseECPnotECC, Seong-10-SAFER, Yoon-11-FREE-p,Jadidi-17-CollaborativeCompression}.  

Ipek et al.~\cite{Ipek-10-DynamicallyReplicatedMemory} and Seong et al.~\cite{Seong-10-SAFER}  assume that writes are spread evenly across the main memory. Their quality metric is the number of writes the memory can receive until the first unrecoverable fault occurs on any of its pages~\cite{Seong-10-SAFER} or until the memory loses all its capacity (each page is deactivated when it reaches its write limit)~\cite{Ipek-10-DynamicallyReplicatedMemory}. They do not relate the number of writes to the time elapsed, and therefore do not need to simulate any application. Yoon et al. propose the same quality metric~\cite{Yoon-11-FREE-p}, but assume that the page write rate is constant, thus expressing memory lifetime in elapsed time, rather than number of writes.

Schechter et al.~\cite{Schechter-10-UseECPnotECC} and Jadidi et al.~\cite{Jadidi-17-CollaborativeCompression} simulate a workload on a system whose main memory has no faulty cells. The former to obtain frequency of writes and the latter to obtain traces of memory accesses. Schechter et al.  evenly distribute the number of writes among all live pages and calculate the bitcell that will fail first~\cite{Schechter-10-UseECPnotECC}. When the number of faulty cells in a page reaches a threshold, the page is deactivated and its writes are distributed evenly among the remaining live pages. Jadidi et al. use the trace of writes to main memory to accumulate the number of writes to each bitcell, deactivating them when they reach their maximum number.~\cite{Jadidi-17-CollaborativeCompression}. The simulation is repeated several times, until the memory loses half of its capacity. They measure the lifetime by counting the number of times the trace is reinjected. Now, from the execution time of the detailed simulation that produced the trace they can calculate a lifespan in terms of elapsed time, but always assuming that performance does not vary with memory degradation.

In summary, to date no procedure capable of accurately estimating the simultaneous degradation of capacity and performance over time has been proposed.


The essence of the problem is as follows. Memory cells age with writes. And as memory degrades performance and write rates also change. But its detailed simulation, cycle by cycle, requires a time that would far exceed the lifetime of the system under study.

\begin{figure}[ht]
    \centering
    \input{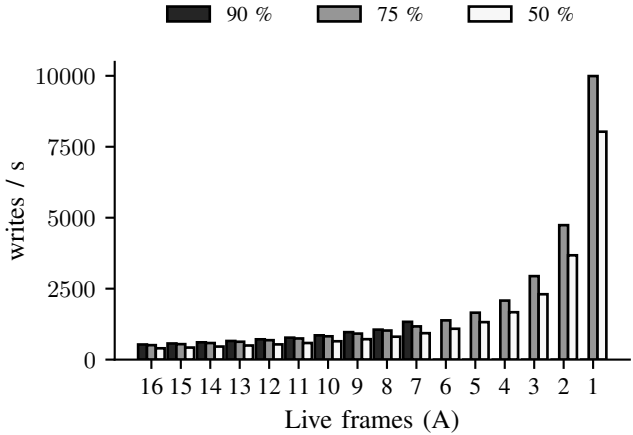}
    \caption{Average write rate per frame in sets with $A$ live frames as a function of  capacity (90\%, 75\%, and 50\%).}  
    \label{fig:wb_barplot}
\end{figure}

Focusing on the cache, on the one hand its degradation leads to an increase in the miss rate, which results in a loss of performance which in turn results in a decrease of the cache write rate. On the other hand, if a certain cache set degrades more than others, it is not correct to equally distribute the write rate of the whole cache in the new associativity configuration.
Let's quantify how the write rate per frame may change as the NV-LLC degrades. Figure \ref{fig:wb_barplot} shows the average write rate per frame in a 16MB, 16-way frame-disabling NV-LLC at various aging stages (see Section \ref{sec:methodology} for the simulated system details). Each bar depicts the average write rate of all frames belonging to a given group of sets, namely the sets with $A$ live frames in a degraded NV-LLC with 90\%, 75\% and 50\% effective capacity, respectively.

At 90\% capacity, all sets have between 16 and 7 live frames, but as the effective capacity reduces to 75\%, more degraded sets, which have only 6 to 1 live frames, also appear. 
Regardless of the effective capacity, as $A$ decreases, the write rate per frame increases noticeably. This increase in write rate has two causes: 
i) the miss rate increases in the sets with fewer live frames and therefore those sets experience a higher write rate, and 
ii) the write rate per set will be spread over fewer live frames. 
On the other hand, when considering the reduction in effective capacity from 75\% to 50\%, a decrease in the write rate per frame is observed for any value of $A$, which is due to a noticeable decrease in system performance.

Furthermore, this non-uniform degradation may affect differently the threads sharing the NV-LLC, selectively reducing the IPC of some of them and changing the pattern of writes in the entire cache. The existence of compression further complicates the modeling, as the data set referenced by each thread may have different compression capabilities that will wear cache bytes unevenly.  In short, a single simulation cannot capture the complexity of all these interactions.

\textbf{Contributions.} 
Accurately addressing this feedback between degradation and performance loss over the lifetime of the NV-LLC is the second problem we tackle in this work.
We provide a forecast procedure to estimate the evolution over time of any metric of interest linked to the LLC (effective capacity, miss rate, IPC, etc.), from the time it starts operating until its storage capacity is exhausted. 

The forecast relies on a sequence of epochs that sample the lifetime of the cache. Each epoch starts with a performance simulation and ends with an aging prediction. The \textit{performance simulation} is carried out with cycle detail on a snapshot of the cache at a particular aging stage and obtains performance metrics (miss rate, IPC, etc.) and in particular, all the write rate statistics needed to feed the aging prediction. The \textit{aging prediction} removes from operation the bytes or frames that die, according to the bitcell endurance model, the cache organization and the write rate statistics received. At the end of each aging prediction phase, a new cache snapshot is generated, with lower capacity than the previous one.

Thus, the performance and capacity forecasts  take into account the interaction between the workload and the non-uniform degradation of the NV-LLC in its multiple dimensions (bank, set, way, byte). It can be applied to a wide range of NV main memory or cache designs, although in this paper we have focused on L2C2 and related alternatives, taking into account replacement and operation with compressed blocks and degraded frames under different redundancy schemes.
Of course performance or capacity forecasts are useful for research purposes, but also can be an industry tool to estimate the life cycle of a NV memory and provide customers with a clear commitment to lifespan and performance.

The rest of the paper is organised as follows. Section~\ref{sec:background} lays the groundwork for NV-LLCs. Section~\ref{sec:nvllc} describes L2C2, a byte-level fault-tolerant cache capable of handling compressed blocks, showing the storage overhead, the detailed design of the block read and write hardware, and the latency penalty incurred in the block read service. In Section~\ref{sec:forecast}, our forecast procedure is conducted on systems with frame disabling and byte disabling with compression. In Section~\ref{sec:validation} we demonstrate the validity of the forecast procedure. Section~\ref{sec:evaluation} evaluates the degradation of L2C2 over time and compares it to various NV-LLC configurations. Finally, Section~\ref{sec:conclusions} concludes this study.

%% file: II-Background.tex
\section{Background}
\label{sec:background}
This section briefly reviews the background regarding the bitcell resilience model, data compression in the context of NV technologies, with emphasis on BDI compression, and finally, the addition of redundant capacity. The reader familiar with these concepts can skip this section without loss of continuity.
\subsection{Bitcell endurance model}

Writing 0 or 1 to an NVM bitcell requires to invest some energy for a time period to alter the value of a physical property in one of the bitcell circuit materials, whose structure, components, dimensions and interface are critical to the proper functionality of the memory~\cite{Mittal-16-ReliabilityTradeOffs, Qureshi-11-PCM, Salehi-17-STT-RAM-Survey}.
Write operations, besides being more costly in time and energy than read operations, eventually degrade bitcells, which render to lose its storage capacity. In this context, the bitcell endurance is defined as the number of writes the bitcell will withstand before it breaks down and loses its storage capability. Bitcell endurance can be approximated by a normal probability distribution of mean $\mu$ = $10^k$ writes and coefficient of variation cv = $\frac{\sigma}{\mu}$, usually between 0.2 and 0.3~\cite{Cintra-13-CharacterizingProcessVariation, Farbeh-16-FloatingECC, Ipek-10-DynamicallyReplicatedMemory, Schechter-10-UseECPnotECC, Seong-10-SAFER, Yoon-11-FREE-p}.
The coefficient of variation reflects the variability in the manufacturing process. 
The  endurance figures are different for each technology and depend on the manufacturer and the target market. For instance, STT-RAM endurance is subject to some design parameters tradeoffs such as retention time, area, power efficiency and read/write latency~\cite{Golonzka-18-MRAMEmbedded, Natsui-20-DualPortSOT-MRAM}.
It is therefore not surprising to find in the literature STT-RAM endurance values from $10^6$ for embedded systems  or IoT applications~\cite{Chih-20-STT-RAM-Endurance2, Golonzka-18-MRAMEmbedded, Lee-18-STT-RAM-Endurance3, Wei-19-STT-RAM-Endurance1} up to $10^{12}$ for general purpose microprocessors~\cite{Farbeh-16-FloatingECC, Huai-08-STT-RAM, Wang-13-I2WAP}.

\subsection{Data compression}
Data compression reduces the block size. This is beneficial in the NVM context because it allows fewer bits to be written and consequently extends the lifetime of the main memory or cache~\cite{Choi-14-AdaptiveCacheCompression, Dgien-14-CompressionArchitecture, Jadidi-17-CollaborativeCompression, Palangappa-18-CASTLE}. Yet, compression has another benefit in the context of a byte-level fault tolerant NV cache such as L2C2: it allows cache frames with dead bytes to hold blocks if compression is high enough~\cite{Ferreron-15-Concertina, Jadidi-17-CollaborativeCompression}.
Any compression mechanism that achieves wide coverage even at the cost of a moderate compression ratio can be useful, so that a large percentage of blocks, once compressed, can be stored in degraded cache frames. On the other hand, the decompression latency must be very low in terms of processor cycles, since decompression is on the critical path of the block service and may affect system performance.

The chosen mechanism is \textit{Base-Delta Immediate (BDI)}, as it achieves high coverage, fast decompression (1 cycle) and a substantial compression ratio \cite{Pekhimenko-12-BaseDeltaImmediate}.
BDI is based on value locality, i.e. on the similarity between the values stored within a block.
It assumes that a 64-byte block is a set of fixed-size values, either 8 8-byte values, 16 4-byte values, or 32 2-byte values.
It determines whether the values can be represented more compactly  as a \textit{Base} value and a series of arithmetic differences (\textit{Deltas}) with respect to that base. 

A block can be compressed with several Base~+~Delta combinations which are computed in parallel. An example with 14 BDI Compression Encodings (CE) is shown in Table \ref{table:compressionclasses}, along with the size values for the Base, Delta and the total compressed size.
Thus, the compression mechanism  chooses for each block the compression encoding (Base + Delta combination) that achieves the highest compression ratio.

\begin{table}[htb]
\caption{BDI compression encodings and their sizes, in Bytes.}
\begin{center}
{\footnotesize
\begin{tabular}{|c|c|c|c||c|c|c|c|}\hline
\textbf{Name} & \textbf{Base} & \textbf{Delta} & \textbf{Size} & \textbf{Name} & \textbf{Base} & \textbf{Delta} & \textbf{Size}\\\hline \hline
All Zeros & 0 & 0 & 0 & B2$\Delta$1 & 2 & 1 & 37 \\\hline
Rep. V(8) & 8 & 0 & 8 & B8$\Delta$4 & 8 & 4 & 37 \\\hline
B8$\Delta$1 & 8 & 1 & 16 & *B8$\Delta$5 & 8 & 5 & 44 \\\hline
B4$\Delta$1 & 4 & 1 & 21 & *B4$\Delta$3 & 4 & 3 & 51 \\\hline
B8$\Delta$2 & 8 & 2 & 23 & *B8$\Delta$6 & 8 & 6 & 51 \\\hline
B8$\Delta$3 & 8 & 3 & 30 & *B8$\Delta$7 & 8 & 7 & 58 \\\hline
B4$\Delta$2 & 4 & 2 & 36 & Uncomp. & - & - & 64 \\\hline
\end{tabular}
}
\label{table:compressionclasses}
\end{center}
\end{table}

\subsection{Addition of redundant capacity}

The reliability of the NV-LLC can be improved by adding redundant capacity.
This can be done by using classical error detection and correction (ECC) codes or more sophisticated techniques~\cite{Ipek-10-DynamicallyReplicatedMemory, Schechter-10-UseECPnotECC, Seong-10-SAFER, Yoon-11-FREE-p}.
The maximum number of bit errors that can be detected and corrected is limited by the available area and energy budget. For instance, Schechter et al. propose ECP~\cite{Schechter-10-UseECPnotECC}, an ECC mechanism that encodes the location of defective bitcells and assigns healthy ones to replace them.

However, in order to further increase reliability, a substantial portion of the redundant capacity could be dedicated to the replacement or expansion of the rated cache capacity stated in the commercial specification. Both alternatives will be evaluated later in this paper.

%% file: III-L2C2.tex
\section{Last-Level Compressed-Contents NV Cache}
\label{sec:nvllc}

This section describes the basic organization of L2C2, also showing the adaptation of BDI compression, metadata layout, the details of block rearrangement and replacement, and how to add redundant capacity.

\subsection{Basic organization}
\subsubsection{Content management between the private L1/L2 levels and the shared L2C2}
\label{sss:content_management}

\begin{figure}[ht]
    \centering
    \includegraphics[width=0.7\linewidth]{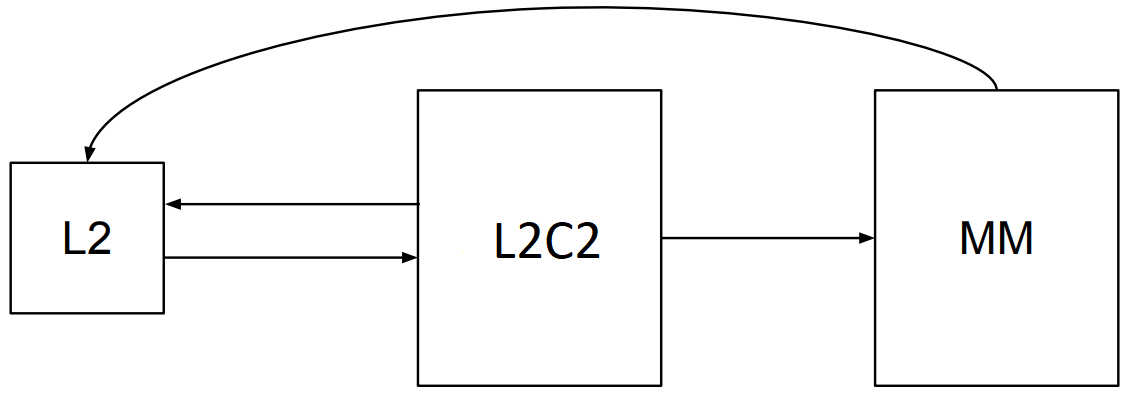}%
    \caption{Block flow diagram of  non-inclusive model.}
    \label{fig:coherence}
\end{figure}

Non-inclusive hierarchies have shown to be specially useful to avoid superfluous block insertions in the LLC~\cite{Cheng-16-LAP}.
Therefore, a non-inclusive organization is used to minimize writes in L2C2, see Figure \ref{fig:coherence}.
A block enters L2C2 by effect of a replacement in L2, provided that the block was not already in L2C2. In case of a write miss in L1 and L2, and a hit in L2C2, the corresponding block is brought to L1/L2 and invalidated in L2C2. Note that in this case, leaving the block in L2C2 does not make sense, because it will eventually have to be written back to L2C2 when it is evicted from L2.

To select the victim block, L2C2 takes into account the recency order according to the following rules: 1) inserted blocks are placed in an LRU list at the MRU position (lowest replacement priority), 2) a read hit in L2C2 places the block to the MRU position, and 3) replacement of a clean block in the private caches is communicated to L2C2; in case such a block is present, it is also placed at the MRU position. 

However, if the LRU cache frame does not have sufficient capacity for the incoming compressed block, it cannot be used as a victim. Then there are two possibilities, either to search in order from least to most recent for the first frame with sufficient capacity (LRU-Fit policy) or to choose the frame with the smallest possible capacity, and if there are several with the same capacity, the LRU one (LRU-Best-Fit policy). Ferrerón et al. test both alternatives and choose LRU-Fit for its better performance~\cite{Ferreron-15-Concertina}, but since in their context writes do not produce degradation, the LRU-Best-Fit policy could be advantageous for the L2C2 design. LRU-Best-Fit avoids writes on the highest capacity frames, and therefore poorly compressible blocks would see their residency opportunities increase. Therefore, in Section~\ref{ss:fit_bestfit} the two policies will be confronted.

\subsubsection{Bitcell fault detection}
Memory cells lose their retention capacity after a certain number of writes. It is therefore essential to handle these permanent faults without losing information.
We assume a SECDED mechanism, able to correct a single-bit error and detect up to two. 
We assume that this ECC mechanism, upon detecting and correcting a single bit fault, triggers an Operating System exception, notifying the identity of the faulty byte\cite{Yoon-11-FREE-p}. Then,  in order to prevent a second (uncorrectable) error from arising within the same region, the exception routine will disable the appropriate region, a whole frame using frame disabling, or a byte in L2C2. Note that  ECC support is already present in many current cache designs; AMD Zen SRAM LLCs, for instance, provide DECTED~\cite{Suggs-20-amd}.

\subsubsection{Wear-leveling mechanism}
\label{subsec:wearleveling}
Writing compressed blocks in a frame is a new source of imbalance in the wear of the cells acting within the frame itself. As we will quantify, if, for example, compressed blocks are always stored from the beginning of the frame, the first bytes of the frame will receive more writes than the last ones. 

Therefore, an intra-frame wear-leveling mechanism is needed to evenly distribute the writes within the frame. We assume a global counter modulo the cache frame size~\cite{Jadidi-17-CollaborativeCompression}. Blocks are stored in the frames starting from the byte indicated by this global counter and using the frame as a circular buffer. Each time the value of the counter is changed, the entire cache must be flushed, but since this must be done every few days or weeks, the impact on performance is negligible. The details and the extension of the mechanism to degraded frames can be found in Section~\ref{subsubsec:blockrearrangement}.

\subsection{BDI adaptation}
Pekhimenko et al. focus their application on achieving a large average compression ratio and therefore dispense with compression encodings with small compression ratios~\cite{Pekhimenko-12-BaseDeltaImmediate}, those marked with an~* in Table~\ref{table:compressionclasses}.
However, L2C2 incorporates them, because in this way frames with few defective bytes will be able to store low compression blocks and thus performance increases noticeably~\cite{Ferreron-15-Concertina}. 

\begin{figure}[ht]
    \centering
    \input{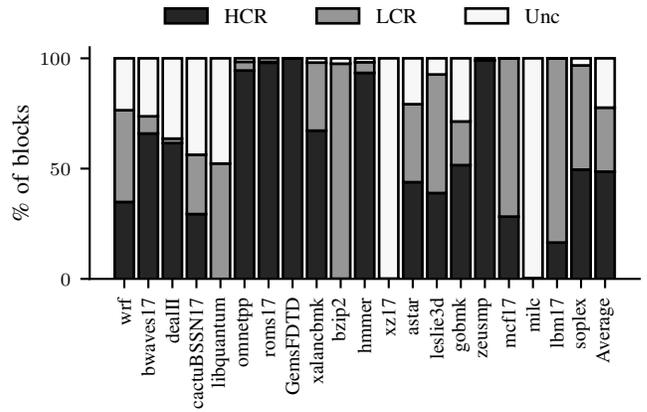}
    \caption{Block classification regarding its compression ratio for the selected SPEC CPU 2006 and 2017 applications.}
    \label{fig:BDICoverage}
\end{figure}

To quantify the importance of such low compression blocks, Figure~\ref{fig:BDICoverage} shows a classification of all blocks written in L2C2 according to the achieved BDI compression ratio for the SPEC CPU 2006 and 2017 applications used in this work. On average, 22\% of the blocks written are uncompressible (Unc), 29\% have low compression ratio (LCR) (compressed block size $>$ 37) and 49\% have high compression ratio (HCR) (compressed block size $\leq$ 37). 
For instance, if all frames in an L2C2 cache have a faulty byte, and the compression mechanism does not use the low-compression ratio encodings, the chance to store 29\% of the blocks would be lost.

\subsection{L2C2 metadata}

\begin{figure}[ht]
    \centering
    \includegraphics[width=\linewidth]{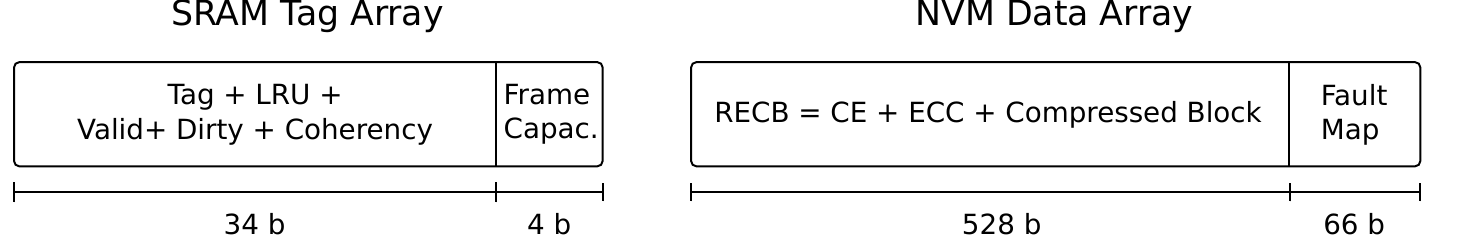}%
    \caption{Layout of a frame entry in the SRAM tag and NVM data arrays.}
    \label{fig:metadata}
\end{figure}

The tag array undergoes the most write requests as it must keep the coherence and replacement states up to date. Should these bit cells fail, the entire data frame should be deactivated. Therefore, we  assume the tag array is built with SRAM technology, free of wear by writing. Our proposal only adds a 4-bit field to store the frame capacity to each tag array entry. This frame capacity is represented in terms of the largest compression encoding the frame can allocate (see Figure \ref{fig:metadata}).

The data array is built using NVM technology. Each frame must have a capacity of 66 bytes: 64 data bytes plus one or two metadata bytes: up to 11 ECC bits and 4 bits representing the compression encoding (CE) of the data block. In addition, a fault bitmap is needed next to the data array to identify faulty bytes. This fault bitmap requires 66 bits for each frame. During the life of the frame this bitmap will experience at most 66 write requests, so it can also be implemented with NVM technology.

\subsection{Block processing: \\ \hspace*{5mm}compression, ECC, replacement and rearrangement}
\label{subsec:nvllcarchitecture}

\subsubsection{L2C2 miss, block writing}
\label{subsubsec:blockwriting}

\begin{figure}[ht]
    \centering
    \includegraphics[width=\linewidth]{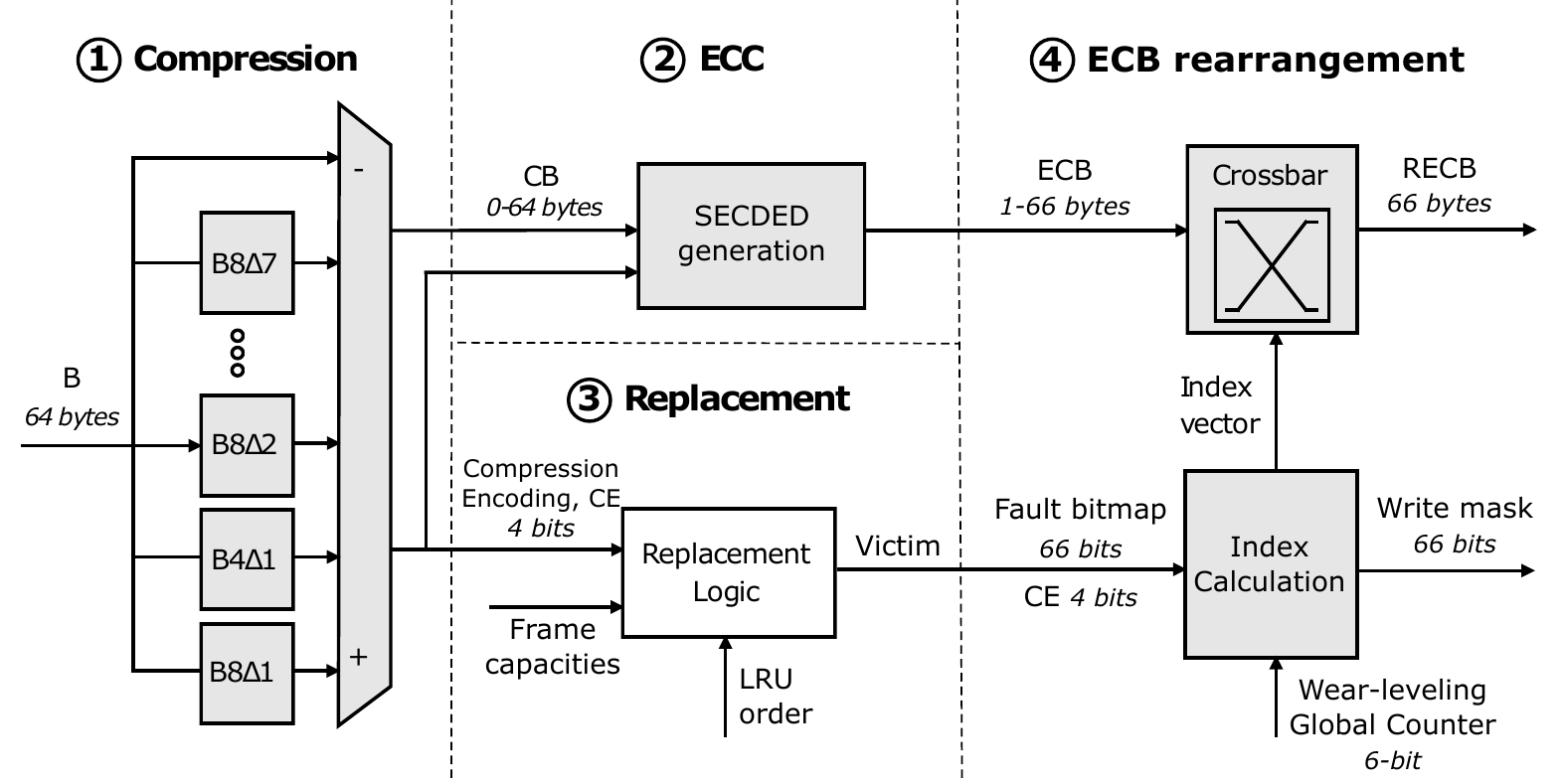}%
    \caption{Flow of writing a block in L2C2 and components involved.}
    \label{fig:writeblock}
\end{figure}

Figure~\ref{fig:writeblock} shows the components involved in the processing of a block B to be written in L2C2, from  compression to rearrangement. In Figures~\ref{fig:writeblock} and \ref{fig:readblock} the shaded boxes represent what is new and/or has been modified to the Concertina proposal~\cite{Ferreron-15-Concertina}.

First, the BDI compression units receive the block B (64~B) [\circled{1}~Compression]. The result of each compression unit is a) whether the block is compressible or not and, if so, b) the compressed block. As a result, the compressed block with the highest compression ratio (CB, 0-64~B) is selected and the corresponding compression encoding (CE, 4~b) is reported.

Next, the ECC bits corresponding to CB are calculated [\circled{2}~ECC]. The ECC mechanism selected in particular is orthogonal to our proposal. As mentioned above, we assume SECDED protection, which means an overhead between 2 and 11 bits encoded in a field of one to two bytes. We call ECB the concatenation of CB and SECDED bits, whose length ranges from 1 to 66 bytes. The length of this ECB determines the minimum capacity a cache frame must have to accommodate the block. 

The replacement logic selects the victim block among the frames with the required minimum capacity [\circled{3} Replacement]. For this, the replacement logic considers
the CE of the incoming block B along with the capacities and  LRU order of the frames still alive in the involved cache set.

Every frame has an associated fault bitmap that points out the faulty bytes (66~b). This fault bitmap information is initialized to '1's indicating that all bytes in a frame are non-defective. In addition, the byte number from which to start writing the frame is reported by the Global Counter (GC, values 0-65). According to the GC and CE values and the fault bitmap,  the block is rearranged for selective writing (RECB, 1-66 bytes) under a write mask (66~b) [\circled{4} Block rearrangement]. The next subsection 3) details the rearrangement logic (ECB or RECB Block rearrangement for L2C2 write or read, respectively).

\subsubsection{L2C2 hit, block reading}
\label{subsubsec:blockreading}

\begin{figure}[ht]
    \centering
    \includegraphics[width=0.9\linewidth]{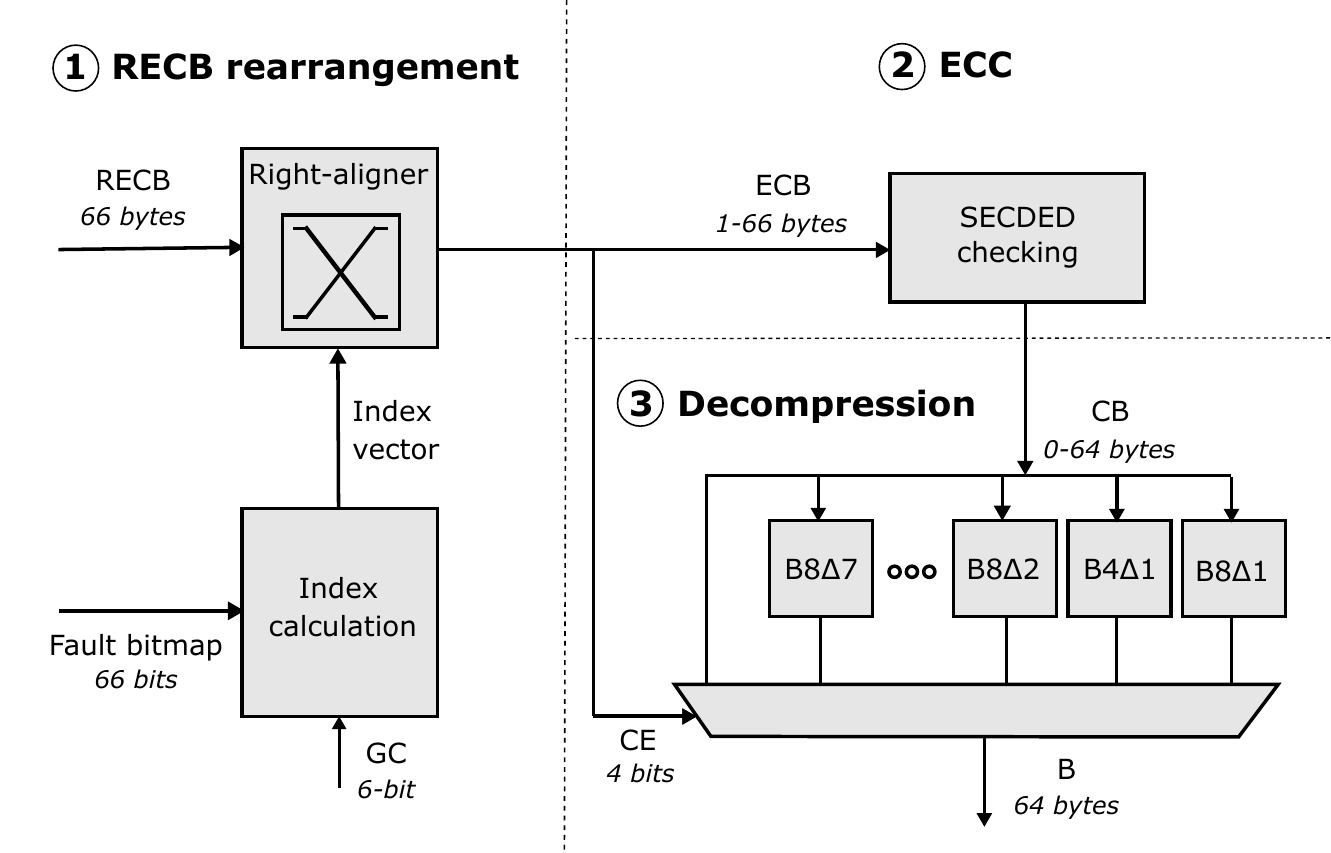}%
    \caption{Flow of reading a block in L2C2 and components involved.}
    \label{fig:readblock}
\end{figure}

Similarly but in the opposite order, Figure \ref{fig:readblock} summarises the read flow of an L2C2 block. First, the block is rearranged using as input RECB, the fault bitmap and the GC value. Then, the ECC of ECB is checked, and from CB and CE the uncompressed B block is obtained and forwarded to L2/L1 [\circled{3} Decompression].

\subsubsection{Rearrangement logic}
\label{subsubsec:blockrearrangement}
The rearrangement logic is composed of two elements: Index Calculation and Crossbar. The index calculation determines the mapping from ECB bytes to RECB bytes (L2C2 write) or conversely, from RECB bytes to ECB bytes (L2C2 read). The crossbar moves bytes from the input ports to the output ports.

\begin{figure}[ht]
    \centering
    \includegraphics[width=\linewidth]{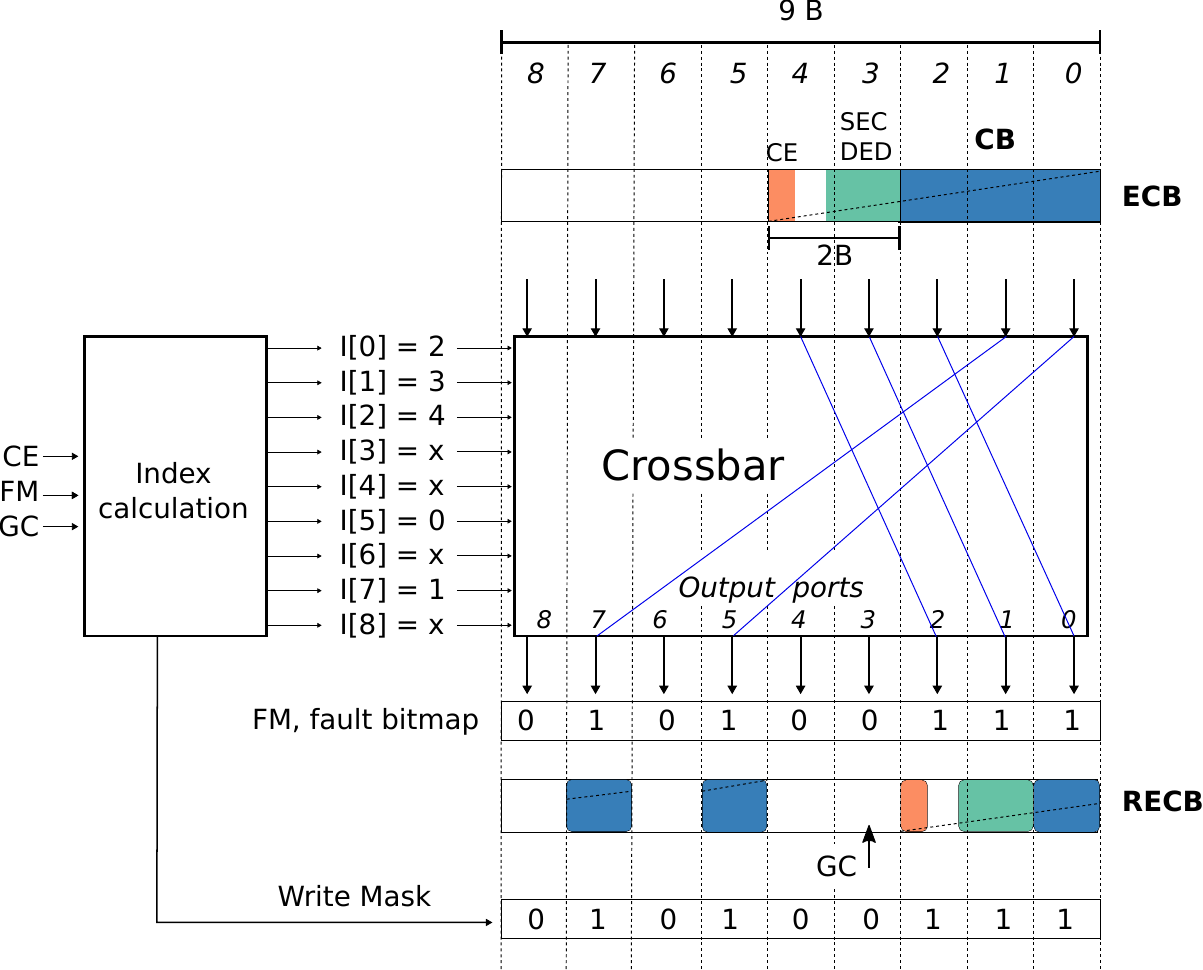}
    \caption{Example of ECB rearranging to write a 9-byte frame.}
    \label{fig:ECB2RECB_example}
\end{figure}

Figure~\ref{fig:ECB2RECB_example} shows an example of rearranging an ECB from the fault bitmap (FM) and the GC and CE values. When writing a frame into L2C2, RECB is an ECB rearrangement consisting of a right rotation starting from the GC value and skipping the faulty bytes. Afterwards, the write is selectively performed on the bytes indicated by the computed write mask. 

Algorithm 1 describes the index calculation for writing N-byte frames. It takes as inputs the fault bitmap (FM) corresponding to the destination frame and the values of the global counter (GC) and compression encoding (CE). 
The outputs are the write mask and the index vector I[N~=~\textit{frame size}] that controls the output ports of the crossbar. For example, I[7]~=~1 means that byte 1 of ECB will appear in the output port 7 of the crossbar, see Figure~\ref{fig:ECB2RECB_example}.

The first for loop (line 2) calculates indexes without considering the global counter value. That is, assuming that ECB is to be rearranged starting with the byte zero of the destination frame. Note that the calculation of each iteration uses the result of the previous one. This implies using N adders in series. Alternatively, our implementation uses a tree of adders, which reduces the computation time to that of log2 (N) adders in series. Each adder uses log2 (N) bits at most.

The two next loops (lines 5 and 6) adjust indexes considering the global counter value. Now, the iterations within each loop are independent and can be calculated in parallel, so the calculation time of the two iterations is that corresponding to two adders in series. 

Finally, the last loop (line 7) calculates the write mask. This loop can be synthesized with an array of 64 7-bit comparators. These comparators act on the calculated indexes and their operation can overlap with the crossbar traversal.

\begin{algorithm}[ht]
\SetKwInput{KwInput}{Input}                
\SetKwInput{KwOutput}{Output}              
\DontPrintSemicolon
  
  \KwInput{
    \;
    FM: N-bit vector fault bitmap  \;
    GC: global counter \;
    size: ECB size, computed from CE \;
    ~~~~~~$0 \leq GC, size \leq N-1$\;
    }
  \KwOutput{\;
  I[N]: N crossbar output port indexes
  \;
  WM[N]: write mask N-bit vector
  \;\;}

  I[0] = 0\;
  \lFor{i=1; i$<$N; i++}{
    I[i] = I[i-1] + FM[i-1];
  }
   T = I[N-1] + FM[N-1];
   
   GCI = I[GC];
   
  \lFor{i=0; i$<$N; i++}{
    I[i] = I[i] - GCI;
  }
  \lFor{i=0; i$<$GC; i++}{
    I[i] = I[i] + T;
  }
  
  \lFor{i=0; i$<$N; i++}  {
	if I[I] $<$ size \&\& FM[I] == 1 then WM[I] = 1
	else WM[I] = 0;}
  
\caption{ECB~$\rightarrow$~RECB Index Calculation}
\end{algorithm}

The index vector is calculated when writing and reading in the same way. In the write circuit, the crossbar is an array of multiplexers governed directly by the index vector.
In the read circuit, the crossbar acts as right-aligner and is more complex. Our implementation assumes NxN comparators of log2 (N) bits and N output multiplexers of N bytes to 1 byte with decoded control. The decoded control of the multiplexer that produces the byte $i$ is generated by N comparators between the value $i$ and the N elements of the index vector.

\textbf{VLSI implementation.} To put into context costs and delays of the rearrangement logic we assume an L2C2 built with 22nm STT-RAM technology, the largest scale of integration available in the NVsim tool  \cite{Dong-12-NVSim}. 
Table \ref{table:nvllccost} shows area, latency and power of the tag SRAM array and the data STT-RAM array that support the 4MB cache banks we are going to use in the experimental section.

\begin{table}[htb]
\caption{Hardware cost comparison.}
\begin{center}
{\footnotesize
\begin{tabular}{|c|c|c|c|c|} \cline{2-5}
\multicolumn{1}{c|}{} & SRAM  & STT-RAM  & ECB & RECB\\
\multicolumn{1}{c|}{} & Tag Array & Data Array  & $\rightarrow$RECB & $\rightarrow$ECB\\
\multicolumn{1}{c|}{} &  22 nm & 22 nm & 16 nm & 16 nm\\\hline
Area (mm$^2$) & 0.116 & 0.74 & 0.021 & 0.025\\\hline
Latency (ns) & 0.28 & 2.41 & 0.33 & 0.38\\\hline
Dynamic read & \multirow{2}{*}{0.17} & \multirow{2}{*}{6.87} & \multirow{2}{*}{-} & \multirow{2}{*}{0.49}\\
power (mW) & & & & \\\hline
Dynamic write & \multirow{2}{*}{0.16} & \multirow{2}{*}{18.64} & \multirow{2}{*}{0.61} & \multirow{2}{*}{-}\\
power (mW) & & & & \\\hline
Static P (mW) & 109 & 338 & 0.53 & 0.7\\\hline
\end{tabular}
}
\label{table:nvllccost}
\end{center}
\end{table}

Both ECB and RECB rearrangement logic are outside the L2C2 core, but the latter is located in the critical path of block delivery to L1/L2. In order to quantify their physical features both have been specified, simulated and laid out with the Synopsys Design Compiler R-2020.09-SP2 and Synopsys IC Compiler R-2020.09-SP2. Due to the lack of a 22nm library, we used the SAED16nm FinFET Low-Vt technology in worst case condition (typical-typical, 125 ºC and  0.8 volts). These tools allowed us to estimate post-layout costs in terms of area, latency and power consumption. Dynamic power values were calculated from activity data obtained from our workload simulations. The latency of the RECB~$\rightarrow$~ECB logic (0.38 ns) plus the delay and setup times of the input and output registers, respectively, can be estimated at about two cycles at 3.5 GHz.
That is, rearranging and decompression increases the L2C2 load-use latency, with respect to a frame-disabling cache, from 30 to 32 cycles, a 6.7\%.

In summary, looking at the figures as a whole,  the overhead seems to be affordable on all metrics. Regarding storage costs, Table~\ref{table:systemscost} also provides a comparison between all the evaluated cache candidates.

\subsection{L2C2+N: adding redundant capacity to L2C2}
\label{ss:redundancy}

Providing L2C2 with a few spare bytes in each frame could be very convenient since it would allow to continue working without loss of performance after the failure of several bytes of each cache frame.

The design presented so far allows to add N spare bytes in a very straightforward way: just increase each frame from 66 to 66+N bytes in the data array, and also increase the bitmaps from 66 to 66+N bits. In addition, the rearrangement logic has to be extended to handle 66+N byte blocks, and the Global Counter has to count modulo 66+N.
Without further changes, the wear-leveling logic will take care of distributing the writes among the 66+N bytes available. A frame will only start to impose performance constraints when its effective capacity falls below 66 bytes.

%% file: IV-Forecasting.tex
\section{Forecast procedure}
\label{sec:forecast}

This section describes a procedure to forecast the capacity and performance evolution of  an NV-LLC
through time, from its initial, fully operational condition, until its complete exhaustion. Without loss of generality the procedure assumes byte granularity, but extending it to other sizes is straightforward. The maximum number of writes supported by bitcells is modelled by a normal distribution. 

The forecast is driven by a detailed, cycle-by-cycle simulation of a workload that can be multiprogrammed, parallel, or a mix of both execution modes. In this paper we opted for a multiprogrammed workload, but the other alternatives can be simulated in exactly the same way.

The forecast procedure determines the live byte configuration in discrete steps of capacity loss, which we call epochs. An epoch starts with a detailed Simulation phase, where performance and write rate measurements are extracted, and continues with a Prediction phase where each byte that fails is disabled and the remaining number of writes of those that are still live is updated.

To the best of our knowledge, this is the first NV-LLC capacity and performance forecast procedure proposed so far.

\subsection{Data structures supporting the forecast procedure}
\label{subsec:forecast_maps}
For each byte of the data array it is necessary to keep track of two key attributes, namely the number of per-byte Remaining Writes and the experienced per-byte Write Rate. These attributes are represented in two data structures, called maps and abbreviated as \emph{RW map} and \emph{WR map}, respectively.

\begin{figure}[ht]
    \centering
    \subfloat[RW map]{\label{subfig:rwmap}
    \includegraphics[width=0.46\linewidth]{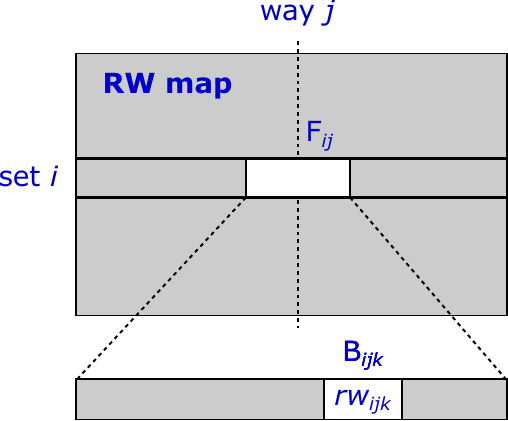}%
    }
    \hfill
    \subfloat[WR map]{\label{subfig:wbmap}
    \includegraphics[width=0.46\linewidth]{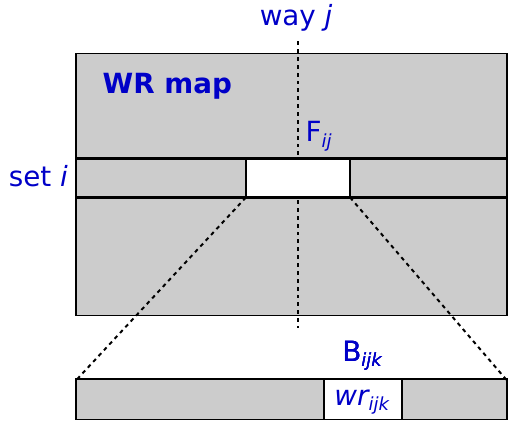}%
    }
    \caption{Per-byte Remaining Writes  and Write Rate maps.
    }
    \label{fig:rwwbmaps_compresion}
\end{figure}

\textbf{RW map}. Each entry of \emph{RW map} holds the number of remaining writes $rw_{ijk}$ of byte $B_{ijk}$ (set $i$, way $j$, byte $k$), see Figure \ref{subfig:rwmap}. \emph{RW map} is initialised according to the statistical endurance model of the memory technology used, as in~\cite{Cintra-13-CharacterizingProcessVariation, Farbeh-16-FloatingECC, Ipek-10-DynamicallyReplicatedMemory, Schechter-10-UseECPnotECC, Seong-10-SAFER, Yoon-11-FREE-p}.

Once the \textit{RW map} is initialized, it would be sufficient to simulate the NV-LLC with the desired workload and update the map on each write, subtracting in all the live bytes of the frame being written. When any byte of the cache reaches its maximum number of writes ($rw_{ijk}= 0$), the corresponding cache region is disabled, the whole frame with frame disabling or the single byte with byte disabling. Then, the simulation would continue with the degraded system.

The simulation should be detailed, cycle-accurate, so that the progressive degradation is reflected in the miss rate and write rate of the remaining healthy regions. However, this naive approach is not feasible, since at detailed simulation speed only a few milliseconds of forecast could be attained.

\textbf{WR map}. An alternative approach, which is nearly as accurate, but with lower simulation cost is the following. After a suitable simulation time we write down in a \textit{WR map}, the write rate per byte $wr_{ijk}$, 
see Fig. \ref{subfig:wbmap}. On the assumption that these per-byte write rates remain constant as long as no further byte is disabled, we can compute the \textit{predicted lifetime} ($PLT$) of each byte $B_{ijk}$ as:

$$ PLT(B_{ijk}) = \frac{rw_{ijk}}{wr_{ijk}} $$

We can use $PLT$ to predict the next byte that becomes faulty. 

\subsection{Basis of the forecast Procedure}
\label{subsec:forecast_basis}

\begin{figure}[ht]
    \centering
    \includegraphics[width=\linewidth]{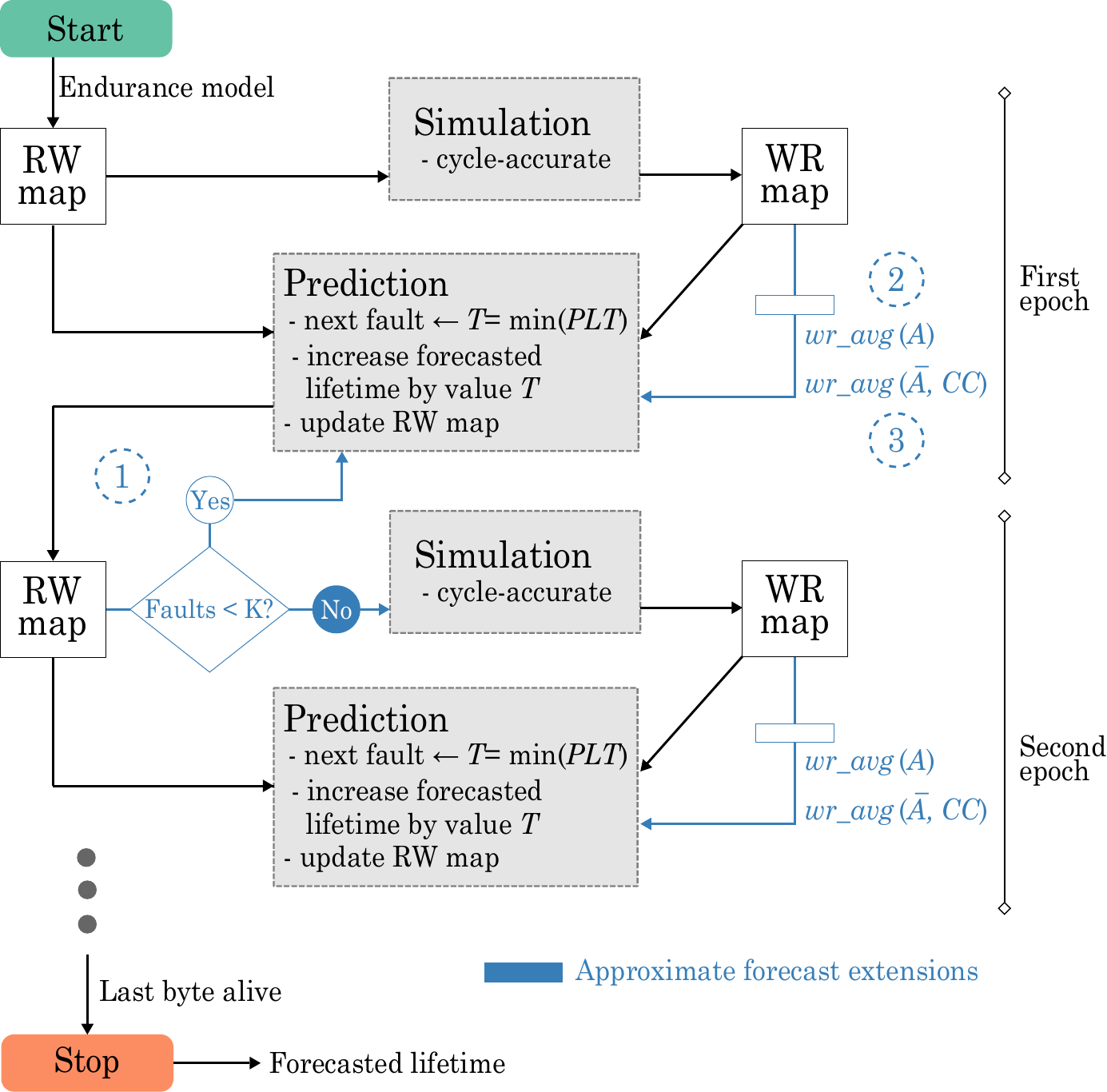}
    \caption{Forecast procedure diagram. Basic procedure and approximate extensions (in blue).}
    \label{fig:forecast}
\end{figure}

The lifetime of an NV-LLC can be forecast using the procedure outlined with black lines in Figure \ref{fig:forecast}. 

The \textit{RW map} is first initialized taking samples from a normal statistical distribution of the maximum number of writes a bitcell can endure. Forecast then proceeds through successive \emph{epochs} which consist of a Simulation phase followed by a Prediction phase.

The Simulation phase requires the development of a microarchitectural LLC model that allows to dynamically configure in each set a variable associativity and, if applicable, a variable number of bytes per frame. So, the simulation will take into account the regions that are still alive, according to the \textit{RW map}, run the workload for a suitable number of cycles, compute the write rates in each byte, and finally update the \textit{WR map}.

The Prediction phase combines the values of both maps to calculate $PLT(B_{ijk})$, selecting the byte with the lowest remaining lifetime, $T=min( PLT(B_{ijk}) )$. The prediction consists of advancing the forecasted lifetime by exactly that value $T$. 
To do so, it is sufficient to subtract from the number of  remaining writes in each byte of the cache, the number of writes that would have occurred in that byte in a time T 
($\forall~ijk : rw_{ijk} = rw_{ijk} - T * wr_{ijk}$).
In this way the next simulation will be performed with the corresponding region disabled, cache frame or byte, so that the behavior of the LLC will take into account the degradation experienced in the data array.  

Forecast advances through single-prediction epochs until all bytes in the cache are disabled. Each epoch adds the variable time $T$ to the NV-LLC lifetime, which depends on the initial \textit{RW map} and the write rate variation.  Although the Prediction phase is computationally very light, this alternative approach still requires as many simulations as there are bytes in the cache, and it is also not affordable considering the runtime required for a detailed simulation.

\definecolor{MiAzul}{RGB}{55,127,184}

To decrease the number of simulations we propose an approximate procedure that extends the forecast duration within each epoch. This approximate forecast procedure acts as follows. In each epoch, the simulation phase does not change: it receives an \textit{RW map} and obtains the corresponding \textit{WR map}. However, the prediction phase has an \textit{extension} of $K$ consecutive predictions, corresponding to the failure of $K$ bytes. After every prediction step the \textit{RW~map} is updated; see \textcolor{MiAzul}{\circled{1}} in Figure \ref{fig:forecast}.

The challenge now is that, as bytes die during the multiple-prediction epoch, the values of the \textit{WR map} may not reflect the effect of the progressive degradation of the NV-LLC during the epoch.

When a cache byte dies there may be a tiny decrease in hit rate and system performance, which may result in tiny changes of the byte write rate across all cache frames. Our model does not take this reduction into account during the Prediction phase within each epoch.
However, if we focus on a shrinking cache set, i.e. one in which a byte has just been disabled, the new write rate in the frames of that set can increase significantly. This effect is evident with frame disabling, see Figure~\ref{fig:wb_barplot}, but  occurs equally with byte disabling. 
Consequently, to increase the epoch extension without introducing significant error, a model is needed to approximate new write rates as bytes fail during prediction. 

Without loss of generality, a uniform distribution of writes among cache sets is assumed in this paper; see Section~\ref{sec:Specific_Situations} for a more general discussion.
Accordingly, the write rate on the bytes of a cache set whose  \textit{health state} has just degraded after a prediction step can be computed by the average write rate of all the bytes belonging to the sets that \textit{were already} in that degraded health state during the simulation. As will be seen below, the health state of a cache set is defined differently for frame- and byte-disabling caches.

\subsection{Approximate forecast procedure
\\ \hspace*{4mm} for frame disabling}
\label{subsec:forecast_app}


In frame disabling, all bytes in a frame receive the same write rate, and it matches the write rate in the frame. Therefore, the \textit{WR map} stores information at frame granularity.

Under the assumption of a uniform distribution of references across sets, for NV-LLCs with frame disabling, the health state of a set can be defined simply as its $A$ number of live frames, with $A$ between one and the initial associativity.

At the end of a Simulation phase, $wr\_avg(A)$, the average write rate per byte in sets with $A$ live frames, is computed from the \emph{WR map}; see \textcolor{MiAzul}{\circled{2}} in Figure \ref{fig:forecast}. Thus, during the Prediction phase the write rate applied to the bytes of a frame changes as the health state of its set changes. That is, while a set has $A$ live frames, the prediction calculations age its bytes with $wr\_avg(A)$, but when one of them dies, the aging will be performed with $wr\_avg(A-1)$.

Note that in the Prediction phase, after disabling a certain number of frames,  sets with a value of $A$ not yet simulated may appear. For instance, let us focus on the black distribution of write rates per frame we showed in Figure~\ref{fig:wb_barplot}. It corresponds to the Simulation phase of an epoch that starts with 
90\% effective capacity. In that epoch, the Prediction phase handles sets with 7 or more live frames. But before reaching $K$ predictions a byte belonging to a set with $A=7$ may die, appearing a new health state, that of the sets with $A=6$ for which there are no available write rate data yet. To cope with these cases, we can stop the prediction, thus ending the epoch prematurely and starting a new simulation. Alternatively, to keep low the number of simulations, we can continue the prediction, also allowing some more error and apply the previous value $wr\_avg(7)$. In this work, we will adopt this second approach.

\subsection{Approximate forecast procedure
\\ \hspace*{4mm} for byte disabling plus Compression}
\label{subsec:forecast_app_com}

Unlike frame disabling, in a cache with byte disabling and compression, such as L2C2, a write to a frame does not always imply a write to all the bytes of the frame and therefore the write rate to the bytes of a frame is lower than the write rate to the frame. The wear-leveling mechanism ensures an even distribution of writes among the live bytes of a frame. Consequently, during prediction, we can assume that the write rate on all live bytes of a frame is equal, and is calculated as the average of the write rates on all of them. 

Moreover, a fault in one or more bytes of a L2C2 frame does not preclude storing blocks, as long as their compressed size is appropriate. 
Now the health state of its sets is more diverse than in frame disabling: at any given time there are not only alive and dead frames, but frames with a very diverse range of effective capacities.

The number of faulty bytes in a frame limits the compression encodings it can accommodate. A frame with a certain effective capacity is associated with a compression class (CC) if it can accommodate compressed blocks of size CC or smaller. For example, a frame with 3 defective bytes has an effective capacity of 61 bytes, which accommodates blocks of any compression encoding except those of size 64 bytes (see Table~\ref{table:compressionclasses} in page 4) and thus it is associated with CC~=~58.



In this context, the prediction of write rate per byte is more complex. For example, think of a set that has only one frame of CC~=~64. All non-compressible blocks will end up in that single frame, which can become a hot spot for writes within the set. But, in another cache set with a majority of frames with CC~=~64, the write rate of the set will be distributed in a substantially equal way among frames.

With this, our Prediction phase will assume that the write rate a byte receives depends on the CC of its frame as well as on the CCs of the rest of the frames in the same cache set.
Therefore, now the health state of a set is abstracted as a 12-tuple $\bar{A}$. It aggregates the compression classes to which each frame belongs to. For instance, a set with tuple $\bar{A}$ = (0, 0, 0, 0, 0, 0, 0, 0, 0, 0, 1, 15) has one frame with CC~=~58 and 15 frames with CC~=~64.

Thus, during prediction, the aging write rate to consider for the bytes of a given frame $F_{ij}$ will depend on its compression class $CC$ and the health state (tuple $\bar{A}$) of the set that contains $F_{ij}$: $wr\_avg(\bar{A}, CC)$.

More specifically,
\begin{align*}
wr\_avg(\bar{A},CC) &= average(wr_{ijk})\\
\forall~ij~ | \textup{ i) } &F_{ij} \in \textup{ set with tuple } \bar{A}\\
\textup{ii) } &F_{ij} \in \textup{ compression class } CC
\end{align*}

Each time a byte is disabled in a frame $F_{ij}$, CC of the frame and $\bar{A}$ of the set are recomputed.
Thereafter, $wr_{ij}$, the aging write rate of $F_{ij}$ with compression class CC,  is approximated by $wr\_avg(\bar{A}, CC)$; see \textcolor{MiAzul}{\circled{3}} in Figure \ref{fig:forecast}.

As in frame disabling, as faults are predicted in succession, sets with a tuple value $\bar{A}$ not yet simulated may appear.
For instance, suppose a set with the same 12-tuple as before: one frame associated with $CC=58$, and 15 frames associated with $CC=64$:
$$ \bar{A}_1 = (0, 0, 0, 0, 0, 0, 0, 0, 0, 0, 1, 15). $$
Suppose a byte of one of the fifteen frames with $CC=64$ fails during the Prediction phase. Now the tuple modeling the set becomes:
$$ \bar{A}_2 = (0, 0, 0, 0, 0, 0, 0, 0, 0, 0, 2, 14). $$

But if in the epoch Simulation no $\bar{A}_2$ tuple was tracked, the values of $wr\_avg(\bar{A}_2, -)$ are unknown.
As in frame disabling, in this work we chose to  continue the prediction, tolerating some more error and using for that set the previous values of $wr\_avg(\bar{A}_1, -)$ as an approximation of $wr\_avg(\bar{A}_2, -)$.

%% file: V-Methodology.tex
\section{Methodology}
\label{sec:methodology}

Details of the multicore system modeled for the cycle-by-cycle simulation phase of each epoch are shown in Table~\ref{table:systemspecification}. It consists of 4 cores, each with two private cache levels L1 and L2, split into instructions and data.
In addition, there is a third cache level (L2C2) which is shared, non-inclusive and distributed in four banks among the cores. The coherence protocol is directory-based MOESI, and the interconnection network is a crossbar connecting the L2 private levels, the banks of the LLC and the directory. The main memory controller is located next to the directory.

\begin{table}[htb]
\caption{System specification.}
\begin{center}
{\footnotesize
\begin{tabular}{|c||l|}\hline
Cores & 4, ARMv8, out-of-order (up to 8 inst/cycle), 3.5 GHz.\\\hline
Coherence & MOESI, directory distributed among LLC banks.\\
Protocol & 64 B data blocks in all levels.\\\hline
\multirow{2}{*}{L1} & Private, 32 KB D, 32 KB I, 4 ways, LRU.\\
 & 3-cycles load-use delay. Fetch on write miss.\\\hline
\multirow{2}{*}{L2} & Private, L1-inclusive, 128 KB D, 128 KB I, 16 ways, LRU.\\
 & 11-cycles load-use delay. Fetch on write miss.\\\hline
 & Shared, non-inclusive, 4 banks, 4MB/bank, 16 ways, LRU.\\ 
STT-RAM & Load-use delay: 30-cycles frame disabling; 32-cycles L2C2.\\
NV-LLC & Frames protected by SECDED. \\
 & Baseline endurance: mean $10^{11}$ wr., $cv$ = 0.2, 0.25, and 0.3. 
 \\\hline
Main & 1 memory controller, DDR4.\\
Memory & 1 channel, 8GB/channel (1200 MHz)\\\hline
NoC & Crossbar between L2C2 banks and L2s. 32 B flits.\\\hline
\end{tabular}
}
\label{table:systemspecification}
\end{center}
\end{table}

We use Gem5 \cite{Lowe-20-gem5} along with the Ruby memory subsystem and Garnet interconnection network. In addition, we use NVSim for the L2C2 latency estimations \cite{Dong-12-NVSim}. The workload consists of 10 mixes randomly built by SPEC CPU 2006 and 2017 benchmarks \cite{Henning-06-SPEC, Bucek-18-SPECCPU2017}, leaving aside applications with very little activity on the LLC~\cite{Navarro-19-SPECCharacterization}. 
Fast-forwarding is performed for the first two billion instructions and then 200M cycles are simulated in detail.
Table~\ref{table:mixes} shows the applications that make up each mix along with the LLC MPKI of the mix, computed by dividing total cache misses by total number of instructions executed by all applications in the mix. Besides, the top ten memory intensive applications, in terms of accesses per kilo instruction, APKI, are superscripted.

The 10 mixes are run in the simulation phase of each epoch to obtain the \textit{WR map} for that epoch. The write rate in each byte of the cache is calculated as the average obtained for the 10 mixes.

\begin{table}[htb]
\caption{Selected SPEC 2006 and SPEC 2017 applications, with suffixes 06 and 17, respectively and their MPKI. Superscript indicating top-10 memory intensive applications.}
\begin{center}
{\footnotesize
\begin{tabular}{|r|l|r|}\hline
mix & Applications & MPKI\\\hline\hline
\#1 & zeusmp06 gobmk06 dealII06 bzip206$^7$ & 1.4\\\hline
\#2 & hmmer06 bzip206$^7$ wrf06 roms17$^9$ & 2.6\\\hline
\#3 & zeusmp06 cactuBSSN17$^1$ hmmer06 soplex06 & 6.1\\\hline
\#4 & omnetpp06 astar06 milc06 libquantum06$^4$ & 4.9\\\hline
\#5 & xalancbmk06$^{10}$ leslie3d06$^3$ bwaves17$^6$ mcf17$^8$ & 10.4\\\hline
\#6 & lbm17$^5$ xz17 GemsFDTD06$^2$ wrf06 & 6.6\\\hline
\#7 & cactuBSSN17$^1$ dealII06 libquantum06$^4$ xalancbmk06$^{10}$ & 7.3\\\hline
\#8 & gobmk06 milc06 mcf17$^8$ lbm17$^5$ & 6.0\\\hline
\#9 & xz17 astar06 bwaves17$^6$ soplex06 & 3.5\\\hline
\#10 & GemsFDTD06$^2$ omnetpp06 roms17$^9$ leslie3d06$^3$ & 10.6\\\hline
\end{tabular}
}
\label{table:mixes}
\end{center}
\end{table}

%% file: VI-Validation.tex
\section{Forecast Validation, Cost, and Specific Situations}
\label{sec:validation}
To validate the forecast procedure it would be necessary to contrast its projections with data from the operation of real NVM caches as they age with a known workload. But unfortunately there is no such information in the public literature. Therefore, in this section we provide tests of the correctness of the assumed hypotheses as a function of the number of epochs employed, evaluating the tradeoff between accuracy and the time spent in the forecast procedure. Finally, we outline forecast alternatives for situations in which some underlying assumptions are not met. 

\subsection{Validation}
\label{sec:validation_cost}

As discussed in Sections~\ref{subsec:forecast_app} and~\ref{subsec:forecast_app_com}, the main source of forecast inaccuracy lies in the Prediction phase, where it is necessary to approximate the write rate of health states that have not yet appeared in the Simulation phase. Forecast with epochs of small extension involves little approximation and may improve the quality of the forecast, but increases  its computational cost. 

To explore this tradeoff between quality and cost, several experiments have been performed, using epochs of different extension in each experiment, which predict a certain cache degradation. Specifically, we predict how much time elapses until 50\% of the cache, $T_{50C}$, degrades. The 50\% degradation is a common case study \cite{Schechter-10-UseECPnotECC,Yoon-11-FREE-p, Jadidi-17-CollaborativeCompression}, and in our experiments we will also focus on it, but any other percentage, including 100\%, corresponding to the total degradation could be used.

A different number of epochs of constant extension is used in each experiment.  The epoch extension is the number $k$ of consecutive predictions disabling frames or bytes, depending on the cache model, and is calculated by simply dividing 50\% of the cache size, measured in frames or bytes, by the number of epochs.

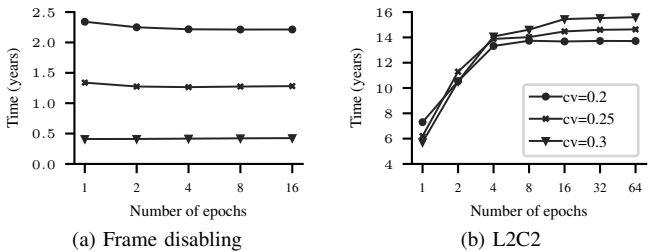
\begin{figure}[ht]
    \centering
    \subfloat[Frame disabling]{\label{subfig:CEbd}
        \input{Figuras/CE_FD.pgf}
    }
    \hfill
    \subfloat[L2C2]{\label{subfig:CEcompression}
        \input{Figuras/CE_BDI.pgf}
    }
    \caption{Forecasted $T_{50C}$ (in years) as a function of the number of epochs for frame disabling and L2C2 caches, with three coefficients of variation: $cv = 0.2$, 0.25, and 0.3.
    }
    \label{fig:convergenceexperiment}
\end{figure}

The Y-axes in Figure \ref{fig:convergenceexperiment} shows $T_{50C}$ as a function of the number of epochs for frame disabling and three L2C2s built with bitcells of different manufacturing variabilities.

As can be seen, the forecast of $T_{50C}$ converges as the number of epochs increases for all coefficients of variation. Using a number of epochs greater than or equal to 8  and 16, $T_{50C}$ varies less than 0.8 and 1.1\% for frame disabling ($k$=16384 frames) and L2C2 ($k$=524288 bytes), respectively.

Previous works performs a single simulation from a fully operational NV memory to obtain the write rate data~\cite{Seong-10-SAFER, Ipek-10-DynamicallyReplicatedMemory, Schechter-10-UseECPnotECC, Jadidi-17-CollaborativeCompression}. From this data, they compute the time at which a bitcell dies, and then recalculate the write rate analytically. In this sense, this methodology is similar to ours when a single epoch is used. But, as Figure~\ref{fig:convergenceexperiment} shows, in both cases but specially for compression, using a single epoch incurs in a non-negligible error as $cv$ grows.

Finally, in order to prove that different \textit{RW maps} do not lead to inconsistent results, five different random seeds have been used.
The seeds are used to initialize different \textit{RW maps} for the three values of $cv$ and  forecast is performed for all of them.
Again, the convergence metric is $T_{50C}$ in a 16-epoch forecast of an L2C2.
The standard deviation of the different forecasted times is below 2\% of the arithmetic mean.


\subsection{Computational cost}
\label{sec:Forecasting_cost}
As shown in Figure~\ref{fig:convergenceexperiment}, after a certain number of epochs the forecast error is negligible, but its computational cost increases almost linearly. Table~\ref{table:computationtime} shows the maximum elapsed times in individual Prediction and Simulation phases, as well as the total, for a forecast reaching up to 50\% capacity degradation in an L2C2. These numbers have been obtained in a multicore AMD EPYC 7662 at 2GHz server with 100GB of main memory.

In view of the costs, in the rest of the paper we will perform 16-epoch forecast for all the experimented systems.

\begin{table}[htb]
\caption{Maximum elapsed times vs. \# epochs to forecast L2C2 from start to 50\% capacity.}
\begin{center}
{
\begin{tabular}{|c|c|c|c|}
\hline
\#epochs & \begin{tabular}[c]{@{}c@{}}Simulation\\ phase, minutes.\end{tabular} & \begin{tabular}[c]{@{}c@{}}Prediction\\ phase, minutes.\end{tabular} & \begin{tabular}[c]{@{}c@{}}Forecast, \\ days.\end{tabular} \\ \hline
16 & 210 &  38 & 2.8\\ \hline
32 & 210 &  21 & 5.1\\ \hline
64 & 210 &  11 & 9.8\\ \hline
\end{tabular}
}
\label{table:computationtime}
\end{center}
\end{table}

\subsection{Specific situations}
\label{sec:Specific_Situations}

In all the experiments performed in this work, our forecast procedure assumes a uniform distribution of writes among sets. This condition is met in most systems either because the workload is diverse over time and produces an even distribution, or because the cache incorporates good wear-leveling mechanisms among sets.

However, in some scenarios it may be important to take non-uniformity into account. As an example, we can think of an embedded system that always runs the same applications. Here the distribution of accesses to the cache sets may well be non-uniform, encouraging the design and comparison of mechanisms to even out the wear between sets.

Certainly, our forecast procedure could also be applied in this context, although the model that approximates the new write rates during the Prediction phase would have to be modified. In particular, the new model could no longer use the average write rate of all frames belonging to sets with a given health state. An alternative could be to obtain the approximation from the distribution of write rates of those frames.
We think such specialized forecast is entirely feasible, but it is beyond the scope of this paper.

%% file: Figuras/CE_FD.pgf
\begingroup%
\makeatletter%
\begin{pgfpicture}%
\pgfpathrectangle{\pgfpointorigin}{\pgfqpoint{1.591882in}{1.118911in}}%
\pgfusepath{use as bounding box, clip}%
\begin{pgfscope}%
\pgfsetbuttcap%
\pgfsetmiterjoin%
\definecolor{currentfill}{rgb}{1.000000,1.000000,1.000000}%
\pgfsetfillcolor{currentfill}%
\pgfsetlinewidth{0.000000pt}%
\definecolor{currentstroke}{rgb}{1.000000,1.000000,1.000000}%
\pgfsetstrokecolor{currentstroke}%
\pgfsetdash{}{0pt}%
\pgfpathmoveto{\pgfqpoint{0.000000in}{0.000000in}}%
\pgfpathlineto{\pgfqpoint{1.591882in}{0.000000in}}%
\pgfpathlineto{\pgfqpoint{1.591882in}{1.118911in}}%
\pgfpathlineto{\pgfqpoint{0.000000in}{1.118911in}}%
\pgfpathclose%
\pgfusepath{fill}%
\end{pgfscope}%
\begin{pgfscope}%
\pgfsetbuttcap%
\pgfsetmiterjoin%
\definecolor{currentfill}{rgb}{1.000000,1.000000,1.000000}%
\pgfsetfillcolor{currentfill}%
\pgfsetlinewidth{0.000000pt}%
\definecolor{currentstroke}{rgb}{0.000000,0.000000,0.000000}%
\pgfsetstrokecolor{currentstroke}%
\pgfsetstrokeopacity{0.000000}%
\pgfsetdash{}{0pt}%
\pgfpathmoveto{\pgfqpoint{0.362463in}{0.288580in}}%
\pgfpathlineto{\pgfqpoint{1.556882in}{0.288580in}}%
\pgfpathlineto{\pgfqpoint{1.556882in}{1.083911in}}%
\pgfpathlineto{\pgfqpoint{0.362463in}{1.083911in}}%
\pgfpathclose%
\pgfusepath{fill}%
\end{pgfscope}%
\begin{pgfscope}%
\pgfsetbuttcap%
\pgfsetroundjoin%
\definecolor{currentfill}{rgb}{0.000000,0.000000,0.000000}%
\pgfsetfillcolor{currentfill}%
\pgfsetlinewidth{0.803000pt}%
\definecolor{currentstroke}{rgb}{0.000000,0.000000,0.000000}%
\pgfsetstrokecolor{currentstroke}%
\pgfsetdash{}{0pt}%
\pgfsys@defobject{currentmarker}{\pgfqpoint{0.000000in}{-0.048611in}}{\pgfqpoint{0.000000in}{0.000000in}}{%
\pgfpathmoveto{\pgfqpoint{0.000000in}{0.000000in}}%
\pgfpathlineto{\pgfqpoint{0.000000in}{-0.048611in}}%
\pgfusepath{stroke,fill}%
}%
\begin{pgfscope}%
\pgfsys@transformshift{0.416755in}{0.288580in}%
\pgfsys@useobject{currentmarker}{}%
\end{pgfscope}%
\end{pgfscope}%
\begin{pgfscope}%
\pgftext[x=0.416755in,y=0.191358in,,top]{\rmfamily\fontsize{5.000000}{6.000000}\selectfont 1}%
\end{pgfscope}%
\begin{pgfscope}%
\pgfsetbuttcap%
\pgfsetroundjoin%
\definecolor{currentfill}{rgb}{0.000000,0.000000,0.000000}%
\pgfsetfillcolor{currentfill}%
\pgfsetlinewidth{0.803000pt}%
\definecolor{currentstroke}{rgb}{0.000000,0.000000,0.000000}%
\pgfsetstrokecolor{currentstroke}%
\pgfsetdash{}{0pt}%
\pgfsys@defobject{currentmarker}{\pgfqpoint{0.000000in}{-0.048611in}}{\pgfqpoint{0.000000in}{0.000000in}}{%
\pgfpathmoveto{\pgfqpoint{0.000000in}{0.000000in}}%
\pgfpathlineto{\pgfqpoint{0.000000in}{-0.048611in}}%
\pgfusepath{stroke,fill}%
}%
\begin{pgfscope}%
\pgfsys@transformshift{0.688214in}{0.288580in}%
\pgfsys@useobject{currentmarker}{}%
\end{pgfscope}%
\end{pgfscope}%
\begin{pgfscope}%
\pgftext[x=0.688214in,y=0.191358in,,top]{\rmfamily\fontsize{5.000000}{6.000000}\selectfont 2}%
\end{pgfscope}%
\begin{pgfscope}%
\pgfsetbuttcap%
\pgfsetroundjoin%
\definecolor{currentfill}{rgb}{0.000000,0.000000,0.000000}%
\pgfsetfillcolor{currentfill}%
\pgfsetlinewidth{0.803000pt}%
\definecolor{currentstroke}{rgb}{0.000000,0.000000,0.000000}%
\pgfsetstrokecolor{currentstroke}%
\pgfsetdash{}{0pt}%
\pgfsys@defobject{currentmarker}{\pgfqpoint{0.000000in}{-0.048611in}}{\pgfqpoint{0.000000in}{0.000000in}}{%
\pgfpathmoveto{\pgfqpoint{0.000000in}{0.000000in}}%
\pgfpathlineto{\pgfqpoint{0.000000in}{-0.048611in}}%
\pgfusepath{stroke,fill}%
}%
\begin{pgfscope}%
\pgfsys@transformshift{0.959673in}{0.288580in}%
\pgfsys@useobject{currentmarker}{}%
\end{pgfscope}%
\end{pgfscope}%
\begin{pgfscope}%
\pgftext[x=0.959673in,y=0.191358in,,top]{\rmfamily\fontsize{5.000000}{6.000000}\selectfont 4}%
\end{pgfscope}%
\begin{pgfscope}%
\pgfsetbuttcap%
\pgfsetroundjoin%
\definecolor{currentfill}{rgb}{0.000000,0.000000,0.000000}%
\pgfsetfillcolor{currentfill}%
\pgfsetlinewidth{0.803000pt}%
\definecolor{currentstroke}{rgb}{0.000000,0.000000,0.000000}%
\pgfsetstrokecolor{currentstroke}%
\pgfsetdash{}{0pt}%
\pgfsys@defobject{currentmarker}{\pgfqpoint{0.000000in}{-0.048611in}}{\pgfqpoint{0.000000in}{0.000000in}}{%
\pgfpathmoveto{\pgfqpoint{0.000000in}{0.000000in}}%
\pgfpathlineto{\pgfqpoint{0.000000in}{-0.048611in}}%
\pgfusepath{stroke,fill}%
}%
\begin{pgfscope}%
\pgfsys@transformshift{1.231131in}{0.288580in}%
\pgfsys@useobject{currentmarker}{}%
\end{pgfscope}%
\end{pgfscope}%
\begin{pgfscope}%
\pgftext[x=1.231131in,y=0.191358in,,top]{\rmfamily\fontsize{5.000000}{6.000000}\selectfont 8}%
\end{pgfscope}%
\begin{pgfscope}%
\pgfsetbuttcap%
\pgfsetroundjoin%
\definecolor{currentfill}{rgb}{0.000000,0.000000,0.000000}%
\pgfsetfillcolor{currentfill}%
\pgfsetlinewidth{0.803000pt}%
\definecolor{currentstroke}{rgb}{0.000000,0.000000,0.000000}%
\pgfsetstrokecolor{currentstroke}%
\pgfsetdash{}{0pt}%
\pgfsys@defobject{currentmarker}{\pgfqpoint{0.000000in}{-0.048611in}}{\pgfqpoint{0.000000in}{0.000000in}}{%
\pgfpathmoveto{\pgfqpoint{0.000000in}{0.000000in}}%
\pgfpathlineto{\pgfqpoint{0.000000in}{-0.048611in}}%
\pgfusepath{stroke,fill}%
}%
\begin{pgfscope}%
\pgfsys@transformshift{1.502590in}{0.288580in}%
\pgfsys@useobject{currentmarker}{}%
\end{pgfscope}%
\end{pgfscope}%
\begin{pgfscope}%
\pgftext[x=1.502590in,y=0.191358in,,top]{\rmfamily\fontsize{5.000000}{6.000000}\selectfont 16}%
\end{pgfscope}%
\begin{pgfscope}%
\pgftext[x=0.959673in,y=0.074074in,,top]{\rmfamily\fontsize{6.000000}{7.200000}\selectfont Number of epochs}%
\end{pgfscope}%
\begin{pgfscope}%
\pgfsetbuttcap%
\pgfsetroundjoin%
\definecolor{currentfill}{rgb}{0.000000,0.000000,0.000000}%
\pgfsetfillcolor{currentfill}%
\pgfsetlinewidth{0.803000pt}%
\definecolor{currentstroke}{rgb}{0.000000,0.000000,0.000000}%
\pgfsetstrokecolor{currentstroke}%
\pgfsetdash{}{0pt}%
\pgfsys@defobject{currentmarker}{\pgfqpoint{-0.048611in}{0.000000in}}{\pgfqpoint{0.000000in}{0.000000in}}{%
\pgfpathmoveto{\pgfqpoint{0.000000in}{0.000000in}}%
\pgfpathlineto{\pgfqpoint{-0.048611in}{0.000000in}}%
\pgfusepath{stroke,fill}%
}%
\begin{pgfscope}%
\pgfsys@transformshift{0.362463in}{0.288580in}%
\pgfsys@useobject{currentmarker}{}%
\end{pgfscope}%
\end{pgfscope}%
\begin{pgfscope}%
\pgftext[x=0.138889in,y=0.264467in,left,base]{\rmfamily\fontsize{5.000000}{6.000000}\selectfont \(\displaystyle 0.0\)}%
\end{pgfscope}%
\begin{pgfscope}%
\pgfsetbuttcap%
\pgfsetroundjoin%
\definecolor{currentfill}{rgb}{0.000000,0.000000,0.000000}%
\pgfsetfillcolor{currentfill}%
\pgfsetlinewidth{0.803000pt}%
\definecolor{currentstroke}{rgb}{0.000000,0.000000,0.000000}%
\pgfsetstrokecolor{currentstroke}%
\pgfsetdash{}{0pt}%
\pgfsys@defobject{currentmarker}{\pgfqpoint{-0.048611in}{0.000000in}}{\pgfqpoint{0.000000in}{0.000000in}}{%
\pgfpathmoveto{\pgfqpoint{0.000000in}{0.000000in}}%
\pgfpathlineto{\pgfqpoint{-0.048611in}{0.000000in}}%
\pgfusepath{stroke,fill}%
}%
\begin{pgfscope}%
\pgfsys@transformshift{0.362463in}{0.447646in}%
\pgfsys@useobject{currentmarker}{}%
\end{pgfscope}%
\end{pgfscope}%
\begin{pgfscope}%
\pgftext[x=0.138889in,y=0.423534in,left,base]{\rmfamily\fontsize{5.000000}{6.000000}\selectfont \(\displaystyle 0.5\)}%
\end{pgfscope}%
\begin{pgfscope}%
\pgfsetbuttcap%
\pgfsetroundjoin%
\definecolor{currentfill}{rgb}{0.000000,0.000000,0.000000}%
\pgfsetfillcolor{currentfill}%
\pgfsetlinewidth{0.803000pt}%
\definecolor{currentstroke}{rgb}{0.000000,0.000000,0.000000}%
\pgfsetstrokecolor{currentstroke}%
\pgfsetdash{}{0pt}%
\pgfsys@defobject{currentmarker}{\pgfqpoint{-0.048611in}{0.000000in}}{\pgfqpoint{0.000000in}{0.000000in}}{%
\pgfpathmoveto{\pgfqpoint{0.000000in}{0.000000in}}%
\pgfpathlineto{\pgfqpoint{-0.048611in}{0.000000in}}%
\pgfusepath{stroke,fill}%
}%
\begin{pgfscope}%
\pgfsys@transformshift{0.362463in}{0.606713in}%
\pgfsys@useobject{currentmarker}{}%
\end{pgfscope}%
\end{pgfscope}%
\begin{pgfscope}%
\pgftext[x=0.138889in,y=0.582600in,left,base]{\rmfamily\fontsize{5.000000}{6.000000}\selectfont \(\displaystyle 1.0\)}%
\end{pgfscope}%
\begin{pgfscope}%
\pgfsetbuttcap%
\pgfsetroundjoin%
\definecolor{currentfill}{rgb}{0.000000,0.000000,0.000000}%
\pgfsetfillcolor{currentfill}%
\pgfsetlinewidth{0.803000pt}%
\definecolor{currentstroke}{rgb}{0.000000,0.000000,0.000000}%
\pgfsetstrokecolor{currentstroke}%
\pgfsetdash{}{0pt}%
\pgfsys@defobject{currentmarker}{\pgfqpoint{-0.048611in}{0.000000in}}{\pgfqpoint{0.000000in}{0.000000in}}{%
\pgfpathmoveto{\pgfqpoint{0.000000in}{0.000000in}}%
\pgfpathlineto{\pgfqpoint{-0.048611in}{0.000000in}}%
\pgfusepath{stroke,fill}%
}%
\begin{pgfscope}%
\pgfsys@transformshift{0.362463in}{0.765779in}%
\pgfsys@useobject{currentmarker}{}%
\end{pgfscope}%
\end{pgfscope}%
\begin{pgfscope}%
\pgftext[x=0.138889in,y=0.741666in,left,base]{\rmfamily\fontsize{5.000000}{6.000000}\selectfont \(\displaystyle 1.5\)}%
\end{pgfscope}%
\begin{pgfscope}%
\pgfsetbuttcap%
\pgfsetroundjoin%
\definecolor{currentfill}{rgb}{0.000000,0.000000,0.000000}%
\pgfsetfillcolor{currentfill}%
\pgfsetlinewidth{0.803000pt}%
\definecolor{currentstroke}{rgb}{0.000000,0.000000,0.000000}%
\pgfsetstrokecolor{currentstroke}%
\pgfsetdash{}{0pt}%
\pgfsys@defobject{currentmarker}{\pgfqpoint{-0.048611in}{0.000000in}}{\pgfqpoint{0.000000in}{0.000000in}}{%
\pgfpathmoveto{\pgfqpoint{0.000000in}{0.000000in}}%
\pgfpathlineto{\pgfqpoint{-0.048611in}{0.000000in}}%
\pgfusepath{stroke,fill}%
}%
\begin{pgfscope}%
\pgfsys@transformshift{0.362463in}{0.924845in}%
\pgfsys@useobject{currentmarker}{}%
\end{pgfscope}%
\end{pgfscope}%
\begin{pgfscope}%
\pgftext[x=0.138889in,y=0.900733in,left,base]{\rmfamily\fontsize{5.000000}{6.000000}\selectfont \(\displaystyle 2.0\)}%
\end{pgfscope}%
\begin{pgfscope}%
\pgfsetbuttcap%
\pgfsetroundjoin%
\definecolor{currentfill}{rgb}{0.000000,0.000000,0.000000}%
\pgfsetfillcolor{currentfill}%
\pgfsetlinewidth{0.803000pt}%
\definecolor{currentstroke}{rgb}{0.000000,0.000000,0.000000}%
\pgfsetstrokecolor{currentstroke}%
\pgfsetdash{}{0pt}%
\pgfsys@defobject{currentmarker}{\pgfqpoint{-0.048611in}{0.000000in}}{\pgfqpoint{0.000000in}{0.000000in}}{%
\pgfpathmoveto{\pgfqpoint{0.000000in}{0.000000in}}%
\pgfpathlineto{\pgfqpoint{-0.048611in}{0.000000in}}%
\pgfusepath{stroke,fill}%
}%
\begin{pgfscope}%
\pgfsys@transformshift{0.362463in}{1.083911in}%
\pgfsys@useobject{currentmarker}{}%
\end{pgfscope}%
\end{pgfscope}%
\begin{pgfscope}%
\pgftext[x=0.138889in,y=1.059799in,left,base]{\rmfamily\fontsize{5.000000}{6.000000}\selectfont \(\displaystyle 2.5\)}%
\end{pgfscope}%
\begin{pgfscope}%
\pgftext[x=0.083333in,y=0.686246in,,bottom,rotate=90.000000]{\rmfamily\fontsize{6.000000}{7.200000}\selectfont Time (years)}%
\end{pgfscope}%
\begin{pgfscope}%
\pgfpathrectangle{\pgfqpoint{0.362463in}{0.288580in}}{\pgfqpoint{1.194419in}{0.795331in}}%
\pgfusepath{clip}%
\pgfsetrectcap%
\pgfsetroundjoin%
\pgfsetlinewidth{0.752812pt}%
\definecolor{currentstroke}{rgb}{0.145098,0.145098,0.145098}%
\pgfsetstrokecolor{currentstroke}%
\pgfsetdash{}{0pt}%
\pgfpathmoveto{\pgfqpoint{0.416755in}{1.033454in}}%
\pgfpathlineto{\pgfqpoint{0.688214in}{1.004375in}}%
\pgfpathlineto{\pgfqpoint{0.959673in}{0.993891in}}%
\pgfpathlineto{\pgfqpoint{1.231131in}{0.992509in}}%
\pgfpathlineto{\pgfqpoint{1.502590in}{0.992771in}}%
\pgfusepath{stroke}%
\end{pgfscope}%
\begin{pgfscope}%
\pgfpathrectangle{\pgfqpoint{0.362463in}{0.288580in}}{\pgfqpoint{1.194419in}{0.795331in}}%
\pgfusepath{clip}%
\pgfsetbuttcap%
\pgfsetroundjoin%
\definecolor{currentfill}{rgb}{0.145098,0.145098,0.145098}%
\pgfsetfillcolor{currentfill}%
\pgfsetlinewidth{1.003750pt}%
\definecolor{currentstroke}{rgb}{0.145098,0.145098,0.145098}%
\pgfsetstrokecolor{currentstroke}%
\pgfsetdash{}{0pt}%
\pgfsys@defobject{currentmarker}{\pgfqpoint{-0.013889in}{-0.013889in}}{\pgfqpoint{0.013889in}{0.013889in}}{%
\pgfpathmoveto{\pgfqpoint{0.000000in}{-0.013889in}}%
\pgfpathcurveto{\pgfqpoint{0.003683in}{-0.013889in}}{\pgfqpoint{0.007216in}{-0.012425in}}{\pgfqpoint{0.009821in}{-0.009821in}}%
\pgfpathcurveto{\pgfqpoint{0.012425in}{-0.007216in}}{\pgfqpoint{0.013889in}{-0.003683in}}{\pgfqpoint{0.013889in}{0.000000in}}%
\pgfpathcurveto{\pgfqpoint{0.013889in}{0.003683in}}{\pgfqpoint{0.012425in}{0.007216in}}{\pgfqpoint{0.009821in}{0.009821in}}%
\pgfpathcurveto{\pgfqpoint{0.007216in}{0.012425in}}{\pgfqpoint{0.003683in}{0.013889in}}{\pgfqpoint{0.000000in}{0.013889in}}%
\pgfpathcurveto{\pgfqpoint{-0.003683in}{0.013889in}}{\pgfqpoint{-0.007216in}{0.012425in}}{\pgfqpoint{-0.009821in}{0.009821in}}%
\pgfpathcurveto{\pgfqpoint{-0.012425in}{0.007216in}}{\pgfqpoint{-0.013889in}{0.003683in}}{\pgfqpoint{-0.013889in}{0.000000in}}%
\pgfpathcurveto{\pgfqpoint{-0.013889in}{-0.003683in}}{\pgfqpoint{-0.012425in}{-0.007216in}}{\pgfqpoint{-0.009821in}{-0.009821in}}%
\pgfpathcurveto{\pgfqpoint{-0.007216in}{-0.012425in}}{\pgfqpoint{-0.003683in}{-0.013889in}}{\pgfqpoint{0.000000in}{-0.013889in}}%
\pgfpathclose%
\pgfusepath{stroke,fill}%
}%
\begin{pgfscope}%
\pgfsys@transformshift{0.416755in}{1.033454in}%
\pgfsys@useobject{currentmarker}{}%
\end{pgfscope}%
\begin{pgfscope}%
\pgfsys@transformshift{0.688214in}{1.004375in}%
\pgfsys@useobject{currentmarker}{}%
\end{pgfscope}%
\begin{pgfscope}%
\pgfsys@transformshift{0.959673in}{0.993891in}%
\pgfsys@useobject{currentmarker}{}%
\end{pgfscope}%
\begin{pgfscope}%
\pgfsys@transformshift{1.231131in}{0.992509in}%
\pgfsys@useobject{currentmarker}{}%
\end{pgfscope}%
\begin{pgfscope}%
\pgfsys@transformshift{1.502590in}{0.992771in}%
\pgfsys@useobject{currentmarker}{}%
\end{pgfscope}%
\end{pgfscope}%
\begin{pgfscope}%
\pgfpathrectangle{\pgfqpoint{0.362463in}{0.288580in}}{\pgfqpoint{1.194419in}{0.795331in}}%
\pgfusepath{clip}%
\pgfsetrectcap%
\pgfsetroundjoin%
\pgfsetlinewidth{0.752812pt}%
\definecolor{currentstroke}{rgb}{0.145098,0.145098,0.145098}%
\pgfsetstrokecolor{currentstroke}%
\pgfsetdash{}{0pt}%
\pgfpathmoveto{\pgfqpoint{0.416755in}{0.714492in}}%
\pgfpathlineto{\pgfqpoint{0.688214in}{0.694045in}}%
\pgfpathlineto{\pgfqpoint{0.959673in}{0.690916in}}%
\pgfpathlineto{\pgfqpoint{1.231131in}{0.693947in}}%
\pgfpathlineto{\pgfqpoint{1.502590in}{0.696113in}}%
\pgfusepath{stroke}%
\end{pgfscope}%
\begin{pgfscope}%
\pgfpathrectangle{\pgfqpoint{0.362463in}{0.288580in}}{\pgfqpoint{1.194419in}{0.795331in}}%
\pgfusepath{clip}%
\pgfsetbuttcap%
\pgfsetroundjoin%
\definecolor{currentfill}{rgb}{0.145098,0.145098,0.145098}%
\pgfsetfillcolor{currentfill}%
\pgfsetlinewidth{1.003750pt}%
\definecolor{currentstroke}{rgb}{0.145098,0.145098,0.145098}%
\pgfsetstrokecolor{currentstroke}%
\pgfsetdash{}{0pt}%
\pgfsys@defobject{currentmarker}{\pgfqpoint{-0.013889in}{-0.013889in}}{\pgfqpoint{0.013889in}{0.013889in}}{%
\pgfpathmoveto{\pgfqpoint{-0.013889in}{-0.013889in}}%
\pgfpathlineto{\pgfqpoint{0.013889in}{0.013889in}}%
\pgfpathmoveto{\pgfqpoint{-0.013889in}{0.013889in}}%
\pgfpathlineto{\pgfqpoint{0.013889in}{-0.013889in}}%
\pgfusepath{stroke,fill}%
}%
\begin{pgfscope}%
\pgfsys@transformshift{0.416755in}{0.714492in}%
\pgfsys@useobject{currentmarker}{}%
\end{pgfscope}%
\begin{pgfscope}%
\pgfsys@transformshift{0.688214in}{0.694045in}%
\pgfsys@useobject{currentmarker}{}%
\end{pgfscope}%
\begin{pgfscope}%
\pgfsys@transformshift{0.959673in}{0.690916in}%
\pgfsys@useobject{currentmarker}{}%
\end{pgfscope}%
\begin{pgfscope}%
\pgfsys@transformshift{1.231131in}{0.693947in}%
\pgfsys@useobject{currentmarker}{}%
\end{pgfscope}%
\begin{pgfscope}%
\pgfsys@transformshift{1.502590in}{0.696113in}%
\pgfsys@useobject{currentmarker}{}%
\end{pgfscope}%
\end{pgfscope}%
\begin{pgfscope}%
\pgfpathrectangle{\pgfqpoint{0.362463in}{0.288580in}}{\pgfqpoint{1.194419in}{0.795331in}}%
\pgfusepath{clip}%
\pgfsetrectcap%
\pgfsetroundjoin%
\pgfsetlinewidth{0.752812pt}%
\definecolor{currentstroke}{rgb}{0.145098,0.145098,0.145098}%
\pgfsetstrokecolor{currentstroke}%
\pgfsetdash{}{0pt}%
\pgfpathmoveto{\pgfqpoint{0.416755in}{0.419027in}}%
\pgfpathlineto{\pgfqpoint{0.688214in}{0.419468in}}%
\pgfpathlineto{\pgfqpoint{0.959673in}{0.421060in}}%
\pgfpathlineto{\pgfqpoint{1.231131in}{0.422725in}}%
\pgfpathlineto{\pgfqpoint{1.502590in}{0.423776in}}%
\pgfusepath{stroke}%
\end{pgfscope}%
\begin{pgfscope}%
\pgfpathrectangle{\pgfqpoint{0.362463in}{0.288580in}}{\pgfqpoint{1.194419in}{0.795331in}}%
\pgfusepath{clip}%
\pgfsetbuttcap%
\pgfsetmiterjoin%
\definecolor{currentfill}{rgb}{0.145098,0.145098,0.145098}%
\pgfsetfillcolor{currentfill}%
\pgfsetlinewidth{1.003750pt}%
\definecolor{currentstroke}{rgb}{0.145098,0.145098,0.145098}%
\pgfsetstrokecolor{currentstroke}%
\pgfsetdash{}{0pt}%
\pgfsys@defobject{currentmarker}{\pgfqpoint{-0.013889in}{-0.013889in}}{\pgfqpoint{0.013889in}{0.013889in}}{%
\pgfpathmoveto{\pgfqpoint{-0.000000in}{-0.013889in}}%
\pgfpathlineto{\pgfqpoint{0.013889in}{0.013889in}}%
\pgfpathlineto{\pgfqpoint{-0.013889in}{0.013889in}}%
\pgfpathclose%
\pgfusepath{stroke,fill}%
}%
\begin{pgfscope}%
\pgfsys@transformshift{0.416755in}{0.419027in}%
\pgfsys@useobject{currentmarker}{}%
\end{pgfscope}%
\begin{pgfscope}%
\pgfsys@transformshift{0.688214in}{0.419468in}%
\pgfsys@useobject{currentmarker}{}%
\end{pgfscope}%
\begin{pgfscope}%
\pgfsys@transformshift{0.959673in}{0.421060in}%
\pgfsys@useobject{currentmarker}{}%
\end{pgfscope}%
\begin{pgfscope}%
\pgfsys@transformshift{1.231131in}{0.422725in}%
\pgfsys@useobject{currentmarker}{}%
\end{pgfscope}%
\begin{pgfscope}%
\pgfsys@transformshift{1.502590in}{0.423776in}%
\pgfsys@useobject{currentmarker}{}%
\end{pgfscope}%
\end{pgfscope}%
\begin{pgfscope}%
\pgfsetrectcap%
\pgfsetmiterjoin%
\pgfsetlinewidth{0.803000pt}%
\definecolor{currentstroke}{rgb}{0.000000,0.000000,0.000000}%
\pgfsetstrokecolor{currentstroke}%
\pgfsetdash{}{0pt}%
\pgfpathmoveto{\pgfqpoint{0.362463in}{0.288580in}}%
\pgfpathlineto{\pgfqpoint{0.362463in}{1.083911in}}%
\pgfusepath{stroke}%
\end{pgfscope}%
\begin{pgfscope}%
\pgfsetrectcap%
\pgfsetmiterjoin%
\pgfsetlinewidth{0.803000pt}%
\definecolor{currentstroke}{rgb}{0.000000,0.000000,0.000000}%
\pgfsetstrokecolor{currentstroke}%
\pgfsetdash{}{0pt}%
\pgfpathmoveto{\pgfqpoint{0.362463in}{0.288580in}}%
\pgfpathlineto{\pgfqpoint{1.556882in}{0.288580in}}%
\pgfusepath{stroke}%
\end{pgfscope}%
\end{pgfpicture}%
\makeatother%
\endgroup%

%% file: Figuras/CE_BDI.pgf
\begingroup%
\makeatletter%
\begin{pgfpicture}%
\pgfpathrectangle{\pgfpointorigin}{\pgfqpoint{1.591882in}{1.118911in}}%
\pgfusepath{use as bounding box, clip}%
\begin{pgfscope}%
\pgfsetbuttcap%
\pgfsetmiterjoin%
\definecolor{currentfill}{rgb}{1.000000,1.000000,1.000000}%
\pgfsetfillcolor{currentfill}%
\pgfsetlinewidth{0.000000pt}%
\definecolor{currentstroke}{rgb}{1.000000,1.000000,1.000000}%
\pgfsetstrokecolor{currentstroke}%
\pgfsetdash{}{0pt}%
\pgfpathmoveto{\pgfqpoint{0.000000in}{0.000000in}}%
\pgfpathlineto{\pgfqpoint{1.591882in}{0.000000in}}%
\pgfpathlineto{\pgfqpoint{1.591882in}{1.118911in}}%
\pgfpathlineto{\pgfqpoint{0.000000in}{1.118911in}}%
\pgfpathclose%
\pgfusepath{fill}%
\end{pgfscope}%
\begin{pgfscope}%
\pgfsetbuttcap%
\pgfsetmiterjoin%
\definecolor{currentfill}{rgb}{1.000000,1.000000,1.000000}%
\pgfsetfillcolor{currentfill}%
\pgfsetlinewidth{0.000000pt}%
\definecolor{currentstroke}{rgb}{0.000000,0.000000,0.000000}%
\pgfsetstrokecolor{currentstroke}%
\pgfsetstrokeopacity{0.000000}%
\pgfsetdash{}{0pt}%
\pgfpathmoveto{\pgfqpoint{0.330634in}{0.288580in}}%
\pgfpathlineto{\pgfqpoint{1.556882in}{0.288580in}}%
\pgfpathlineto{\pgfqpoint{1.556882in}{1.083911in}}%
\pgfpathlineto{\pgfqpoint{0.330634in}{1.083911in}}%
\pgfpathclose%
\pgfusepath{fill}%
\end{pgfscope}%
\begin{pgfscope}%
\pgfsetbuttcap%
\pgfsetroundjoin%
\definecolor{currentfill}{rgb}{0.000000,0.000000,0.000000}%
\pgfsetfillcolor{currentfill}%
\pgfsetlinewidth{0.803000pt}%
\definecolor{currentstroke}{rgb}{0.000000,0.000000,0.000000}%
\pgfsetstrokecolor{currentstroke}%
\pgfsetdash{}{0pt}%
\pgfsys@defobject{currentmarker}{\pgfqpoint{0.000000in}{-0.048611in}}{\pgfqpoint{0.000000in}{0.000000in}}{%
\pgfpathmoveto{\pgfqpoint{0.000000in}{0.000000in}}%
\pgfpathlineto{\pgfqpoint{0.000000in}{-0.048611in}}%
\pgfusepath{stroke,fill}%
}%
\begin{pgfscope}%
\pgfsys@transformshift{0.386373in}{0.288580in}%
\pgfsys@useobject{currentmarker}{}%
\end{pgfscope}%
\end{pgfscope}%
\begin{pgfscope}%
\pgftext[x=0.386373in,y=0.191358in,,top]{\rmfamily\fontsize{5.000000}{6.000000}\selectfont 1}%
\end{pgfscope}%
\begin{pgfscope}%
\pgfsetbuttcap%
\pgfsetroundjoin%
\definecolor{currentfill}{rgb}{0.000000,0.000000,0.000000}%
\pgfsetfillcolor{currentfill}%
\pgfsetlinewidth{0.803000pt}%
\definecolor{currentstroke}{rgb}{0.000000,0.000000,0.000000}%
\pgfsetstrokecolor{currentstroke}%
\pgfsetdash{}{0pt}%
\pgfsys@defobject{currentmarker}{\pgfqpoint{0.000000in}{-0.048611in}}{\pgfqpoint{0.000000in}{0.000000in}}{%
\pgfpathmoveto{\pgfqpoint{0.000000in}{0.000000in}}%
\pgfpathlineto{\pgfqpoint{0.000000in}{-0.048611in}}%
\pgfusepath{stroke,fill}%
}%
\begin{pgfscope}%
\pgfsys@transformshift{0.572168in}{0.288580in}%
\pgfsys@useobject{currentmarker}{}%
\end{pgfscope}%
\end{pgfscope}%
\begin{pgfscope}%
\pgftext[x=0.572168in,y=0.191358in,,top]{\rmfamily\fontsize{5.000000}{6.000000}\selectfont 2}%
\end{pgfscope}%
\begin{pgfscope}%
\pgfsetbuttcap%
\pgfsetroundjoin%
\definecolor{currentfill}{rgb}{0.000000,0.000000,0.000000}%
\pgfsetfillcolor{currentfill}%
\pgfsetlinewidth{0.803000pt}%
\definecolor{currentstroke}{rgb}{0.000000,0.000000,0.000000}%
\pgfsetstrokecolor{currentstroke}%
\pgfsetdash{}{0pt}%
\pgfsys@defobject{currentmarker}{\pgfqpoint{0.000000in}{-0.048611in}}{\pgfqpoint{0.000000in}{0.000000in}}{%
\pgfpathmoveto{\pgfqpoint{0.000000in}{0.000000in}}%
\pgfpathlineto{\pgfqpoint{0.000000in}{-0.048611in}}%
\pgfusepath{stroke,fill}%
}%
\begin{pgfscope}%
\pgfsys@transformshift{0.757963in}{0.288580in}%
\pgfsys@useobject{currentmarker}{}%
\end{pgfscope}%
\end{pgfscope}%
\begin{pgfscope}%
\pgftext[x=0.757963in,y=0.191358in,,top]{\rmfamily\fontsize{5.000000}{6.000000}\selectfont 4}%
\end{pgfscope}%
\begin{pgfscope}%
\pgfsetbuttcap%
\pgfsetroundjoin%
\definecolor{currentfill}{rgb}{0.000000,0.000000,0.000000}%
\pgfsetfillcolor{currentfill}%
\pgfsetlinewidth{0.803000pt}%
\definecolor{currentstroke}{rgb}{0.000000,0.000000,0.000000}%
\pgfsetstrokecolor{currentstroke}%
\pgfsetdash{}{0pt}%
\pgfsys@defobject{currentmarker}{\pgfqpoint{0.000000in}{-0.048611in}}{\pgfqpoint{0.000000in}{0.000000in}}{%
\pgfpathmoveto{\pgfqpoint{0.000000in}{0.000000in}}%
\pgfpathlineto{\pgfqpoint{0.000000in}{-0.048611in}}%
\pgfusepath{stroke,fill}%
}%
\begin{pgfscope}%
\pgfsys@transformshift{0.943758in}{0.288580in}%
\pgfsys@useobject{currentmarker}{}%
\end{pgfscope}%
\end{pgfscope}%
\begin{pgfscope}%
\pgftext[x=0.943758in,y=0.191358in,,top]{\rmfamily\fontsize{5.000000}{6.000000}\selectfont 8}%
\end{pgfscope}%
\begin{pgfscope}%
\pgfsetbuttcap%
\pgfsetroundjoin%
\definecolor{currentfill}{rgb}{0.000000,0.000000,0.000000}%
\pgfsetfillcolor{currentfill}%
\pgfsetlinewidth{0.803000pt}%
\definecolor{currentstroke}{rgb}{0.000000,0.000000,0.000000}%
\pgfsetstrokecolor{currentstroke}%
\pgfsetdash{}{0pt}%
\pgfsys@defobject{currentmarker}{\pgfqpoint{0.000000in}{-0.048611in}}{\pgfqpoint{0.000000in}{0.000000in}}{%
\pgfpathmoveto{\pgfqpoint{0.000000in}{0.000000in}}%
\pgfpathlineto{\pgfqpoint{0.000000in}{-0.048611in}}%
\pgfusepath{stroke,fill}%
}%
\begin{pgfscope}%
\pgfsys@transformshift{1.129553in}{0.288580in}%
\pgfsys@useobject{currentmarker}{}%
\end{pgfscope}%
\end{pgfscope}%
\begin{pgfscope}%
\pgftext[x=1.129553in,y=0.191358in,,top]{\rmfamily\fontsize{5.000000}{6.000000}\selectfont 16}%
\end{pgfscope}%
\begin{pgfscope}%
\pgfsetbuttcap%
\pgfsetroundjoin%
\definecolor{currentfill}{rgb}{0.000000,0.000000,0.000000}%
\pgfsetfillcolor{currentfill}%
\pgfsetlinewidth{0.803000pt}%
\definecolor{currentstroke}{rgb}{0.000000,0.000000,0.000000}%
\pgfsetstrokecolor{currentstroke}%
\pgfsetdash{}{0pt}%
\pgfsys@defobject{currentmarker}{\pgfqpoint{0.000000in}{-0.048611in}}{\pgfqpoint{0.000000in}{0.000000in}}{%
\pgfpathmoveto{\pgfqpoint{0.000000in}{0.000000in}}%
\pgfpathlineto{\pgfqpoint{0.000000in}{-0.048611in}}%
\pgfusepath{stroke,fill}%
}%
\begin{pgfscope}%
\pgfsys@transformshift{1.315348in}{0.288580in}%
\pgfsys@useobject{currentmarker}{}%
\end{pgfscope}%
\end{pgfscope}%
\begin{pgfscope}%
\pgftext[x=1.315348in,y=0.191358in,,top]{\rmfamily\fontsize{5.000000}{6.000000}\selectfont 32}%
\end{pgfscope}%
\begin{pgfscope}%
\pgfsetbuttcap%
\pgfsetroundjoin%
\definecolor{currentfill}{rgb}{0.000000,0.000000,0.000000}%
\pgfsetfillcolor{currentfill}%
\pgfsetlinewidth{0.803000pt}%
\definecolor{currentstroke}{rgb}{0.000000,0.000000,0.000000}%
\pgfsetstrokecolor{currentstroke}%
\pgfsetdash{}{0pt}%
\pgfsys@defobject{currentmarker}{\pgfqpoint{0.000000in}{-0.048611in}}{\pgfqpoint{0.000000in}{0.000000in}}{%
\pgfpathmoveto{\pgfqpoint{0.000000in}{0.000000in}}%
\pgfpathlineto{\pgfqpoint{0.000000in}{-0.048611in}}%
\pgfusepath{stroke,fill}%
}%
\begin{pgfscope}%
\pgfsys@transformshift{1.501143in}{0.288580in}%
\pgfsys@useobject{currentmarker}{}%
\end{pgfscope}%
\end{pgfscope}%
\begin{pgfscope}%
\pgftext[x=1.501143in,y=0.191358in,,top]{\rmfamily\fontsize{5.000000}{6.000000}\selectfont 64}%
\end{pgfscope}%
\begin{pgfscope}%
\pgftext[x=0.943758in,y=0.074074in,,top]{\rmfamily\fontsize{6.000000}{7.200000}\selectfont Number of epochs}%
\end{pgfscope}%
\begin{pgfscope}%
\pgfsetbuttcap%
\pgfsetroundjoin%
\definecolor{currentfill}{rgb}{0.000000,0.000000,0.000000}%
\pgfsetfillcolor{currentfill}%
\pgfsetlinewidth{0.803000pt}%
\definecolor{currentstroke}{rgb}{0.000000,0.000000,0.000000}%
\pgfsetstrokecolor{currentstroke}%
\pgfsetdash{}{0pt}%
\pgfsys@defobject{currentmarker}{\pgfqpoint{-0.048611in}{0.000000in}}{\pgfqpoint{0.000000in}{0.000000in}}{%
\pgfpathmoveto{\pgfqpoint{0.000000in}{0.000000in}}%
\pgfpathlineto{\pgfqpoint{-0.048611in}{0.000000in}}%
\pgfusepath{stroke,fill}%
}%
\begin{pgfscope}%
\pgfsys@transformshift{0.330634in}{0.288580in}%
\pgfsys@useobject{currentmarker}{}%
\end{pgfscope}%
\end{pgfscope}%
\begin{pgfscope}%
\pgftext[x=0.186150in,y=0.264467in,left,base]{\rmfamily\fontsize{5.000000}{6.000000}\selectfont \(\displaystyle 4\)}%
\end{pgfscope}%
\begin{pgfscope}%
\pgfsetbuttcap%
\pgfsetroundjoin%
\definecolor{currentfill}{rgb}{0.000000,0.000000,0.000000}%
\pgfsetfillcolor{currentfill}%
\pgfsetlinewidth{0.803000pt}%
\definecolor{currentstroke}{rgb}{0.000000,0.000000,0.000000}%
\pgfsetstrokecolor{currentstroke}%
\pgfsetdash{}{0pt}%
\pgfsys@defobject{currentmarker}{\pgfqpoint{-0.048611in}{0.000000in}}{\pgfqpoint{0.000000in}{0.000000in}}{%
\pgfpathmoveto{\pgfqpoint{0.000000in}{0.000000in}}%
\pgfpathlineto{\pgfqpoint{-0.048611in}{0.000000in}}%
\pgfusepath{stroke,fill}%
}%
\begin{pgfscope}%
\pgfsys@transformshift{0.330634in}{0.421135in}%
\pgfsys@useobject{currentmarker}{}%
\end{pgfscope}%
\end{pgfscope}%
\begin{pgfscope}%
\pgftext[x=0.186150in,y=0.397023in,left,base]{\rmfamily\fontsize{5.000000}{6.000000}\selectfont \(\displaystyle 6\)}%
\end{pgfscope}%
\begin{pgfscope}%
\pgfsetbuttcap%
\pgfsetroundjoin%
\definecolor{currentfill}{rgb}{0.000000,0.000000,0.000000}%
\pgfsetfillcolor{currentfill}%
\pgfsetlinewidth{0.803000pt}%
\definecolor{currentstroke}{rgb}{0.000000,0.000000,0.000000}%
\pgfsetstrokecolor{currentstroke}%
\pgfsetdash{}{0pt}%
\pgfsys@defobject{currentmarker}{\pgfqpoint{-0.048611in}{0.000000in}}{\pgfqpoint{0.000000in}{0.000000in}}{%
\pgfpathmoveto{\pgfqpoint{0.000000in}{0.000000in}}%
\pgfpathlineto{\pgfqpoint{-0.048611in}{0.000000in}}%
\pgfusepath{stroke,fill}%
}%
\begin{pgfscope}%
\pgfsys@transformshift{0.330634in}{0.553690in}%
\pgfsys@useobject{currentmarker}{}%
\end{pgfscope}%
\end{pgfscope}%
\begin{pgfscope}%
\pgftext[x=0.186150in,y=0.529578in,left,base]{\rmfamily\fontsize{5.000000}{6.000000}\selectfont \(\displaystyle 8\)}%
\end{pgfscope}%
\begin{pgfscope}%
\pgfsetbuttcap%
\pgfsetroundjoin%
\definecolor{currentfill}{rgb}{0.000000,0.000000,0.000000}%
\pgfsetfillcolor{currentfill}%
\pgfsetlinewidth{0.803000pt}%
\definecolor{currentstroke}{rgb}{0.000000,0.000000,0.000000}%
\pgfsetstrokecolor{currentstroke}%
\pgfsetdash{}{0pt}%
\pgfsys@defobject{currentmarker}{\pgfqpoint{-0.048611in}{0.000000in}}{\pgfqpoint{0.000000in}{0.000000in}}{%
\pgfpathmoveto{\pgfqpoint{0.000000in}{0.000000in}}%
\pgfpathlineto{\pgfqpoint{-0.048611in}{0.000000in}}%
\pgfusepath{stroke,fill}%
}%
\begin{pgfscope}%
\pgfsys@transformshift{0.330634in}{0.686246in}%
\pgfsys@useobject{currentmarker}{}%
\end{pgfscope}%
\end{pgfscope}%
\begin{pgfscope}%
\pgftext[x=0.138889in,y=0.662133in,left,base]{\rmfamily\fontsize{5.000000}{6.000000}\selectfont \(\displaystyle 10\)}%
\end{pgfscope}%
\begin{pgfscope}%
\pgfsetbuttcap%
\pgfsetroundjoin%
\definecolor{currentfill}{rgb}{0.000000,0.000000,0.000000}%
\pgfsetfillcolor{currentfill}%
\pgfsetlinewidth{0.803000pt}%
\definecolor{currentstroke}{rgb}{0.000000,0.000000,0.000000}%
\pgfsetstrokecolor{currentstroke}%
\pgfsetdash{}{0pt}%
\pgfsys@defobject{currentmarker}{\pgfqpoint{-0.048611in}{0.000000in}}{\pgfqpoint{0.000000in}{0.000000in}}{%
\pgfpathmoveto{\pgfqpoint{0.000000in}{0.000000in}}%
\pgfpathlineto{\pgfqpoint{-0.048611in}{0.000000in}}%
\pgfusepath{stroke,fill}%
}%
\begin{pgfscope}%
\pgfsys@transformshift{0.330634in}{0.818801in}%
\pgfsys@useobject{currentmarker}{}%
\end{pgfscope}%
\end{pgfscope}%
\begin{pgfscope}%
\pgftext[x=0.138889in,y=0.794688in,left,base]{\rmfamily\fontsize{5.000000}{6.000000}\selectfont \(\displaystyle 12\)}%
\end{pgfscope}%
\begin{pgfscope}%
\pgfsetbuttcap%
\pgfsetroundjoin%
\definecolor{currentfill}{rgb}{0.000000,0.000000,0.000000}%
\pgfsetfillcolor{currentfill}%
\pgfsetlinewidth{0.803000pt}%
\definecolor{currentstroke}{rgb}{0.000000,0.000000,0.000000}%
\pgfsetstrokecolor{currentstroke}%
\pgfsetdash{}{0pt}%
\pgfsys@defobject{currentmarker}{\pgfqpoint{-0.048611in}{0.000000in}}{\pgfqpoint{0.000000in}{0.000000in}}{%
\pgfpathmoveto{\pgfqpoint{0.000000in}{0.000000in}}%
\pgfpathlineto{\pgfqpoint{-0.048611in}{0.000000in}}%
\pgfusepath{stroke,fill}%
}%
\begin{pgfscope}%
\pgfsys@transformshift{0.330634in}{0.951356in}%
\pgfsys@useobject{currentmarker}{}%
\end{pgfscope}%
\end{pgfscope}%
\begin{pgfscope}%
\pgftext[x=0.138889in,y=0.927244in,left,base]{\rmfamily\fontsize{5.000000}{6.000000}\selectfont \(\displaystyle 14\)}%
\end{pgfscope}%
\begin{pgfscope}%
\pgfsetbuttcap%
\pgfsetroundjoin%
\definecolor{currentfill}{rgb}{0.000000,0.000000,0.000000}%
\pgfsetfillcolor{currentfill}%
\pgfsetlinewidth{0.803000pt}%
\definecolor{currentstroke}{rgb}{0.000000,0.000000,0.000000}%
\pgfsetstrokecolor{currentstroke}%
\pgfsetdash{}{0pt}%
\pgfsys@defobject{currentmarker}{\pgfqpoint{-0.048611in}{0.000000in}}{\pgfqpoint{0.000000in}{0.000000in}}{%
\pgfpathmoveto{\pgfqpoint{0.000000in}{0.000000in}}%
\pgfpathlineto{\pgfqpoint{-0.048611in}{0.000000in}}%
\pgfusepath{stroke,fill}%
}%
\begin{pgfscope}%
\pgfsys@transformshift{0.330634in}{1.083911in}%
\pgfsys@useobject{currentmarker}{}%
\end{pgfscope}%
\end{pgfscope}%
\begin{pgfscope}%
\pgftext[x=0.138889in,y=1.059799in,left,base]{\rmfamily\fontsize{5.000000}{6.000000}\selectfont \(\displaystyle 16\)}%
\end{pgfscope}%
\begin{pgfscope}%
\pgftext[x=0.083333in,y=0.686246in,,bottom,rotate=90.000000]{\rmfamily\fontsize{6.000000}{7.200000}\selectfont Time (years)}%
\end{pgfscope}%
\begin{pgfscope}%
\pgfpathrectangle{\pgfqpoint{0.330634in}{0.288580in}}{\pgfqpoint{1.226248in}{0.795331in}}%
\pgfusepath{clip}%
\pgfsetrectcap%
\pgfsetroundjoin%
\pgfsetlinewidth{0.752812pt}%
\definecolor{currentstroke}{rgb}{0.145098,0.145098,0.145098}%
\pgfsetstrokecolor{currentstroke}%
\pgfsetdash{}{0pt}%
\pgfpathmoveto{\pgfqpoint{0.386373in}{0.507224in}}%
\pgfpathlineto{\pgfqpoint{0.572168in}{0.721813in}}%
\pgfpathlineto{\pgfqpoint{0.757963in}{0.906382in}}%
\pgfpathlineto{\pgfqpoint{0.943758in}{0.933856in}}%
\pgfpathlineto{\pgfqpoint{1.129553in}{0.930016in}}%
\pgfpathlineto{\pgfqpoint{1.315348in}{0.932887in}}%
\pgfpathlineto{\pgfqpoint{1.501143in}{0.932308in}}%
\pgfusepath{stroke}%
\end{pgfscope}%
\begin{pgfscope}%
\pgfpathrectangle{\pgfqpoint{0.330634in}{0.288580in}}{\pgfqpoint{1.226248in}{0.795331in}}%
\pgfusepath{clip}%
\pgfsetbuttcap%
\pgfsetroundjoin%
\definecolor{currentfill}{rgb}{0.145098,0.145098,0.145098}%
\pgfsetfillcolor{currentfill}%
\pgfsetlinewidth{1.003750pt}%
\definecolor{currentstroke}{rgb}{0.145098,0.145098,0.145098}%
\pgfsetstrokecolor{currentstroke}%
\pgfsetdash{}{0pt}%
\pgfsys@defobject{currentmarker}{\pgfqpoint{-0.013889in}{-0.013889in}}{\pgfqpoint{0.013889in}{0.013889in}}{%
\pgfpathmoveto{\pgfqpoint{0.000000in}{-0.013889in}}%
\pgfpathcurveto{\pgfqpoint{0.003683in}{-0.013889in}}{\pgfqpoint{0.007216in}{-0.012425in}}{\pgfqpoint{0.009821in}{-0.009821in}}%
\pgfpathcurveto{\pgfqpoint{0.012425in}{-0.007216in}}{\pgfqpoint{0.013889in}{-0.003683in}}{\pgfqpoint{0.013889in}{0.000000in}}%
\pgfpathcurveto{\pgfqpoint{0.013889in}{0.003683in}}{\pgfqpoint{0.012425in}{0.007216in}}{\pgfqpoint{0.009821in}{0.009821in}}%
\pgfpathcurveto{\pgfqpoint{0.007216in}{0.012425in}}{\pgfqpoint{0.003683in}{0.013889in}}{\pgfqpoint{0.000000in}{0.013889in}}%
\pgfpathcurveto{\pgfqpoint{-0.003683in}{0.013889in}}{\pgfqpoint{-0.007216in}{0.012425in}}{\pgfqpoint{-0.009821in}{0.009821in}}%
\pgfpathcurveto{\pgfqpoint{-0.012425in}{0.007216in}}{\pgfqpoint{-0.013889in}{0.003683in}}{\pgfqpoint{-0.013889in}{0.000000in}}%
\pgfpathcurveto{\pgfqpoint{-0.013889in}{-0.003683in}}{\pgfqpoint{-0.012425in}{-0.007216in}}{\pgfqpoint{-0.009821in}{-0.009821in}}%
\pgfpathcurveto{\pgfqpoint{-0.007216in}{-0.012425in}}{\pgfqpoint{-0.003683in}{-0.013889in}}{\pgfqpoint{0.000000in}{-0.013889in}}%
\pgfpathclose%
\pgfusepath{stroke,fill}%
}%
\begin{pgfscope}%
\pgfsys@transformshift{0.386373in}{0.507224in}%
\pgfsys@useobject{currentmarker}{}%
\end{pgfscope}%
\begin{pgfscope}%
\pgfsys@transformshift{0.572168in}{0.721813in}%
\pgfsys@useobject{currentmarker}{}%
\end{pgfscope}%
\begin{pgfscope}%
\pgfsys@transformshift{0.757963in}{0.906382in}%
\pgfsys@useobject{currentmarker}{}%
\end{pgfscope}%
\begin{pgfscope}%
\pgfsys@transformshift{0.943758in}{0.933856in}%
\pgfsys@useobject{currentmarker}{}%
\end{pgfscope}%
\begin{pgfscope}%
\pgfsys@transformshift{1.129553in}{0.930016in}%
\pgfsys@useobject{currentmarker}{}%
\end{pgfscope}%
\begin{pgfscope}%
\pgfsys@transformshift{1.315348in}{0.932887in}%
\pgfsys@useobject{currentmarker}{}%
\end{pgfscope}%
\begin{pgfscope}%
\pgfsys@transformshift{1.501143in}{0.932308in}%
\pgfsys@useobject{currentmarker}{}%
\end{pgfscope}%
\end{pgfscope}%
\begin{pgfscope}%
\pgfpathrectangle{\pgfqpoint{0.330634in}{0.288580in}}{\pgfqpoint{1.226248in}{0.795331in}}%
\pgfusepath{clip}%
\pgfsetrectcap%
\pgfsetroundjoin%
\pgfsetlinewidth{0.752812pt}%
\definecolor{currentstroke}{rgb}{0.145098,0.145098,0.145098}%
\pgfsetstrokecolor{currentstroke}%
\pgfsetdash{}{0pt}%
\pgfpathmoveto{\pgfqpoint{0.386373in}{0.434740in}}%
\pgfpathlineto{\pgfqpoint{0.572168in}{0.771663in}}%
\pgfpathlineto{\pgfqpoint{0.757963in}{0.943360in}}%
\pgfpathlineto{\pgfqpoint{0.943758in}{0.952787in}}%
\pgfpathlineto{\pgfqpoint{1.129553in}{0.982630in}}%
\pgfpathlineto{\pgfqpoint{1.315348in}{0.991022in}}%
\pgfpathlineto{\pgfqpoint{1.501143in}{0.993229in}}%
\pgfusepath{stroke}%
\end{pgfscope}%
\begin{pgfscope}%
\pgfpathrectangle{\pgfqpoint{0.330634in}{0.288580in}}{\pgfqpoint{1.226248in}{0.795331in}}%
\pgfusepath{clip}%
\pgfsetbuttcap%
\pgfsetroundjoin%
\definecolor{currentfill}{rgb}{0.145098,0.145098,0.145098}%
\pgfsetfillcolor{currentfill}%
\pgfsetlinewidth{1.003750pt}%
\definecolor{currentstroke}{rgb}{0.145098,0.145098,0.145098}%
\pgfsetstrokecolor{currentstroke}%
\pgfsetdash{}{0pt}%
\pgfsys@defobject{currentmarker}{\pgfqpoint{-0.013889in}{-0.013889in}}{\pgfqpoint{0.013889in}{0.013889in}}{%
\pgfpathmoveto{\pgfqpoint{-0.013889in}{-0.013889in}}%
\pgfpathlineto{\pgfqpoint{0.013889in}{0.013889in}}%
\pgfpathmoveto{\pgfqpoint{-0.013889in}{0.013889in}}%
\pgfpathlineto{\pgfqpoint{0.013889in}{-0.013889in}}%
\pgfusepath{stroke,fill}%
}%
\begin{pgfscope}%
\pgfsys@transformshift{0.386373in}{0.434740in}%
\pgfsys@useobject{currentmarker}{}%
\end{pgfscope}%
\begin{pgfscope}%
\pgfsys@transformshift{0.572168in}{0.771663in}%
\pgfsys@useobject{currentmarker}{}%
\end{pgfscope}%
\begin{pgfscope}%
\pgfsys@transformshift{0.757963in}{0.943360in}%
\pgfsys@useobject{currentmarker}{}%
\end{pgfscope}%
\begin{pgfscope}%
\pgfsys@transformshift{0.943758in}{0.952787in}%
\pgfsys@useobject{currentmarker}{}%
\end{pgfscope}%
\begin{pgfscope}%
\pgfsys@transformshift{1.129553in}{0.982630in}%
\pgfsys@useobject{currentmarker}{}%
\end{pgfscope}%
\begin{pgfscope}%
\pgfsys@transformshift{1.315348in}{0.991022in}%
\pgfsys@useobject{currentmarker}{}%
\end{pgfscope}%
\begin{pgfscope}%
\pgfsys@transformshift{1.501143in}{0.993229in}%
\pgfsys@useobject{currentmarker}{}%
\end{pgfscope}%
\end{pgfscope}%
\begin{pgfscope}%
\pgfpathrectangle{\pgfqpoint{0.330634in}{0.288580in}}{\pgfqpoint{1.226248in}{0.795331in}}%
\pgfusepath{clip}%
\pgfsetrectcap%
\pgfsetroundjoin%
\pgfsetlinewidth{0.752812pt}%
\definecolor{currentstroke}{rgb}{0.145098,0.145098,0.145098}%
\pgfsetstrokecolor{currentstroke}%
\pgfsetdash{}{0pt}%
\pgfpathmoveto{\pgfqpoint{0.386373in}{0.402293in}}%
\pgfpathlineto{\pgfqpoint{0.572168in}{0.720799in}}%
\pgfpathlineto{\pgfqpoint{0.757963in}{0.956219in}}%
\pgfpathlineto{\pgfqpoint{0.943758in}{0.991793in}}%
\pgfpathlineto{\pgfqpoint{1.129553in}{1.047256in}}%
\pgfpathlineto{\pgfqpoint{1.315348in}{1.052558in}}%
\pgfpathlineto{\pgfqpoint{1.501143in}{1.057146in}}%
\pgfusepath{stroke}%
\end{pgfscope}%
\begin{pgfscope}%
\pgfpathrectangle{\pgfqpoint{0.330634in}{0.288580in}}{\pgfqpoint{1.226248in}{0.795331in}}%
\pgfusepath{clip}%
\pgfsetbuttcap%
\pgfsetmiterjoin%
\definecolor{currentfill}{rgb}{0.145098,0.145098,0.145098}%
\pgfsetfillcolor{currentfill}%
\pgfsetlinewidth{1.003750pt}%
\definecolor{currentstroke}{rgb}{0.145098,0.145098,0.145098}%
\pgfsetstrokecolor{currentstroke}%
\pgfsetdash{}{0pt}%
\pgfsys@defobject{currentmarker}{\pgfqpoint{-0.013889in}{-0.013889in}}{\pgfqpoint{0.013889in}{0.013889in}}{%
\pgfpathmoveto{\pgfqpoint{-0.000000in}{-0.013889in}}%
\pgfpathlineto{\pgfqpoint{0.013889in}{0.013889in}}%
\pgfpathlineto{\pgfqpoint{-0.013889in}{0.013889in}}%
\pgfpathclose%
\pgfusepath{stroke,fill}%
}%
\begin{pgfscope}%
\pgfsys@transformshift{0.386373in}{0.402293in}%
\pgfsys@useobject{currentmarker}{}%
\end{pgfscope}%
\begin{pgfscope}%
\pgfsys@transformshift{0.572168in}{0.720799in}%
\pgfsys@useobject{currentmarker}{}%
\end{pgfscope}%
\begin{pgfscope}%
\pgfsys@transformshift{0.757963in}{0.956219in}%
\pgfsys@useobject{currentmarker}{}%
\end{pgfscope}%
\begin{pgfscope}%
\pgfsys@transformshift{0.943758in}{0.991793in}%
\pgfsys@useobject{currentmarker}{}%
\end{pgfscope}%
\begin{pgfscope}%
\pgfsys@transformshift{1.129553in}{1.047256in}%
\pgfsys@useobject{currentmarker}{}%
\end{pgfscope}%
\begin{pgfscope}%
\pgfsys@transformshift{1.315348in}{1.052558in}%
\pgfsys@useobject{currentmarker}{}%
\end{pgfscope}%
\begin{pgfscope}%
\pgfsys@transformshift{1.501143in}{1.057146in}%
\pgfsys@useobject{currentmarker}{}%
\end{pgfscope}%
\end{pgfscope}%
\begin{pgfscope}%
\pgfsetrectcap%
\pgfsetmiterjoin%
\pgfsetlinewidth{0.803000pt}%
\definecolor{currentstroke}{rgb}{0.000000,0.000000,0.000000}%
\pgfsetstrokecolor{currentstroke}%
\pgfsetdash{}{0pt}%
\pgfpathmoveto{\pgfqpoint{0.330634in}{0.288580in}}%
\pgfpathlineto{\pgfqpoint{0.330634in}{1.083911in}}%
\pgfusepath{stroke}%
\end{pgfscope}%
\begin{pgfscope}%
\pgfsetrectcap%
\pgfsetmiterjoin%
\pgfsetlinewidth{0.803000pt}%
\definecolor{currentstroke}{rgb}{0.000000,0.000000,0.000000}%
\pgfsetstrokecolor{currentstroke}%
\pgfsetdash{}{0pt}%
\pgfpathmoveto{\pgfqpoint{0.330634in}{0.288580in}}%
\pgfpathlineto{\pgfqpoint{1.556882in}{0.288580in}}%
\pgfusepath{stroke}%
\end{pgfscope}%
\begin{pgfscope}%
\pgfsetbuttcap%
\pgfsetmiterjoin%
\definecolor{currentfill}{rgb}{1.000000,1.000000,1.000000}%
\pgfsetfillcolor{currentfill}%
\pgfsetfillopacity{0.800000}%
\pgfsetlinewidth{1.003750pt}%
\definecolor{currentstroke}{rgb}{0.800000,0.800000,0.800000}%
\pgfsetstrokecolor{currentstroke}%
\pgfsetstrokeopacity{0.800000}%
\pgfsetdash{}{0pt}%
\pgfpathmoveto{\pgfqpoint{0.931038in}{0.330247in}}%
\pgfpathlineto{\pgfqpoint{1.498549in}{0.330247in}}%
\pgfpathquadraticcurveto{\pgfqpoint{1.515215in}{0.330247in}}{\pgfqpoint{1.515215in}{0.346913in}}%
\pgfpathlineto{\pgfqpoint{1.515215in}{0.687191in}}%
\pgfpathquadraticcurveto{\pgfqpoint{1.515215in}{0.703858in}}{\pgfqpoint{1.498549in}{0.703858in}}%
\pgfpathlineto{\pgfqpoint{0.931038in}{0.703858in}}%
\pgfpathquadraticcurveto{\pgfqpoint{0.914371in}{0.703858in}}{\pgfqpoint{0.914371in}{0.687191in}}%
\pgfpathlineto{\pgfqpoint{0.914371in}{0.346913in}}%
\pgfpathquadraticcurveto{\pgfqpoint{0.914371in}{0.330247in}}{\pgfqpoint{0.931038in}{0.330247in}}%
\pgfpathclose%
\pgfusepath{stroke,fill}%
\end{pgfscope}%
\begin{pgfscope}%
\pgfsetrectcap%
\pgfsetroundjoin%
\pgfsetlinewidth{0.752812pt}%
\definecolor{currentstroke}{rgb}{0.145098,0.145098,0.145098}%
\pgfsetstrokecolor{currentstroke}%
\pgfsetdash{}{0pt}%
\pgfpathmoveto{\pgfqpoint{0.947704in}{0.641358in}}%
\pgfpathlineto{\pgfqpoint{1.114371in}{0.641358in}}%
\pgfusepath{stroke}%
\end{pgfscope}%
\begin{pgfscope}%
\pgfsetbuttcap%
\pgfsetroundjoin%
\definecolor{currentfill}{rgb}{0.145098,0.145098,0.145098}%
\pgfsetfillcolor{currentfill}%
\pgfsetlinewidth{1.003750pt}%
\definecolor{currentstroke}{rgb}{0.145098,0.145098,0.145098}%
\pgfsetstrokecolor{currentstroke}%
\pgfsetdash{}{0pt}%
\pgfsys@defobject{currentmarker}{\pgfqpoint{-0.013889in}{-0.013889in}}{\pgfqpoint{0.013889in}{0.013889in}}{%
\pgfpathmoveto{\pgfqpoint{0.000000in}{-0.013889in}}%
\pgfpathcurveto{\pgfqpoint{0.003683in}{-0.013889in}}{\pgfqpoint{0.007216in}{-0.012425in}}{\pgfqpoint{0.009821in}{-0.009821in}}%
\pgfpathcurveto{\pgfqpoint{0.012425in}{-0.007216in}}{\pgfqpoint{0.013889in}{-0.003683in}}{\pgfqpoint{0.013889in}{0.000000in}}%
\pgfpathcurveto{\pgfqpoint{0.013889in}{0.003683in}}{\pgfqpoint{0.012425in}{0.007216in}}{\pgfqpoint{0.009821in}{0.009821in}}%
\pgfpathcurveto{\pgfqpoint{0.007216in}{0.012425in}}{\pgfqpoint{0.003683in}{0.013889in}}{\pgfqpoint{0.000000in}{0.013889in}}%
\pgfpathcurveto{\pgfqpoint{-0.003683in}{0.013889in}}{\pgfqpoint{-0.007216in}{0.012425in}}{\pgfqpoint{-0.009821in}{0.009821in}}%
\pgfpathcurveto{\pgfqpoint{-0.012425in}{0.007216in}}{\pgfqpoint{-0.013889in}{0.003683in}}{\pgfqpoint{-0.013889in}{0.000000in}}%
\pgfpathcurveto{\pgfqpoint{-0.013889in}{-0.003683in}}{\pgfqpoint{-0.012425in}{-0.007216in}}{\pgfqpoint{-0.009821in}{-0.009821in}}%
\pgfpathcurveto{\pgfqpoint{-0.007216in}{-0.012425in}}{\pgfqpoint{-0.003683in}{-0.013889in}}{\pgfqpoint{0.000000in}{-0.013889in}}%
\pgfpathclose%
\pgfusepath{stroke,fill}%
}%
\begin{pgfscope}%
\pgfsys@transformshift{1.031038in}{0.641358in}%
\pgfsys@useobject{currentmarker}{}%
\end{pgfscope}%
\end{pgfscope}%
\begin{pgfscope}%
\pgftext[x=1.122704in,y=0.612191in,left,base]{\rmfamily\fontsize{6.000000}{7.200000}\selectfont cv=0.2}%
\end{pgfscope}%
\begin{pgfscope}%
\pgfsetrectcap%
\pgfsetroundjoin%
\pgfsetlinewidth{0.752812pt}%
\definecolor{currentstroke}{rgb}{0.145098,0.145098,0.145098}%
\pgfsetstrokecolor{currentstroke}%
\pgfsetdash{}{0pt}%
\pgfpathmoveto{\pgfqpoint{0.947704in}{0.525154in}}%
\pgfpathlineto{\pgfqpoint{1.114371in}{0.525154in}}%
\pgfusepath{stroke}%
\end{pgfscope}%
\begin{pgfscope}%
\pgfsetbuttcap%
\pgfsetroundjoin%
\definecolor{currentfill}{rgb}{0.145098,0.145098,0.145098}%
\pgfsetfillcolor{currentfill}%
\pgfsetlinewidth{1.003750pt}%
\definecolor{currentstroke}{rgb}{0.145098,0.145098,0.145098}%
\pgfsetstrokecolor{currentstroke}%
\pgfsetdash{}{0pt}%
\pgfsys@defobject{currentmarker}{\pgfqpoint{-0.013889in}{-0.013889in}}{\pgfqpoint{0.013889in}{0.013889in}}{%
\pgfpathmoveto{\pgfqpoint{-0.013889in}{-0.013889in}}%
\pgfpathlineto{\pgfqpoint{0.013889in}{0.013889in}}%
\pgfpathmoveto{\pgfqpoint{-0.013889in}{0.013889in}}%
\pgfpathlineto{\pgfqpoint{0.013889in}{-0.013889in}}%
\pgfusepath{stroke,fill}%
}%
\begin{pgfscope}%
\pgfsys@transformshift{1.031038in}{0.525154in}%
\pgfsys@useobject{currentmarker}{}%
\end{pgfscope}%
\end{pgfscope}%
\begin{pgfscope}%
\pgftext[x=1.122704in,y=0.495988in,left,base]{\rmfamily\fontsize{6.000000}{7.200000}\selectfont cv=0.25}%
\end{pgfscope}%
\begin{pgfscope}%
\pgfsetrectcap%
\pgfsetroundjoin%
\pgfsetlinewidth{0.752812pt}%
\definecolor{currentstroke}{rgb}{0.145098,0.145098,0.145098}%
\pgfsetstrokecolor{currentstroke}%
\pgfsetdash{}{0pt}%
\pgfpathmoveto{\pgfqpoint{0.947704in}{0.408950in}}%
\pgfpathlineto{\pgfqpoint{1.114371in}{0.408950in}}%
\pgfusepath{stroke}%
\end{pgfscope}%
\begin{pgfscope}%
\pgfsetbuttcap%
\pgfsetmiterjoin%
\definecolor{currentfill}{rgb}{0.145098,0.145098,0.145098}%
\pgfsetfillcolor{currentfill}%
\pgfsetlinewidth{1.003750pt}%
\definecolor{currentstroke}{rgb}{0.145098,0.145098,0.145098}%
\pgfsetstrokecolor{currentstroke}%
\pgfsetdash{}{0pt}%
\pgfsys@defobject{currentmarker}{\pgfqpoint{-0.013889in}{-0.013889in}}{\pgfqpoint{0.013889in}{0.013889in}}{%
\pgfpathmoveto{\pgfqpoint{-0.000000in}{-0.013889in}}%
\pgfpathlineto{\pgfqpoint{0.013889in}{0.013889in}}%
\pgfpathlineto{\pgfqpoint{-0.013889in}{0.013889in}}%
\pgfpathclose%
\pgfusepath{stroke,fill}%
}%
\begin{pgfscope}%
\pgfsys@transformshift{1.031038in}{0.408950in}%
\pgfsys@useobject{currentmarker}{}%
\end{pgfscope}%
\end{pgfscope}%
\begin{pgfscope}%
\pgftext[x=1.122704in,y=0.379784in,left,base]{\rmfamily\fontsize{6.000000}{7.200000}\selectfont cv=0.3}%
\end{pgfscope}%
\end{pgfpicture}%
\makeatother%
\endgroup%

%% file: VII-Evaluation.tex
\section{Evaluation}
\label{sec:evaluation}

This section shows the evolution of capacity and performance for several NV-LLC organizations, from 100\% to 50\% capacity, along with experiments on wear leveling, replacement, cache size and workload. 
For all tested organizations, the forecast procedure uses 16 epochs of constant extension.

We analyze four NV-LLCs candidates, two based on frame disabling and two on byte disabling plus compression:

\begin{itemize}
    \item \textbf{Frame disabling cache (FD)}. 
    A bitcell failure is just handled by disabling the corresponding frame \cite{Chang-07-BD1, Wuu-05-Asynchronous}.

    \item \textbf{Frame disabling cache with ECP6 (FD+6)}. Frame endurance is increased allowing the failure of up to six bitcells. After the seventh failure, the frame is disabled, because an eighth failure would no longer be recoverable~\cite{Schechter-10-UseECPnotECC}. This is achieved by adding six ECPs per frame to the base SECDED mechanism.
    
    \item \textbf{L2C2}. A bitcell failure is  handled by disabling the corresponding byte. Cache blocks are stored compressed with BDI. It has an intra-frame wear-leveling mechanism and an LRU-Fit replacement policy; see Section \ref{sec:nvllc}.
    
    \item \textbf{L2C2+6}. An L2C2 with 6 spare bytes per cache frame; see Section \ref{ss:redundancy}.
\end{itemize}
    Two variations of L2C2 are also tested;
\begin{itemize}
    \item \textbf{L2C2-NWL}. It is an L2C2 \emph{without} the intra-frame wear-leveling mechanism. The Index Calculation circuit has less complexity;  see Section \ref{subsubsec:blockrearrangement}. Writing always starts at the least significant live byte of the frame.
    
    \item \textbf{L2C2-BF}. It is an L2C2 with  LRU-Best-Fit replacement policy instead of LRU-Fit.
\end{itemize}

Table \ref{table:systemscost} shows the number of storage bits per frame of the tag and data arrays, along with the percentage increments with respect to FD.

\begin{table}[htb]
\caption{Frame costs in bits. Percentage overhead relative to FD.}
\begin{center}
{
\begin{tabular}{|c|c|c|c|c|} \cline{2-5}
\multicolumn{1}{c|}{} & \multicolumn{2}{c|}{SRAM Tag Array} & \multicolumn{2}{c|}{STT-RAM Data Array} \\\cline{2-5}
\multicolumn{1}{c|}{} & Bits & Overhead, \% & Bits & Overhead, \% \\\hline
FD & 34 & - & 529 & - \\\hline
FD+6 & 34 & 0 & 595 & 12.5\\\hline
L2C2 & 38 & 11.8 & 594 & 12.3\\\hline
L2C2+6 & 38 & 11.8 & 648 & 22.5\\\hline
\end{tabular}
}
\label{table:systemscost}
\end{center}
\end{table}

\subsection{Lifetime}

\begin{figure*}[ht]
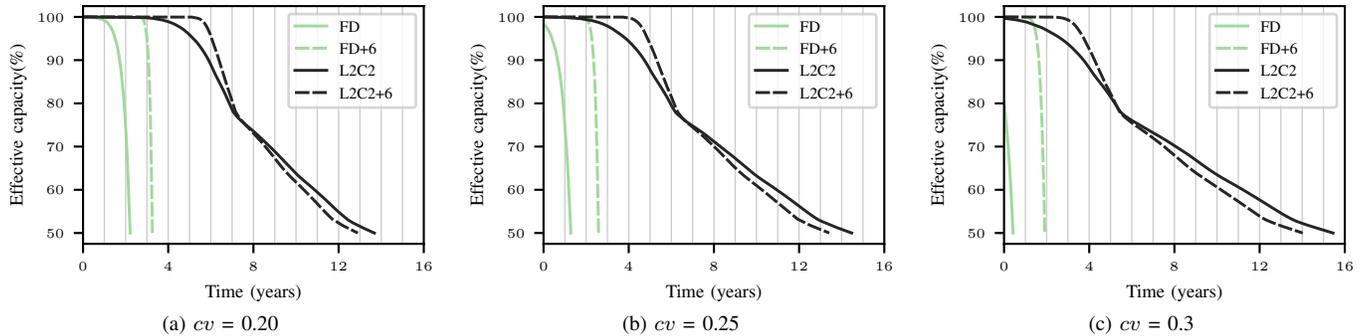

    \centering
    \subfloat[$cv$ = 0.20]{\label{subfig:lifetime_020}
        \input{Figuras/lifetime_cov020.pgf}
    }
    \hfill
    \subfloat[$cv$ = 0.25]{\label{subfig:lifetime_025}
        \input{Figuras/lifetime_cov025.pgf}
    }
    \hfill
    \subfloat[$cv$ = 0.3]{\label{subfig:lifetime_030}
        \input{Figuras/lifetime_cov030.pgf}
    }
    \caption{Effective capacity evolution over time until 50\% of capacity is lost.
    }
    \label{fig:lifetime_ev}
\end{figure*}

Figure \ref{fig:lifetime_ev} shows forecasts of capacity degradation, from start-up until 50\% of effective capacity is lost, considering bitcells with increasing manufacturing variabilities for the four NV-LLC candidates.
The \textit{effective} capacity shown on the Y-axis is the one contributing to cache block storage. For example, L2C2+6 has 100\% effective capacity as long as its nominal 16MB capacity is available, regardless of whether or not the spare bytes are coming into play.
Besides, Table~\ref{table:t50c_I50C_I5y_cv}a shows $T_{50C}$, the time required to lose 50\% of the nominal cache capacity.

\begin{figure*}[ht]
    \centering
    \subfloat[$cv$ = 0.2]
    {\label{subfig:ipc_020}
        \input{Figuras/ipc_cov020.pgf}
    }
    \hfill
    \subfloat[$cv$ = 0.25]
    {\label{subfig:ipc_025}
        \input{Figuras/ipc_cov025.pgf}
    }
    \hfill
    \subfloat[$cv$ = 0.3]
    {\label{subfig:ipc_030}
        \input{Figuras/ipc_cov030.pgf}
    }
    \caption{Normalized IPC evolution over time until 50\% of capacity is lost.
    }
    \label{fig:ipc_ev}
\end{figure*}
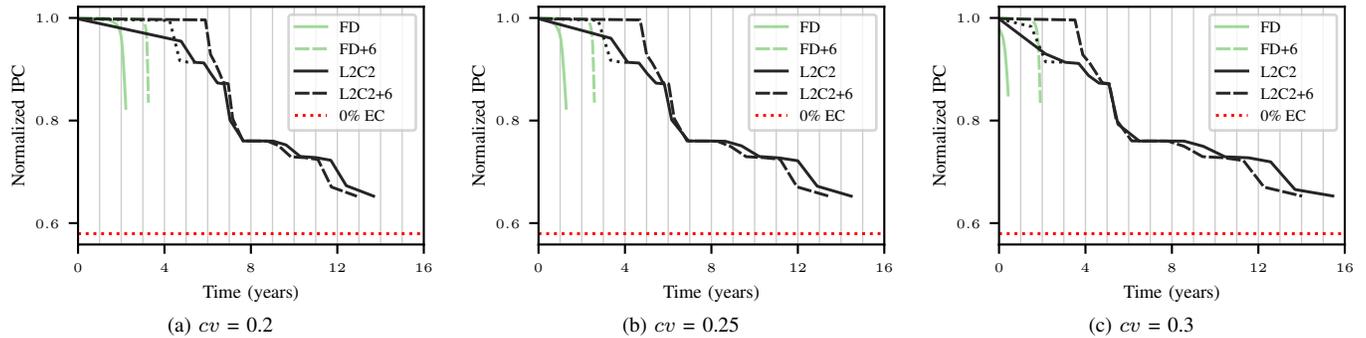

 


First of all, it can be seen that FD manufactured with high variability starts with an effective capacity that may be well below the nominal one; i.e., an FD with $cv=0.3$ starts operating with less than 80\% of nominal capacity because many frames come out of production with defective bitcells; see Figure~\ref{fig:lifetime_ev}c. FD+6, in contrast, completely solves this problem by adding redundancy. 
On the other hand, $T_{50C}$ decreases markedly for FD and FD+6 as the manufacturing variability increases, while for L2C2 and L2C2+6 it is the other way around; see Table~\ref{table:t50c_I50C_I5y_cv}a. This is due to the byte-level disabling capability of L2C2, which tolerates early byte failures and takes advantage of the later ones.

Second, compared to the sharp drop observed in frame-disabling caches, the byte-disabling ones show a much more progressive degradation of capacity, resulting in a longer $T_{50C}$. L2C2 is the longest lived cache, in terms of $T_{50C}$ from 13.7 to 15.4 years, and FD the least, from 2.2 to 0.42 years, depending on $cv$. L2C2+6 lasts a little less than L2C2, but it is the one that maintains the nominal capacity for the longest time, namely $T_{99C}$, between 5.6 and 3.1 years, depending on $cv$; see Table~\ref{table:t50c_I50C_I5y_cv}b.
As an example, in terms of $T_{50C}$, see Table~\ref{table:t50c_I50C_I5y_cv}a, L2C2 is alive 6, 11 and 37 times longer than FD for $cv$ values of 0.2, 0.25 and 0.3, respectively.

Third, as time goes by, and contrary to expectations, the effective capacity of L2C2+6 is no longer greater than that of L2C2, with the curves intersecting at around 7.5-5.5 years, depending on $cv$. As will be seen in the next subsection the explanation is as follows: before the curves cross, the L2C2+6 system maintains a higher IPC, which implies a higher write rate and a consequent earlier degradation.

\begin{table}[htb]
\centering
\caption{$T_{50C}$, $T_{99C}$, and $T_{99P}$ in years; $I_{50C|5y}$ in instructions.}
\label{table:t50c_I50C_I5y_cv}
\begin{tabular}{cccclccc}
\cline{2-4} \cline{6-8}
\multicolumn{1}{c|}{} & \multicolumn{3}{c|}{$T_{50C}$, years.} & \multicolumn{1}{l|}{} & \multicolumn{3}{c|}{$T_{99C}$, years.} 

\\ \cline{1-4} \cline{6-8}
\multicolumn{1}{|c|}{\textit{cv}} & \multicolumn{1}{c|}{\textit{0.2}} & \multicolumn{1}{c|}{\textit{0.25}} & \multicolumn{1}{c|}{\textit{0.3}} & \multicolumn{1}{l|}{} & \multicolumn{1}{c|}{\textit{0.2}} & \multicolumn{1}{c|}{\textit{0.25}} & \multicolumn{1}{c|}{\textit{0.3}} \\ 

\cline{1-4} \cline{6-8}
\multicolumn{1}{|c|}{FD} & 
\multicolumn{1}{c|}{2.2} & 
\multicolumn{1}{c|}{1.3} & 
\multicolumn{1}{c|}{0.42} & 

\multicolumn{1}{l|}{} &
\multicolumn{1}{c|}{1.1} &
\multicolumn{1}{c|}{-} &
\multicolumn{1}{c|}{-} \\

\cline{1-4} \cline{6-8}
\multicolumn{1}{|c|}{FD+6} &
\multicolumn{1}{c|}{3.3} &
\multicolumn{1}{c|}{2.6} &
\multicolumn{1}{c|}{1.9} & 

\multicolumn{1}{l|}{} &
\multicolumn{1}{c|}{2.9} &
\multicolumn{1}{c|}{2.1} &
\multicolumn{1}{c|}{1.3} \\

\cline{1-4} \cline{6-8}
\multicolumn{1}{|c|}{L2C2} &
\multicolumn{1}{c|}{13.7} &
\multicolumn{1}{c|}{14.5} &
\multicolumn{1}{c|}{15.4} &

\multicolumn{1}{l|}{} &
\multicolumn{1}{c|}{3.9} &
\multicolumn{1}{c|}{2.4} &
\multicolumn{1}{c|}{0.9} \\

\cline{1-4} \cline{6-8}
\multicolumn{1}{|c|}{L2C2+6} &
\multicolumn{1}{c|}{12.9} &
\multicolumn{1}{c|}{13.4} &
\multicolumn{1}{c|}{14.0} &

\multicolumn{1}{l|}{} &
\multicolumn{1}{c|}{5.6} &
\multicolumn{1}{c|}{4.4} &
\multicolumn{1}{c|}{3.1} \\

\cline{1-4} \cline{6-8}
 & \multicolumn{3}{c}{(a)} & & \multicolumn{3}{c}{(b)} \\
 
\multicolumn{1}{l}{} & \multicolumn{1}{l}{} & \multicolumn{1}{l}{} & \multicolumn{1}{l}{} & & \multicolumn{1}{l}{} & \multicolumn{1}{l}{} & \multicolumn{1}{l}{} \\ \cline{2-4} \cline{6-8} 
\multicolumn{1}{l|}{} & \multicolumn{3}{c|}{$T_{99P}$, years.} & \multicolumn{1}{l|}{} & \multicolumn{3}{c|}{$I_{50C|5y}$, instr. $\times10^{18}$} \\ \cline{1-4} \cline{6-8} 
\multicolumn{1}{|c|}{\textit{cv}} & \multicolumn{1}{c|}{\textit{0.2}} & \multicolumn{1}{c|}{\textit{0.25}} & \multicolumn{1}{c|}{\textit{0.3}} & \multicolumn{1}{l|}{} & \multicolumn{1}{c|}{\textit{0.2}} & \multicolumn{1}{c|}{\textit{0.25}} & \multicolumn{1}{c|}{\textit{0.3}} \\ \cline{1-4} \cline{6-8} 
\multicolumn{1}{|c|}{FD} &
\multicolumn{1}{c|}{1.7} &
\multicolumn{1}{c|}{0.65} &
\multicolumn{1}{c|}{-} & 

\multicolumn{1}{l|}{} &
\multicolumn{1}{c|}{0.89} &
\multicolumn{1}{c|}{0.51} &
\multicolumn{1}{c|}{0.16} \\

\cline{1-4} \cline{6-8} 
\multicolumn{1}{|c|}{FD+6}  &
\multicolumn{1}{c|}{3.1} &
\multicolumn{1}{c|}{2.4} &
\multicolumn{1}{c|}{1.6} &

\multicolumn{1}{l|}{} &
\multicolumn{1}{c|}{1.32} &
\multicolumn{1}{c|}{1.05} &
\multicolumn{1}{c|}{0.77} \\

\cline{1-4} \cline{6-8} 
\multicolumn{1}{|c|}{L2C2} &
\multicolumn{1}{c|}{4.3} & 
\multicolumn{1}{c|}{2.8} &
\multicolumn{1}{c|}{0.82} & 

\multicolumn{1}{l|}{} &
\multicolumn{1}{c|}{2.01} &
\multicolumn{1}{c|}{1.96} &
\multicolumn{1}{c|}{1.90} \\ 

\cline{1-4} \cline{6-8} 
\multicolumn{1}{|c|}{L2C2+6} &
\multicolumn{1}{c|}{5.9} &
\multicolumn{1}{c|}{4.7} &
\multicolumn{1}{c|}{3.5} &

\multicolumn{1}{l|}{} &
\multicolumn{1}{c|}{2.03} &
\multicolumn{1}{c|}{2.03} &
\multicolumn{1}{c|}{1.98} \\ 

\cline{1-4} \cline{6-8} 
\multicolumn{1}{l}{} & \multicolumn{3}{c}{(c)} & & \multicolumn{3}{c}{(d)} 
\end{tabular}
\end{table}

\subsection{Performance}

Figure \ref{fig:ipc_ev} shows the IPC forecast over time from start-up until 50\% of effective capacity is lost for the NV-LLC candidates. The IPCs have been normalized to the IPC of a system with an NV-LLC with all bitcells operational. The bottom dotted red line (0\%  EC) represents the IPC of a system with a fully impaired NV-LLC, i.e. with zero effective capacity. 

Before going into the analysis, notice the forecast procedure only provides IPC values after the simulation phase of each epoch, corresponding to the health state computed by the previous epoch. Intermediate IPC values within epochs are obtained by linear interpolation.
However, should it be necessary to have a more precise IPC within an epoch, it is sufficient to halve the epoch extension once or several times.
For instance, we observe that the first L2C2 epochs are long in forecasted time \textit{and} produce a significant drop in IPC.
To obtain more detail on the IPC loss during that period, the extension of the first two epochs has been halved, resulting in the new IPC values depicted in the dashed black lines in Figure \ref{fig:ipc_ev}.

Four observations can be highlighted from the curves.

First, after losing 50\% of capacity, the IPC with frame-disabling caches is around 20\% higher than that of byte-disabling.
This is because having 50\% effective capacity with FD or FD+6 implies that 50\% of frames can store any block, whereas with L2C2 and L2C2+6, it implies that the capacity of all frames has been reduced and therefore some blocks cannot be stored in any frame.

Second, consistent with the effective capacity forecasts, the IPC degrades later and more gradually in L2C2 and L2C2+6.
The steps seen in their lines correspond to periods in which the possibility of storing blocks of a given compression encoding has been lost.

Third, the crossings in the IPC and capacity curves occur at the same times. After these crossings, L2C2 performs slightly better and lasts slightly longer than L2C2+6.  The reason is to be found in the first 4-6 years of operation of L2C2+6 at maximum performance, years that, compared to L2C2, cause a higher write wear.


And fourth, in the first years of operation L2C2+6 keeps the maximum performance, L2C2 loses it progressively, and FD and FD+6 loses it abruptly. The index $T_{99P}$, the time during which performance holds above 99\% of the maximum allows to quantify these facts; see Table~\ref{table:t50c_I50C_I5y_cv}c. L2C2+6 excels at $T_{99P}$ for all $cv$ values, with L2C2 in second place, except for $cv=0.3$, where FD+6 is better.

From the above analysis, L2C2+6 seems to be the best candidate, followed by L2C2, and at some distance FD+6. 

To get more insight, we propose to measure the work performed by the different organizations using the aggregate number of instructions executed by the four cores, with an utilization of 100\%, until a certain wear-out condition is reached. We calculate this value with the integral of the CPI curve. Since according to Belkhir et al. the average lifetime of a server is three to five years~\cite{Belkhir-18-Assessing}, we propose the index $I_{50C|5y}$ which measures the number of instructions executed until 50\% of the capacity is exhausted or until five years have elapsed, whichever is earlier; see Table~\ref{table:t50c_I50C_I5y_cv}d.

Regarding this index, we can say that the increase in manufacturing variability is very bad for frame disabling, with reductions of 82 and 42\% of $I_{50C|5y}$ in FD and FD+6, going from $cv$ 0.2 to 0.3. In contrast, that same increase in $cv$ slightly reduces $I_{50C|5y}$ in L2C2 and L2C2+6 by 5.5 and 2.5\%, respectively. 

In short, L2C2+6 offers the best performance in all indexes, with an additional storage cost over L2C2 and FD+6 of less than 10\%. The second option, cheaper but with less performance, is L2C2, which requires about 12.3\% more data array storage than FD, the base option without redundancy.


\subsection{Intra-frame wear-leveling impact on lifetime}
\label{ss:eval_wl}

In this experiment we aim to see the importance of the intra-frame wear-leveling mechanism. 
In an L2C2 without intra-frame wear-leveling
there is an imbalance between the number of writes that receive the low order bytes and the high order bytes of a frame. Concretely, higher order bytes will not be written if the block compressed to some extent. This imbalance will make lower order bytes receive more write operations than higher order ones so they will become faulty before than in a cache with intra-frame wear-leveling.

To model the L2C2 without wear-leveling, L2C2-NWL, the write rate is not averaged across the bytes of a frame. Thus, the rate used to age a byte depends not only on the compression class of the frame it belongs to, $CC$, and the health state of the set, $\bar{A}$, but also on the position the byte occupies among the live bytes in the frame.

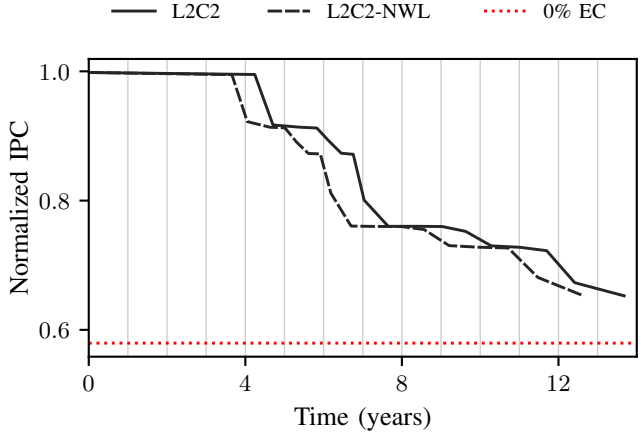
\begin{figure}[ht]
    \centering
    \input{Figuras/ipc_nwl.pgf}
    \caption{IPC evolution until losing 50\% of  capacity of an L2C2 without intra-frame wear-leveling mechanism, L2C2-NWL, for cv = 0.2.}
    \label{fig:nwl}
\end{figure}

Figure \ref{fig:nwl} shows the IPC evolution until the NV-LLC loses 50\% of its effective capacity for $cv=0.2$. IPC of L2C2-NWL starts dropping at 3.7 years while L2C2 IPC drops at 4.3 years (16\% later). This temporal shift linking points of equal performance is evident throughout the duration studied, being around one year on many occasions.

\subsection{Fit vs. Best-Fit replacement}
\label{ss:fit_bestfit}

\begin{figure}[ht]
    \centering
    \input{Figuras/ipc_best_fit.pgf}
    \caption{IPC evolution until losing 50\% of  capacity of an L2C2 with LRU-Best-Fit replacement policy, L2C2-BF, for $cv$ = 0.2.}
    \label{fig:bf}
\end{figure}
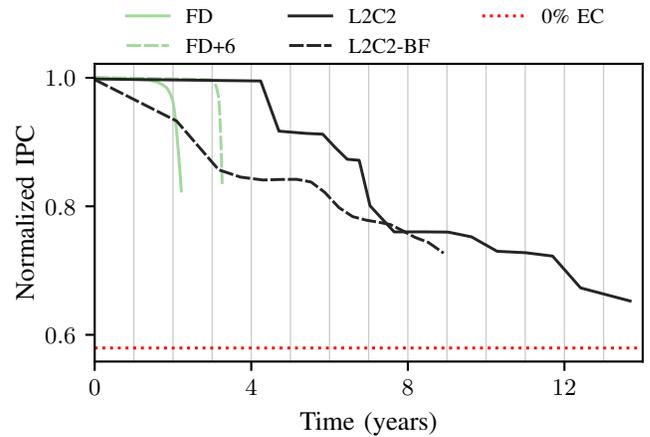

In L2C2 an alternative replacement policy to LRU-Fit is LRU-Best-Fit, which consists of choosing the smallest LRU frame capable of holding the incoming compressed block; see L2C2-BF in Figure~\ref{fig:bf}.  In principle, LRU-Best-Fit could be advantageous since it would preserve frames with larger capacity from writes, allowing in the long term the hosting of blocks with low compression capacity; see Section~\ref{sss:content_management}.  However, L2C2-BF takes 8.9 years to lose 50\% of its capacity, while L2C2 reaches the same loss at 13.7 years, i.e. 54\% longer. Besides, the IPC drop L2C2-BF experiences at the early stages (0-2 years) is even more pronounced than that of FD. The explanation for both effects is that when the first frame in a set experiences the first byte failure, all the compressible blocks  addressed to this set, 78\% of the total, will be allocated to this recently degraded frame; see Figure~\ref{fig:BDICoverage} in page 6. This incurring in substantial conflict misses that degrade performance. 

\subsection{Sensitivity analysis}

\begin{figure*}[ht]
    \centering
    \subfloat[32MB caches.]{\label{subfig:ipc_32MB}
        \input{Figuras/ipc_32MB.pgf}
    }
    \hfill
    \subfloat[8 cores.]{\label{subfig:ipc_8c}
        \input{Figuras/ipc_8c.pgf}
    }
    \hfill
    \subfloat[Memory-intensive programs.]{\label{subfig:ipc_mi}
        \input{Figuras/ipc_MI.pgf}
    }
    \caption{IPC evolution until losing 50\% of  capacity of FD and L2C2 for $cv = 0.2$, doubling cache size (a), doubling the number of cores while keeping the same 4MB/core (b), and considering the most memory-intensive programs (c).}
    \label{fig:ipc_sa}
\end{figure*}
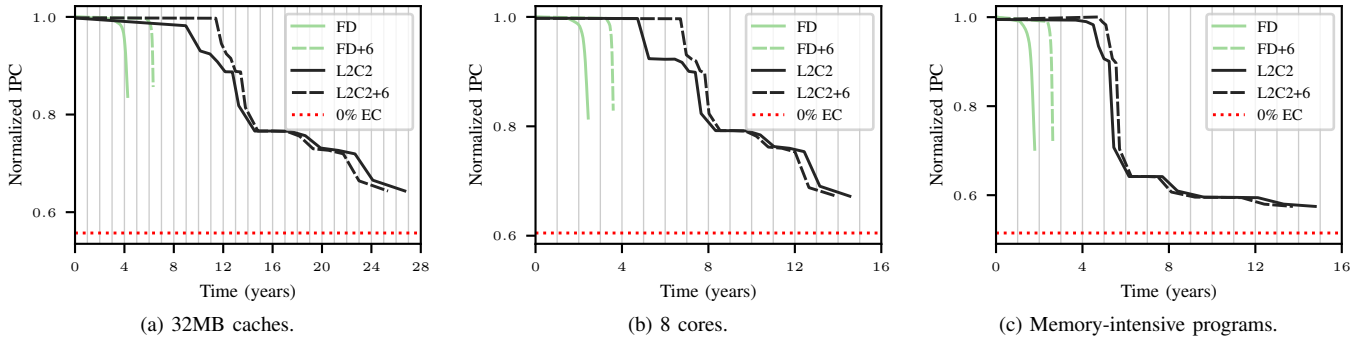

To further add generality to the results presented so far, we elaborate on three aspects; see Figure \ref{fig:ipc_sa}.
First, the LLC bank size is increased from 4 to 8 MB per bank.
Second, the system is scaled by a factor of 2, going from 4 to 8 cores, from 4 to 8 banks of NV-LLC and from 1 to 2 main memory controllers.
And third, the workload mixes are changed, including only the top ten memory intensive applications; see applications with superscript in the Table~\ref{table:mixes}.

Doubling cache capacity with the same number of cores extends performance over time to a similar amount across all cache organizations. For example, for L2C2+6, $T_{50C}$ goes from 12.9 to 25.3 years when increasing size from 16 to 32 MB; see Figure~\ref{subfig:ipc_020} vs. Figure~\ref{subfig:ipc_32MB}.

By scaling the system, simultaneously doubling number of cores, cache size and memory bandwidth, the performance-time curves for all cache organizations maintain their shape; see Figure~\ref{subfig:ipc_020} vs. Figure~\ref{subfig:ipc_8c}. This is an expected conclusion, which reinforces the possibility of incorporating L2C2-type caches in future generations of on-chip multiprocessors.

When considering more memory intensive applications a first observation is that the performance at full capacity exhaustion is lower, which indicates, not surprisingly, a higher dependence of performance on the quality of the memory hierarchy; see the red baselines (0\% EC) in Figures~\ref{subfig:ipc_mi} and \ref{subfig:ipc_020}. In addition, the performance drop is sharper and occurs earlier. For example, for L2C2+6 the first drop is one year earlier and the relative IPC drops from 0.76 to 0.64. Again it can be reasoned that applications that exhibit intensive LLC usage are more sensitive to capacity loss, so overall system performance is more affected.

In summary, this sensitivity analysis shows that both the results and the forecast procedure itself are consistent when varying two significant dimensions, capacity and workload.

\subsection{Technological projections of lifetime and performance of NV-LLCs}
\label{technological-projection-of-the-lifetime}

As we have explained so far, the forward-looking behavior of an NV-LLC can be estimated by applying a forecast procedure that has three key elements, namely, a statistical model of bitcell write endurance, a detailed simulation model of the NV-LLC organization, and a workload. 
In principle, a new forecast with a change in any of these three elements requires a feasible, but high computation time. 

All the results so far have been obtained for baseline bitcells with given endurances modelled with mean $\mu=10^{11}$ and ${cv}=0.2-0.3$.
To obtain results concerning other bitcell endurances and/or NV-LLC latencies, of course the whole forecast procedure can be repeated, creating new $RW maps$ and changing the latencies in the simulation model.

However, as long as the NV-LLC latencies are assumed constant, it is possible to take advantage of the properties of the linear transformation of Gaussian distributions to reuse the forecast data and obtain projections for other NVM technologies with a different bitcell write endurance values.

Specifically, if an NV-LLC is built with an improved technology, which offers the same cache latencies, but uses bitcells with $k$ times more endurance ($\mu_{i}=k \cdot\mu_{b},\ \sigma_{i}=k\cdot\sigma_{b}$), new capacity and IPC indices as a function of time can be calculated as follows:

\begin{itemize}
    \item Cap. improved bitcells ($t$) = Cap. baseline bitcells ($\frac{t}{k}$)
    \item IPC improved bitcells ($t$) = IPC baseline bitcells ($\frac{t}{k}$)
\end{itemize}

That is, new indexes with improved bitcells at time ($t$) can be obtained from the forecast made with baseline bitcells at an earlier time ($\frac{t}{k}$); see Appendix A.

Thus, from a few reference forecasts, many technology projections can be obtained. Table~\ref{table:lifetime} is an example that focuses on two arbitrary, but interesting, indices: $T_{90C}$ and $T_{90P}$, calculated from the central column forecasts for $\mu=10^{11}$ and $cv=0.2-0.3$.  $T_{90C}$ and $T_{90P}$ are the elapsed times to reduce the rated capacity and performance, measured in IPC, to 90\% of the initial values, respectively.
Note that the values of $T_{90C}$ and $T_{90P}$ scale linearly with the value of $\mu$. That is, the value of $T_{90C}$ for $\mu=10^{12}$ is equal to 10 times the value of $T_{90C}$ for $\mu=10^{11}$.
As it can be seen, $T_{90C}$ always trails $T_{90P}$ and L2C2+6 is the best cache organization. 

These types of indices can serve as a basis for signing a Service Level Agreement, SLA, with prospective customers.  It is plausible to think that a manufacturer can have a portfolio of NVM qualities and technologies and that a customer can choose the product with the best performance/cost ratio for her/his needs. 
For example, for a smartphone projected for a daily usage of 6 hours at 100\% and with an average product life of 1.8 years~\cite{Belkhir-18-Assessing} several cache organizations and manufacturing variabilities of those shown in the  $\mu=10^{11}$ writes/bitcell columns may fit. These figures could be representative of a technology with moderate write endurance, but comparatively inexpensive.




\begin{table}[htb]
\caption{$T_{90C}$ and $T_{90P}$ for FD+6, L2C2 and L2C2+6, varying $cv$ and $\mu$; $m$= months, $y$= years.}
\begin{center}
{
\footnotesize
\begin{tabular}{|c|c|r|r|r|r|r|r|} \cline{3-8}
\multicolumn{1}{c}{} & \multicolumn{1}{c|}{} & \multicolumn{6}{c|}{Mean number of writes to fail, $\mu$.}\\\hline
\multirow{2}{*}{$cv$} & \multirow{2}{*}{Cache} & \multicolumn{2}{c|}{$10^{10}$} & \multicolumn{2}{c|}{$10^{11}$} & \multicolumn{2}{c|}{$10^{12}$}\\\cline{3-8}
 & & \multicolumn{1}{c|}{$T_{90C}$} & \multicolumn{1}{c|}{$T_{90P}$} & \multicolumn{1}{c|}{$T_{90C}$} & \multicolumn{1}{c|}{$T_{90P}$} & \multicolumn{1}{c|}{$T_{90C}$} & \multicolumn{1}{c|}{$T_{90P}$}\\\hline
\multirow{3}{*}{0.2} & FD+6 & 3.7m & 3.9m & 3.1y & 3.2y & 30,7y & 32.2y\\\cline{2-8}
 & L2C2 & 7.1m & 7.2m & 5.9y & 6.0y & 58,9y & 60.1y\\\cline{2-8}
 & L2C2+6 & 7.7m & 7.8m & 6.4y & 6.5y & 63.8y & 64.7y\\\hline
\multirow{3}{*}{0.25} & FD+6 & 2.8m & 3.1m & 2.4y & 2.5y & 23.5y & 25.4y\\\cline{2-8}
 & L2C2 & 5.7m & 5.8m & 4.7y & 4.9y & 47.3y & 48.7y\\\cline{2-8}
 & L2C2+6 & 6.4m & 6.5m & 5.3y & 5.4y & 53.1y & 54.2y\\\hline
\multirow{3}{*}{0.3} & FD+6 & 2.0m & 2.2m & 1.6y & 1.9y & 16.4y & 18.6y\\\cline{2-8}
 & L2C2 & 4.5m & 4.7m & 3.7y & 3.9y & 37.3y & 39.1y\\\cline{2-8}
 & L2C2+6 & 5.1m & 5.3m & 4.3y & 4.4y & 42.5y & 43.8y\\\hline
\multicolumn{2}{c|}{} & \multicolumn{2}{c|}{Projections} & \multicolumn{2}{c|}{Forecast} & \multicolumn{2}{c|}{Projections} \\\cline{3-8}
\end{tabular}
}
\label{table:lifetime}
\end{center}
\end{table}

%% file: Figuras/lifetime_cov020.pgf
\begingroup%
\makeatletter%
\begin{pgfpicture}%
\pgfpathrectangle{\pgfpointorigin}{\pgfqpoint{2.222509in}{1.591882in}}%
\pgfusepath{use as bounding box, clip}%
\begin{pgfscope}%
\pgfsetbuttcap%
\pgfsetmiterjoin%
\definecolor{currentfill}{rgb}{1.000000,1.000000,1.000000}%
\pgfsetfillcolor{currentfill}%
\pgfsetlinewidth{0.000000pt}%
\definecolor{currentstroke}{rgb}{1.000000,1.000000,1.000000}%
\pgfsetstrokecolor{currentstroke}%
\pgfsetdash{}{0pt}%
\pgfpathmoveto{\pgfqpoint{0.000000in}{0.000000in}}%
\pgfpathlineto{\pgfqpoint{2.222509in}{0.000000in}}%
\pgfpathlineto{\pgfqpoint{2.222509in}{1.591882in}}%
\pgfpathlineto{\pgfqpoint{0.000000in}{1.591882in}}%
\pgfpathclose%
\pgfusepath{fill}%
\end{pgfscope}%
\begin{pgfscope}%
\pgfsetbuttcap%
\pgfsetmiterjoin%
\definecolor{currentfill}{rgb}{1.000000,1.000000,1.000000}%
\pgfsetfillcolor{currentfill}%
\pgfsetlinewidth{0.000000pt}%
\definecolor{currentstroke}{rgb}{0.000000,0.000000,0.000000}%
\pgfsetstrokecolor{currentstroke}%
\pgfsetstrokeopacity{0.000000}%
\pgfsetdash{}{0pt}%
\pgfpathmoveto{\pgfqpoint{0.391785in}{0.311728in}}%
\pgfpathlineto{\pgfqpoint{2.175248in}{0.311728in}}%
\pgfpathlineto{\pgfqpoint{2.175248in}{1.556882in}}%
\pgfpathlineto{\pgfqpoint{0.391785in}{1.556882in}}%
\pgfpathclose%
\pgfusepath{fill}%
\end{pgfscope}%
\begin{pgfscope}%
\pgfsetbuttcap%
\pgfsetroundjoin%
\definecolor{currentfill}{rgb}{0.000000,0.000000,0.000000}%
\pgfsetfillcolor{currentfill}%
\pgfsetlinewidth{0.803000pt}%
\definecolor{currentstroke}{rgb}{0.000000,0.000000,0.000000}%
\pgfsetstrokecolor{currentstroke}%
\pgfsetdash{}{0pt}%
\pgfsys@defobject{currentmarker}{\pgfqpoint{0.000000in}{-0.048611in}}{\pgfqpoint{0.000000in}{0.000000in}}{%
\pgfpathmoveto{\pgfqpoint{0.000000in}{0.000000in}}%
\pgfpathlineto{\pgfqpoint{0.000000in}{-0.048611in}}%
\pgfusepath{stroke,fill}%
}%
\begin{pgfscope}%
\pgfsys@transformshift{0.391785in}{0.311728in}%
\pgfsys@useobject{currentmarker}{}%
\end{pgfscope}%
\end{pgfscope}%
\begin{pgfscope}%
\pgftext[x=0.391785in,y=0.214506in,,top]{\rmfamily\fontsize{5.000000}{6.000000}\selectfont \(\displaystyle 0\)}%
\end{pgfscope}%
\begin{pgfscope}%
\pgfsetbuttcap%
\pgfsetroundjoin%
\definecolor{currentfill}{rgb}{0.000000,0.000000,0.000000}%
\pgfsetfillcolor{currentfill}%
\pgfsetlinewidth{0.803000pt}%
\definecolor{currentstroke}{rgb}{0.000000,0.000000,0.000000}%
\pgfsetstrokecolor{currentstroke}%
\pgfsetdash{}{0pt}%
\pgfsys@defobject{currentmarker}{\pgfqpoint{0.000000in}{-0.048611in}}{\pgfqpoint{0.000000in}{0.000000in}}{%
\pgfpathmoveto{\pgfqpoint{0.000000in}{0.000000in}}%
\pgfpathlineto{\pgfqpoint{0.000000in}{-0.048611in}}%
\pgfusepath{stroke,fill}%
}%
\begin{pgfscope}%
\pgfsys@transformshift{0.837650in}{0.311728in}%
\pgfsys@useobject{currentmarker}{}%
\end{pgfscope}%
\end{pgfscope}%
\begin{pgfscope}%
\pgftext[x=0.837650in,y=0.214506in,,top]{\rmfamily\fontsize{5.000000}{6.000000}\selectfont \(\displaystyle 4\)}%
\end{pgfscope}%
\begin{pgfscope}%
\pgfsetbuttcap%
\pgfsetroundjoin%
\definecolor{currentfill}{rgb}{0.000000,0.000000,0.000000}%
\pgfsetfillcolor{currentfill}%
\pgfsetlinewidth{0.803000pt}%
\definecolor{currentstroke}{rgb}{0.000000,0.000000,0.000000}%
\pgfsetstrokecolor{currentstroke}%
\pgfsetdash{}{0pt}%
\pgfsys@defobject{currentmarker}{\pgfqpoint{0.000000in}{-0.048611in}}{\pgfqpoint{0.000000in}{0.000000in}}{%
\pgfpathmoveto{\pgfqpoint{0.000000in}{0.000000in}}%
\pgfpathlineto{\pgfqpoint{0.000000in}{-0.048611in}}%
\pgfusepath{stroke,fill}%
}%
\begin{pgfscope}%
\pgfsys@transformshift{1.283516in}{0.311728in}%
\pgfsys@useobject{currentmarker}{}%
\end{pgfscope}%
\end{pgfscope}%
\begin{pgfscope}%
\pgftext[x=1.283516in,y=0.214506in,,top]{\rmfamily\fontsize{5.000000}{6.000000}\selectfont \(\displaystyle 8\)}%
\end{pgfscope}%
\begin{pgfscope}%
\pgfsetbuttcap%
\pgfsetroundjoin%
\definecolor{currentfill}{rgb}{0.000000,0.000000,0.000000}%
\pgfsetfillcolor{currentfill}%
\pgfsetlinewidth{0.803000pt}%
\definecolor{currentstroke}{rgb}{0.000000,0.000000,0.000000}%
\pgfsetstrokecolor{currentstroke}%
\pgfsetdash{}{0pt}%
\pgfsys@defobject{currentmarker}{\pgfqpoint{0.000000in}{-0.048611in}}{\pgfqpoint{0.000000in}{0.000000in}}{%
\pgfpathmoveto{\pgfqpoint{0.000000in}{0.000000in}}%
\pgfpathlineto{\pgfqpoint{0.000000in}{-0.048611in}}%
\pgfusepath{stroke,fill}%
}%
\begin{pgfscope}%
\pgfsys@transformshift{1.729382in}{0.311728in}%
\pgfsys@useobject{currentmarker}{}%
\end{pgfscope}%
\end{pgfscope}%
\begin{pgfscope}%
\pgftext[x=1.729382in,y=0.214506in,,top]{\rmfamily\fontsize{5.000000}{6.000000}\selectfont \(\displaystyle 12\)}%
\end{pgfscope}%
\begin{pgfscope}%
\pgfsetbuttcap%
\pgfsetroundjoin%
\definecolor{currentfill}{rgb}{0.000000,0.000000,0.000000}%
\pgfsetfillcolor{currentfill}%
\pgfsetlinewidth{0.803000pt}%
\definecolor{currentstroke}{rgb}{0.000000,0.000000,0.000000}%
\pgfsetstrokecolor{currentstroke}%
\pgfsetdash{}{0pt}%
\pgfsys@defobject{currentmarker}{\pgfqpoint{0.000000in}{-0.048611in}}{\pgfqpoint{0.000000in}{0.000000in}}{%
\pgfpathmoveto{\pgfqpoint{0.000000in}{0.000000in}}%
\pgfpathlineto{\pgfqpoint{0.000000in}{-0.048611in}}%
\pgfusepath{stroke,fill}%
}%
\begin{pgfscope}%
\pgfsys@transformshift{2.175248in}{0.311728in}%
\pgfsys@useobject{currentmarker}{}%
\end{pgfscope}%
\end{pgfscope}%
\begin{pgfscope}%
\pgftext[x=2.175248in,y=0.214506in,,top]{\rmfamily\fontsize{5.000000}{6.000000}\selectfont \(\displaystyle 16\)}%
\end{pgfscope}%
\begin{pgfscope}%
\pgftext[x=1.283516in,y=0.097222in,,top]{\rmfamily\fontsize{7.000000}{8.400000}\selectfont Time (years)}%
\end{pgfscope}%
\begin{pgfscope}%
\pgfsetbuttcap%
\pgfsetroundjoin%
\definecolor{currentfill}{rgb}{0.000000,0.000000,0.000000}%
\pgfsetfillcolor{currentfill}%
\pgfsetlinewidth{0.803000pt}%
\definecolor{currentstroke}{rgb}{0.000000,0.000000,0.000000}%
\pgfsetstrokecolor{currentstroke}%
\pgfsetdash{}{0pt}%
\pgfsys@defobject{currentmarker}{\pgfqpoint{-0.048611in}{0.000000in}}{\pgfqpoint{0.000000in}{0.000000in}}{%
\pgfpathmoveto{\pgfqpoint{0.000000in}{0.000000in}}%
\pgfpathlineto{\pgfqpoint{-0.048611in}{0.000000in}}%
\pgfusepath{stroke,fill}%
}%
\begin{pgfscope}%
\pgfsys@transformshift{0.391785in}{0.368326in}%
\pgfsys@useobject{currentmarker}{}%
\end{pgfscope}%
\end{pgfscope}%
\begin{pgfscope}%
\pgftext[x=0.200039in,y=0.344213in,left,base]{\rmfamily\fontsize{5.000000}{6.000000}\selectfont \(\displaystyle 50\)}%
\end{pgfscope}%
\begin{pgfscope}%
\pgfsetbuttcap%
\pgfsetroundjoin%
\definecolor{currentfill}{rgb}{0.000000,0.000000,0.000000}%
\pgfsetfillcolor{currentfill}%
\pgfsetlinewidth{0.803000pt}%
\definecolor{currentstroke}{rgb}{0.000000,0.000000,0.000000}%
\pgfsetstrokecolor{currentstroke}%
\pgfsetdash{}{0pt}%
\pgfsys@defobject{currentmarker}{\pgfqpoint{-0.048611in}{0.000000in}}{\pgfqpoint{0.000000in}{0.000000in}}{%
\pgfpathmoveto{\pgfqpoint{0.000000in}{0.000000in}}%
\pgfpathlineto{\pgfqpoint{-0.048611in}{0.000000in}}%
\pgfusepath{stroke,fill}%
}%
\begin{pgfscope}%
\pgfsys@transformshift{0.391785in}{0.594717in}%
\pgfsys@useobject{currentmarker}{}%
\end{pgfscope}%
\end{pgfscope}%
\begin{pgfscope}%
\pgftext[x=0.200039in,y=0.570605in,left,base]{\rmfamily\fontsize{5.000000}{6.000000}\selectfont \(\displaystyle 60\)}%
\end{pgfscope}%
\begin{pgfscope}%
\pgfsetbuttcap%
\pgfsetroundjoin%
\definecolor{currentfill}{rgb}{0.000000,0.000000,0.000000}%
\pgfsetfillcolor{currentfill}%
\pgfsetlinewidth{0.803000pt}%
\definecolor{currentstroke}{rgb}{0.000000,0.000000,0.000000}%
\pgfsetstrokecolor{currentstroke}%
\pgfsetdash{}{0pt}%
\pgfsys@defobject{currentmarker}{\pgfqpoint{-0.048611in}{0.000000in}}{\pgfqpoint{0.000000in}{0.000000in}}{%
\pgfpathmoveto{\pgfqpoint{0.000000in}{0.000000in}}%
\pgfpathlineto{\pgfqpoint{-0.048611in}{0.000000in}}%
\pgfusepath{stroke,fill}%
}%
\begin{pgfscope}%
\pgfsys@transformshift{0.391785in}{0.821109in}%
\pgfsys@useobject{currentmarker}{}%
\end{pgfscope}%
\end{pgfscope}%
\begin{pgfscope}%
\pgftext[x=0.200039in,y=0.796997in,left,base]{\rmfamily\fontsize{5.000000}{6.000000}\selectfont \(\displaystyle 70\)}%
\end{pgfscope}%
\begin{pgfscope}%
\pgfsetbuttcap%
\pgfsetroundjoin%
\definecolor{currentfill}{rgb}{0.000000,0.000000,0.000000}%
\pgfsetfillcolor{currentfill}%
\pgfsetlinewidth{0.803000pt}%
\definecolor{currentstroke}{rgb}{0.000000,0.000000,0.000000}%
\pgfsetstrokecolor{currentstroke}%
\pgfsetdash{}{0pt}%
\pgfsys@defobject{currentmarker}{\pgfqpoint{-0.048611in}{0.000000in}}{\pgfqpoint{0.000000in}{0.000000in}}{%
\pgfpathmoveto{\pgfqpoint{0.000000in}{0.000000in}}%
\pgfpathlineto{\pgfqpoint{-0.048611in}{0.000000in}}%
\pgfusepath{stroke,fill}%
}%
\begin{pgfscope}%
\pgfsys@transformshift{0.391785in}{1.047501in}%
\pgfsys@useobject{currentmarker}{}%
\end{pgfscope}%
\end{pgfscope}%
\begin{pgfscope}%
\pgftext[x=0.200039in,y=1.023388in,left,base]{\rmfamily\fontsize{5.000000}{6.000000}\selectfont \(\displaystyle 80\)}%
\end{pgfscope}%
\begin{pgfscope}%
\pgfsetbuttcap%
\pgfsetroundjoin%
\definecolor{currentfill}{rgb}{0.000000,0.000000,0.000000}%
\pgfsetfillcolor{currentfill}%
\pgfsetlinewidth{0.803000pt}%
\definecolor{currentstroke}{rgb}{0.000000,0.000000,0.000000}%
\pgfsetstrokecolor{currentstroke}%
\pgfsetdash{}{0pt}%
\pgfsys@defobject{currentmarker}{\pgfqpoint{-0.048611in}{0.000000in}}{\pgfqpoint{0.000000in}{0.000000in}}{%
\pgfpathmoveto{\pgfqpoint{0.000000in}{0.000000in}}%
\pgfpathlineto{\pgfqpoint{-0.048611in}{0.000000in}}%
\pgfusepath{stroke,fill}%
}%
\begin{pgfscope}%
\pgfsys@transformshift{0.391785in}{1.273892in}%
\pgfsys@useobject{currentmarker}{}%
\end{pgfscope}%
\end{pgfscope}%
\begin{pgfscope}%
\pgftext[x=0.200039in,y=1.249780in,left,base]{\rmfamily\fontsize{5.000000}{6.000000}\selectfont \(\displaystyle 90\)}%
\end{pgfscope}%
\begin{pgfscope}%
\pgfsetbuttcap%
\pgfsetroundjoin%
\definecolor{currentfill}{rgb}{0.000000,0.000000,0.000000}%
\pgfsetfillcolor{currentfill}%
\pgfsetlinewidth{0.803000pt}%
\definecolor{currentstroke}{rgb}{0.000000,0.000000,0.000000}%
\pgfsetstrokecolor{currentstroke}%
\pgfsetdash{}{0pt}%
\pgfsys@defobject{currentmarker}{\pgfqpoint{-0.048611in}{0.000000in}}{\pgfqpoint{0.000000in}{0.000000in}}{%
\pgfpathmoveto{\pgfqpoint{0.000000in}{0.000000in}}%
\pgfpathlineto{\pgfqpoint{-0.048611in}{0.000000in}}%
\pgfusepath{stroke,fill}%
}%
\begin{pgfscope}%
\pgfsys@transformshift{0.391785in}{1.500284in}%
\pgfsys@useobject{currentmarker}{}%
\end{pgfscope}%
\end{pgfscope}%
\begin{pgfscope}%
\pgftext[x=0.152778in,y=1.476171in,left,base]{\rmfamily\fontsize{5.000000}{6.000000}\selectfont \(\displaystyle 100\)}%
\end{pgfscope}%
\begin{pgfscope}%
\pgftext[x=0.097222in,y=0.934305in,,bottom,rotate=90.000000]{\rmfamily\fontsize{7.000000}{8.400000}\selectfont Effective capacity(\%)}%
\end{pgfscope}%
\begin{pgfscope}%
\pgfpathrectangle{\pgfqpoint{0.391785in}{0.311728in}}{\pgfqpoint{1.783463in}{1.245154in}}%
\pgfusepath{clip}%
\pgfsetrectcap%
\pgfsetroundjoin%
\pgfsetlinewidth{0.501875pt}%
\definecolor{currentstroke}{rgb}{0.800000,0.800000,0.800000}%
\pgfsetstrokecolor{currentstroke}%
\pgfsetdash{}{0pt}%
\pgfpathmoveto{\pgfqpoint{0.503251in}{0.311728in}}%
\pgfpathlineto{\pgfqpoint{0.503251in}{1.556882in}}%
\pgfusepath{stroke}%
\end{pgfscope}%
\begin{pgfscope}%
\pgfpathrectangle{\pgfqpoint{0.391785in}{0.311728in}}{\pgfqpoint{1.783463in}{1.245154in}}%
\pgfusepath{clip}%
\pgfsetrectcap%
\pgfsetroundjoin%
\pgfsetlinewidth{0.501875pt}%
\definecolor{currentstroke}{rgb}{0.800000,0.800000,0.800000}%
\pgfsetstrokecolor{currentstroke}%
\pgfsetdash{}{0pt}%
\pgfpathmoveto{\pgfqpoint{0.614718in}{0.311728in}}%
\pgfpathlineto{\pgfqpoint{0.614718in}{1.556882in}}%
\pgfusepath{stroke}%
\end{pgfscope}%
\begin{pgfscope}%
\pgfpathrectangle{\pgfqpoint{0.391785in}{0.311728in}}{\pgfqpoint{1.783463in}{1.245154in}}%
\pgfusepath{clip}%
\pgfsetrectcap%
\pgfsetroundjoin%
\pgfsetlinewidth{0.501875pt}%
\definecolor{currentstroke}{rgb}{0.800000,0.800000,0.800000}%
\pgfsetstrokecolor{currentstroke}%
\pgfsetdash{}{0pt}%
\pgfpathmoveto{\pgfqpoint{0.726184in}{0.311728in}}%
\pgfpathlineto{\pgfqpoint{0.726184in}{1.556882in}}%
\pgfusepath{stroke}%
\end{pgfscope}%
\begin{pgfscope}%
\pgfpathrectangle{\pgfqpoint{0.391785in}{0.311728in}}{\pgfqpoint{1.783463in}{1.245154in}}%
\pgfusepath{clip}%
\pgfsetrectcap%
\pgfsetroundjoin%
\pgfsetlinewidth{0.501875pt}%
\definecolor{currentstroke}{rgb}{0.800000,0.800000,0.800000}%
\pgfsetstrokecolor{currentstroke}%
\pgfsetdash{}{0pt}%
\pgfpathmoveto{\pgfqpoint{0.837650in}{0.311728in}}%
\pgfpathlineto{\pgfqpoint{0.837650in}{1.556882in}}%
\pgfusepath{stroke}%
\end{pgfscope}%
\begin{pgfscope}%
\pgfpathrectangle{\pgfqpoint{0.391785in}{0.311728in}}{\pgfqpoint{1.783463in}{1.245154in}}%
\pgfusepath{clip}%
\pgfsetrectcap%
\pgfsetroundjoin%
\pgfsetlinewidth{0.501875pt}%
\definecolor{currentstroke}{rgb}{0.800000,0.800000,0.800000}%
\pgfsetstrokecolor{currentstroke}%
\pgfsetdash{}{0pt}%
\pgfpathmoveto{\pgfqpoint{0.949117in}{0.311728in}}%
\pgfpathlineto{\pgfqpoint{0.949117in}{1.556882in}}%
\pgfusepath{stroke}%
\end{pgfscope}%
\begin{pgfscope}%
\pgfpathrectangle{\pgfqpoint{0.391785in}{0.311728in}}{\pgfqpoint{1.783463in}{1.245154in}}%
\pgfusepath{clip}%
\pgfsetrectcap%
\pgfsetroundjoin%
\pgfsetlinewidth{0.501875pt}%
\definecolor{currentstroke}{rgb}{0.800000,0.800000,0.800000}%
\pgfsetstrokecolor{currentstroke}%
\pgfsetdash{}{0pt}%
\pgfpathmoveto{\pgfqpoint{1.060583in}{0.311728in}}%
\pgfpathlineto{\pgfqpoint{1.060583in}{1.556882in}}%
\pgfusepath{stroke}%
\end{pgfscope}%
\begin{pgfscope}%
\pgfpathrectangle{\pgfqpoint{0.391785in}{0.311728in}}{\pgfqpoint{1.783463in}{1.245154in}}%
\pgfusepath{clip}%
\pgfsetrectcap%
\pgfsetroundjoin%
\pgfsetlinewidth{0.501875pt}%
\definecolor{currentstroke}{rgb}{0.800000,0.800000,0.800000}%
\pgfsetstrokecolor{currentstroke}%
\pgfsetdash{}{0pt}%
\pgfpathmoveto{\pgfqpoint{1.172050in}{0.311728in}}%
\pgfpathlineto{\pgfqpoint{1.172050in}{1.556882in}}%
\pgfusepath{stroke}%
\end{pgfscope}%
\begin{pgfscope}%
\pgfpathrectangle{\pgfqpoint{0.391785in}{0.311728in}}{\pgfqpoint{1.783463in}{1.245154in}}%
\pgfusepath{clip}%
\pgfsetrectcap%
\pgfsetroundjoin%
\pgfsetlinewidth{0.501875pt}%
\definecolor{currentstroke}{rgb}{0.800000,0.800000,0.800000}%
\pgfsetstrokecolor{currentstroke}%
\pgfsetdash{}{0pt}%
\pgfpathmoveto{\pgfqpoint{1.283516in}{0.311728in}}%
\pgfpathlineto{\pgfqpoint{1.283516in}{1.556882in}}%
\pgfusepath{stroke}%
\end{pgfscope}%
\begin{pgfscope}%
\pgfpathrectangle{\pgfqpoint{0.391785in}{0.311728in}}{\pgfqpoint{1.783463in}{1.245154in}}%
\pgfusepath{clip}%
\pgfsetrectcap%
\pgfsetroundjoin%
\pgfsetlinewidth{0.501875pt}%
\definecolor{currentstroke}{rgb}{0.800000,0.800000,0.800000}%
\pgfsetstrokecolor{currentstroke}%
\pgfsetdash{}{0pt}%
\pgfpathmoveto{\pgfqpoint{1.394983in}{0.311728in}}%
\pgfpathlineto{\pgfqpoint{1.394983in}{1.556882in}}%
\pgfusepath{stroke}%
\end{pgfscope}%
\begin{pgfscope}%
\pgfpathrectangle{\pgfqpoint{0.391785in}{0.311728in}}{\pgfqpoint{1.783463in}{1.245154in}}%
\pgfusepath{clip}%
\pgfsetrectcap%
\pgfsetroundjoin%
\pgfsetlinewidth{0.501875pt}%
\definecolor{currentstroke}{rgb}{0.800000,0.800000,0.800000}%
\pgfsetstrokecolor{currentstroke}%
\pgfsetdash{}{0pt}%
\pgfpathmoveto{\pgfqpoint{1.506449in}{0.311728in}}%
\pgfpathlineto{\pgfqpoint{1.506449in}{1.556882in}}%
\pgfusepath{stroke}%
\end{pgfscope}%
\begin{pgfscope}%
\pgfpathrectangle{\pgfqpoint{0.391785in}{0.311728in}}{\pgfqpoint{1.783463in}{1.245154in}}%
\pgfusepath{clip}%
\pgfsetrectcap%
\pgfsetroundjoin%
\pgfsetlinewidth{0.501875pt}%
\definecolor{currentstroke}{rgb}{0.800000,0.800000,0.800000}%
\pgfsetstrokecolor{currentstroke}%
\pgfsetdash{}{0pt}%
\pgfpathmoveto{\pgfqpoint{1.617916in}{0.311728in}}%
\pgfpathlineto{\pgfqpoint{1.617916in}{1.556882in}}%
\pgfusepath{stroke}%
\end{pgfscope}%
\begin{pgfscope}%
\pgfpathrectangle{\pgfqpoint{0.391785in}{0.311728in}}{\pgfqpoint{1.783463in}{1.245154in}}%
\pgfusepath{clip}%
\pgfsetrectcap%
\pgfsetroundjoin%
\pgfsetlinewidth{0.501875pt}%
\definecolor{currentstroke}{rgb}{0.800000,0.800000,0.800000}%
\pgfsetstrokecolor{currentstroke}%
\pgfsetdash{}{0pt}%
\pgfpathmoveto{\pgfqpoint{1.729382in}{0.311728in}}%
\pgfpathlineto{\pgfqpoint{1.729382in}{1.556882in}}%
\pgfusepath{stroke}%
\end{pgfscope}%
\begin{pgfscope}%
\pgfpathrectangle{\pgfqpoint{0.391785in}{0.311728in}}{\pgfqpoint{1.783463in}{1.245154in}}%
\pgfusepath{clip}%
\pgfsetrectcap%
\pgfsetroundjoin%
\pgfsetlinewidth{0.501875pt}%
\definecolor{currentstroke}{rgb}{0.800000,0.800000,0.800000}%
\pgfsetstrokecolor{currentstroke}%
\pgfsetdash{}{0pt}%
\pgfpathmoveto{\pgfqpoint{1.840848in}{0.311728in}}%
\pgfpathlineto{\pgfqpoint{1.840848in}{1.556882in}}%
\pgfusepath{stroke}%
\end{pgfscope}%
\begin{pgfscope}%
\pgfpathrectangle{\pgfqpoint{0.391785in}{0.311728in}}{\pgfqpoint{1.783463in}{1.245154in}}%
\pgfusepath{clip}%
\pgfsetrectcap%
\pgfsetroundjoin%
\pgfsetlinewidth{0.501875pt}%
\definecolor{currentstroke}{rgb}{0.800000,0.800000,0.800000}%
\pgfsetstrokecolor{currentstroke}%
\pgfsetdash{}{0pt}%
\pgfpathmoveto{\pgfqpoint{1.952315in}{0.311728in}}%
\pgfpathlineto{\pgfqpoint{1.952315in}{1.556882in}}%
\pgfusepath{stroke}%
\end{pgfscope}%
\begin{pgfscope}%
\pgfpathrectangle{\pgfqpoint{0.391785in}{0.311728in}}{\pgfqpoint{1.783463in}{1.245154in}}%
\pgfusepath{clip}%
\pgfsetrectcap%
\pgfsetroundjoin%
\pgfsetlinewidth{0.501875pt}%
\definecolor{currentstroke}{rgb}{0.800000,0.800000,0.800000}%
\pgfsetstrokecolor{currentstroke}%
\pgfsetdash{}{0pt}%
\pgfpathmoveto{\pgfqpoint{2.063781in}{0.311728in}}%
\pgfpathlineto{\pgfqpoint{2.063781in}{1.556882in}}%
\pgfusepath{stroke}%
\end{pgfscope}%
\begin{pgfscope}%
\pgfpathrectangle{\pgfqpoint{0.391785in}{0.311728in}}{\pgfqpoint{1.783463in}{1.245154in}}%
\pgfusepath{clip}%
\pgfsetrectcap%
\pgfsetroundjoin%
\pgfsetlinewidth{1.104125pt}%
\definecolor{currentstroke}{rgb}{0.631373,0.850980,0.607843}%
\pgfsetstrokecolor{currentstroke}%
\pgfsetdash{}{0pt}%
\pgfpathmoveto{\pgfqpoint{0.391785in}{1.499921in}}%
\pgfpathlineto{\pgfqpoint{0.393291in}{1.499895in}}%
\pgfpathlineto{\pgfqpoint{0.432606in}{1.498626in}}%
\pgfpathlineto{\pgfqpoint{0.455104in}{1.496717in}}%
\pgfpathlineto{\pgfqpoint{0.494127in}{1.486613in}}%
\pgfpathlineto{\pgfqpoint{0.495330in}{1.486147in}}%
\pgfpathlineto{\pgfqpoint{0.498131in}{1.484696in}}%
\pgfpathlineto{\pgfqpoint{0.505038in}{1.480792in}}%
\pgfpathlineto{\pgfqpoint{0.506629in}{1.479808in}}%
\pgfpathlineto{\pgfqpoint{0.510421in}{1.477225in}}%
\pgfpathlineto{\pgfqpoint{0.511652in}{1.476284in}}%
\pgfpathlineto{\pgfqpoint{0.515776in}{1.472804in}}%
\pgfpathlineto{\pgfqpoint{0.518310in}{1.470412in}}%
\pgfpathlineto{\pgfqpoint{0.526387in}{1.461784in}}%
\pgfpathlineto{\pgfqpoint{0.528293in}{1.459331in}}%
\pgfpathlineto{\pgfqpoint{0.529438in}{1.457690in}}%
\pgfpathlineto{\pgfqpoint{0.530611in}{1.456015in}}%
\pgfpathlineto{\pgfqpoint{0.532501in}{1.453174in}}%
\pgfpathlineto{\pgfqpoint{0.535597in}{1.448182in}}%
\pgfpathlineto{\pgfqpoint{0.541218in}{1.438060in}}%
\pgfpathlineto{\pgfqpoint{0.543676in}{1.433242in}}%
\pgfpathlineto{\pgfqpoint{0.544562in}{1.431514in}}%
\pgfpathlineto{\pgfqpoint{0.545702in}{1.428863in}}%
\pgfpathlineto{\pgfqpoint{0.547864in}{1.423837in}}%
\pgfpathlineto{\pgfqpoint{0.548486in}{1.422230in}}%
\pgfpathlineto{\pgfqpoint{0.550151in}{1.417705in}}%
\pgfpathlineto{\pgfqpoint{0.551528in}{1.413879in}}%
\pgfpathlineto{\pgfqpoint{0.554002in}{1.406746in}}%
\pgfpathlineto{\pgfqpoint{0.554619in}{1.404880in}}%
\pgfpathlineto{\pgfqpoint{0.567685in}{1.359817in}}%
\pgfpathlineto{\pgfqpoint{0.569066in}{1.353910in}}%
\pgfpathlineto{\pgfqpoint{0.573076in}{1.334824in}}%
\pgfpathlineto{\pgfqpoint{0.574904in}{1.325005in}}%
\pgfpathlineto{\pgfqpoint{0.575259in}{1.323122in}}%
\pgfpathlineto{\pgfqpoint{0.582734in}{1.279293in}}%
\pgfpathlineto{\pgfqpoint{0.592319in}{1.205117in}}%
\pgfpathlineto{\pgfqpoint{0.604624in}{1.075290in}}%
\pgfpathlineto{\pgfqpoint{0.605811in}{1.059745in}}%
\pgfpathlineto{\pgfqpoint{0.606412in}{1.051929in}}%
\pgfpathlineto{\pgfqpoint{0.608318in}{1.026012in}}%
\pgfpathlineto{\pgfqpoint{0.609135in}{1.014682in}}%
\pgfpathlineto{\pgfqpoint{0.610347in}{0.996882in}}%
\pgfpathlineto{\pgfqpoint{0.615697in}{0.910823in}}%
\pgfpathlineto{\pgfqpoint{0.615976in}{0.905840in}}%
\pgfpathlineto{\pgfqpoint{0.617455in}{0.878861in}}%
\pgfpathlineto{\pgfqpoint{0.618335in}{0.861735in}}%
\pgfpathlineto{\pgfqpoint{0.619703in}{0.835205in}}%
\pgfpathlineto{\pgfqpoint{0.621118in}{0.806447in}}%
\pgfpathlineto{\pgfqpoint{0.621513in}{0.797759in}}%
\pgfpathlineto{\pgfqpoint{0.638517in}{0.368334in}}%
\pgfpathlineto{\pgfqpoint{0.638517in}{0.368334in}}%
\pgfusepath{stroke}%
\end{pgfscope}%
\begin{pgfscope}%
\pgfpathrectangle{\pgfqpoint{0.391785in}{0.311728in}}{\pgfqpoint{1.783463in}{1.245154in}}%
\pgfusepath{clip}%
\pgfsetbuttcap%
\pgfsetroundjoin%
\pgfsetlinewidth{1.104125pt}%
\definecolor{currentstroke}{rgb}{0.631373,0.850980,0.607843}%
\pgfsetstrokecolor{currentstroke}%
\pgfsetdash{{5.500000pt}{1.100000pt}}{0.000000pt}%
\pgfpathmoveto{\pgfqpoint{0.391785in}{1.500284in}}%
\pgfpathlineto{\pgfqpoint{0.687398in}{1.499170in}}%
\pgfpathlineto{\pgfqpoint{0.695605in}{1.496890in}}%
\pgfpathlineto{\pgfqpoint{0.699020in}{1.495059in}}%
\pgfpathlineto{\pgfqpoint{0.700728in}{1.493582in}}%
\pgfpathlineto{\pgfqpoint{0.702054in}{1.492365in}}%
\pgfpathlineto{\pgfqpoint{0.704333in}{1.489782in}}%
\pgfpathlineto{\pgfqpoint{0.706615in}{1.486302in}}%
\pgfpathlineto{\pgfqpoint{0.707938in}{1.483970in}}%
\pgfpathlineto{\pgfqpoint{0.710315in}{1.479238in}}%
\pgfpathlineto{\pgfqpoint{0.712704in}{1.472700in}}%
\pgfpathlineto{\pgfqpoint{0.713203in}{1.471111in}}%
\pgfpathlineto{\pgfqpoint{0.714717in}{1.465549in}}%
\pgfpathlineto{\pgfqpoint{0.715890in}{1.461093in}}%
\pgfpathlineto{\pgfqpoint{0.716421in}{1.458874in}}%
\pgfpathlineto{\pgfqpoint{0.725062in}{1.403559in}}%
\pgfpathlineto{\pgfqpoint{0.726184in}{1.392315in}}%
\pgfpathlineto{\pgfqpoint{0.726825in}{1.385786in}}%
\pgfpathlineto{\pgfqpoint{0.728023in}{1.371338in}}%
\pgfpathlineto{\pgfqpoint{0.728568in}{1.364645in}}%
\pgfpathlineto{\pgfqpoint{0.736745in}{1.220498in}}%
\pgfpathlineto{\pgfqpoint{0.738042in}{1.187586in}}%
\pgfpathlineto{\pgfqpoint{0.738527in}{1.174735in}}%
\pgfpathlineto{\pgfqpoint{0.739900in}{1.135286in}}%
\pgfpathlineto{\pgfqpoint{0.740546in}{1.114861in}}%
\pgfpathlineto{\pgfqpoint{0.743025in}{1.029449in}}%
\pgfpathlineto{\pgfqpoint{0.743315in}{1.018248in}}%
\pgfpathlineto{\pgfqpoint{0.744231in}{0.981812in}}%
\pgfpathlineto{\pgfqpoint{0.745375in}{0.931567in}}%
\pgfpathlineto{\pgfqpoint{0.753124in}{0.487859in}}%
\pgfpathlineto{\pgfqpoint{0.754843in}{0.368334in}}%
\pgfpathlineto{\pgfqpoint{0.754843in}{0.368334in}}%
\pgfusepath{stroke}%
\end{pgfscope}%
\begin{pgfscope}%
\pgfpathrectangle{\pgfqpoint{0.391785in}{0.311728in}}{\pgfqpoint{1.783463in}{1.245154in}}%
\pgfusepath{clip}%
\pgfsetrectcap%
\pgfsetroundjoin%
\pgfsetlinewidth{1.104125pt}%
\definecolor{currentstroke}{rgb}{0.145098,0.145098,0.145098}%
\pgfsetstrokecolor{currentstroke}%
\pgfsetdash{}{0pt}%
\pgfpathmoveto{\pgfqpoint{0.391785in}{1.500278in}}%
\pgfpathlineto{\pgfqpoint{0.706523in}{1.496918in}}%
\pgfpathlineto{\pgfqpoint{0.712609in}{1.496543in}}%
\pgfpathlineto{\pgfqpoint{0.740421in}{1.494345in}}%
\pgfpathlineto{\pgfqpoint{0.749621in}{1.493405in}}%
\pgfpathlineto{\pgfqpoint{0.761779in}{1.491968in}}%
\pgfpathlineto{\pgfqpoint{0.808501in}{1.483603in}}%
\pgfpathlineto{\pgfqpoint{0.811117in}{1.482969in}}%
\pgfpathlineto{\pgfqpoint{0.833589in}{1.476652in}}%
\pgfpathlineto{\pgfqpoint{0.835820in}{1.475926in}}%
\pgfpathlineto{\pgfqpoint{0.840626in}{1.474320in}}%
\pgfpathlineto{\pgfqpoint{0.861331in}{1.466340in}}%
\pgfpathlineto{\pgfqpoint{0.862478in}{1.465826in}}%
\pgfpathlineto{\pgfqpoint{0.867156in}{1.463703in}}%
\pgfpathlineto{\pgfqpoint{0.868724in}{1.462972in}}%
\pgfpathlineto{\pgfqpoint{0.873939in}{1.460505in}}%
\pgfpathlineto{\pgfqpoint{0.875463in}{1.459753in}}%
\pgfpathlineto{\pgfqpoint{0.899397in}{1.446377in}}%
\pgfpathlineto{\pgfqpoint{0.900700in}{1.445554in}}%
\pgfpathlineto{\pgfqpoint{0.903328in}{1.443904in}}%
\pgfpathlineto{\pgfqpoint{0.904516in}{1.443120in}}%
\pgfpathlineto{\pgfqpoint{0.909117in}{1.440042in}}%
\pgfpathlineto{\pgfqpoint{0.911360in}{1.438504in}}%
\pgfpathlineto{\pgfqpoint{0.912898in}{1.437401in}}%
\pgfpathlineto{\pgfqpoint{0.917538in}{1.434041in}}%
\pgfpathlineto{\pgfqpoint{0.918813in}{1.433092in}}%
\pgfpathlineto{\pgfqpoint{0.921197in}{1.431258in}}%
\pgfpathlineto{\pgfqpoint{0.923262in}{1.429661in}}%
\pgfpathlineto{\pgfqpoint{0.929017in}{1.424664in}}%
\pgfpathlineto{\pgfqpoint{0.933017in}{1.421256in}}%
\pgfpathlineto{\pgfqpoint{0.936499in}{1.418274in}}%
\pgfpathlineto{\pgfqpoint{0.938151in}{1.416857in}}%
\pgfpathlineto{\pgfqpoint{0.940734in}{1.414642in}}%
\pgfpathlineto{\pgfqpoint{0.942015in}{1.413544in}}%
\pgfpathlineto{\pgfqpoint{0.943577in}{1.412198in}}%
\pgfpathlineto{\pgfqpoint{0.985844in}{1.369121in}}%
\pgfpathlineto{\pgfqpoint{0.987140in}{1.367545in}}%
\pgfpathlineto{\pgfqpoint{0.991564in}{1.361934in}}%
\pgfpathlineto{\pgfqpoint{0.993123in}{1.359890in}}%
\pgfpathlineto{\pgfqpoint{0.994805in}{1.357710in}}%
\pgfpathlineto{\pgfqpoint{0.997304in}{1.354509in}}%
\pgfpathlineto{\pgfqpoint{0.999426in}{1.351749in}}%
\pgfpathlineto{\pgfqpoint{1.015154in}{1.329851in}}%
\pgfpathlineto{\pgfqpoint{1.020951in}{1.321135in}}%
\pgfpathlineto{\pgfqpoint{1.021959in}{1.319569in}}%
\pgfpathlineto{\pgfqpoint{1.026549in}{1.312325in}}%
\pgfpathlineto{\pgfqpoint{1.027535in}{1.310750in}}%
\pgfpathlineto{\pgfqpoint{1.030452in}{1.305990in}}%
\pgfpathlineto{\pgfqpoint{1.032185in}{1.303169in}}%
\pgfpathlineto{\pgfqpoint{1.033745in}{1.300602in}}%
\pgfpathlineto{\pgfqpoint{1.034736in}{1.298931in}}%
\pgfpathlineto{\pgfqpoint{1.036758in}{1.295523in}}%
\pgfpathlineto{\pgfqpoint{1.037803in}{1.293744in}}%
\pgfpathlineto{\pgfqpoint{1.038977in}{1.291780in}}%
\pgfpathlineto{\pgfqpoint{1.043570in}{1.283769in}}%
\pgfpathlineto{\pgfqpoint{1.044654in}{1.281765in}}%
\pgfpathlineto{\pgfqpoint{1.049075in}{1.273398in}}%
\pgfpathlineto{\pgfqpoint{1.050698in}{1.270296in}}%
\pgfpathlineto{\pgfqpoint{1.060168in}{1.251204in}}%
\pgfpathlineto{\pgfqpoint{1.066070in}{1.238544in}}%
\pgfpathlineto{\pgfqpoint{1.067036in}{1.236436in}}%
\pgfpathlineto{\pgfqpoint{1.073372in}{1.222317in}}%
\pgfpathlineto{\pgfqpoint{1.074671in}{1.219378in}}%
\pgfpathlineto{\pgfqpoint{1.077027in}{1.214252in}}%
\pgfpathlineto{\pgfqpoint{1.078400in}{1.211290in}}%
\pgfpathlineto{\pgfqpoint{1.080587in}{1.206678in}}%
\pgfpathlineto{\pgfqpoint{1.083334in}{1.200949in}}%
\pgfpathlineto{\pgfqpoint{1.084509in}{1.198490in}}%
\pgfpathlineto{\pgfqpoint{1.086591in}{1.194253in}}%
\pgfpathlineto{\pgfqpoint{1.087447in}{1.192504in}}%
\pgfpathlineto{\pgfqpoint{1.118755in}{1.131145in}}%
\pgfpathlineto{\pgfqpoint{1.120165in}{1.128350in}}%
\pgfpathlineto{\pgfqpoint{1.136473in}{1.094475in}}%
\pgfpathlineto{\pgfqpoint{1.137590in}{1.092068in}}%
\pgfpathlineto{\pgfqpoint{1.177229in}{1.003274in}}%
\pgfpathlineto{\pgfqpoint{1.179771in}{0.999882in}}%
\pgfpathlineto{\pgfqpoint{1.180892in}{0.998397in}}%
\pgfpathlineto{\pgfqpoint{1.183504in}{0.995055in}}%
\pgfpathlineto{\pgfqpoint{1.188357in}{0.989019in}}%
\pgfpathlineto{\pgfqpoint{1.189400in}{0.987745in}}%
\pgfpathlineto{\pgfqpoint{1.201219in}{0.974289in}}%
\pgfpathlineto{\pgfqpoint{1.202454in}{0.972993in}}%
\pgfpathlineto{\pgfqpoint{1.204642in}{0.970725in}}%
\pgfpathlineto{\pgfqpoint{1.206723in}{0.968575in}}%
\pgfpathlineto{\pgfqpoint{1.210916in}{0.964398in}}%
\pgfpathlineto{\pgfqpoint{1.212255in}{0.963098in}}%
\pgfpathlineto{\pgfqpoint{1.214309in}{0.961147in}}%
\pgfpathlineto{\pgfqpoint{1.217450in}{0.958171in}}%
\pgfpathlineto{\pgfqpoint{1.218627in}{0.957085in}}%
\pgfpathlineto{\pgfqpoint{1.220527in}{0.955326in}}%
\pgfpathlineto{\pgfqpoint{1.223235in}{0.952911in}}%
\pgfpathlineto{\pgfqpoint{1.225269in}{0.951104in}}%
\pgfpathlineto{\pgfqpoint{1.227160in}{0.949415in}}%
\pgfpathlineto{\pgfqpoint{1.229563in}{0.947296in}}%
\pgfpathlineto{\pgfqpoint{1.230811in}{0.946225in}}%
\pgfpathlineto{\pgfqpoint{1.247095in}{0.932249in}}%
\pgfpathlineto{\pgfqpoint{1.249051in}{0.930614in}}%
\pgfpathlineto{\pgfqpoint{1.254558in}{0.926053in}}%
\pgfpathlineto{\pgfqpoint{1.257119in}{0.923920in}}%
\pgfpathlineto{\pgfqpoint{1.259889in}{0.921640in}}%
\pgfpathlineto{\pgfqpoint{1.276546in}{0.907514in}}%
\pgfpathlineto{\pgfqpoint{1.277928in}{0.906322in}}%
\pgfpathlineto{\pgfqpoint{1.279777in}{0.904728in}}%
\pgfpathlineto{\pgfqpoint{1.281295in}{0.903401in}}%
\pgfpathlineto{\pgfqpoint{1.283338in}{0.901642in}}%
\pgfpathlineto{\pgfqpoint{1.284866in}{0.900349in}}%
\pgfpathlineto{\pgfqpoint{1.303124in}{0.883933in}}%
\pgfpathlineto{\pgfqpoint{1.304274in}{0.882880in}}%
\pgfpathlineto{\pgfqpoint{1.312435in}{0.875275in}}%
\pgfpathlineto{\pgfqpoint{1.314453in}{0.873364in}}%
\pgfpathlineto{\pgfqpoint{1.316136in}{0.871761in}}%
\pgfpathlineto{\pgfqpoint{1.317688in}{0.870247in}}%
\pgfpathlineto{\pgfqpoint{1.321115in}{0.866926in}}%
\pgfpathlineto{\pgfqpoint{1.323505in}{0.864583in}}%
\pgfpathlineto{\pgfqpoint{1.325480in}{0.862706in}}%
\pgfpathlineto{\pgfqpoint{1.326831in}{0.861483in}}%
\pgfpathlineto{\pgfqpoint{1.328910in}{0.859577in}}%
\pgfpathlineto{\pgfqpoint{1.332836in}{0.855933in}}%
\pgfpathlineto{\pgfqpoint{1.336049in}{0.852942in}}%
\pgfpathlineto{\pgfqpoint{1.337184in}{0.851867in}}%
\pgfpathlineto{\pgfqpoint{1.340569in}{0.848674in}}%
\pgfpathlineto{\pgfqpoint{1.342060in}{0.847277in}}%
\pgfpathlineto{\pgfqpoint{1.343450in}{0.845955in}}%
\pgfpathlineto{\pgfqpoint{1.346186in}{0.843360in}}%
\pgfpathlineto{\pgfqpoint{1.347675in}{0.841940in}}%
\pgfpathlineto{\pgfqpoint{1.350337in}{0.839431in}}%
\pgfpathlineto{\pgfqpoint{1.352684in}{0.837211in}}%
\pgfpathlineto{\pgfqpoint{1.354360in}{0.835601in}}%
\pgfpathlineto{\pgfqpoint{1.361680in}{0.828579in}}%
\pgfpathlineto{\pgfqpoint{1.363957in}{0.826408in}}%
\pgfpathlineto{\pgfqpoint{1.365329in}{0.825079in}}%
\pgfpathlineto{\pgfqpoint{1.367854in}{0.822661in}}%
\pgfpathlineto{\pgfqpoint{1.369091in}{0.821479in}}%
\pgfpathlineto{\pgfqpoint{1.370555in}{0.820035in}}%
\pgfpathlineto{\pgfqpoint{1.373696in}{0.816969in}}%
\pgfpathlineto{\pgfqpoint{1.375706in}{0.815005in}}%
\pgfpathlineto{\pgfqpoint{1.377265in}{0.813481in}}%
\pgfpathlineto{\pgfqpoint{1.378641in}{0.812122in}}%
\pgfpathlineto{\pgfqpoint{1.380041in}{0.810749in}}%
\pgfpathlineto{\pgfqpoint{1.390810in}{0.800141in}}%
\pgfpathlineto{\pgfqpoint{1.396192in}{0.794839in}}%
\pgfpathlineto{\pgfqpoint{1.398032in}{0.793033in}}%
\pgfpathlineto{\pgfqpoint{1.402265in}{0.788682in}}%
\pgfpathlineto{\pgfqpoint{1.403600in}{0.787320in}}%
\pgfpathlineto{\pgfqpoint{1.409471in}{0.781254in}}%
\pgfpathlineto{\pgfqpoint{1.411778in}{0.778848in}}%
\pgfpathlineto{\pgfqpoint{1.415670in}{0.774789in}}%
\pgfpathlineto{\pgfqpoint{1.420309in}{0.769956in}}%
\pgfpathlineto{\pgfqpoint{1.425294in}{0.764673in}}%
\pgfpathlineto{\pgfqpoint{1.427415in}{0.762429in}}%
\pgfpathlineto{\pgfqpoint{1.429992in}{0.759687in}}%
\pgfpathlineto{\pgfqpoint{1.435330in}{0.753989in}}%
\pgfpathlineto{\pgfqpoint{1.436825in}{0.752391in}}%
\pgfpathlineto{\pgfqpoint{1.439017in}{0.750034in}}%
\pgfpathlineto{\pgfqpoint{1.440439in}{0.748488in}}%
\pgfpathlineto{\pgfqpoint{1.444407in}{0.744226in}}%
\pgfpathlineto{\pgfqpoint{1.515361in}{0.671414in}}%
\pgfpathlineto{\pgfqpoint{1.516924in}{0.669958in}}%
\pgfpathlineto{\pgfqpoint{1.518331in}{0.668632in}}%
\pgfpathlineto{\pgfqpoint{1.520870in}{0.666325in}}%
\pgfpathlineto{\pgfqpoint{1.522304in}{0.665007in}}%
\pgfpathlineto{\pgfqpoint{1.527154in}{0.660605in}}%
\pgfpathlineto{\pgfqpoint{1.528448in}{0.659452in}}%
\pgfpathlineto{\pgfqpoint{1.530434in}{0.657692in}}%
\pgfpathlineto{\pgfqpoint{1.532052in}{0.656231in}}%
\pgfpathlineto{\pgfqpoint{1.543435in}{0.646341in}}%
\pgfpathlineto{\pgfqpoint{1.544583in}{0.645369in}}%
\pgfpathlineto{\pgfqpoint{1.546483in}{0.643743in}}%
\pgfpathlineto{\pgfqpoint{1.548976in}{0.641632in}}%
\pgfpathlineto{\pgfqpoint{1.550840in}{0.640042in}}%
\pgfpathlineto{\pgfqpoint{1.551966in}{0.639109in}}%
\pgfpathlineto{\pgfqpoint{1.555958in}{0.635760in}}%
\pgfpathlineto{\pgfqpoint{1.557143in}{0.634756in}}%
\pgfpathlineto{\pgfqpoint{1.559756in}{0.632553in}}%
\pgfpathlineto{\pgfqpoint{1.561109in}{0.631386in}}%
\pgfpathlineto{\pgfqpoint{1.564444in}{0.628558in}}%
\pgfpathlineto{\pgfqpoint{1.565612in}{0.627557in}}%
\pgfpathlineto{\pgfqpoint{1.569612in}{0.624150in}}%
\pgfpathlineto{\pgfqpoint{1.572347in}{0.621822in}}%
\pgfpathlineto{\pgfqpoint{1.573518in}{0.620827in}}%
\pgfpathlineto{\pgfqpoint{1.576180in}{0.618533in}}%
\pgfpathlineto{\pgfqpoint{1.577349in}{0.617556in}}%
\pgfpathlineto{\pgfqpoint{1.579103in}{0.616044in}}%
\pgfpathlineto{\pgfqpoint{1.591736in}{0.605180in}}%
\pgfpathlineto{\pgfqpoint{1.593367in}{0.603730in}}%
\pgfpathlineto{\pgfqpoint{1.595906in}{0.601516in}}%
\pgfpathlineto{\pgfqpoint{1.600304in}{0.597671in}}%
\pgfpathlineto{\pgfqpoint{1.602383in}{0.595826in}}%
\pgfpathlineto{\pgfqpoint{1.606403in}{0.592244in}}%
\pgfpathlineto{\pgfqpoint{1.607519in}{0.591249in}}%
\pgfpathlineto{\pgfqpoint{1.610549in}{0.588548in}}%
\pgfpathlineto{\pgfqpoint{1.643026in}{0.559397in}}%
\pgfpathlineto{\pgfqpoint{1.645034in}{0.557549in}}%
\pgfpathlineto{\pgfqpoint{1.647112in}{0.555625in}}%
\pgfpathlineto{\pgfqpoint{1.648898in}{0.553986in}}%
\pgfpathlineto{\pgfqpoint{1.650055in}{0.552942in}}%
\pgfpathlineto{\pgfqpoint{1.651827in}{0.551321in}}%
\pgfpathlineto{\pgfqpoint{1.653903in}{0.549420in}}%
\pgfpathlineto{\pgfqpoint{1.656171in}{0.547333in}}%
\pgfpathlineto{\pgfqpoint{1.659372in}{0.544347in}}%
\pgfpathlineto{\pgfqpoint{1.661034in}{0.542764in}}%
\pgfpathlineto{\pgfqpoint{1.668350in}{0.535904in}}%
\pgfpathlineto{\pgfqpoint{1.670689in}{0.533730in}}%
\pgfpathlineto{\pgfqpoint{1.672595in}{0.531956in}}%
\pgfpathlineto{\pgfqpoint{1.674991in}{0.529700in}}%
\pgfpathlineto{\pgfqpoint{1.676726in}{0.528054in}}%
\pgfpathlineto{\pgfqpoint{1.678232in}{0.526630in}}%
\pgfpathlineto{\pgfqpoint{1.680782in}{0.524197in}}%
\pgfpathlineto{\pgfqpoint{1.682298in}{0.522737in}}%
\pgfpathlineto{\pgfqpoint{1.694571in}{0.511018in}}%
\pgfpathlineto{\pgfqpoint{1.696251in}{0.509401in}}%
\pgfpathlineto{\pgfqpoint{1.698653in}{0.507168in}}%
\pgfpathlineto{\pgfqpoint{1.701667in}{0.504363in}}%
\pgfpathlineto{\pgfqpoint{1.703480in}{0.502674in}}%
\pgfpathlineto{\pgfqpoint{1.706744in}{0.499594in}}%
\pgfpathlineto{\pgfqpoint{1.709503in}{0.497008in}}%
\pgfpathlineto{\pgfqpoint{1.711203in}{0.495407in}}%
\pgfpathlineto{\pgfqpoint{1.712357in}{0.494320in}}%
\pgfpathlineto{\pgfqpoint{1.729540in}{0.478358in}}%
\pgfpathlineto{\pgfqpoint{1.731230in}{0.476815in}}%
\pgfpathlineto{\pgfqpoint{1.777832in}{0.437562in}}%
\pgfpathlineto{\pgfqpoint{1.780448in}{0.435926in}}%
\pgfpathlineto{\pgfqpoint{1.788234in}{0.431170in}}%
\pgfpathlineto{\pgfqpoint{1.790434in}{0.429870in}}%
\pgfpathlineto{\pgfqpoint{1.793946in}{0.427801in}}%
\pgfpathlineto{\pgfqpoint{1.795797in}{0.426733in}}%
\pgfpathlineto{\pgfqpoint{1.799607in}{0.424540in}}%
\pgfpathlineto{\pgfqpoint{1.801843in}{0.423268in}}%
\pgfpathlineto{\pgfqpoint{1.805282in}{0.421353in}}%
\pgfpathlineto{\pgfqpoint{1.806435in}{0.420717in}}%
\pgfpathlineto{\pgfqpoint{1.812361in}{0.417496in}}%
\pgfpathlineto{\pgfqpoint{1.865218in}{0.391557in}}%
\pgfpathlineto{\pgfqpoint{1.868592in}{0.390017in}}%
\pgfpathlineto{\pgfqpoint{1.869973in}{0.389391in}}%
\pgfpathlineto{\pgfqpoint{1.887026in}{0.381611in}}%
\pgfpathlineto{\pgfqpoint{1.890043in}{0.380255in}}%
\pgfpathlineto{\pgfqpoint{1.891367in}{0.379638in}}%
\pgfpathlineto{\pgfqpoint{1.894502in}{0.378225in}}%
\pgfpathlineto{\pgfqpoint{1.898965in}{0.376206in}}%
\pgfpathlineto{\pgfqpoint{1.909182in}{0.371602in}}%
\pgfpathlineto{\pgfqpoint{1.916425in}{0.368326in}}%
\pgfpathlineto{\pgfqpoint{1.916425in}{0.368326in}}%
\pgfusepath{stroke}%
\end{pgfscope}%
\begin{pgfscope}%
\pgfpathrectangle{\pgfqpoint{0.391785in}{0.311728in}}{\pgfqpoint{1.783463in}{1.245154in}}%
\pgfusepath{clip}%
\pgfsetbuttcap%
\pgfsetroundjoin%
\pgfsetlinewidth{1.104125pt}%
\definecolor{currentstroke}{rgb}{0.145098,0.145098,0.145098}%
\pgfsetstrokecolor{currentstroke}%
\pgfsetdash{{5.500000pt}{1.100000pt}}{0.000000pt}%
\pgfpathmoveto{\pgfqpoint{0.391785in}{1.500284in}}%
\pgfpathlineto{\pgfqpoint{0.962006in}{1.499173in}}%
\pgfpathlineto{\pgfqpoint{0.996203in}{1.492344in}}%
\pgfpathlineto{\pgfqpoint{1.003263in}{1.488989in}}%
\pgfpathlineto{\pgfqpoint{1.004503in}{1.488281in}}%
\pgfpathlineto{\pgfqpoint{1.011945in}{1.483278in}}%
\pgfpathlineto{\pgfqpoint{1.025466in}{1.469942in}}%
\pgfpathlineto{\pgfqpoint{1.027820in}{1.466941in}}%
\pgfpathlineto{\pgfqpoint{1.029742in}{1.464328in}}%
\pgfpathlineto{\pgfqpoint{1.030910in}{1.462660in}}%
\pgfpathlineto{\pgfqpoint{1.038140in}{1.450995in}}%
\pgfpathlineto{\pgfqpoint{1.040073in}{1.447461in}}%
\pgfpathlineto{\pgfqpoint{1.040906in}{1.445869in}}%
\pgfpathlineto{\pgfqpoint{1.042451in}{1.442832in}}%
\pgfpathlineto{\pgfqpoint{1.043538in}{1.440653in}}%
\pgfpathlineto{\pgfqpoint{1.044796in}{1.438014in}}%
\pgfpathlineto{\pgfqpoint{1.045821in}{1.435816in}}%
\pgfpathlineto{\pgfqpoint{1.047275in}{1.432603in}}%
\pgfpathlineto{\pgfqpoint{1.056234in}{1.412612in}}%
\pgfpathlineto{\pgfqpoint{1.057535in}{1.409424in}}%
\pgfpathlineto{\pgfqpoint{1.096972in}{1.294864in}}%
\pgfpathlineto{\pgfqpoint{1.100581in}{1.283263in}}%
\pgfpathlineto{\pgfqpoint{1.101242in}{1.281127in}}%
\pgfpathlineto{\pgfqpoint{1.148728in}{1.128640in}}%
\pgfpathlineto{\pgfqpoint{1.149398in}{1.126716in}}%
\pgfpathlineto{\pgfqpoint{1.150685in}{1.123048in}}%
\pgfpathlineto{\pgfqpoint{1.151880in}{1.119619in}}%
\pgfpathlineto{\pgfqpoint{1.154759in}{1.111297in}}%
\pgfpathlineto{\pgfqpoint{1.156296in}{1.106860in}}%
\pgfpathlineto{\pgfqpoint{1.156891in}{1.105120in}}%
\pgfpathlineto{\pgfqpoint{1.158198in}{1.101267in}}%
\pgfpathlineto{\pgfqpoint{1.162698in}{1.087777in}}%
\pgfpathlineto{\pgfqpoint{1.164016in}{1.083755in}}%
\pgfpathlineto{\pgfqpoint{1.167058in}{1.074479in}}%
\pgfpathlineto{\pgfqpoint{1.168322in}{1.070604in}}%
\pgfpathlineto{\pgfqpoint{1.169925in}{1.065685in}}%
\pgfpathlineto{\pgfqpoint{1.187448in}{1.011101in}}%
\pgfpathlineto{\pgfqpoint{1.187919in}{1.009723in}}%
\pgfpathlineto{\pgfqpoint{1.189400in}{1.005432in}}%
\pgfpathlineto{\pgfqpoint{1.190943in}{1.002517in}}%
\pgfpathlineto{\pgfqpoint{1.192420in}{0.999908in}}%
\pgfpathlineto{\pgfqpoint{1.193827in}{0.997426in}}%
\pgfpathlineto{\pgfqpoint{1.195035in}{0.995340in}}%
\pgfpathlineto{\pgfqpoint{1.197995in}{0.990396in}}%
\pgfpathlineto{\pgfqpoint{1.199413in}{0.988097in}}%
\pgfpathlineto{\pgfqpoint{1.203385in}{0.981876in}}%
\pgfpathlineto{\pgfqpoint{1.204682in}{0.979915in}}%
\pgfpathlineto{\pgfqpoint{1.206581in}{0.977124in}}%
\pgfpathlineto{\pgfqpoint{1.208220in}{0.974779in}}%
\pgfpathlineto{\pgfqpoint{1.211462in}{0.970290in}}%
\pgfpathlineto{\pgfqpoint{1.219512in}{0.960119in}}%
\pgfpathlineto{\pgfqpoint{1.220958in}{0.958414in}}%
\pgfpathlineto{\pgfqpoint{1.222116in}{0.957057in}}%
\pgfpathlineto{\pgfqpoint{1.224565in}{0.954217in}}%
\pgfpathlineto{\pgfqpoint{1.226126in}{0.952444in}}%
\pgfpathlineto{\pgfqpoint{1.228831in}{0.949419in}}%
\pgfpathlineto{\pgfqpoint{1.233744in}{0.944001in}}%
\pgfpathlineto{\pgfqpoint{1.234859in}{0.942791in}}%
\pgfpathlineto{\pgfqpoint{1.236127in}{0.941407in}}%
\pgfpathlineto{\pgfqpoint{1.238617in}{0.938706in}}%
\pgfpathlineto{\pgfqpoint{1.240466in}{0.936684in}}%
\pgfpathlineto{\pgfqpoint{1.241841in}{0.935207in}}%
\pgfpathlineto{\pgfqpoint{1.243947in}{0.932963in}}%
\pgfpathlineto{\pgfqpoint{1.245465in}{0.931394in}}%
\pgfpathlineto{\pgfqpoint{1.247418in}{0.929406in}}%
\pgfpathlineto{\pgfqpoint{1.249652in}{0.927088in}}%
\pgfpathlineto{\pgfqpoint{1.253837in}{0.922747in}}%
\pgfpathlineto{\pgfqpoint{1.255054in}{0.921463in}}%
\pgfpathlineto{\pgfqpoint{1.287978in}{0.886329in}}%
\pgfpathlineto{\pgfqpoint{1.289485in}{0.884696in}}%
\pgfpathlineto{\pgfqpoint{1.291235in}{0.882803in}}%
\pgfpathlineto{\pgfqpoint{1.327689in}{0.842271in}}%
\pgfpathlineto{\pgfqpoint{1.332188in}{0.837024in}}%
\pgfpathlineto{\pgfqpoint{1.333797in}{0.835149in}}%
\pgfpathlineto{\pgfqpoint{1.335726in}{0.832870in}}%
\pgfpathlineto{\pgfqpoint{1.339451in}{0.828477in}}%
\pgfpathlineto{\pgfqpoint{1.341706in}{0.825792in}}%
\pgfpathlineto{\pgfqpoint{1.343094in}{0.824117in}}%
\pgfpathlineto{\pgfqpoint{1.344243in}{0.822757in}}%
\pgfpathlineto{\pgfqpoint{1.347361in}{0.819049in}}%
\pgfpathlineto{\pgfqpoint{1.351598in}{0.813984in}}%
\pgfpathlineto{\pgfqpoint{1.353890in}{0.811180in}}%
\pgfpathlineto{\pgfqpoint{1.355288in}{0.809476in}}%
\pgfpathlineto{\pgfqpoint{1.356790in}{0.807617in}}%
\pgfpathlineto{\pgfqpoint{1.363343in}{0.799500in}}%
\pgfpathlineto{\pgfqpoint{1.365307in}{0.797019in}}%
\pgfpathlineto{\pgfqpoint{1.367896in}{0.793785in}}%
\pgfpathlineto{\pgfqpoint{1.369272in}{0.792066in}}%
\pgfpathlineto{\pgfqpoint{1.378195in}{0.781045in}}%
\pgfpathlineto{\pgfqpoint{1.380064in}{0.778695in}}%
\pgfpathlineto{\pgfqpoint{1.381985in}{0.776288in}}%
\pgfpathlineto{\pgfqpoint{1.383224in}{0.774711in}}%
\pgfpathlineto{\pgfqpoint{1.384579in}{0.773032in}}%
\pgfpathlineto{\pgfqpoint{1.387337in}{0.769555in}}%
\pgfpathlineto{\pgfqpoint{1.408624in}{0.742499in}}%
\pgfpathlineto{\pgfqpoint{1.409981in}{0.740777in}}%
\pgfpathlineto{\pgfqpoint{1.413390in}{0.736492in}}%
\pgfpathlineto{\pgfqpoint{1.414938in}{0.734568in}}%
\pgfpathlineto{\pgfqpoint{1.420166in}{0.728119in}}%
\pgfpathlineto{\pgfqpoint{1.422783in}{0.724955in}}%
\pgfpathlineto{\pgfqpoint{1.435915in}{0.708748in}}%
\pgfpathlineto{\pgfqpoint{1.437008in}{0.707423in}}%
\pgfpathlineto{\pgfqpoint{1.473517in}{0.667142in}}%
\pgfpathlineto{\pgfqpoint{1.477731in}{0.662892in}}%
\pgfpathlineto{\pgfqpoint{1.479686in}{0.660987in}}%
\pgfpathlineto{\pgfqpoint{1.482458in}{0.658264in}}%
\pgfpathlineto{\pgfqpoint{1.483588in}{0.657158in}}%
\pgfpathlineto{\pgfqpoint{1.485024in}{0.655761in}}%
\pgfpathlineto{\pgfqpoint{1.488217in}{0.652641in}}%
\pgfpathlineto{\pgfqpoint{1.490183in}{0.650754in}}%
\pgfpathlineto{\pgfqpoint{1.491861in}{0.649125in}}%
\pgfpathlineto{\pgfqpoint{1.493217in}{0.647819in}}%
\pgfpathlineto{\pgfqpoint{1.497768in}{0.643444in}}%
\pgfpathlineto{\pgfqpoint{1.499286in}{0.641968in}}%
\pgfpathlineto{\pgfqpoint{1.504460in}{0.636957in}}%
\pgfpathlineto{\pgfqpoint{1.505909in}{0.635556in}}%
\pgfpathlineto{\pgfqpoint{1.516256in}{0.625437in}}%
\pgfpathlineto{\pgfqpoint{1.529829in}{0.612141in}}%
\pgfpathlineto{\pgfqpoint{1.532296in}{0.609681in}}%
\pgfpathlineto{\pgfqpoint{1.534816in}{0.607228in}}%
\pgfpathlineto{\pgfqpoint{1.536678in}{0.605391in}}%
\pgfpathlineto{\pgfqpoint{1.539295in}{0.602811in}}%
\pgfpathlineto{\pgfqpoint{1.546350in}{0.595744in}}%
\pgfpathlineto{\pgfqpoint{1.547615in}{0.594470in}}%
\pgfpathlineto{\pgfqpoint{1.569377in}{0.572344in}}%
\pgfpathlineto{\pgfqpoint{1.570869in}{0.570832in}}%
\pgfpathlineto{\pgfqpoint{1.576361in}{0.565191in}}%
\pgfpathlineto{\pgfqpoint{1.595786in}{0.545193in}}%
\pgfpathlineto{\pgfqpoint{1.597704in}{0.543191in}}%
\pgfpathlineto{\pgfqpoint{1.598854in}{0.542007in}}%
\pgfpathlineto{\pgfqpoint{1.600404in}{0.540395in}}%
\pgfpathlineto{\pgfqpoint{1.601531in}{0.539203in}}%
\pgfpathlineto{\pgfqpoint{1.605863in}{0.534653in}}%
\pgfpathlineto{\pgfqpoint{1.607087in}{0.533368in}}%
\pgfpathlineto{\pgfqpoint{1.608394in}{0.532017in}}%
\pgfpathlineto{\pgfqpoint{1.610191in}{0.530136in}}%
\pgfpathlineto{\pgfqpoint{1.612203in}{0.527994in}}%
\pgfpathlineto{\pgfqpoint{1.613533in}{0.526598in}}%
\pgfpathlineto{\pgfqpoint{1.616138in}{0.523803in}}%
\pgfpathlineto{\pgfqpoint{1.617662in}{0.522188in}}%
\pgfpathlineto{\pgfqpoint{1.664166in}{0.472644in}}%
\pgfpathlineto{\pgfqpoint{1.666267in}{0.470451in}}%
\pgfpathlineto{\pgfqpoint{1.667875in}{0.468768in}}%
\pgfpathlineto{\pgfqpoint{1.669846in}{0.466728in}}%
\pgfpathlineto{\pgfqpoint{1.670975in}{0.465565in}}%
\pgfpathlineto{\pgfqpoint{1.672679in}{0.463836in}}%
\pgfpathlineto{\pgfqpoint{1.691109in}{0.445996in}}%
\pgfpathlineto{\pgfqpoint{1.693425in}{0.443918in}}%
\pgfpathlineto{\pgfqpoint{1.694966in}{0.442541in}}%
\pgfpathlineto{\pgfqpoint{1.749160in}{0.408744in}}%
\pgfpathlineto{\pgfqpoint{1.751785in}{0.407360in}}%
\pgfpathlineto{\pgfqpoint{1.753149in}{0.406631in}}%
\pgfpathlineto{\pgfqpoint{1.756206in}{0.405021in}}%
\pgfpathlineto{\pgfqpoint{1.759628in}{0.403241in}}%
\pgfpathlineto{\pgfqpoint{1.761502in}{0.402282in}}%
\pgfpathlineto{\pgfqpoint{1.764467in}{0.400770in}}%
\pgfpathlineto{\pgfqpoint{1.771263in}{0.397319in}}%
\pgfpathlineto{\pgfqpoint{1.774585in}{0.395623in}}%
\pgfpathlineto{\pgfqpoint{1.782033in}{0.391958in}}%
\pgfpathlineto{\pgfqpoint{1.785248in}{0.390373in}}%
\pgfpathlineto{\pgfqpoint{1.786404in}{0.389798in}}%
\pgfpathlineto{\pgfqpoint{1.788292in}{0.388855in}}%
\pgfpathlineto{\pgfqpoint{1.791274in}{0.387398in}}%
\pgfpathlineto{\pgfqpoint{1.795419in}{0.385381in}}%
\pgfpathlineto{\pgfqpoint{1.800860in}{0.382736in}}%
\pgfpathlineto{\pgfqpoint{1.801974in}{0.382199in}}%
\pgfpathlineto{\pgfqpoint{1.822613in}{0.372056in}}%
\pgfpathlineto{\pgfqpoint{1.826136in}{0.370328in}}%
\pgfpathlineto{\pgfqpoint{1.828455in}{0.369185in}}%
\pgfpathlineto{\pgfqpoint{1.830171in}{0.368326in}}%
\pgfpathlineto{\pgfqpoint{1.830171in}{0.368326in}}%
\pgfusepath{stroke}%
\end{pgfscope}%
\begin{pgfscope}%
\pgfsetrectcap%
\pgfsetmiterjoin%
\pgfsetlinewidth{0.803000pt}%
\definecolor{currentstroke}{rgb}{0.000000,0.000000,0.000000}%
\pgfsetstrokecolor{currentstroke}%
\pgfsetdash{}{0pt}%
\pgfpathmoveto{\pgfqpoint{0.391785in}{0.311728in}}%
\pgfpathlineto{\pgfqpoint{0.391785in}{1.556882in}}%
\pgfusepath{stroke}%
\end{pgfscope}%
\begin{pgfscope}%
\pgfsetrectcap%
\pgfsetmiterjoin%
\pgfsetlinewidth{0.803000pt}%
\definecolor{currentstroke}{rgb}{0.000000,0.000000,0.000000}%
\pgfsetstrokecolor{currentstroke}%
\pgfsetdash{}{0pt}%
\pgfpathmoveto{\pgfqpoint{2.175248in}{0.311728in}}%
\pgfpathlineto{\pgfqpoint{2.175248in}{1.556882in}}%
\pgfusepath{stroke}%
\end{pgfscope}%
\begin{pgfscope}%
\pgfsetrectcap%
\pgfsetmiterjoin%
\pgfsetlinewidth{0.803000pt}%
\definecolor{currentstroke}{rgb}{0.000000,0.000000,0.000000}%
\pgfsetstrokecolor{currentstroke}%
\pgfsetdash{}{0pt}%
\pgfpathmoveto{\pgfqpoint{0.391785in}{0.311728in}}%
\pgfpathlineto{\pgfqpoint{2.175248in}{0.311728in}}%
\pgfusepath{stroke}%
\end{pgfscope}%
\begin{pgfscope}%
\pgfsetrectcap%
\pgfsetmiterjoin%
\pgfsetlinewidth{0.803000pt}%
\definecolor{currentstroke}{rgb}{0.000000,0.000000,0.000000}%
\pgfsetstrokecolor{currentstroke}%
\pgfsetdash{}{0pt}%
\pgfpathmoveto{\pgfqpoint{0.391785in}{1.556882in}}%
\pgfpathlineto{\pgfqpoint{2.175248in}{1.556882in}}%
\pgfusepath{stroke}%
\end{pgfscope}%
\begin{pgfscope}%
\pgfsetbuttcap%
\pgfsetmiterjoin%
\definecolor{currentfill}{rgb}{1.000000,1.000000,1.000000}%
\pgfsetfillcolor{currentfill}%
\pgfsetfillopacity{0.800000}%
\pgfsetlinewidth{1.003750pt}%
\definecolor{currentstroke}{rgb}{0.800000,0.800000,0.800000}%
\pgfsetstrokecolor{currentstroke}%
\pgfsetstrokeopacity{0.800000}%
\pgfsetdash{}{0pt}%
\pgfpathmoveto{\pgfqpoint{1.484203in}{1.025400in}}%
\pgfpathlineto{\pgfqpoint{2.116914in}{1.025400in}}%
\pgfpathquadraticcurveto{\pgfqpoint{2.133581in}{1.025400in}}{\pgfqpoint{2.133581in}{1.042067in}}%
\pgfpathlineto{\pgfqpoint{2.133581in}{1.498549in}}%
\pgfpathquadraticcurveto{\pgfqpoint{2.133581in}{1.515215in}}{\pgfqpoint{2.116914in}{1.515215in}}%
\pgfpathlineto{\pgfqpoint{1.484203in}{1.515215in}}%
\pgfpathquadraticcurveto{\pgfqpoint{1.467536in}{1.515215in}}{\pgfqpoint{1.467536in}{1.498549in}}%
\pgfpathlineto{\pgfqpoint{1.467536in}{1.042067in}}%
\pgfpathquadraticcurveto{\pgfqpoint{1.467536in}{1.025400in}}{\pgfqpoint{1.484203in}{1.025400in}}%
\pgfpathclose%
\pgfusepath{stroke,fill}%
\end{pgfscope}%
\begin{pgfscope}%
\pgfsetrectcap%
\pgfsetroundjoin%
\pgfsetlinewidth{1.104125pt}%
\definecolor{currentstroke}{rgb}{0.631373,0.850980,0.607843}%
\pgfsetstrokecolor{currentstroke}%
\pgfsetdash{}{0pt}%
\pgfpathmoveto{\pgfqpoint{1.500870in}{1.452715in}}%
\pgfpathlineto{\pgfqpoint{1.667536in}{1.452715in}}%
\pgfusepath{stroke}%
\end{pgfscope}%
\begin{pgfscope}%
\pgftext[x=1.734203in,y=1.423549in,left,base]{\rmfamily\fontsize{6.000000}{7.200000}\selectfont FD}%
\end{pgfscope}%
\begin{pgfscope}%
\pgfsetbuttcap%
\pgfsetroundjoin%
\pgfsetlinewidth{1.104125pt}%
\definecolor{currentstroke}{rgb}{0.631373,0.850980,0.607843}%
\pgfsetstrokecolor{currentstroke}%
\pgfsetdash{{5.500000pt}{1.100000pt}}{0.000000pt}%
\pgfpathmoveto{\pgfqpoint{1.500870in}{1.336511in}}%
\pgfpathlineto{\pgfqpoint{1.667536in}{1.336511in}}%
\pgfusepath{stroke}%
\end{pgfscope}%
\begin{pgfscope}%
\pgftext[x=1.734203in,y=1.307345in,left,base]{\rmfamily\fontsize{6.000000}{7.200000}\selectfont FD+6}%
\end{pgfscope}%
\begin{pgfscope}%
\pgfsetrectcap%
\pgfsetroundjoin%
\pgfsetlinewidth{1.104125pt}%
\definecolor{currentstroke}{rgb}{0.145098,0.145098,0.145098}%
\pgfsetstrokecolor{currentstroke}%
\pgfsetdash{}{0pt}%
\pgfpathmoveto{\pgfqpoint{1.500870in}{1.220308in}}%
\pgfpathlineto{\pgfqpoint{1.667536in}{1.220308in}}%
\pgfusepath{stroke}%
\end{pgfscope}%
\begin{pgfscope}%
\pgftext[x=1.734203in,y=1.191141in,left,base]{\rmfamily\fontsize{6.000000}{7.200000}\selectfont L2C2}%
\end{pgfscope}%
\begin{pgfscope}%
\pgfsetbuttcap%
\pgfsetroundjoin%
\pgfsetlinewidth{1.104125pt}%
\definecolor{currentstroke}{rgb}{0.145098,0.145098,0.145098}%
\pgfsetstrokecolor{currentstroke}%
\pgfsetdash{{5.500000pt}{1.100000pt}}{0.000000pt}%
\pgfpathmoveto{\pgfqpoint{1.500870in}{1.104104in}}%
\pgfpathlineto{\pgfqpoint{1.667536in}{1.104104in}}%
\pgfusepath{stroke}%
\end{pgfscope}%
\begin{pgfscope}%
\pgftext[x=1.734203in,y=1.074937in,left,base]{\rmfamily\fontsize{6.000000}{7.200000}\selectfont L2C2+6}%
\end{pgfscope}%
\end{pgfpicture}%
\makeatother%
\endgroup%

%% file: Figuras/ipc_cov020.pgf
\begingroup%
\makeatletter%
\begin{pgfpicture}%
\pgfpathrectangle{\pgfpointorigin}{\pgfqpoint{2.222509in}{1.591882in}}%
\pgfusepath{use as bounding box, clip}%
\begin{pgfscope}%
\pgfsetbuttcap%
\pgfsetmiterjoin%
\definecolor{currentfill}{rgb}{1.000000,1.000000,1.000000}%
\pgfsetfillcolor{currentfill}%
\pgfsetlinewidth{0.000000pt}%
\definecolor{currentstroke}{rgb}{1.000000,1.000000,1.000000}%
\pgfsetstrokecolor{currentstroke}%
\pgfsetdash{}{0pt}%
\pgfpathmoveto{\pgfqpoint{0.000000in}{0.000000in}}%
\pgfpathlineto{\pgfqpoint{2.222509in}{0.000000in}}%
\pgfpathlineto{\pgfqpoint{2.222509in}{1.591882in}}%
\pgfpathlineto{\pgfqpoint{0.000000in}{1.591882in}}%
\pgfpathclose%
\pgfusepath{fill}%
\end{pgfscope}%
\begin{pgfscope}%
\pgfsetbuttcap%
\pgfsetmiterjoin%
\definecolor{currentfill}{rgb}{1.000000,1.000000,1.000000}%
\pgfsetfillcolor{currentfill}%
\pgfsetlinewidth{0.000000pt}%
\definecolor{currentstroke}{rgb}{0.000000,0.000000,0.000000}%
\pgfsetstrokecolor{currentstroke}%
\pgfsetstrokeopacity{0.000000}%
\pgfsetdash{}{0pt}%
\pgfpathmoveto{\pgfqpoint{0.365549in}{0.311728in}}%
\pgfpathlineto{\pgfqpoint{2.175248in}{0.311728in}}%
\pgfpathlineto{\pgfqpoint{2.175248in}{1.556882in}}%
\pgfpathlineto{\pgfqpoint{0.365549in}{1.556882in}}%
\pgfpathclose%
\pgfusepath{fill}%
\end{pgfscope}%
\begin{pgfscope}%
\pgfsetbuttcap%
\pgfsetroundjoin%
\definecolor{currentfill}{rgb}{0.000000,0.000000,0.000000}%
\pgfsetfillcolor{currentfill}%
\pgfsetlinewidth{0.803000pt}%
\definecolor{currentstroke}{rgb}{0.000000,0.000000,0.000000}%
\pgfsetstrokecolor{currentstroke}%
\pgfsetdash{}{0pt}%
\pgfsys@defobject{currentmarker}{\pgfqpoint{0.000000in}{-0.048611in}}{\pgfqpoint{0.000000in}{0.000000in}}{%
\pgfpathmoveto{\pgfqpoint{0.000000in}{0.000000in}}%
\pgfpathlineto{\pgfqpoint{0.000000in}{-0.048611in}}%
\pgfusepath{stroke,fill}%
}%
\begin{pgfscope}%
\pgfsys@transformshift{0.365549in}{0.311728in}%
\pgfsys@useobject{currentmarker}{}%
\end{pgfscope}%
\end{pgfscope}%
\begin{pgfscope}%
\pgftext[x=0.365549in,y=0.214506in,,top]{\rmfamily\fontsize{5.000000}{6.000000}\selectfont \(\displaystyle 0\)}%
\end{pgfscope}%
\begin{pgfscope}%
\pgfsetbuttcap%
\pgfsetroundjoin%
\definecolor{currentfill}{rgb}{0.000000,0.000000,0.000000}%
\pgfsetfillcolor{currentfill}%
\pgfsetlinewidth{0.803000pt}%
\definecolor{currentstroke}{rgb}{0.000000,0.000000,0.000000}%
\pgfsetstrokecolor{currentstroke}%
\pgfsetdash{}{0pt}%
\pgfsys@defobject{currentmarker}{\pgfqpoint{0.000000in}{-0.048611in}}{\pgfqpoint{0.000000in}{0.000000in}}{%
\pgfpathmoveto{\pgfqpoint{0.000000in}{0.000000in}}%
\pgfpathlineto{\pgfqpoint{0.000000in}{-0.048611in}}%
\pgfusepath{stroke,fill}%
}%
\begin{pgfscope}%
\pgfsys@transformshift{0.817974in}{0.311728in}%
\pgfsys@useobject{currentmarker}{}%
\end{pgfscope}%
\end{pgfscope}%
\begin{pgfscope}%
\pgftext[x=0.817974in,y=0.214506in,,top]{\rmfamily\fontsize{5.000000}{6.000000}\selectfont \(\displaystyle 4\)}%
\end{pgfscope}%
\begin{pgfscope}%
\pgfsetbuttcap%
\pgfsetroundjoin%
\definecolor{currentfill}{rgb}{0.000000,0.000000,0.000000}%
\pgfsetfillcolor{currentfill}%
\pgfsetlinewidth{0.803000pt}%
\definecolor{currentstroke}{rgb}{0.000000,0.000000,0.000000}%
\pgfsetstrokecolor{currentstroke}%
\pgfsetdash{}{0pt}%
\pgfsys@defobject{currentmarker}{\pgfqpoint{0.000000in}{-0.048611in}}{\pgfqpoint{0.000000in}{0.000000in}}{%
\pgfpathmoveto{\pgfqpoint{0.000000in}{0.000000in}}%
\pgfpathlineto{\pgfqpoint{0.000000in}{-0.048611in}}%
\pgfusepath{stroke,fill}%
}%
\begin{pgfscope}%
\pgfsys@transformshift{1.270398in}{0.311728in}%
\pgfsys@useobject{currentmarker}{}%
\end{pgfscope}%
\end{pgfscope}%
\begin{pgfscope}%
\pgftext[x=1.270398in,y=0.214506in,,top]{\rmfamily\fontsize{5.000000}{6.000000}\selectfont \(\displaystyle 8\)}%
\end{pgfscope}%
\begin{pgfscope}%
\pgfsetbuttcap%
\pgfsetroundjoin%
\definecolor{currentfill}{rgb}{0.000000,0.000000,0.000000}%
\pgfsetfillcolor{currentfill}%
\pgfsetlinewidth{0.803000pt}%
\definecolor{currentstroke}{rgb}{0.000000,0.000000,0.000000}%
\pgfsetstrokecolor{currentstroke}%
\pgfsetdash{}{0pt}%
\pgfsys@defobject{currentmarker}{\pgfqpoint{0.000000in}{-0.048611in}}{\pgfqpoint{0.000000in}{0.000000in}}{%
\pgfpathmoveto{\pgfqpoint{0.000000in}{0.000000in}}%
\pgfpathlineto{\pgfqpoint{0.000000in}{-0.048611in}}%
\pgfusepath{stroke,fill}%
}%
\begin{pgfscope}%
\pgfsys@transformshift{1.722823in}{0.311728in}%
\pgfsys@useobject{currentmarker}{}%
\end{pgfscope}%
\end{pgfscope}%
\begin{pgfscope}%
\pgftext[x=1.722823in,y=0.214506in,,top]{\rmfamily\fontsize{5.000000}{6.000000}\selectfont \(\displaystyle 12\)}%
\end{pgfscope}%
\begin{pgfscope}%
\pgfsetbuttcap%
\pgfsetroundjoin%
\definecolor{currentfill}{rgb}{0.000000,0.000000,0.000000}%
\pgfsetfillcolor{currentfill}%
\pgfsetlinewidth{0.803000pt}%
\definecolor{currentstroke}{rgb}{0.000000,0.000000,0.000000}%
\pgfsetstrokecolor{currentstroke}%
\pgfsetdash{}{0pt}%
\pgfsys@defobject{currentmarker}{\pgfqpoint{0.000000in}{-0.048611in}}{\pgfqpoint{0.000000in}{0.000000in}}{%
\pgfpathmoveto{\pgfqpoint{0.000000in}{0.000000in}}%
\pgfpathlineto{\pgfqpoint{0.000000in}{-0.048611in}}%
\pgfusepath{stroke,fill}%
}%
\begin{pgfscope}%
\pgfsys@transformshift{2.175248in}{0.311728in}%
\pgfsys@useobject{currentmarker}{}%
\end{pgfscope}%
\end{pgfscope}%
\begin{pgfscope}%
\pgftext[x=2.175248in,y=0.214506in,,top]{\rmfamily\fontsize{5.000000}{6.000000}\selectfont \(\displaystyle 16\)}%
\end{pgfscope}%
\begin{pgfscope}%
\pgftext[x=1.270398in,y=0.097222in,,top]{\rmfamily\fontsize{7.000000}{8.400000}\selectfont Time (years)}%
\end{pgfscope}%
\begin{pgfscope}%
\pgfsetbuttcap%
\pgfsetroundjoin%
\definecolor{currentfill}{rgb}{0.000000,0.000000,0.000000}%
\pgfsetfillcolor{currentfill}%
\pgfsetlinewidth{0.803000pt}%
\definecolor{currentstroke}{rgb}{0.000000,0.000000,0.000000}%
\pgfsetstrokecolor{currentstroke}%
\pgfsetdash{}{0pt}%
\pgfsys@defobject{currentmarker}{\pgfqpoint{-0.048611in}{0.000000in}}{\pgfqpoint{0.000000in}{0.000000in}}{%
\pgfpathmoveto{\pgfqpoint{0.000000in}{0.000000in}}%
\pgfpathlineto{\pgfqpoint{-0.048611in}{0.000000in}}%
\pgfusepath{stroke,fill}%
}%
\begin{pgfscope}%
\pgfsys@transformshift{0.365549in}{0.423571in}%
\pgfsys@useobject{currentmarker}{}%
\end{pgfscope}%
\end{pgfscope}%
\begin{pgfscope}%
\pgftext[x=0.141975in,y=0.399459in,left,base]{\rmfamily\fontsize{5.000000}{6.000000}\selectfont \(\displaystyle 0.6\)}%
\end{pgfscope}%
\begin{pgfscope}%
\pgfsetbuttcap%
\pgfsetroundjoin%
\definecolor{currentfill}{rgb}{0.000000,0.000000,0.000000}%
\pgfsetfillcolor{currentfill}%
\pgfsetlinewidth{0.803000pt}%
\definecolor{currentstroke}{rgb}{0.000000,0.000000,0.000000}%
\pgfsetstrokecolor{currentstroke}%
\pgfsetdash{}{0pt}%
\pgfsys@defobject{currentmarker}{\pgfqpoint{-0.048611in}{0.000000in}}{\pgfqpoint{0.000000in}{0.000000in}}{%
\pgfpathmoveto{\pgfqpoint{0.000000in}{0.000000in}}%
\pgfpathlineto{\pgfqpoint{-0.048611in}{0.000000in}}%
\pgfusepath{stroke,fill}%
}%
\begin{pgfscope}%
\pgfsys@transformshift{0.365549in}{0.960959in}%
\pgfsys@useobject{currentmarker}{}%
\end{pgfscope}%
\end{pgfscope}%
\begin{pgfscope}%
\pgftext[x=0.141975in,y=0.936847in,left,base]{\rmfamily\fontsize{5.000000}{6.000000}\selectfont \(\displaystyle 0.8\)}%
\end{pgfscope}%
\begin{pgfscope}%
\pgfsetbuttcap%
\pgfsetroundjoin%
\definecolor{currentfill}{rgb}{0.000000,0.000000,0.000000}%
\pgfsetfillcolor{currentfill}%
\pgfsetlinewidth{0.803000pt}%
\definecolor{currentstroke}{rgb}{0.000000,0.000000,0.000000}%
\pgfsetstrokecolor{currentstroke}%
\pgfsetdash{}{0pt}%
\pgfsys@defobject{currentmarker}{\pgfqpoint{-0.048611in}{0.000000in}}{\pgfqpoint{0.000000in}{0.000000in}}{%
\pgfpathmoveto{\pgfqpoint{0.000000in}{0.000000in}}%
\pgfpathlineto{\pgfqpoint{-0.048611in}{0.000000in}}%
\pgfusepath{stroke,fill}%
}%
\begin{pgfscope}%
\pgfsys@transformshift{0.365549in}{1.498347in}%
\pgfsys@useobject{currentmarker}{}%
\end{pgfscope}%
\end{pgfscope}%
\begin{pgfscope}%
\pgftext[x=0.141975in,y=1.474234in,left,base]{\rmfamily\fontsize{5.000000}{6.000000}\selectfont \(\displaystyle 1.0\)}%
\end{pgfscope}%
\begin{pgfscope}%
\pgftext[x=0.086419in,y=0.934305in,,bottom,rotate=90.000000]{\rmfamily\fontsize{7.000000}{8.400000}\selectfont Normalized IPC}%
\end{pgfscope}%
\begin{pgfscope}%
\pgfpathrectangle{\pgfqpoint{0.365549in}{0.311728in}}{\pgfqpoint{1.809698in}{1.245154in}}%
\pgfusepath{clip}%
\pgfsetrectcap%
\pgfsetroundjoin%
\pgfsetlinewidth{0.501875pt}%
\definecolor{currentstroke}{rgb}{0.800000,0.800000,0.800000}%
\pgfsetstrokecolor{currentstroke}%
\pgfsetdash{}{0pt}%
\pgfpathmoveto{\pgfqpoint{0.478655in}{0.311728in}}%
\pgfpathlineto{\pgfqpoint{0.478655in}{1.556882in}}%
\pgfusepath{stroke}%
\end{pgfscope}%
\begin{pgfscope}%
\pgfpathrectangle{\pgfqpoint{0.365549in}{0.311728in}}{\pgfqpoint{1.809698in}{1.245154in}}%
\pgfusepath{clip}%
\pgfsetrectcap%
\pgfsetroundjoin%
\pgfsetlinewidth{0.501875pt}%
\definecolor{currentstroke}{rgb}{0.800000,0.800000,0.800000}%
\pgfsetstrokecolor{currentstroke}%
\pgfsetdash{}{0pt}%
\pgfpathmoveto{\pgfqpoint{0.591762in}{0.311728in}}%
\pgfpathlineto{\pgfqpoint{0.591762in}{1.556882in}}%
\pgfusepath{stroke}%
\end{pgfscope}%
\begin{pgfscope}%
\pgfpathrectangle{\pgfqpoint{0.365549in}{0.311728in}}{\pgfqpoint{1.809698in}{1.245154in}}%
\pgfusepath{clip}%
\pgfsetrectcap%
\pgfsetroundjoin%
\pgfsetlinewidth{0.501875pt}%
\definecolor{currentstroke}{rgb}{0.800000,0.800000,0.800000}%
\pgfsetstrokecolor{currentstroke}%
\pgfsetdash{}{0pt}%
\pgfpathmoveto{\pgfqpoint{0.704868in}{0.311728in}}%
\pgfpathlineto{\pgfqpoint{0.704868in}{1.556882in}}%
\pgfusepath{stroke}%
\end{pgfscope}%
\begin{pgfscope}%
\pgfpathrectangle{\pgfqpoint{0.365549in}{0.311728in}}{\pgfqpoint{1.809698in}{1.245154in}}%
\pgfusepath{clip}%
\pgfsetrectcap%
\pgfsetroundjoin%
\pgfsetlinewidth{0.501875pt}%
\definecolor{currentstroke}{rgb}{0.800000,0.800000,0.800000}%
\pgfsetstrokecolor{currentstroke}%
\pgfsetdash{}{0pt}%
\pgfpathmoveto{\pgfqpoint{0.817974in}{0.311728in}}%
\pgfpathlineto{\pgfqpoint{0.817974in}{1.556882in}}%
\pgfusepath{stroke}%
\end{pgfscope}%
\begin{pgfscope}%
\pgfpathrectangle{\pgfqpoint{0.365549in}{0.311728in}}{\pgfqpoint{1.809698in}{1.245154in}}%
\pgfusepath{clip}%
\pgfsetrectcap%
\pgfsetroundjoin%
\pgfsetlinewidth{0.501875pt}%
\definecolor{currentstroke}{rgb}{0.800000,0.800000,0.800000}%
\pgfsetstrokecolor{currentstroke}%
\pgfsetdash{}{0pt}%
\pgfpathmoveto{\pgfqpoint{0.931080in}{0.311728in}}%
\pgfpathlineto{\pgfqpoint{0.931080in}{1.556882in}}%
\pgfusepath{stroke}%
\end{pgfscope}%
\begin{pgfscope}%
\pgfpathrectangle{\pgfqpoint{0.365549in}{0.311728in}}{\pgfqpoint{1.809698in}{1.245154in}}%
\pgfusepath{clip}%
\pgfsetrectcap%
\pgfsetroundjoin%
\pgfsetlinewidth{0.501875pt}%
\definecolor{currentstroke}{rgb}{0.800000,0.800000,0.800000}%
\pgfsetstrokecolor{currentstroke}%
\pgfsetdash{}{0pt}%
\pgfpathmoveto{\pgfqpoint{1.044186in}{0.311728in}}%
\pgfpathlineto{\pgfqpoint{1.044186in}{1.556882in}}%
\pgfusepath{stroke}%
\end{pgfscope}%
\begin{pgfscope}%
\pgfpathrectangle{\pgfqpoint{0.365549in}{0.311728in}}{\pgfqpoint{1.809698in}{1.245154in}}%
\pgfusepath{clip}%
\pgfsetrectcap%
\pgfsetroundjoin%
\pgfsetlinewidth{0.501875pt}%
\definecolor{currentstroke}{rgb}{0.800000,0.800000,0.800000}%
\pgfsetstrokecolor{currentstroke}%
\pgfsetdash{}{0pt}%
\pgfpathmoveto{\pgfqpoint{1.157292in}{0.311728in}}%
\pgfpathlineto{\pgfqpoint{1.157292in}{1.556882in}}%
\pgfusepath{stroke}%
\end{pgfscope}%
\begin{pgfscope}%
\pgfpathrectangle{\pgfqpoint{0.365549in}{0.311728in}}{\pgfqpoint{1.809698in}{1.245154in}}%
\pgfusepath{clip}%
\pgfsetrectcap%
\pgfsetroundjoin%
\pgfsetlinewidth{0.501875pt}%
\definecolor{currentstroke}{rgb}{0.800000,0.800000,0.800000}%
\pgfsetstrokecolor{currentstroke}%
\pgfsetdash{}{0pt}%
\pgfpathmoveto{\pgfqpoint{1.270398in}{0.311728in}}%
\pgfpathlineto{\pgfqpoint{1.270398in}{1.556882in}}%
\pgfusepath{stroke}%
\end{pgfscope}%
\begin{pgfscope}%
\pgfpathrectangle{\pgfqpoint{0.365549in}{0.311728in}}{\pgfqpoint{1.809698in}{1.245154in}}%
\pgfusepath{clip}%
\pgfsetrectcap%
\pgfsetroundjoin%
\pgfsetlinewidth{0.501875pt}%
\definecolor{currentstroke}{rgb}{0.800000,0.800000,0.800000}%
\pgfsetstrokecolor{currentstroke}%
\pgfsetdash{}{0pt}%
\pgfpathmoveto{\pgfqpoint{1.383505in}{0.311728in}}%
\pgfpathlineto{\pgfqpoint{1.383505in}{1.556882in}}%
\pgfusepath{stroke}%
\end{pgfscope}%
\begin{pgfscope}%
\pgfpathrectangle{\pgfqpoint{0.365549in}{0.311728in}}{\pgfqpoint{1.809698in}{1.245154in}}%
\pgfusepath{clip}%
\pgfsetrectcap%
\pgfsetroundjoin%
\pgfsetlinewidth{0.501875pt}%
\definecolor{currentstroke}{rgb}{0.800000,0.800000,0.800000}%
\pgfsetstrokecolor{currentstroke}%
\pgfsetdash{}{0pt}%
\pgfpathmoveto{\pgfqpoint{1.496611in}{0.311728in}}%
\pgfpathlineto{\pgfqpoint{1.496611in}{1.556882in}}%
\pgfusepath{stroke}%
\end{pgfscope}%
\begin{pgfscope}%
\pgfpathrectangle{\pgfqpoint{0.365549in}{0.311728in}}{\pgfqpoint{1.809698in}{1.245154in}}%
\pgfusepath{clip}%
\pgfsetrectcap%
\pgfsetroundjoin%
\pgfsetlinewidth{0.501875pt}%
\definecolor{currentstroke}{rgb}{0.800000,0.800000,0.800000}%
\pgfsetstrokecolor{currentstroke}%
\pgfsetdash{}{0pt}%
\pgfpathmoveto{\pgfqpoint{1.609717in}{0.311728in}}%
\pgfpathlineto{\pgfqpoint{1.609717in}{1.556882in}}%
\pgfusepath{stroke}%
\end{pgfscope}%
\begin{pgfscope}%
\pgfpathrectangle{\pgfqpoint{0.365549in}{0.311728in}}{\pgfqpoint{1.809698in}{1.245154in}}%
\pgfusepath{clip}%
\pgfsetrectcap%
\pgfsetroundjoin%
\pgfsetlinewidth{0.501875pt}%
\definecolor{currentstroke}{rgb}{0.800000,0.800000,0.800000}%
\pgfsetstrokecolor{currentstroke}%
\pgfsetdash{}{0pt}%
\pgfpathmoveto{\pgfqpoint{1.722823in}{0.311728in}}%
\pgfpathlineto{\pgfqpoint{1.722823in}{1.556882in}}%
\pgfusepath{stroke}%
\end{pgfscope}%
\begin{pgfscope}%
\pgfpathrectangle{\pgfqpoint{0.365549in}{0.311728in}}{\pgfqpoint{1.809698in}{1.245154in}}%
\pgfusepath{clip}%
\pgfsetrectcap%
\pgfsetroundjoin%
\pgfsetlinewidth{0.501875pt}%
\definecolor{currentstroke}{rgb}{0.800000,0.800000,0.800000}%
\pgfsetstrokecolor{currentstroke}%
\pgfsetdash{}{0pt}%
\pgfpathmoveto{\pgfqpoint{1.835929in}{0.311728in}}%
\pgfpathlineto{\pgfqpoint{1.835929in}{1.556882in}}%
\pgfusepath{stroke}%
\end{pgfscope}%
\begin{pgfscope}%
\pgfpathrectangle{\pgfqpoint{0.365549in}{0.311728in}}{\pgfqpoint{1.809698in}{1.245154in}}%
\pgfusepath{clip}%
\pgfsetrectcap%
\pgfsetroundjoin%
\pgfsetlinewidth{0.501875pt}%
\definecolor{currentstroke}{rgb}{0.800000,0.800000,0.800000}%
\pgfsetstrokecolor{currentstroke}%
\pgfsetdash{}{0pt}%
\pgfpathmoveto{\pgfqpoint{1.949035in}{0.311728in}}%
\pgfpathlineto{\pgfqpoint{1.949035in}{1.556882in}}%
\pgfusepath{stroke}%
\end{pgfscope}%
\begin{pgfscope}%
\pgfpathrectangle{\pgfqpoint{0.365549in}{0.311728in}}{\pgfqpoint{1.809698in}{1.245154in}}%
\pgfusepath{clip}%
\pgfsetrectcap%
\pgfsetroundjoin%
\pgfsetlinewidth{0.501875pt}%
\definecolor{currentstroke}{rgb}{0.800000,0.800000,0.800000}%
\pgfsetstrokecolor{currentstroke}%
\pgfsetdash{}{0pt}%
\pgfpathmoveto{\pgfqpoint{2.062142in}{0.311728in}}%
\pgfpathlineto{\pgfqpoint{2.062142in}{1.556882in}}%
\pgfusepath{stroke}%
\end{pgfscope}%
\begin{pgfscope}%
\pgfpathrectangle{\pgfqpoint{0.365549in}{0.311728in}}{\pgfqpoint{1.809698in}{1.245154in}}%
\pgfusepath{clip}%
\pgfsetrectcap%
\pgfsetroundjoin%
\pgfsetlinewidth{1.104125pt}%
\definecolor{currentstroke}{rgb}{0.631373,0.850980,0.607843}%
\pgfsetstrokecolor{currentstroke}%
\pgfsetdash{}{0pt}%
\pgfpathmoveto{\pgfqpoint{0.365549in}{1.498706in}}%
\pgfpathlineto{\pgfqpoint{0.521626in}{1.488733in}}%
\pgfpathlineto{\pgfqpoint{0.544352in}{1.481558in}}%
\pgfpathlineto{\pgfqpoint{0.558045in}{1.472303in}}%
\pgfpathlineto{\pgfqpoint{0.567678in}{1.463908in}}%
\pgfpathlineto{\pgfqpoint{0.575257in}{1.453218in}}%
\pgfpathlineto{\pgfqpoint{0.581507in}{1.438223in}}%
\pgfpathlineto{\pgfqpoint{0.586802in}{1.422653in}}%
\pgfpathlineto{\pgfqpoint{0.591381in}{1.400125in}}%
\pgfpathlineto{\pgfqpoint{0.595368in}{1.365758in}}%
\pgfpathlineto{\pgfqpoint{0.598904in}{1.324216in}}%
\pgfpathlineto{\pgfqpoint{0.602150in}{1.277867in}}%
\pgfpathlineto{\pgfqpoint{0.605191in}{1.224774in}}%
\pgfpathlineto{\pgfqpoint{0.608047in}{1.178712in}}%
\pgfpathlineto{\pgfqpoint{0.610765in}{1.125763in}}%
\pgfpathlineto{\pgfqpoint{0.613410in}{1.077262in}}%
\pgfpathlineto{\pgfqpoint{0.615911in}{1.025030in}}%
\pgfusepath{stroke}%
\end{pgfscope}%
\begin{pgfscope}%
\pgfpathrectangle{\pgfqpoint{0.365549in}{0.311728in}}{\pgfqpoint{1.809698in}{1.245154in}}%
\pgfusepath{clip}%
\pgfsetbuttcap%
\pgfsetroundjoin%
\pgfsetlinewidth{1.104125pt}%
\definecolor{currentstroke}{rgb}{0.631373,0.850980,0.607843}%
\pgfsetstrokecolor{currentstroke}%
\pgfsetdash{{5.500000pt}{1.100000pt}}{0.000000pt}%
\pgfpathmoveto{\pgfqpoint{0.365549in}{1.500284in}}%
\pgfpathlineto{\pgfqpoint{0.700427in}{1.490957in}}%
\pgfpathlineto{\pgfqpoint{0.707757in}{1.482562in}}%
\pgfpathlineto{\pgfqpoint{0.712348in}{1.472877in}}%
\pgfpathlineto{\pgfqpoint{0.715710in}{1.465343in}}%
\pgfpathlineto{\pgfqpoint{0.718395in}{1.452931in}}%
\pgfpathlineto{\pgfqpoint{0.720636in}{1.439442in}}%
\pgfpathlineto{\pgfqpoint{0.722593in}{1.423658in}}%
\pgfpathlineto{\pgfqpoint{0.724277in}{1.404501in}}%
\pgfpathlineto{\pgfqpoint{0.725772in}{1.380035in}}%
\pgfpathlineto{\pgfqpoint{0.727153in}{1.343157in}}%
\pgfpathlineto{\pgfqpoint{0.728446in}{1.301687in}}%
\pgfpathlineto{\pgfqpoint{0.729657in}{1.252325in}}%
\pgfpathlineto{\pgfqpoint{0.730799in}{1.206550in}}%
\pgfpathlineto{\pgfqpoint{0.731879in}{1.154820in}}%
\pgfpathlineto{\pgfqpoint{0.732910in}{1.106535in}}%
\pgfpathlineto{\pgfqpoint{0.733948in}{1.054231in}}%
\pgfusepath{stroke}%
\end{pgfscope}%
\begin{pgfscope}%
\pgfpathrectangle{\pgfqpoint{0.365549in}{0.311728in}}{\pgfqpoint{1.809698in}{1.245154in}}%
\pgfusepath{clip}%
\pgfsetrectcap%
\pgfsetroundjoin%
\pgfsetlinewidth{1.104125pt}%
\definecolor{currentstroke}{rgb}{0.145098,0.145098,0.145098}%
\pgfsetstrokecolor{currentstroke}%
\pgfsetdash{}{0pt}%
\pgfpathmoveto{\pgfqpoint{0.365549in}{1.493540in}}%
\pgfpathlineto{\pgfqpoint{0.905009in}{1.376592in}}%
\pgfpathlineto{\pgfqpoint{0.976587in}{1.265957in}}%
\pgfpathlineto{\pgfqpoint{1.024538in}{1.262729in}}%
\pgfpathlineto{\pgfqpoint{1.059419in}{1.208774in}}%
\pgfpathlineto{\pgfqpoint{1.095183in}{1.157762in}}%
\pgfpathlineto{\pgfqpoint{1.129808in}{1.153457in}}%
\pgfpathlineto{\pgfqpoint{1.161214in}{0.963112in}}%
\pgfpathlineto{\pgfqpoint{1.230907in}{0.853912in}}%
\pgfpathlineto{\pgfqpoint{1.312048in}{0.854127in}}%
\pgfpathlineto{\pgfqpoint{1.386827in}{0.853123in}}%
\pgfpathlineto{\pgfqpoint{1.454578in}{0.833034in}}%
\pgfpathlineto{\pgfqpoint{1.528213in}{0.772909in}}%
\pgfpathlineto{\pgfqpoint{1.611304in}{0.766811in}}%
\pgfpathlineto{\pgfqpoint{1.688752in}{0.752748in}}%
\pgfpathlineto{\pgfqpoint{1.769560in}{0.619800in}}%
\pgfpathlineto{\pgfqpoint{1.912618in}{0.564985in}}%
\pgfusepath{stroke}%
\end{pgfscope}%
\begin{pgfscope}%
\pgfpathrectangle{\pgfqpoint{0.365549in}{0.311728in}}{\pgfqpoint{1.809698in}{1.245154in}}%
\pgfusepath{clip}%
\pgfsetbuttcap%
\pgfsetroundjoin%
\pgfsetlinewidth{1.104125pt}%
\definecolor{currentstroke}{rgb}{0.145098,0.145098,0.145098}%
\pgfsetstrokecolor{currentstroke}%
\pgfsetdash{{5.500000pt}{1.100000pt}}{0.000000pt}%
\pgfpathmoveto{\pgfqpoint{0.365549in}{1.494329in}}%
\pgfpathlineto{\pgfqpoint{1.032067in}{1.488015in}}%
\pgfpathlineto{\pgfqpoint{1.059251in}{1.306638in}}%
\pgfpathlineto{\pgfqpoint{1.083248in}{1.261724in}}%
\pgfpathlineto{\pgfqpoint{1.104349in}{1.214299in}}%
\pgfpathlineto{\pgfqpoint{1.127210in}{1.158408in}}%
\pgfpathlineto{\pgfqpoint{1.151784in}{1.153744in}}%
\pgfpathlineto{\pgfqpoint{1.175037in}{0.970430in}}%
\pgfpathlineto{\pgfqpoint{1.228942in}{0.854414in}}%
\pgfpathlineto{\pgfqpoint{1.296223in}{0.853984in}}%
\pgfpathlineto{\pgfqpoint{1.356804in}{0.853410in}}%
\pgfpathlineto{\pgfqpoint{1.414148in}{0.830092in}}%
\pgfpathlineto{\pgfqpoint{1.479522in}{0.770972in}}%
\pgfpathlineto{\pgfqpoint{1.552233in}{0.766667in}}%
\pgfpathlineto{\pgfqpoint{1.621206in}{0.757053in}}%
\pgfpathlineto{\pgfqpoint{1.691939in}{0.612769in}}%
\pgfpathlineto{\pgfqpoint{1.825095in}{0.564483in}}%
\pgfusepath{stroke}%
\end{pgfscope}%
\begin{pgfscope}%
\pgfpathrectangle{\pgfqpoint{0.365549in}{0.311728in}}{\pgfqpoint{1.809698in}{1.245154in}}%
\pgfusepath{clip}%
\pgfsetbuttcap%
\pgfsetroundjoin%
\pgfsetlinewidth{1.104125pt}%
\definecolor{currentstroke}{rgb}{1.000000,0.000000,0.000000}%
\pgfsetstrokecolor{currentstroke}%
\pgfsetdash{{1.100000pt}{1.815000pt}}{0.000000pt}%
\pgfpathmoveto{\pgfqpoint{0.365549in}{0.368326in}}%
\pgfpathlineto{\pgfqpoint{2.175248in}{0.368326in}}%
\pgfusepath{stroke}%
\end{pgfscope}%
\begin{pgfscope}%
\pgfpathrectangle{\pgfqpoint{0.365549in}{0.311728in}}{\pgfqpoint{1.809698in}{1.245154in}}%
\pgfusepath{clip}%
\pgfsetbuttcap%
\pgfsetroundjoin%
\pgfsetlinewidth{1.104125pt}%
\definecolor{currentstroke}{rgb}{0.145098,0.145098,0.145098}%
\pgfsetstrokecolor{currentstroke}%
\pgfsetdash{{1.100000pt}{1.815000pt}}{0.000000pt}%
\pgfpathmoveto{\pgfqpoint{0.365549in}{1.493540in}}%
\pgfpathlineto{\pgfqpoint{0.845184in}{1.485217in}}%
\pgfpathlineto{\pgfqpoint{0.897912in}{1.275356in}}%
\pgfpathlineto{\pgfqpoint{0.942132in}{1.266316in}}%
\pgfpathlineto{\pgfqpoint{0.974766in}{1.265455in}}%
\pgfusepath{stroke}%
\end{pgfscope}%
\begin{pgfscope}%
\pgfsetrectcap%
\pgfsetmiterjoin%
\pgfsetlinewidth{0.803000pt}%
\definecolor{currentstroke}{rgb}{0.000000,0.000000,0.000000}%
\pgfsetstrokecolor{currentstroke}%
\pgfsetdash{}{0pt}%
\pgfpathmoveto{\pgfqpoint{0.365549in}{0.311728in}}%
\pgfpathlineto{\pgfqpoint{0.365549in}{1.556882in}}%
\pgfusepath{stroke}%
\end{pgfscope}%
\begin{pgfscope}%
\pgfsetrectcap%
\pgfsetmiterjoin%
\pgfsetlinewidth{0.803000pt}%
\definecolor{currentstroke}{rgb}{0.000000,0.000000,0.000000}%
\pgfsetstrokecolor{currentstroke}%
\pgfsetdash{}{0pt}%
\pgfpathmoveto{\pgfqpoint{2.175248in}{0.311728in}}%
\pgfpathlineto{\pgfqpoint{2.175248in}{1.556882in}}%
\pgfusepath{stroke}%
\end{pgfscope}%
\begin{pgfscope}%
\pgfsetrectcap%
\pgfsetmiterjoin%
\pgfsetlinewidth{0.803000pt}%
\definecolor{currentstroke}{rgb}{0.000000,0.000000,0.000000}%
\pgfsetstrokecolor{currentstroke}%
\pgfsetdash{}{0pt}%
\pgfpathmoveto{\pgfqpoint{0.365549in}{0.311728in}}%
\pgfpathlineto{\pgfqpoint{2.175248in}{0.311728in}}%
\pgfusepath{stroke}%
\end{pgfscope}%
\begin{pgfscope}%
\pgfsetrectcap%
\pgfsetmiterjoin%
\pgfsetlinewidth{0.803000pt}%
\definecolor{currentstroke}{rgb}{0.000000,0.000000,0.000000}%
\pgfsetstrokecolor{currentstroke}%
\pgfsetdash{}{0pt}%
\pgfpathmoveto{\pgfqpoint{0.365549in}{1.556882in}}%
\pgfpathlineto{\pgfqpoint{2.175248in}{1.556882in}}%
\pgfusepath{stroke}%
\end{pgfscope}%
\begin{pgfscope}%
\pgfsetbuttcap%
\pgfsetmiterjoin%
\definecolor{currentfill}{rgb}{1.000000,1.000000,1.000000}%
\pgfsetfillcolor{currentfill}%
\pgfsetfillopacity{0.800000}%
\pgfsetlinewidth{1.003750pt}%
\definecolor{currentstroke}{rgb}{0.800000,0.800000,0.800000}%
\pgfsetstrokecolor{currentstroke}%
\pgfsetstrokeopacity{0.800000}%
\pgfsetdash{}{0pt}%
\pgfpathmoveto{\pgfqpoint{1.484203in}{0.909196in}}%
\pgfpathlineto{\pgfqpoint{2.116914in}{0.909196in}}%
\pgfpathquadraticcurveto{\pgfqpoint{2.133581in}{0.909196in}}{\pgfqpoint{2.133581in}{0.925863in}}%
\pgfpathlineto{\pgfqpoint{2.133581in}{1.498549in}}%
\pgfpathquadraticcurveto{\pgfqpoint{2.133581in}{1.515215in}}{\pgfqpoint{2.116914in}{1.515215in}}%
\pgfpathlineto{\pgfqpoint{1.484203in}{1.515215in}}%
\pgfpathquadraticcurveto{\pgfqpoint{1.467536in}{1.515215in}}{\pgfqpoint{1.467536in}{1.498549in}}%
\pgfpathlineto{\pgfqpoint{1.467536in}{0.925863in}}%
\pgfpathquadraticcurveto{\pgfqpoint{1.467536in}{0.909196in}}{\pgfqpoint{1.484203in}{0.909196in}}%
\pgfpathclose%
\pgfusepath{stroke,fill}%
\end{pgfscope}%
\begin{pgfscope}%
\pgfsetrectcap%
\pgfsetroundjoin%
\pgfsetlinewidth{1.104125pt}%
\definecolor{currentstroke}{rgb}{0.631373,0.850980,0.607843}%
\pgfsetstrokecolor{currentstroke}%
\pgfsetdash{}{0pt}%
\pgfpathmoveto{\pgfqpoint{1.500870in}{1.452715in}}%
\pgfpathlineto{\pgfqpoint{1.667536in}{1.452715in}}%
\pgfusepath{stroke}%
\end{pgfscope}%
\begin{pgfscope}%
\pgftext[x=1.734203in,y=1.423549in,left,base]{\rmfamily\fontsize{6.000000}{7.200000}\selectfont FD}%
\end{pgfscope}%
\begin{pgfscope}%
\pgfsetbuttcap%
\pgfsetroundjoin%
\pgfsetlinewidth{1.104125pt}%
\definecolor{currentstroke}{rgb}{0.631373,0.850980,0.607843}%
\pgfsetstrokecolor{currentstroke}%
\pgfsetdash{{5.500000pt}{1.100000pt}}{0.000000pt}%
\pgfpathmoveto{\pgfqpoint{1.500870in}{1.336511in}}%
\pgfpathlineto{\pgfqpoint{1.667536in}{1.336511in}}%
\pgfusepath{stroke}%
\end{pgfscope}%
\begin{pgfscope}%
\pgftext[x=1.734203in,y=1.307345in,left,base]{\rmfamily\fontsize{6.000000}{7.200000}\selectfont FD+6}%
\end{pgfscope}%
\begin{pgfscope}%
\pgfsetrectcap%
\pgfsetroundjoin%
\pgfsetlinewidth{1.104125pt}%
\definecolor{currentstroke}{rgb}{0.145098,0.145098,0.145098}%
\pgfsetstrokecolor{currentstroke}%
\pgfsetdash{}{0pt}%
\pgfpathmoveto{\pgfqpoint{1.500870in}{1.220308in}}%
\pgfpathlineto{\pgfqpoint{1.667536in}{1.220308in}}%
\pgfusepath{stroke}%
\end{pgfscope}%
\begin{pgfscope}%
\pgftext[x=1.734203in,y=1.191141in,left,base]{\rmfamily\fontsize{6.000000}{7.200000}\selectfont L2C2}%
\end{pgfscope}%
\begin{pgfscope}%
\pgfsetbuttcap%
\pgfsetroundjoin%
\pgfsetlinewidth{1.104125pt}%
\definecolor{currentstroke}{rgb}{0.145098,0.145098,0.145098}%
\pgfsetstrokecolor{currentstroke}%
\pgfsetdash{{5.500000pt}{1.100000pt}}{0.000000pt}%
\pgfpathmoveto{\pgfqpoint{1.500870in}{1.104104in}}%
\pgfpathlineto{\pgfqpoint{1.667536in}{1.104104in}}%
\pgfusepath{stroke}%
\end{pgfscope}%
\begin{pgfscope}%
\pgftext[x=1.734203in,y=1.074937in,left,base]{\rmfamily\fontsize{6.000000}{7.200000}\selectfont L2C2+6}%
\end{pgfscope}%
\begin{pgfscope}%
\pgfsetbuttcap%
\pgfsetroundjoin%
\pgfsetlinewidth{1.104125pt}%
\definecolor{currentstroke}{rgb}{1.000000,0.000000,0.000000}%
\pgfsetstrokecolor{currentstroke}%
\pgfsetdash{{1.100000pt}{1.815000pt}}{0.000000pt}%
\pgfpathmoveto{\pgfqpoint{1.500870in}{0.987900in}}%
\pgfpathlineto{\pgfqpoint{1.667536in}{0.987900in}}%
\pgfusepath{stroke}%
\end{pgfscope}%
\begin{pgfscope}%
\pgftext[x=1.734203in,y=0.958734in,left,base]{\rmfamily\fontsize{6.000000}{7.200000}\selectfont 0\% EC}%
\end{pgfscope}%
\end{pgfpicture}%
\makeatother%
\endgroup%

%% file: Figuras/ipc_cov025.pgf
\begingroup%
\makeatletter%
\begin{pgfpicture}%
\pgfpathrectangle{\pgfpointorigin}{\pgfqpoint{2.222509in}{1.591882in}}%
\pgfusepath{use as bounding box, clip}%
\begin{pgfscope}%
\pgfsetbuttcap%
\pgfsetmiterjoin%
\definecolor{currentfill}{rgb}{1.000000,1.000000,1.000000}%
\pgfsetfillcolor{currentfill}%
\pgfsetlinewidth{0.000000pt}%
\definecolor{currentstroke}{rgb}{1.000000,1.000000,1.000000}%
\pgfsetstrokecolor{currentstroke}%
\pgfsetdash{}{0pt}%
\pgfpathmoveto{\pgfqpoint{0.000000in}{0.000000in}}%
\pgfpathlineto{\pgfqpoint{2.222509in}{0.000000in}}%
\pgfpathlineto{\pgfqpoint{2.222509in}{1.591882in}}%
\pgfpathlineto{\pgfqpoint{0.000000in}{1.591882in}}%
\pgfpathclose%
\pgfusepath{fill}%
\end{pgfscope}%
\begin{pgfscope}%
\pgfsetbuttcap%
\pgfsetmiterjoin%
\definecolor{currentfill}{rgb}{1.000000,1.000000,1.000000}%
\pgfsetfillcolor{currentfill}%
\pgfsetlinewidth{0.000000pt}%
\definecolor{currentstroke}{rgb}{0.000000,0.000000,0.000000}%
\pgfsetstrokecolor{currentstroke}%
\pgfsetstrokeopacity{0.000000}%
\pgfsetdash{}{0pt}%
\pgfpathmoveto{\pgfqpoint{0.365549in}{0.311728in}}%
\pgfpathlineto{\pgfqpoint{2.175248in}{0.311728in}}%
\pgfpathlineto{\pgfqpoint{2.175248in}{1.556882in}}%
\pgfpathlineto{\pgfqpoint{0.365549in}{1.556882in}}%
\pgfpathclose%
\pgfusepath{fill}%
\end{pgfscope}%
\begin{pgfscope}%
\pgfsetbuttcap%
\pgfsetroundjoin%
\definecolor{currentfill}{rgb}{0.000000,0.000000,0.000000}%
\pgfsetfillcolor{currentfill}%
\pgfsetlinewidth{0.803000pt}%
\definecolor{currentstroke}{rgb}{0.000000,0.000000,0.000000}%
\pgfsetstrokecolor{currentstroke}%
\pgfsetdash{}{0pt}%
\pgfsys@defobject{currentmarker}{\pgfqpoint{0.000000in}{-0.048611in}}{\pgfqpoint{0.000000in}{0.000000in}}{%
\pgfpathmoveto{\pgfqpoint{0.000000in}{0.000000in}}%
\pgfpathlineto{\pgfqpoint{0.000000in}{-0.048611in}}%
\pgfusepath{stroke,fill}%
}%
\begin{pgfscope}%
\pgfsys@transformshift{0.365549in}{0.311728in}%
\pgfsys@useobject{currentmarker}{}%
\end{pgfscope}%
\end{pgfscope}%
\begin{pgfscope}%
\pgftext[x=0.365549in,y=0.214506in,,top]{\rmfamily\fontsize{5.000000}{6.000000}\selectfont \(\displaystyle 0\)}%
\end{pgfscope}%
\begin{pgfscope}%
\pgfsetbuttcap%
\pgfsetroundjoin%
\definecolor{currentfill}{rgb}{0.000000,0.000000,0.000000}%
\pgfsetfillcolor{currentfill}%
\pgfsetlinewidth{0.803000pt}%
\definecolor{currentstroke}{rgb}{0.000000,0.000000,0.000000}%
\pgfsetstrokecolor{currentstroke}%
\pgfsetdash{}{0pt}%
\pgfsys@defobject{currentmarker}{\pgfqpoint{0.000000in}{-0.048611in}}{\pgfqpoint{0.000000in}{0.000000in}}{%
\pgfpathmoveto{\pgfqpoint{0.000000in}{0.000000in}}%
\pgfpathlineto{\pgfqpoint{0.000000in}{-0.048611in}}%
\pgfusepath{stroke,fill}%
}%
\begin{pgfscope}%
\pgfsys@transformshift{0.817974in}{0.311728in}%
\pgfsys@useobject{currentmarker}{}%
\end{pgfscope}%
\end{pgfscope}%
\begin{pgfscope}%
\pgftext[x=0.817974in,y=0.214506in,,top]{\rmfamily\fontsize{5.000000}{6.000000}\selectfont \(\displaystyle 4\)}%
\end{pgfscope}%
\begin{pgfscope}%
\pgfsetbuttcap%
\pgfsetroundjoin%
\definecolor{currentfill}{rgb}{0.000000,0.000000,0.000000}%
\pgfsetfillcolor{currentfill}%
\pgfsetlinewidth{0.803000pt}%
\definecolor{currentstroke}{rgb}{0.000000,0.000000,0.000000}%
\pgfsetstrokecolor{currentstroke}%
\pgfsetdash{}{0pt}%
\pgfsys@defobject{currentmarker}{\pgfqpoint{0.000000in}{-0.048611in}}{\pgfqpoint{0.000000in}{0.000000in}}{%
\pgfpathmoveto{\pgfqpoint{0.000000in}{0.000000in}}%
\pgfpathlineto{\pgfqpoint{0.000000in}{-0.048611in}}%
\pgfusepath{stroke,fill}%
}%
\begin{pgfscope}%
\pgfsys@transformshift{1.270398in}{0.311728in}%
\pgfsys@useobject{currentmarker}{}%
\end{pgfscope}%
\end{pgfscope}%
\begin{pgfscope}%
\pgftext[x=1.270398in,y=0.214506in,,top]{\rmfamily\fontsize{5.000000}{6.000000}\selectfont \(\displaystyle 8\)}%
\end{pgfscope}%
\begin{pgfscope}%
\pgfsetbuttcap%
\pgfsetroundjoin%
\definecolor{currentfill}{rgb}{0.000000,0.000000,0.000000}%
\pgfsetfillcolor{currentfill}%
\pgfsetlinewidth{0.803000pt}%
\definecolor{currentstroke}{rgb}{0.000000,0.000000,0.000000}%
\pgfsetstrokecolor{currentstroke}%
\pgfsetdash{}{0pt}%
\pgfsys@defobject{currentmarker}{\pgfqpoint{0.000000in}{-0.048611in}}{\pgfqpoint{0.000000in}{0.000000in}}{%
\pgfpathmoveto{\pgfqpoint{0.000000in}{0.000000in}}%
\pgfpathlineto{\pgfqpoint{0.000000in}{-0.048611in}}%
\pgfusepath{stroke,fill}%
}%
\begin{pgfscope}%
\pgfsys@transformshift{1.722823in}{0.311728in}%
\pgfsys@useobject{currentmarker}{}%
\end{pgfscope}%
\end{pgfscope}%
\begin{pgfscope}%
\pgftext[x=1.722823in,y=0.214506in,,top]{\rmfamily\fontsize{5.000000}{6.000000}\selectfont \(\displaystyle 12\)}%
\end{pgfscope}%
\begin{pgfscope}%
\pgfsetbuttcap%
\pgfsetroundjoin%
\definecolor{currentfill}{rgb}{0.000000,0.000000,0.000000}%
\pgfsetfillcolor{currentfill}%
\pgfsetlinewidth{0.803000pt}%
\definecolor{currentstroke}{rgb}{0.000000,0.000000,0.000000}%
\pgfsetstrokecolor{currentstroke}%
\pgfsetdash{}{0pt}%
\pgfsys@defobject{currentmarker}{\pgfqpoint{0.000000in}{-0.048611in}}{\pgfqpoint{0.000000in}{0.000000in}}{%
\pgfpathmoveto{\pgfqpoint{0.000000in}{0.000000in}}%
\pgfpathlineto{\pgfqpoint{0.000000in}{-0.048611in}}%
\pgfusepath{stroke,fill}%
}%
\begin{pgfscope}%
\pgfsys@transformshift{2.175248in}{0.311728in}%
\pgfsys@useobject{currentmarker}{}%
\end{pgfscope}%
\end{pgfscope}%
\begin{pgfscope}%
\pgftext[x=2.175248in,y=0.214506in,,top]{\rmfamily\fontsize{5.000000}{6.000000}\selectfont \(\displaystyle 16\)}%
\end{pgfscope}%
\begin{pgfscope}%
\pgftext[x=1.270398in,y=0.097222in,,top]{\rmfamily\fontsize{7.000000}{8.400000}\selectfont Time (years)}%
\end{pgfscope}%
\begin{pgfscope}%
\pgfsetbuttcap%
\pgfsetroundjoin%
\definecolor{currentfill}{rgb}{0.000000,0.000000,0.000000}%
\pgfsetfillcolor{currentfill}%
\pgfsetlinewidth{0.803000pt}%
\definecolor{currentstroke}{rgb}{0.000000,0.000000,0.000000}%
\pgfsetstrokecolor{currentstroke}%
\pgfsetdash{}{0pt}%
\pgfsys@defobject{currentmarker}{\pgfqpoint{-0.048611in}{0.000000in}}{\pgfqpoint{0.000000in}{0.000000in}}{%
\pgfpathmoveto{\pgfqpoint{0.000000in}{0.000000in}}%
\pgfpathlineto{\pgfqpoint{-0.048611in}{0.000000in}}%
\pgfusepath{stroke,fill}%
}%
\begin{pgfscope}%
\pgfsys@transformshift{0.365549in}{0.423571in}%
\pgfsys@useobject{currentmarker}{}%
\end{pgfscope}%
\end{pgfscope}%
\begin{pgfscope}%
\pgftext[x=0.141975in,y=0.399459in,left,base]{\rmfamily\fontsize{5.000000}{6.000000}\selectfont \(\displaystyle 0.6\)}%
\end{pgfscope}%
\begin{pgfscope}%
\pgfsetbuttcap%
\pgfsetroundjoin%
\definecolor{currentfill}{rgb}{0.000000,0.000000,0.000000}%
\pgfsetfillcolor{currentfill}%
\pgfsetlinewidth{0.803000pt}%
\definecolor{currentstroke}{rgb}{0.000000,0.000000,0.000000}%
\pgfsetstrokecolor{currentstroke}%
\pgfsetdash{}{0pt}%
\pgfsys@defobject{currentmarker}{\pgfqpoint{-0.048611in}{0.000000in}}{\pgfqpoint{0.000000in}{0.000000in}}{%
\pgfpathmoveto{\pgfqpoint{0.000000in}{0.000000in}}%
\pgfpathlineto{\pgfqpoint{-0.048611in}{0.000000in}}%
\pgfusepath{stroke,fill}%
}%
\begin{pgfscope}%
\pgfsys@transformshift{0.365549in}{0.960959in}%
\pgfsys@useobject{currentmarker}{}%
\end{pgfscope}%
\end{pgfscope}%
\begin{pgfscope}%
\pgftext[x=0.141975in,y=0.936847in,left,base]{\rmfamily\fontsize{5.000000}{6.000000}\selectfont \(\displaystyle 0.8\)}%
\end{pgfscope}%
\begin{pgfscope}%
\pgfsetbuttcap%
\pgfsetroundjoin%
\definecolor{currentfill}{rgb}{0.000000,0.000000,0.000000}%
\pgfsetfillcolor{currentfill}%
\pgfsetlinewidth{0.803000pt}%
\definecolor{currentstroke}{rgb}{0.000000,0.000000,0.000000}%
\pgfsetstrokecolor{currentstroke}%
\pgfsetdash{}{0pt}%
\pgfsys@defobject{currentmarker}{\pgfqpoint{-0.048611in}{0.000000in}}{\pgfqpoint{0.000000in}{0.000000in}}{%
\pgfpathmoveto{\pgfqpoint{0.000000in}{0.000000in}}%
\pgfpathlineto{\pgfqpoint{-0.048611in}{0.000000in}}%
\pgfusepath{stroke,fill}%
}%
\begin{pgfscope}%
\pgfsys@transformshift{0.365549in}{1.498347in}%
\pgfsys@useobject{currentmarker}{}%
\end{pgfscope}%
\end{pgfscope}%
\begin{pgfscope}%
\pgftext[x=0.141975in,y=1.474234in,left,base]{\rmfamily\fontsize{5.000000}{6.000000}\selectfont \(\displaystyle 1.0\)}%
\end{pgfscope}%
\begin{pgfscope}%
\pgftext[x=0.086419in,y=0.934305in,,bottom,rotate=90.000000]{\rmfamily\fontsize{7.000000}{8.400000}\selectfont Normalized IPC}%
\end{pgfscope}%
\begin{pgfscope}%
\pgfpathrectangle{\pgfqpoint{0.365549in}{0.311728in}}{\pgfqpoint{1.809698in}{1.245154in}}%
\pgfusepath{clip}%
\pgfsetrectcap%
\pgfsetroundjoin%
\pgfsetlinewidth{0.501875pt}%
\definecolor{currentstroke}{rgb}{0.800000,0.800000,0.800000}%
\pgfsetstrokecolor{currentstroke}%
\pgfsetdash{}{0pt}%
\pgfpathmoveto{\pgfqpoint{0.478655in}{0.311728in}}%
\pgfpathlineto{\pgfqpoint{0.478655in}{1.556882in}}%
\pgfusepath{stroke}%
\end{pgfscope}%
\begin{pgfscope}%
\pgfpathrectangle{\pgfqpoint{0.365549in}{0.311728in}}{\pgfqpoint{1.809698in}{1.245154in}}%
\pgfusepath{clip}%
\pgfsetrectcap%
\pgfsetroundjoin%
\pgfsetlinewidth{0.501875pt}%
\definecolor{currentstroke}{rgb}{0.800000,0.800000,0.800000}%
\pgfsetstrokecolor{currentstroke}%
\pgfsetdash{}{0pt}%
\pgfpathmoveto{\pgfqpoint{0.591762in}{0.311728in}}%
\pgfpathlineto{\pgfqpoint{0.591762in}{1.556882in}}%
\pgfusepath{stroke}%
\end{pgfscope}%
\begin{pgfscope}%
\pgfpathrectangle{\pgfqpoint{0.365549in}{0.311728in}}{\pgfqpoint{1.809698in}{1.245154in}}%
\pgfusepath{clip}%
\pgfsetrectcap%
\pgfsetroundjoin%
\pgfsetlinewidth{0.501875pt}%
\definecolor{currentstroke}{rgb}{0.800000,0.800000,0.800000}%
\pgfsetstrokecolor{currentstroke}%
\pgfsetdash{}{0pt}%
\pgfpathmoveto{\pgfqpoint{0.704868in}{0.311728in}}%
\pgfpathlineto{\pgfqpoint{0.704868in}{1.556882in}}%
\pgfusepath{stroke}%
\end{pgfscope}%
\begin{pgfscope}%
\pgfpathrectangle{\pgfqpoint{0.365549in}{0.311728in}}{\pgfqpoint{1.809698in}{1.245154in}}%
\pgfusepath{clip}%
\pgfsetrectcap%
\pgfsetroundjoin%
\pgfsetlinewidth{0.501875pt}%
\definecolor{currentstroke}{rgb}{0.800000,0.800000,0.800000}%
\pgfsetstrokecolor{currentstroke}%
\pgfsetdash{}{0pt}%
\pgfpathmoveto{\pgfqpoint{0.817974in}{0.311728in}}%
\pgfpathlineto{\pgfqpoint{0.817974in}{1.556882in}}%
\pgfusepath{stroke}%
\end{pgfscope}%
\begin{pgfscope}%
\pgfpathrectangle{\pgfqpoint{0.365549in}{0.311728in}}{\pgfqpoint{1.809698in}{1.245154in}}%
\pgfusepath{clip}%
\pgfsetrectcap%
\pgfsetroundjoin%
\pgfsetlinewidth{0.501875pt}%
\definecolor{currentstroke}{rgb}{0.800000,0.800000,0.800000}%
\pgfsetstrokecolor{currentstroke}%
\pgfsetdash{}{0pt}%
\pgfpathmoveto{\pgfqpoint{0.931080in}{0.311728in}}%
\pgfpathlineto{\pgfqpoint{0.931080in}{1.556882in}}%
\pgfusepath{stroke}%
\end{pgfscope}%
\begin{pgfscope}%
\pgfpathrectangle{\pgfqpoint{0.365549in}{0.311728in}}{\pgfqpoint{1.809698in}{1.245154in}}%
\pgfusepath{clip}%
\pgfsetrectcap%
\pgfsetroundjoin%
\pgfsetlinewidth{0.501875pt}%
\definecolor{currentstroke}{rgb}{0.800000,0.800000,0.800000}%
\pgfsetstrokecolor{currentstroke}%
\pgfsetdash{}{0pt}%
\pgfpathmoveto{\pgfqpoint{1.044186in}{0.311728in}}%
\pgfpathlineto{\pgfqpoint{1.044186in}{1.556882in}}%
\pgfusepath{stroke}%
\end{pgfscope}%
\begin{pgfscope}%
\pgfpathrectangle{\pgfqpoint{0.365549in}{0.311728in}}{\pgfqpoint{1.809698in}{1.245154in}}%
\pgfusepath{clip}%
\pgfsetrectcap%
\pgfsetroundjoin%
\pgfsetlinewidth{0.501875pt}%
\definecolor{currentstroke}{rgb}{0.800000,0.800000,0.800000}%
\pgfsetstrokecolor{currentstroke}%
\pgfsetdash{}{0pt}%
\pgfpathmoveto{\pgfqpoint{1.157292in}{0.311728in}}%
\pgfpathlineto{\pgfqpoint{1.157292in}{1.556882in}}%
\pgfusepath{stroke}%
\end{pgfscope}%
\begin{pgfscope}%
\pgfpathrectangle{\pgfqpoint{0.365549in}{0.311728in}}{\pgfqpoint{1.809698in}{1.245154in}}%
\pgfusepath{clip}%
\pgfsetrectcap%
\pgfsetroundjoin%
\pgfsetlinewidth{0.501875pt}%
\definecolor{currentstroke}{rgb}{0.800000,0.800000,0.800000}%
\pgfsetstrokecolor{currentstroke}%
\pgfsetdash{}{0pt}%
\pgfpathmoveto{\pgfqpoint{1.270398in}{0.311728in}}%
\pgfpathlineto{\pgfqpoint{1.270398in}{1.556882in}}%
\pgfusepath{stroke}%
\end{pgfscope}%
\begin{pgfscope}%
\pgfpathrectangle{\pgfqpoint{0.365549in}{0.311728in}}{\pgfqpoint{1.809698in}{1.245154in}}%
\pgfusepath{clip}%
\pgfsetrectcap%
\pgfsetroundjoin%
\pgfsetlinewidth{0.501875pt}%
\definecolor{currentstroke}{rgb}{0.800000,0.800000,0.800000}%
\pgfsetstrokecolor{currentstroke}%
\pgfsetdash{}{0pt}%
\pgfpathmoveto{\pgfqpoint{1.383505in}{0.311728in}}%
\pgfpathlineto{\pgfqpoint{1.383505in}{1.556882in}}%
\pgfusepath{stroke}%
\end{pgfscope}%
\begin{pgfscope}%
\pgfpathrectangle{\pgfqpoint{0.365549in}{0.311728in}}{\pgfqpoint{1.809698in}{1.245154in}}%
\pgfusepath{clip}%
\pgfsetrectcap%
\pgfsetroundjoin%
\pgfsetlinewidth{0.501875pt}%
\definecolor{currentstroke}{rgb}{0.800000,0.800000,0.800000}%
\pgfsetstrokecolor{currentstroke}%
\pgfsetdash{}{0pt}%
\pgfpathmoveto{\pgfqpoint{1.496611in}{0.311728in}}%
\pgfpathlineto{\pgfqpoint{1.496611in}{1.556882in}}%
\pgfusepath{stroke}%
\end{pgfscope}%
\begin{pgfscope}%
\pgfpathrectangle{\pgfqpoint{0.365549in}{0.311728in}}{\pgfqpoint{1.809698in}{1.245154in}}%
\pgfusepath{clip}%
\pgfsetrectcap%
\pgfsetroundjoin%
\pgfsetlinewidth{0.501875pt}%
\definecolor{currentstroke}{rgb}{0.800000,0.800000,0.800000}%
\pgfsetstrokecolor{currentstroke}%
\pgfsetdash{}{0pt}%
\pgfpathmoveto{\pgfqpoint{1.609717in}{0.311728in}}%
\pgfpathlineto{\pgfqpoint{1.609717in}{1.556882in}}%
\pgfusepath{stroke}%
\end{pgfscope}%
\begin{pgfscope}%
\pgfpathrectangle{\pgfqpoint{0.365549in}{0.311728in}}{\pgfqpoint{1.809698in}{1.245154in}}%
\pgfusepath{clip}%
\pgfsetrectcap%
\pgfsetroundjoin%
\pgfsetlinewidth{0.501875pt}%
\definecolor{currentstroke}{rgb}{0.800000,0.800000,0.800000}%
\pgfsetstrokecolor{currentstroke}%
\pgfsetdash{}{0pt}%
\pgfpathmoveto{\pgfqpoint{1.722823in}{0.311728in}}%
\pgfpathlineto{\pgfqpoint{1.722823in}{1.556882in}}%
\pgfusepath{stroke}%
\end{pgfscope}%
\begin{pgfscope}%
\pgfpathrectangle{\pgfqpoint{0.365549in}{0.311728in}}{\pgfqpoint{1.809698in}{1.245154in}}%
\pgfusepath{clip}%
\pgfsetrectcap%
\pgfsetroundjoin%
\pgfsetlinewidth{0.501875pt}%
\definecolor{currentstroke}{rgb}{0.800000,0.800000,0.800000}%
\pgfsetstrokecolor{currentstroke}%
\pgfsetdash{}{0pt}%
\pgfpathmoveto{\pgfqpoint{1.835929in}{0.311728in}}%
\pgfpathlineto{\pgfqpoint{1.835929in}{1.556882in}}%
\pgfusepath{stroke}%
\end{pgfscope}%
\begin{pgfscope}%
\pgfpathrectangle{\pgfqpoint{0.365549in}{0.311728in}}{\pgfqpoint{1.809698in}{1.245154in}}%
\pgfusepath{clip}%
\pgfsetrectcap%
\pgfsetroundjoin%
\pgfsetlinewidth{0.501875pt}%
\definecolor{currentstroke}{rgb}{0.800000,0.800000,0.800000}%
\pgfsetstrokecolor{currentstroke}%
\pgfsetdash{}{0pt}%
\pgfpathmoveto{\pgfqpoint{1.949035in}{0.311728in}}%
\pgfpathlineto{\pgfqpoint{1.949035in}{1.556882in}}%
\pgfusepath{stroke}%
\end{pgfscope}%
\begin{pgfscope}%
\pgfpathrectangle{\pgfqpoint{0.365549in}{0.311728in}}{\pgfqpoint{1.809698in}{1.245154in}}%
\pgfusepath{clip}%
\pgfsetrectcap%
\pgfsetroundjoin%
\pgfsetlinewidth{0.501875pt}%
\definecolor{currentstroke}{rgb}{0.800000,0.800000,0.800000}%
\pgfsetstrokecolor{currentstroke}%
\pgfsetdash{}{0pt}%
\pgfpathmoveto{\pgfqpoint{2.062142in}{0.311728in}}%
\pgfpathlineto{\pgfqpoint{2.062142in}{1.556882in}}%
\pgfusepath{stroke}%
\end{pgfscope}%
\begin{pgfscope}%
\pgfpathrectangle{\pgfqpoint{0.365549in}{0.311728in}}{\pgfqpoint{1.809698in}{1.245154in}}%
\pgfusepath{clip}%
\pgfsetrectcap%
\pgfsetroundjoin%
\pgfsetlinewidth{1.104125pt}%
\definecolor{currentstroke}{rgb}{0.631373,0.850980,0.607843}%
\pgfsetstrokecolor{currentstroke}%
\pgfsetdash{}{0pt}%
\pgfpathmoveto{\pgfqpoint{0.365549in}{1.495979in}}%
\pgfpathlineto{\pgfqpoint{0.408611in}{1.485002in}}%
\pgfpathlineto{\pgfqpoint{0.430265in}{1.477612in}}%
\pgfpathlineto{\pgfqpoint{0.444532in}{1.468500in}}%
\pgfpathlineto{\pgfqpoint{0.455012in}{1.460823in}}%
\pgfpathlineto{\pgfqpoint{0.463654in}{1.446402in}}%
\pgfpathlineto{\pgfqpoint{0.470816in}{1.431191in}}%
\pgfpathlineto{\pgfqpoint{0.477002in}{1.412465in}}%
\pgfpathlineto{\pgfqpoint{0.482393in}{1.386493in}}%
\pgfpathlineto{\pgfqpoint{0.487055in}{1.346242in}}%
\pgfpathlineto{\pgfqpoint{0.491290in}{1.302261in}}%
\pgfpathlineto{\pgfqpoint{0.495170in}{1.251321in}}%
\pgfpathlineto{\pgfqpoint{0.498837in}{1.203680in}}%
\pgfpathlineto{\pgfqpoint{0.502323in}{1.151735in}}%
\pgfpathlineto{\pgfqpoint{0.505676in}{1.102373in}}%
\pgfpathlineto{\pgfqpoint{0.508882in}{1.054087in}}%
\pgfpathlineto{\pgfqpoint{0.510440in}{1.025173in}}%
\pgfusepath{stroke}%
\end{pgfscope}%
\begin{pgfscope}%
\pgfpathrectangle{\pgfqpoint{0.365549in}{0.311728in}}{\pgfqpoint{1.809698in}{1.245154in}}%
\pgfusepath{clip}%
\pgfsetbuttcap%
\pgfsetroundjoin%
\pgfsetlinewidth{1.104125pt}%
\definecolor{currentstroke}{rgb}{0.631373,0.850980,0.607843}%
\pgfsetstrokecolor{currentstroke}%
\pgfsetdash{{5.500000pt}{1.100000pt}}{0.000000pt}%
\pgfpathmoveto{\pgfqpoint{0.365549in}{1.500284in}}%
\pgfpathlineto{\pgfqpoint{0.616004in}{1.490957in}}%
\pgfpathlineto{\pgfqpoint{0.625167in}{1.482562in}}%
\pgfpathlineto{\pgfqpoint{0.630906in}{1.472877in}}%
\pgfpathlineto{\pgfqpoint{0.635109in}{1.465343in}}%
\pgfpathlineto{\pgfqpoint{0.638466in}{1.452931in}}%
\pgfpathlineto{\pgfqpoint{0.641266in}{1.439155in}}%
\pgfpathlineto{\pgfqpoint{0.643714in}{1.422725in}}%
\pgfpathlineto{\pgfqpoint{0.645818in}{1.404430in}}%
\pgfpathlineto{\pgfqpoint{0.647687in}{1.380035in}}%
\pgfpathlineto{\pgfqpoint{0.649413in}{1.343516in}}%
\pgfpathlineto{\pgfqpoint{0.651031in}{1.302620in}}%
\pgfpathlineto{\pgfqpoint{0.652544in}{1.252038in}}%
\pgfpathlineto{\pgfqpoint{0.653970in}{1.206909in}}%
\pgfpathlineto{\pgfqpoint{0.655321in}{1.154677in}}%
\pgfpathlineto{\pgfqpoint{0.656609in}{1.107037in}}%
\pgfpathlineto{\pgfqpoint{0.657906in}{1.054374in}}%
\pgfusepath{stroke}%
\end{pgfscope}%
\begin{pgfscope}%
\pgfpathrectangle{\pgfqpoint{0.365549in}{0.311728in}}{\pgfqpoint{1.809698in}{1.245154in}}%
\pgfusepath{clip}%
\pgfsetrectcap%
\pgfsetroundjoin%
\pgfsetlinewidth{1.104125pt}%
\definecolor{currentstroke}{rgb}{0.145098,0.145098,0.145098}%
\pgfsetstrokecolor{currentstroke}%
\pgfsetdash{}{0pt}%
\pgfpathmoveto{\pgfqpoint{0.365549in}{1.494185in}}%
\pgfpathlineto{\pgfqpoint{0.744857in}{1.391802in}}%
\pgfpathlineto{\pgfqpoint{0.832542in}{1.265814in}}%
\pgfpathlineto{\pgfqpoint{0.891817in}{1.262155in}}%
\pgfpathlineto{\pgfqpoint{0.935276in}{1.205618in}}%
\pgfpathlineto{\pgfqpoint{0.980532in}{1.156901in}}%
\pgfpathlineto{\pgfqpoint{1.023816in}{1.153314in}}%
\pgfpathlineto{\pgfqpoint{1.062619in}{0.964259in}}%
\pgfpathlineto{\pgfqpoint{1.150722in}{0.853769in}}%
\pgfpathlineto{\pgfqpoint{1.250169in}{0.853840in}}%
\pgfpathlineto{\pgfqpoint{1.343261in}{0.852405in}}%
\pgfpathlineto{\pgfqpoint{1.427985in}{0.828370in}}%
\pgfpathlineto{\pgfqpoint{1.520654in}{0.772551in}}%
\pgfpathlineto{\pgfqpoint{1.625156in}{0.765161in}}%
\pgfpathlineto{\pgfqpoint{1.722572in}{0.750381in}}%
\pgfpathlineto{\pgfqpoint{1.824107in}{0.617218in}}%
\pgfpathlineto{\pgfqpoint{2.002406in}{0.564914in}}%
\pgfusepath{stroke}%
\end{pgfscope}%
\begin{pgfscope}%
\pgfpathrectangle{\pgfqpoint{0.365549in}{0.311728in}}{\pgfqpoint{1.809698in}{1.245154in}}%
\pgfusepath{clip}%
\pgfsetbuttcap%
\pgfsetroundjoin%
\pgfsetlinewidth{1.104125pt}%
\definecolor{currentstroke}{rgb}{0.145098,0.145098,0.145098}%
\pgfsetstrokecolor{currentstroke}%
\pgfsetdash{{5.500000pt}{1.100000pt}}{0.000000pt}%
\pgfpathmoveto{\pgfqpoint{0.365549in}{1.494401in}}%
\pgfpathlineto{\pgfqpoint{0.897403in}{1.488015in}}%
\pgfpathlineto{\pgfqpoint{0.931289in}{1.306351in}}%
\pgfpathlineto{\pgfqpoint{0.961294in}{1.261724in}}%
\pgfpathlineto{\pgfqpoint{0.987641in}{1.214586in}}%
\pgfpathlineto{\pgfqpoint{1.016178in}{1.159125in}}%
\pgfpathlineto{\pgfqpoint{1.046892in}{1.152812in}}%
\pgfpathlineto{\pgfqpoint{1.075872in}{0.971506in}}%
\pgfpathlineto{\pgfqpoint{1.142864in}{0.854199in}}%
\pgfpathlineto{\pgfqpoint{1.226439in}{0.853625in}}%
\pgfpathlineto{\pgfqpoint{1.300346in}{0.852979in}}%
\pgfpathlineto{\pgfqpoint{1.370930in}{0.825141in}}%
\pgfpathlineto{\pgfqpoint{1.452046in}{0.771690in}}%
\pgfpathlineto{\pgfqpoint{1.542918in}{0.766739in}}%
\pgfpathlineto{\pgfqpoint{1.629880in}{0.758058in}}%
\pgfpathlineto{\pgfqpoint{1.718420in}{0.613487in}}%
\pgfpathlineto{\pgfqpoint{1.883369in}{0.565344in}}%
\pgfusepath{stroke}%
\end{pgfscope}%
\begin{pgfscope}%
\pgfpathrectangle{\pgfqpoint{0.365549in}{0.311728in}}{\pgfqpoint{1.809698in}{1.245154in}}%
\pgfusepath{clip}%
\pgfsetbuttcap%
\pgfsetroundjoin%
\pgfsetlinewidth{1.104125pt}%
\definecolor{currentstroke}{rgb}{1.000000,0.000000,0.000000}%
\pgfsetstrokecolor{currentstroke}%
\pgfsetdash{{1.100000pt}{1.815000pt}}{0.000000pt}%
\pgfpathmoveto{\pgfqpoint{0.365549in}{0.368326in}}%
\pgfpathlineto{\pgfqpoint{2.175248in}{0.368326in}}%
\pgfusepath{stroke}%
\end{pgfscope}%
\begin{pgfscope}%
\pgfpathrectangle{\pgfqpoint{0.365549in}{0.311728in}}{\pgfqpoint{1.809698in}{1.245154in}}%
\pgfusepath{clip}%
\pgfsetbuttcap%
\pgfsetroundjoin%
\pgfsetlinewidth{1.104125pt}%
\definecolor{currentstroke}{rgb}{0.145098,0.145098,0.145098}%
\pgfsetstrokecolor{currentstroke}%
\pgfsetdash{{1.100000pt}{1.815000pt}}{0.000000pt}%
\pgfpathmoveto{\pgfqpoint{0.365549in}{1.494185in}}%
\pgfpathlineto{\pgfqpoint{0.677901in}{1.486365in}}%
\pgfpathlineto{\pgfqpoint{0.742774in}{1.276576in}}%
\pgfpathlineto{\pgfqpoint{0.797532in}{1.266388in}}%
\pgfpathlineto{\pgfqpoint{0.838087in}{1.265383in}}%
\pgfusepath{stroke}%
\end{pgfscope}%
\begin{pgfscope}%
\pgfsetrectcap%
\pgfsetmiterjoin%
\pgfsetlinewidth{0.803000pt}%
\definecolor{currentstroke}{rgb}{0.000000,0.000000,0.000000}%
\pgfsetstrokecolor{currentstroke}%
\pgfsetdash{}{0pt}%
\pgfpathmoveto{\pgfqpoint{0.365549in}{0.311728in}}%
\pgfpathlineto{\pgfqpoint{0.365549in}{1.556882in}}%
\pgfusepath{stroke}%
\end{pgfscope}%
\begin{pgfscope}%
\pgfsetrectcap%
\pgfsetmiterjoin%
\pgfsetlinewidth{0.803000pt}%
\definecolor{currentstroke}{rgb}{0.000000,0.000000,0.000000}%
\pgfsetstrokecolor{currentstroke}%
\pgfsetdash{}{0pt}%
\pgfpathmoveto{\pgfqpoint{2.175248in}{0.311728in}}%
\pgfpathlineto{\pgfqpoint{2.175248in}{1.556882in}}%
\pgfusepath{stroke}%
\end{pgfscope}%
\begin{pgfscope}%
\pgfsetrectcap%
\pgfsetmiterjoin%
\pgfsetlinewidth{0.803000pt}%
\definecolor{currentstroke}{rgb}{0.000000,0.000000,0.000000}%
\pgfsetstrokecolor{currentstroke}%
\pgfsetdash{}{0pt}%
\pgfpathmoveto{\pgfqpoint{0.365549in}{0.311728in}}%
\pgfpathlineto{\pgfqpoint{2.175248in}{0.311728in}}%
\pgfusepath{stroke}%
\end{pgfscope}%
\begin{pgfscope}%
\pgfsetrectcap%
\pgfsetmiterjoin%
\pgfsetlinewidth{0.803000pt}%
\definecolor{currentstroke}{rgb}{0.000000,0.000000,0.000000}%
\pgfsetstrokecolor{currentstroke}%
\pgfsetdash{}{0pt}%
\pgfpathmoveto{\pgfqpoint{0.365549in}{1.556882in}}%
\pgfpathlineto{\pgfqpoint{2.175248in}{1.556882in}}%
\pgfusepath{stroke}%
\end{pgfscope}%
\begin{pgfscope}%
\pgfsetbuttcap%
\pgfsetmiterjoin%
\definecolor{currentfill}{rgb}{1.000000,1.000000,1.000000}%
\pgfsetfillcolor{currentfill}%
\pgfsetfillopacity{0.800000}%
\pgfsetlinewidth{1.003750pt}%
\definecolor{currentstroke}{rgb}{0.800000,0.800000,0.800000}%
\pgfsetstrokecolor{currentstroke}%
\pgfsetstrokeopacity{0.800000}%
\pgfsetdash{}{0pt}%
\pgfpathmoveto{\pgfqpoint{1.484203in}{0.909196in}}%
\pgfpathlineto{\pgfqpoint{2.116914in}{0.909196in}}%
\pgfpathquadraticcurveto{\pgfqpoint{2.133581in}{0.909196in}}{\pgfqpoint{2.133581in}{0.925863in}}%
\pgfpathlineto{\pgfqpoint{2.133581in}{1.498549in}}%
\pgfpathquadraticcurveto{\pgfqpoint{2.133581in}{1.515215in}}{\pgfqpoint{2.116914in}{1.515215in}}%
\pgfpathlineto{\pgfqpoint{1.484203in}{1.515215in}}%
\pgfpathquadraticcurveto{\pgfqpoint{1.467536in}{1.515215in}}{\pgfqpoint{1.467536in}{1.498549in}}%
\pgfpathlineto{\pgfqpoint{1.467536in}{0.925863in}}%
\pgfpathquadraticcurveto{\pgfqpoint{1.467536in}{0.909196in}}{\pgfqpoint{1.484203in}{0.909196in}}%
\pgfpathclose%
\pgfusepath{stroke,fill}%
\end{pgfscope}%
\begin{pgfscope}%
\pgfsetrectcap%
\pgfsetroundjoin%
\pgfsetlinewidth{1.104125pt}%
\definecolor{currentstroke}{rgb}{0.631373,0.850980,0.607843}%
\pgfsetstrokecolor{currentstroke}%
\pgfsetdash{}{0pt}%
\pgfpathmoveto{\pgfqpoint{1.500870in}{1.452715in}}%
\pgfpathlineto{\pgfqpoint{1.667536in}{1.452715in}}%
\pgfusepath{stroke}%
\end{pgfscope}%
\begin{pgfscope}%
\pgftext[x=1.734203in,y=1.423549in,left,base]{\rmfamily\fontsize{6.000000}{7.200000}\selectfont FD}%
\end{pgfscope}%
\begin{pgfscope}%
\pgfsetbuttcap%
\pgfsetroundjoin%
\pgfsetlinewidth{1.104125pt}%
\definecolor{currentstroke}{rgb}{0.631373,0.850980,0.607843}%
\pgfsetstrokecolor{currentstroke}%
\pgfsetdash{{5.500000pt}{1.100000pt}}{0.000000pt}%
\pgfpathmoveto{\pgfqpoint{1.500870in}{1.336511in}}%
\pgfpathlineto{\pgfqpoint{1.667536in}{1.336511in}}%
\pgfusepath{stroke}%
\end{pgfscope}%
\begin{pgfscope}%
\pgftext[x=1.734203in,y=1.307345in,left,base]{\rmfamily\fontsize{6.000000}{7.200000}\selectfont FD+6}%
\end{pgfscope}%
\begin{pgfscope}%
\pgfsetrectcap%
\pgfsetroundjoin%
\pgfsetlinewidth{1.104125pt}%
\definecolor{currentstroke}{rgb}{0.145098,0.145098,0.145098}%
\pgfsetstrokecolor{currentstroke}%
\pgfsetdash{}{0pt}%
\pgfpathmoveto{\pgfqpoint{1.500870in}{1.220308in}}%
\pgfpathlineto{\pgfqpoint{1.667536in}{1.220308in}}%
\pgfusepath{stroke}%
\end{pgfscope}%
\begin{pgfscope}%
\pgftext[x=1.734203in,y=1.191141in,left,base]{\rmfamily\fontsize{6.000000}{7.200000}\selectfont L2C2}%
\end{pgfscope}%
\begin{pgfscope}%
\pgfsetbuttcap%
\pgfsetroundjoin%
\pgfsetlinewidth{1.104125pt}%
\definecolor{currentstroke}{rgb}{0.145098,0.145098,0.145098}%
\pgfsetstrokecolor{currentstroke}%
\pgfsetdash{{5.500000pt}{1.100000pt}}{0.000000pt}%
\pgfpathmoveto{\pgfqpoint{1.500870in}{1.104104in}}%
\pgfpathlineto{\pgfqpoint{1.667536in}{1.104104in}}%
\pgfusepath{stroke}%
\end{pgfscope}%
\begin{pgfscope}%
\pgftext[x=1.734203in,y=1.074937in,left,base]{\rmfamily\fontsize{6.000000}{7.200000}\selectfont L2C2+6}%
\end{pgfscope}%
\begin{pgfscope}%
\pgfsetbuttcap%
\pgfsetroundjoin%
\pgfsetlinewidth{1.104125pt}%
\definecolor{currentstroke}{rgb}{1.000000,0.000000,0.000000}%
\pgfsetstrokecolor{currentstroke}%
\pgfsetdash{{1.100000pt}{1.815000pt}}{0.000000pt}%
\pgfpathmoveto{\pgfqpoint{1.500870in}{0.987900in}}%
\pgfpathlineto{\pgfqpoint{1.667536in}{0.987900in}}%
\pgfusepath{stroke}%
\end{pgfscope}%
\begin{pgfscope}%
\pgftext[x=1.734203in,y=0.958734in,left,base]{\rmfamily\fontsize{6.000000}{7.200000}\selectfont 0\% EC}%
\end{pgfscope}%
\end{pgfpicture}%
\makeatother%
\endgroup%

%% file: Figuras/ipc_cov030.pgf
\begingroup%
\makeatletter%
\begin{pgfpicture}%
\pgfpathrectangle{\pgfpointorigin}{\pgfqpoint{2.222509in}{1.591882in}}%
\pgfusepath{use as bounding box, clip}%
\begin{pgfscope}%
\pgfsetbuttcap%
\pgfsetmiterjoin%
\definecolor{currentfill}{rgb}{1.000000,1.000000,1.000000}%
\pgfsetfillcolor{currentfill}%
\pgfsetlinewidth{0.000000pt}%
\definecolor{currentstroke}{rgb}{1.000000,1.000000,1.000000}%
\pgfsetstrokecolor{currentstroke}%
\pgfsetdash{}{0pt}%
\pgfpathmoveto{\pgfqpoint{0.000000in}{0.000000in}}%
\pgfpathlineto{\pgfqpoint{2.222509in}{0.000000in}}%
\pgfpathlineto{\pgfqpoint{2.222509in}{1.591882in}}%
\pgfpathlineto{\pgfqpoint{0.000000in}{1.591882in}}%
\pgfpathclose%
\pgfusepath{fill}%
\end{pgfscope}%
\begin{pgfscope}%
\pgfsetbuttcap%
\pgfsetmiterjoin%
\definecolor{currentfill}{rgb}{1.000000,1.000000,1.000000}%
\pgfsetfillcolor{currentfill}%
\pgfsetlinewidth{0.000000pt}%
\definecolor{currentstroke}{rgb}{0.000000,0.000000,0.000000}%
\pgfsetstrokecolor{currentstroke}%
\pgfsetstrokeopacity{0.000000}%
\pgfsetdash{}{0pt}%
\pgfpathmoveto{\pgfqpoint{0.365549in}{0.311728in}}%
\pgfpathlineto{\pgfqpoint{2.175248in}{0.311728in}}%
\pgfpathlineto{\pgfqpoint{2.175248in}{1.556882in}}%
\pgfpathlineto{\pgfqpoint{0.365549in}{1.556882in}}%
\pgfpathclose%
\pgfusepath{fill}%
\end{pgfscope}%
\begin{pgfscope}%
\pgfsetbuttcap%
\pgfsetroundjoin%
\definecolor{currentfill}{rgb}{0.000000,0.000000,0.000000}%
\pgfsetfillcolor{currentfill}%
\pgfsetlinewidth{0.803000pt}%
\definecolor{currentstroke}{rgb}{0.000000,0.000000,0.000000}%
\pgfsetstrokecolor{currentstroke}%
\pgfsetdash{}{0pt}%
\pgfsys@defobject{currentmarker}{\pgfqpoint{0.000000in}{-0.048611in}}{\pgfqpoint{0.000000in}{0.000000in}}{%
\pgfpathmoveto{\pgfqpoint{0.000000in}{0.000000in}}%
\pgfpathlineto{\pgfqpoint{0.000000in}{-0.048611in}}%
\pgfusepath{stroke,fill}%
}%
\begin{pgfscope}%
\pgfsys@transformshift{0.365549in}{0.311728in}%
\pgfsys@useobject{currentmarker}{}%
\end{pgfscope}%
\end{pgfscope}%
\begin{pgfscope}%
\pgftext[x=0.365549in,y=0.214506in,,top]{\rmfamily\fontsize{5.000000}{6.000000}\selectfont \(\displaystyle 0\)}%
\end{pgfscope}%
\begin{pgfscope}%
\pgfsetbuttcap%
\pgfsetroundjoin%
\definecolor{currentfill}{rgb}{0.000000,0.000000,0.000000}%
\pgfsetfillcolor{currentfill}%
\pgfsetlinewidth{0.803000pt}%
\definecolor{currentstroke}{rgb}{0.000000,0.000000,0.000000}%
\pgfsetstrokecolor{currentstroke}%
\pgfsetdash{}{0pt}%
\pgfsys@defobject{currentmarker}{\pgfqpoint{0.000000in}{-0.048611in}}{\pgfqpoint{0.000000in}{0.000000in}}{%
\pgfpathmoveto{\pgfqpoint{0.000000in}{0.000000in}}%
\pgfpathlineto{\pgfqpoint{0.000000in}{-0.048611in}}%
\pgfusepath{stroke,fill}%
}%
\begin{pgfscope}%
\pgfsys@transformshift{0.817974in}{0.311728in}%
\pgfsys@useobject{currentmarker}{}%
\end{pgfscope}%
\end{pgfscope}%
\begin{pgfscope}%
\pgftext[x=0.817974in,y=0.214506in,,top]{\rmfamily\fontsize{5.000000}{6.000000}\selectfont \(\displaystyle 4\)}%
\end{pgfscope}%
\begin{pgfscope}%
\pgfsetbuttcap%
\pgfsetroundjoin%
\definecolor{currentfill}{rgb}{0.000000,0.000000,0.000000}%
\pgfsetfillcolor{currentfill}%
\pgfsetlinewidth{0.803000pt}%
\definecolor{currentstroke}{rgb}{0.000000,0.000000,0.000000}%
\pgfsetstrokecolor{currentstroke}%
\pgfsetdash{}{0pt}%
\pgfsys@defobject{currentmarker}{\pgfqpoint{0.000000in}{-0.048611in}}{\pgfqpoint{0.000000in}{0.000000in}}{%
\pgfpathmoveto{\pgfqpoint{0.000000in}{0.000000in}}%
\pgfpathlineto{\pgfqpoint{0.000000in}{-0.048611in}}%
\pgfusepath{stroke,fill}%
}%
\begin{pgfscope}%
\pgfsys@transformshift{1.270398in}{0.311728in}%
\pgfsys@useobject{currentmarker}{}%
\end{pgfscope}%
\end{pgfscope}%
\begin{pgfscope}%
\pgftext[x=1.270398in,y=0.214506in,,top]{\rmfamily\fontsize{5.000000}{6.000000}\selectfont \(\displaystyle 8\)}%
\end{pgfscope}%
\begin{pgfscope}%
\pgfsetbuttcap%
\pgfsetroundjoin%
\definecolor{currentfill}{rgb}{0.000000,0.000000,0.000000}%
\pgfsetfillcolor{currentfill}%
\pgfsetlinewidth{0.803000pt}%
\definecolor{currentstroke}{rgb}{0.000000,0.000000,0.000000}%
\pgfsetstrokecolor{currentstroke}%
\pgfsetdash{}{0pt}%
\pgfsys@defobject{currentmarker}{\pgfqpoint{0.000000in}{-0.048611in}}{\pgfqpoint{0.000000in}{0.000000in}}{%
\pgfpathmoveto{\pgfqpoint{0.000000in}{0.000000in}}%
\pgfpathlineto{\pgfqpoint{0.000000in}{-0.048611in}}%
\pgfusepath{stroke,fill}%
}%
\begin{pgfscope}%
\pgfsys@transformshift{1.722823in}{0.311728in}%
\pgfsys@useobject{currentmarker}{}%
\end{pgfscope}%
\end{pgfscope}%
\begin{pgfscope}%
\pgftext[x=1.722823in,y=0.214506in,,top]{\rmfamily\fontsize{5.000000}{6.000000}\selectfont \(\displaystyle 12\)}%
\end{pgfscope}%
\begin{pgfscope}%
\pgfsetbuttcap%
\pgfsetroundjoin%
\definecolor{currentfill}{rgb}{0.000000,0.000000,0.000000}%
\pgfsetfillcolor{currentfill}%
\pgfsetlinewidth{0.803000pt}%
\definecolor{currentstroke}{rgb}{0.000000,0.000000,0.000000}%
\pgfsetstrokecolor{currentstroke}%
\pgfsetdash{}{0pt}%
\pgfsys@defobject{currentmarker}{\pgfqpoint{0.000000in}{-0.048611in}}{\pgfqpoint{0.000000in}{0.000000in}}{%
\pgfpathmoveto{\pgfqpoint{0.000000in}{0.000000in}}%
\pgfpathlineto{\pgfqpoint{0.000000in}{-0.048611in}}%
\pgfusepath{stroke,fill}%
}%
\begin{pgfscope}%
\pgfsys@transformshift{2.175248in}{0.311728in}%
\pgfsys@useobject{currentmarker}{}%
\end{pgfscope}%
\end{pgfscope}%
\begin{pgfscope}%
\pgftext[x=2.175248in,y=0.214506in,,top]{\rmfamily\fontsize{5.000000}{6.000000}\selectfont \(\displaystyle 16\)}%
\end{pgfscope}%
\begin{pgfscope}%
\pgftext[x=1.270398in,y=0.097222in,,top]{\rmfamily\fontsize{7.000000}{8.400000}\selectfont Time (years)}%
\end{pgfscope}%
\begin{pgfscope}%
\pgfsetbuttcap%
\pgfsetroundjoin%
\definecolor{currentfill}{rgb}{0.000000,0.000000,0.000000}%
\pgfsetfillcolor{currentfill}%
\pgfsetlinewidth{0.803000pt}%
\definecolor{currentstroke}{rgb}{0.000000,0.000000,0.000000}%
\pgfsetstrokecolor{currentstroke}%
\pgfsetdash{}{0pt}%
\pgfsys@defobject{currentmarker}{\pgfqpoint{-0.048611in}{0.000000in}}{\pgfqpoint{0.000000in}{0.000000in}}{%
\pgfpathmoveto{\pgfqpoint{0.000000in}{0.000000in}}%
\pgfpathlineto{\pgfqpoint{-0.048611in}{0.000000in}}%
\pgfusepath{stroke,fill}%
}%
\begin{pgfscope}%
\pgfsys@transformshift{0.365549in}{0.423571in}%
\pgfsys@useobject{currentmarker}{}%
\end{pgfscope}%
\end{pgfscope}%
\begin{pgfscope}%
\pgftext[x=0.141975in,y=0.399459in,left,base]{\rmfamily\fontsize{5.000000}{6.000000}\selectfont \(\displaystyle 0.6\)}%
\end{pgfscope}%
\begin{pgfscope}%
\pgfsetbuttcap%
\pgfsetroundjoin%
\definecolor{currentfill}{rgb}{0.000000,0.000000,0.000000}%
\pgfsetfillcolor{currentfill}%
\pgfsetlinewidth{0.803000pt}%
\definecolor{currentstroke}{rgb}{0.000000,0.000000,0.000000}%
\pgfsetstrokecolor{currentstroke}%
\pgfsetdash{}{0pt}%
\pgfsys@defobject{currentmarker}{\pgfqpoint{-0.048611in}{0.000000in}}{\pgfqpoint{0.000000in}{0.000000in}}{%
\pgfpathmoveto{\pgfqpoint{0.000000in}{0.000000in}}%
\pgfpathlineto{\pgfqpoint{-0.048611in}{0.000000in}}%
\pgfusepath{stroke,fill}%
}%
\begin{pgfscope}%
\pgfsys@transformshift{0.365549in}{0.960959in}%
\pgfsys@useobject{currentmarker}{}%
\end{pgfscope}%
\end{pgfscope}%
\begin{pgfscope}%
\pgftext[x=0.141975in,y=0.936847in,left,base]{\rmfamily\fontsize{5.000000}{6.000000}\selectfont \(\displaystyle 0.8\)}%
\end{pgfscope}%
\begin{pgfscope}%
\pgfsetbuttcap%
\pgfsetroundjoin%
\definecolor{currentfill}{rgb}{0.000000,0.000000,0.000000}%
\pgfsetfillcolor{currentfill}%
\pgfsetlinewidth{0.803000pt}%
\definecolor{currentstroke}{rgb}{0.000000,0.000000,0.000000}%
\pgfsetstrokecolor{currentstroke}%
\pgfsetdash{}{0pt}%
\pgfsys@defobject{currentmarker}{\pgfqpoint{-0.048611in}{0.000000in}}{\pgfqpoint{0.000000in}{0.000000in}}{%
\pgfpathmoveto{\pgfqpoint{0.000000in}{0.000000in}}%
\pgfpathlineto{\pgfqpoint{-0.048611in}{0.000000in}}%
\pgfusepath{stroke,fill}%
}%
\begin{pgfscope}%
\pgfsys@transformshift{0.365549in}{1.498347in}%
\pgfsys@useobject{currentmarker}{}%
\end{pgfscope}%
\end{pgfscope}%
\begin{pgfscope}%
\pgftext[x=0.141975in,y=1.474234in,left,base]{\rmfamily\fontsize{5.000000}{6.000000}\selectfont \(\displaystyle 1.0\)}%
\end{pgfscope}%
\begin{pgfscope}%
\pgftext[x=0.086419in,y=0.934305in,,bottom,rotate=90.000000]{\rmfamily\fontsize{7.000000}{8.400000}\selectfont Normalized IPC}%
\end{pgfscope}%
\begin{pgfscope}%
\pgfpathrectangle{\pgfqpoint{0.365549in}{0.311728in}}{\pgfqpoint{1.809698in}{1.245154in}}%
\pgfusepath{clip}%
\pgfsetrectcap%
\pgfsetroundjoin%
\pgfsetlinewidth{0.501875pt}%
\definecolor{currentstroke}{rgb}{0.800000,0.800000,0.800000}%
\pgfsetstrokecolor{currentstroke}%
\pgfsetdash{}{0pt}%
\pgfpathmoveto{\pgfqpoint{0.478655in}{0.311728in}}%
\pgfpathlineto{\pgfqpoint{0.478655in}{1.556882in}}%
\pgfusepath{stroke}%
\end{pgfscope}%
\begin{pgfscope}%
\pgfpathrectangle{\pgfqpoint{0.365549in}{0.311728in}}{\pgfqpoint{1.809698in}{1.245154in}}%
\pgfusepath{clip}%
\pgfsetrectcap%
\pgfsetroundjoin%
\pgfsetlinewidth{0.501875pt}%
\definecolor{currentstroke}{rgb}{0.800000,0.800000,0.800000}%
\pgfsetstrokecolor{currentstroke}%
\pgfsetdash{}{0pt}%
\pgfpathmoveto{\pgfqpoint{0.591762in}{0.311728in}}%
\pgfpathlineto{\pgfqpoint{0.591762in}{1.556882in}}%
\pgfusepath{stroke}%
\end{pgfscope}%
\begin{pgfscope}%
\pgfpathrectangle{\pgfqpoint{0.365549in}{0.311728in}}{\pgfqpoint{1.809698in}{1.245154in}}%
\pgfusepath{clip}%
\pgfsetrectcap%
\pgfsetroundjoin%
\pgfsetlinewidth{0.501875pt}%
\definecolor{currentstroke}{rgb}{0.800000,0.800000,0.800000}%
\pgfsetstrokecolor{currentstroke}%
\pgfsetdash{}{0pt}%
\pgfpathmoveto{\pgfqpoint{0.704868in}{0.311728in}}%
\pgfpathlineto{\pgfqpoint{0.704868in}{1.556882in}}%
\pgfusepath{stroke}%
\end{pgfscope}%
\begin{pgfscope}%
\pgfpathrectangle{\pgfqpoint{0.365549in}{0.311728in}}{\pgfqpoint{1.809698in}{1.245154in}}%
\pgfusepath{clip}%
\pgfsetrectcap%
\pgfsetroundjoin%
\pgfsetlinewidth{0.501875pt}%
\definecolor{currentstroke}{rgb}{0.800000,0.800000,0.800000}%
\pgfsetstrokecolor{currentstroke}%
\pgfsetdash{}{0pt}%
\pgfpathmoveto{\pgfqpoint{0.817974in}{0.311728in}}%
\pgfpathlineto{\pgfqpoint{0.817974in}{1.556882in}}%
\pgfusepath{stroke}%
\end{pgfscope}%
\begin{pgfscope}%
\pgfpathrectangle{\pgfqpoint{0.365549in}{0.311728in}}{\pgfqpoint{1.809698in}{1.245154in}}%
\pgfusepath{clip}%
\pgfsetrectcap%
\pgfsetroundjoin%
\pgfsetlinewidth{0.501875pt}%
\definecolor{currentstroke}{rgb}{0.800000,0.800000,0.800000}%
\pgfsetstrokecolor{currentstroke}%
\pgfsetdash{}{0pt}%
\pgfpathmoveto{\pgfqpoint{0.931080in}{0.311728in}}%
\pgfpathlineto{\pgfqpoint{0.931080in}{1.556882in}}%
\pgfusepath{stroke}%
\end{pgfscope}%
\begin{pgfscope}%
\pgfpathrectangle{\pgfqpoint{0.365549in}{0.311728in}}{\pgfqpoint{1.809698in}{1.245154in}}%
\pgfusepath{clip}%
\pgfsetrectcap%
\pgfsetroundjoin%
\pgfsetlinewidth{0.501875pt}%
\definecolor{currentstroke}{rgb}{0.800000,0.800000,0.800000}%
\pgfsetstrokecolor{currentstroke}%
\pgfsetdash{}{0pt}%
\pgfpathmoveto{\pgfqpoint{1.044186in}{0.311728in}}%
\pgfpathlineto{\pgfqpoint{1.044186in}{1.556882in}}%
\pgfusepath{stroke}%
\end{pgfscope}%
\begin{pgfscope}%
\pgfpathrectangle{\pgfqpoint{0.365549in}{0.311728in}}{\pgfqpoint{1.809698in}{1.245154in}}%
\pgfusepath{clip}%
\pgfsetrectcap%
\pgfsetroundjoin%
\pgfsetlinewidth{0.501875pt}%
\definecolor{currentstroke}{rgb}{0.800000,0.800000,0.800000}%
\pgfsetstrokecolor{currentstroke}%
\pgfsetdash{}{0pt}%
\pgfpathmoveto{\pgfqpoint{1.157292in}{0.311728in}}%
\pgfpathlineto{\pgfqpoint{1.157292in}{1.556882in}}%
\pgfusepath{stroke}%
\end{pgfscope}%
\begin{pgfscope}%
\pgfpathrectangle{\pgfqpoint{0.365549in}{0.311728in}}{\pgfqpoint{1.809698in}{1.245154in}}%
\pgfusepath{clip}%
\pgfsetrectcap%
\pgfsetroundjoin%
\pgfsetlinewidth{0.501875pt}%
\definecolor{currentstroke}{rgb}{0.800000,0.800000,0.800000}%
\pgfsetstrokecolor{currentstroke}%
\pgfsetdash{}{0pt}%
\pgfpathmoveto{\pgfqpoint{1.270398in}{0.311728in}}%
\pgfpathlineto{\pgfqpoint{1.270398in}{1.556882in}}%
\pgfusepath{stroke}%
\end{pgfscope}%
\begin{pgfscope}%
\pgfpathrectangle{\pgfqpoint{0.365549in}{0.311728in}}{\pgfqpoint{1.809698in}{1.245154in}}%
\pgfusepath{clip}%
\pgfsetrectcap%
\pgfsetroundjoin%
\pgfsetlinewidth{0.501875pt}%
\definecolor{currentstroke}{rgb}{0.800000,0.800000,0.800000}%
\pgfsetstrokecolor{currentstroke}%
\pgfsetdash{}{0pt}%
\pgfpathmoveto{\pgfqpoint{1.383505in}{0.311728in}}%
\pgfpathlineto{\pgfqpoint{1.383505in}{1.556882in}}%
\pgfusepath{stroke}%
\end{pgfscope}%
\begin{pgfscope}%
\pgfpathrectangle{\pgfqpoint{0.365549in}{0.311728in}}{\pgfqpoint{1.809698in}{1.245154in}}%
\pgfusepath{clip}%
\pgfsetrectcap%
\pgfsetroundjoin%
\pgfsetlinewidth{0.501875pt}%
\definecolor{currentstroke}{rgb}{0.800000,0.800000,0.800000}%
\pgfsetstrokecolor{currentstroke}%
\pgfsetdash{}{0pt}%
\pgfpathmoveto{\pgfqpoint{1.496611in}{0.311728in}}%
\pgfpathlineto{\pgfqpoint{1.496611in}{1.556882in}}%
\pgfusepath{stroke}%
\end{pgfscope}%
\begin{pgfscope}%
\pgfpathrectangle{\pgfqpoint{0.365549in}{0.311728in}}{\pgfqpoint{1.809698in}{1.245154in}}%
\pgfusepath{clip}%
\pgfsetrectcap%
\pgfsetroundjoin%
\pgfsetlinewidth{0.501875pt}%
\definecolor{currentstroke}{rgb}{0.800000,0.800000,0.800000}%
\pgfsetstrokecolor{currentstroke}%
\pgfsetdash{}{0pt}%
\pgfpathmoveto{\pgfqpoint{1.609717in}{0.311728in}}%
\pgfpathlineto{\pgfqpoint{1.609717in}{1.556882in}}%
\pgfusepath{stroke}%
\end{pgfscope}%
\begin{pgfscope}%
\pgfpathrectangle{\pgfqpoint{0.365549in}{0.311728in}}{\pgfqpoint{1.809698in}{1.245154in}}%
\pgfusepath{clip}%
\pgfsetrectcap%
\pgfsetroundjoin%
\pgfsetlinewidth{0.501875pt}%
\definecolor{currentstroke}{rgb}{0.800000,0.800000,0.800000}%
\pgfsetstrokecolor{currentstroke}%
\pgfsetdash{}{0pt}%
\pgfpathmoveto{\pgfqpoint{1.722823in}{0.311728in}}%
\pgfpathlineto{\pgfqpoint{1.722823in}{1.556882in}}%
\pgfusepath{stroke}%
\end{pgfscope}%
\begin{pgfscope}%
\pgfpathrectangle{\pgfqpoint{0.365549in}{0.311728in}}{\pgfqpoint{1.809698in}{1.245154in}}%
\pgfusepath{clip}%
\pgfsetrectcap%
\pgfsetroundjoin%
\pgfsetlinewidth{0.501875pt}%
\definecolor{currentstroke}{rgb}{0.800000,0.800000,0.800000}%
\pgfsetstrokecolor{currentstroke}%
\pgfsetdash{}{0pt}%
\pgfpathmoveto{\pgfqpoint{1.835929in}{0.311728in}}%
\pgfpathlineto{\pgfqpoint{1.835929in}{1.556882in}}%
\pgfusepath{stroke}%
\end{pgfscope}%
\begin{pgfscope}%
\pgfpathrectangle{\pgfqpoint{0.365549in}{0.311728in}}{\pgfqpoint{1.809698in}{1.245154in}}%
\pgfusepath{clip}%
\pgfsetrectcap%
\pgfsetroundjoin%
\pgfsetlinewidth{0.501875pt}%
\definecolor{currentstroke}{rgb}{0.800000,0.800000,0.800000}%
\pgfsetstrokecolor{currentstroke}%
\pgfsetdash{}{0pt}%
\pgfpathmoveto{\pgfqpoint{1.949035in}{0.311728in}}%
\pgfpathlineto{\pgfqpoint{1.949035in}{1.556882in}}%
\pgfusepath{stroke}%
\end{pgfscope}%
\begin{pgfscope}%
\pgfpathrectangle{\pgfqpoint{0.365549in}{0.311728in}}{\pgfqpoint{1.809698in}{1.245154in}}%
\pgfusepath{clip}%
\pgfsetrectcap%
\pgfsetroundjoin%
\pgfsetlinewidth{0.501875pt}%
\definecolor{currentstroke}{rgb}{0.800000,0.800000,0.800000}%
\pgfsetstrokecolor{currentstroke}%
\pgfsetdash{}{0pt}%
\pgfpathmoveto{\pgfqpoint{2.062142in}{0.311728in}}%
\pgfpathlineto{\pgfqpoint{2.062142in}{1.556882in}}%
\pgfusepath{stroke}%
\end{pgfscope}%
\begin{pgfscope}%
\pgfpathrectangle{\pgfqpoint{0.365549in}{0.311728in}}{\pgfqpoint{1.809698in}{1.245154in}}%
\pgfusepath{clip}%
\pgfsetrectcap%
\pgfsetroundjoin%
\pgfsetlinewidth{1.104125pt}%
\definecolor{currentstroke}{rgb}{0.631373,0.850980,0.607843}%
\pgfsetstrokecolor{currentstroke}%
\pgfsetdash{}{0pt}%
\pgfpathmoveto{\pgfqpoint{0.365549in}{1.441595in}}%
\pgfpathlineto{\pgfqpoint{0.373303in}{1.427317in}}%
\pgfpathlineto{\pgfqpoint{0.379933in}{1.411389in}}%
\pgfpathlineto{\pgfqpoint{0.385904in}{1.394313in}}%
\pgfpathlineto{\pgfqpoint{0.391106in}{1.366403in}}%
\pgfpathlineto{\pgfqpoint{0.395822in}{1.333256in}}%
\pgfpathlineto{\pgfqpoint{0.400117in}{1.288701in}}%
\pgfpathlineto{\pgfqpoint{0.404063in}{1.242783in}}%
\pgfpathlineto{\pgfqpoint{0.407797in}{1.192273in}}%
\pgfpathlineto{\pgfqpoint{0.411262in}{1.136597in}}%
\pgfpathlineto{\pgfqpoint{0.413616in}{1.094337in}}%
\pgfusepath{stroke}%
\end{pgfscope}%
\begin{pgfscope}%
\pgfpathrectangle{\pgfqpoint{0.365549in}{0.311728in}}{\pgfqpoint{1.809698in}{1.245154in}}%
\pgfusepath{clip}%
\pgfsetbuttcap%
\pgfsetroundjoin%
\pgfsetlinewidth{1.104125pt}%
\definecolor{currentstroke}{rgb}{0.631373,0.850980,0.607843}%
\pgfsetstrokecolor{currentstroke}%
\pgfsetdash{{5.500000pt}{1.100000pt}}{0.000000pt}%
\pgfpathmoveto{\pgfqpoint{0.365549in}{1.500284in}}%
\pgfpathlineto{\pgfqpoint{0.531582in}{1.490957in}}%
\pgfpathlineto{\pgfqpoint{0.542577in}{1.482419in}}%
\pgfpathlineto{\pgfqpoint{0.549465in}{1.472661in}}%
\pgfpathlineto{\pgfqpoint{0.554508in}{1.465558in}}%
\pgfpathlineto{\pgfqpoint{0.558536in}{1.452357in}}%
\pgfpathlineto{\pgfqpoint{0.561898in}{1.439801in}}%
\pgfpathlineto{\pgfqpoint{0.564834in}{1.423227in}}%
\pgfpathlineto{\pgfqpoint{0.567359in}{1.405721in}}%
\pgfpathlineto{\pgfqpoint{0.569603in}{1.379174in}}%
\pgfpathlineto{\pgfqpoint{0.571672in}{1.344449in}}%
\pgfpathlineto{\pgfqpoint{0.573613in}{1.303409in}}%
\pgfpathlineto{\pgfqpoint{0.575429in}{1.252899in}}%
\pgfpathlineto{\pgfqpoint{0.577141in}{1.206909in}}%
\pgfpathlineto{\pgfqpoint{0.578761in}{1.155251in}}%
\pgfpathlineto{\pgfqpoint{0.580306in}{1.107037in}}%
\pgfpathlineto{\pgfqpoint{0.581863in}{1.053513in}}%
\pgfusepath{stroke}%
\end{pgfscope}%
\begin{pgfscope}%
\pgfpathrectangle{\pgfqpoint{0.365549in}{0.311728in}}{\pgfqpoint{1.809698in}{1.245154in}}%
\pgfusepath{clip}%
\pgfsetrectcap%
\pgfsetroundjoin%
\pgfsetlinewidth{1.104125pt}%
\definecolor{currentstroke}{rgb}{0.145098,0.145098,0.145098}%
\pgfsetstrokecolor{currentstroke}%
\pgfsetdash{}{0pt}%
\pgfpathmoveto{\pgfqpoint{0.365549in}{1.493683in}}%
\pgfpathlineto{\pgfqpoint{0.608248in}{1.310440in}}%
\pgfpathlineto{\pgfqpoint{0.713277in}{1.265814in}}%
\pgfpathlineto{\pgfqpoint{0.782637in}{1.260002in}}%
\pgfpathlineto{\pgfqpoint{0.834331in}{1.196506in}}%
\pgfpathlineto{\pgfqpoint{0.889019in}{1.156901in}}%
\pgfpathlineto{\pgfqpoint{0.940529in}{1.152525in}}%
\pgfpathlineto{\pgfqpoint{0.988212in}{0.941659in}}%
\pgfpathlineto{\pgfqpoint{1.099862in}{0.853553in}}%
\pgfpathlineto{\pgfqpoint{1.222373in}{0.853840in}}%
\pgfpathlineto{\pgfqpoint{1.334423in}{0.853195in}}%
\pgfpathlineto{\pgfqpoint{1.434848in}{0.825859in}}%
\pgfpathlineto{\pgfqpoint{1.548123in}{0.771116in}}%
\pgfpathlineto{\pgfqpoint{1.671326in}{0.765878in}}%
\pgfpathlineto{\pgfqpoint{1.787415in}{0.744569in}}%
\pgfpathlineto{\pgfqpoint{1.914607in}{0.599639in}}%
\pgfpathlineto{\pgfqpoint{2.112693in}{0.565775in}}%
\pgfusepath{stroke}%
\end{pgfscope}%
\begin{pgfscope}%
\pgfpathrectangle{\pgfqpoint{0.365549in}{0.311728in}}{\pgfqpoint{1.809698in}{1.245154in}}%
\pgfusepath{clip}%
\pgfsetbuttcap%
\pgfsetroundjoin%
\pgfsetlinewidth{1.104125pt}%
\definecolor{currentstroke}{rgb}{0.145098,0.145098,0.145098}%
\pgfsetstrokecolor{currentstroke}%
\pgfsetdash{{5.500000pt}{1.100000pt}}{0.000000pt}%
\pgfpathmoveto{\pgfqpoint{0.365549in}{1.494401in}}%
\pgfpathlineto{\pgfqpoint{0.763186in}{1.487800in}}%
\pgfpathlineto{\pgfqpoint{0.803980in}{1.305490in}}%
\pgfpathlineto{\pgfqpoint{0.839935in}{1.262155in}}%
\pgfpathlineto{\pgfqpoint{0.871595in}{1.213438in}}%
\pgfpathlineto{\pgfqpoint{0.905977in}{1.158767in}}%
\pgfpathlineto{\pgfqpoint{0.942908in}{1.153744in}}%
\pgfpathlineto{\pgfqpoint{0.978142in}{0.967201in}}%
\pgfpathlineto{\pgfqpoint{1.059845in}{0.854056in}}%
\pgfpathlineto{\pgfqpoint{1.159980in}{0.854127in}}%
\pgfpathlineto{\pgfqpoint{1.249185in}{0.853482in}}%
\pgfpathlineto{\pgfqpoint{1.334271in}{0.825931in}}%
\pgfpathlineto{\pgfqpoint{1.431038in}{0.771761in}}%
\pgfpathlineto{\pgfqpoint{1.540621in}{0.766237in}}%
\pgfpathlineto{\pgfqpoint{1.644869in}{0.751385in}}%
\pgfpathlineto{\pgfqpoint{1.752599in}{0.612339in}}%
\pgfpathlineto{\pgfqpoint{1.951916in}{0.564985in}}%
\pgfusepath{stroke}%
\end{pgfscope}%
\begin{pgfscope}%
\pgfpathrectangle{\pgfqpoint{0.365549in}{0.311728in}}{\pgfqpoint{1.809698in}{1.245154in}}%
\pgfusepath{clip}%
\pgfsetbuttcap%
\pgfsetroundjoin%
\pgfsetlinewidth{1.104125pt}%
\definecolor{currentstroke}{rgb}{1.000000,0.000000,0.000000}%
\pgfsetstrokecolor{currentstroke}%
\pgfsetdash{{1.100000pt}{1.815000pt}}{0.000000pt}%
\pgfpathmoveto{\pgfqpoint{0.365549in}{0.368326in}}%
\pgfpathlineto{\pgfqpoint{2.175248in}{0.368326in}}%
\pgfusepath{stroke}%
\end{pgfscope}%
\begin{pgfscope}%
\pgfpathrectangle{\pgfqpoint{0.365549in}{0.311728in}}{\pgfqpoint{1.809698in}{1.245154in}}%
\pgfusepath{clip}%
\pgfsetbuttcap%
\pgfsetroundjoin%
\pgfsetlinewidth{1.104125pt}%
\definecolor{currentstroke}{rgb}{0.145098,0.145098,0.145098}%
\pgfsetstrokecolor{currentstroke}%
\pgfsetdash{{1.100000pt}{1.815000pt}}{0.000000pt}%
\pgfpathmoveto{\pgfqpoint{0.365549in}{1.493683in}}%
\pgfpathlineto{\pgfqpoint{0.534191in}{1.453146in}}%
\pgfpathlineto{\pgfqpoint{0.609564in}{1.268325in}}%
\pgfpathlineto{\pgfqpoint{0.671236in}{1.267249in}}%
\pgfpathlineto{\pgfqpoint{0.717557in}{1.266101in}}%
\pgfusepath{stroke}%
\end{pgfscope}%
\begin{pgfscope}%
\pgfsetrectcap%
\pgfsetmiterjoin%
\pgfsetlinewidth{0.803000pt}%
\definecolor{currentstroke}{rgb}{0.000000,0.000000,0.000000}%
\pgfsetstrokecolor{currentstroke}%
\pgfsetdash{}{0pt}%
\pgfpathmoveto{\pgfqpoint{0.365549in}{0.311728in}}%
\pgfpathlineto{\pgfqpoint{0.365549in}{1.556882in}}%
\pgfusepath{stroke}%
\end{pgfscope}%
\begin{pgfscope}%
\pgfsetrectcap%
\pgfsetmiterjoin%
\pgfsetlinewidth{0.803000pt}%
\definecolor{currentstroke}{rgb}{0.000000,0.000000,0.000000}%
\pgfsetstrokecolor{currentstroke}%
\pgfsetdash{}{0pt}%
\pgfpathmoveto{\pgfqpoint{2.175248in}{0.311728in}}%
\pgfpathlineto{\pgfqpoint{2.175248in}{1.556882in}}%
\pgfusepath{stroke}%
\end{pgfscope}%
\begin{pgfscope}%
\pgfsetrectcap%
\pgfsetmiterjoin%
\pgfsetlinewidth{0.803000pt}%
\definecolor{currentstroke}{rgb}{0.000000,0.000000,0.000000}%
\pgfsetstrokecolor{currentstroke}%
\pgfsetdash{}{0pt}%
\pgfpathmoveto{\pgfqpoint{0.365549in}{0.311728in}}%
\pgfpathlineto{\pgfqpoint{2.175248in}{0.311728in}}%
\pgfusepath{stroke}%
\end{pgfscope}%
\begin{pgfscope}%
\pgfsetrectcap%
\pgfsetmiterjoin%
\pgfsetlinewidth{0.803000pt}%
\definecolor{currentstroke}{rgb}{0.000000,0.000000,0.000000}%
\pgfsetstrokecolor{currentstroke}%
\pgfsetdash{}{0pt}%
\pgfpathmoveto{\pgfqpoint{0.365549in}{1.556882in}}%
\pgfpathlineto{\pgfqpoint{2.175248in}{1.556882in}}%
\pgfusepath{stroke}%
\end{pgfscope}%
\begin{pgfscope}%
\pgfsetbuttcap%
\pgfsetmiterjoin%
\definecolor{currentfill}{rgb}{1.000000,1.000000,1.000000}%
\pgfsetfillcolor{currentfill}%
\pgfsetfillopacity{0.800000}%
\pgfsetlinewidth{1.003750pt}%
\definecolor{currentstroke}{rgb}{0.800000,0.800000,0.800000}%
\pgfsetstrokecolor{currentstroke}%
\pgfsetstrokeopacity{0.800000}%
\pgfsetdash{}{0pt}%
\pgfpathmoveto{\pgfqpoint{1.484203in}{0.909196in}}%
\pgfpathlineto{\pgfqpoint{2.116914in}{0.909196in}}%
\pgfpathquadraticcurveto{\pgfqpoint{2.133581in}{0.909196in}}{\pgfqpoint{2.133581in}{0.925863in}}%
\pgfpathlineto{\pgfqpoint{2.133581in}{1.498549in}}%
\pgfpathquadraticcurveto{\pgfqpoint{2.133581in}{1.515215in}}{\pgfqpoint{2.116914in}{1.515215in}}%
\pgfpathlineto{\pgfqpoint{1.484203in}{1.515215in}}%
\pgfpathquadraticcurveto{\pgfqpoint{1.467536in}{1.515215in}}{\pgfqpoint{1.467536in}{1.498549in}}%
\pgfpathlineto{\pgfqpoint{1.467536in}{0.925863in}}%
\pgfpathquadraticcurveto{\pgfqpoint{1.467536in}{0.909196in}}{\pgfqpoint{1.484203in}{0.909196in}}%
\pgfpathclose%
\pgfusepath{stroke,fill}%
\end{pgfscope}%
\begin{pgfscope}%
\pgfsetrectcap%
\pgfsetroundjoin%
\pgfsetlinewidth{1.104125pt}%
\definecolor{currentstroke}{rgb}{0.631373,0.850980,0.607843}%
\pgfsetstrokecolor{currentstroke}%
\pgfsetdash{}{0pt}%
\pgfpathmoveto{\pgfqpoint{1.500870in}{1.452715in}}%
\pgfpathlineto{\pgfqpoint{1.667536in}{1.452715in}}%
\pgfusepath{stroke}%
\end{pgfscope}%
\begin{pgfscope}%
\pgftext[x=1.734203in,y=1.423549in,left,base]{\rmfamily\fontsize{6.000000}{7.200000}\selectfont FD}%
\end{pgfscope}%
\begin{pgfscope}%
\pgfsetbuttcap%
\pgfsetroundjoin%
\pgfsetlinewidth{1.104125pt}%
\definecolor{currentstroke}{rgb}{0.631373,0.850980,0.607843}%
\pgfsetstrokecolor{currentstroke}%
\pgfsetdash{{5.500000pt}{1.100000pt}}{0.000000pt}%
\pgfpathmoveto{\pgfqpoint{1.500870in}{1.336511in}}%
\pgfpathlineto{\pgfqpoint{1.667536in}{1.336511in}}%
\pgfusepath{stroke}%
\end{pgfscope}%
\begin{pgfscope}%
\pgftext[x=1.734203in,y=1.307345in,left,base]{\rmfamily\fontsize{6.000000}{7.200000}\selectfont FD+6}%
\end{pgfscope}%
\begin{pgfscope}%
\pgfsetrectcap%
\pgfsetroundjoin%
\pgfsetlinewidth{1.104125pt}%
\definecolor{currentstroke}{rgb}{0.145098,0.145098,0.145098}%
\pgfsetstrokecolor{currentstroke}%
\pgfsetdash{}{0pt}%
\pgfpathmoveto{\pgfqpoint{1.500870in}{1.220308in}}%
\pgfpathlineto{\pgfqpoint{1.667536in}{1.220308in}}%
\pgfusepath{stroke}%
\end{pgfscope}%
\begin{pgfscope}%
\pgftext[x=1.734203in,y=1.191141in,left,base]{\rmfamily\fontsize{6.000000}{7.200000}\selectfont L2C2}%
\end{pgfscope}%
\begin{pgfscope}%
\pgfsetbuttcap%
\pgfsetroundjoin%
\pgfsetlinewidth{1.104125pt}%
\definecolor{currentstroke}{rgb}{0.145098,0.145098,0.145098}%
\pgfsetstrokecolor{currentstroke}%
\pgfsetdash{{5.500000pt}{1.100000pt}}{0.000000pt}%
\pgfpathmoveto{\pgfqpoint{1.500870in}{1.104104in}}%
\pgfpathlineto{\pgfqpoint{1.667536in}{1.104104in}}%
\pgfusepath{stroke}%
\end{pgfscope}%
\begin{pgfscope}%
\pgftext[x=1.734203in,y=1.074937in,left,base]{\rmfamily\fontsize{6.000000}{7.200000}\selectfont L2C2+6}%
\end{pgfscope}%
\begin{pgfscope}%
\pgfsetbuttcap%
\pgfsetroundjoin%
\pgfsetlinewidth{1.104125pt}%
\definecolor{currentstroke}{rgb}{1.000000,0.000000,0.000000}%
\pgfsetstrokecolor{currentstroke}%
\pgfsetdash{{1.100000pt}{1.815000pt}}{0.000000pt}%
\pgfpathmoveto{\pgfqpoint{1.500870in}{0.987900in}}%
\pgfpathlineto{\pgfqpoint{1.667536in}{0.987900in}}%
\pgfusepath{stroke}%
\end{pgfscope}%
\begin{pgfscope}%
\pgftext[x=1.734203in,y=0.958734in,left,base]{\rmfamily\fontsize{6.000000}{7.200000}\selectfont 0\% EC}%
\end{pgfscope}%
\end{pgfpicture}%
\makeatother%
\endgroup%

%% file: Figuras/ipc_nwl.pgf
\begingroup%
\makeatletter%
\begin{pgfpicture}%
\pgfpathrectangle{\pgfpointorigin}{\pgfqpoint{3.343624in}{2.296532in}}%
\pgfusepath{use as bounding box, clip}%
\begin{pgfscope}%
\pgfsetbuttcap%
\pgfsetmiterjoin%
\definecolor{currentfill}{rgb}{1.000000,1.000000,1.000000}%
\pgfsetfillcolor{currentfill}%
\pgfsetlinewidth{0.000000pt}%
\definecolor{currentstroke}{rgb}{1.000000,1.000000,1.000000}%
\pgfsetstrokecolor{currentstroke}%
\pgfsetdash{}{0pt}%
\pgfpathmoveto{\pgfqpoint{0.000000in}{0.000000in}}%
\pgfpathlineto{\pgfqpoint{3.343624in}{0.000000in}}%
\pgfpathlineto{\pgfqpoint{3.343624in}{2.296532in}}%
\pgfpathlineto{\pgfqpoint{0.000000in}{2.296532in}}%
\pgfpathclose%
\pgfusepath{fill}%
\end{pgfscope}%
\begin{pgfscope}%
\pgfsetbuttcap%
\pgfsetmiterjoin%
\definecolor{currentfill}{rgb}{1.000000,1.000000,1.000000}%
\pgfsetfillcolor{currentfill}%
\pgfsetlinewidth{0.000000pt}%
\definecolor{currentstroke}{rgb}{0.000000,0.000000,0.000000}%
\pgfsetstrokecolor{currentstroke}%
\pgfsetstrokeopacity{0.000000}%
\pgfsetdash{}{0pt}%
\pgfpathmoveto{\pgfqpoint{0.440392in}{0.402778in}}%
\pgfpathlineto{\pgfqpoint{3.308624in}{0.402778in}}%
\pgfpathlineto{\pgfqpoint{3.308624in}{1.962226in}}%
\pgfpathlineto{\pgfqpoint{0.440392in}{1.962226in}}%
\pgfpathclose%
\pgfusepath{fill}%
\end{pgfscope}%
\begin{pgfscope}%
\pgfsetbuttcap%
\pgfsetroundjoin%
\definecolor{currentfill}{rgb}{0.000000,0.000000,0.000000}%
\pgfsetfillcolor{currentfill}%
\pgfsetlinewidth{0.803000pt}%
\definecolor{currentstroke}{rgb}{0.000000,0.000000,0.000000}%
\pgfsetstrokecolor{currentstroke}%
\pgfsetdash{}{0pt}%
\pgfsys@defobject{currentmarker}{\pgfqpoint{0.000000in}{-0.048611in}}{\pgfqpoint{0.000000in}{0.000000in}}{%
\pgfpathmoveto{\pgfqpoint{0.000000in}{0.000000in}}%
\pgfpathlineto{\pgfqpoint{0.000000in}{-0.048611in}}%
\pgfusepath{stroke,fill}%
}%
\begin{pgfscope}%
\pgfsys@transformshift{0.440392in}{0.402778in}%
\pgfsys@useobject{currentmarker}{}%
\end{pgfscope}%
\end{pgfscope}%
\begin{pgfscope}%
\pgftext[x=0.440392in,y=0.305556in,,top]{\rmfamily\fontsize{9.000000}{10.800000}\selectfont \(\displaystyle 0\)}%
\end{pgfscope}%
\begin{pgfscope}%
\pgfsetbuttcap%
\pgfsetroundjoin%
\definecolor{currentfill}{rgb}{0.000000,0.000000,0.000000}%
\pgfsetfillcolor{currentfill}%
\pgfsetlinewidth{0.803000pt}%
\definecolor{currentstroke}{rgb}{0.000000,0.000000,0.000000}%
\pgfsetstrokecolor{currentstroke}%
\pgfsetdash{}{0pt}%
\pgfsys@defobject{currentmarker}{\pgfqpoint{0.000000in}{-0.048611in}}{\pgfqpoint{0.000000in}{0.000000in}}{%
\pgfpathmoveto{\pgfqpoint{0.000000in}{0.000000in}}%
\pgfpathlineto{\pgfqpoint{0.000000in}{-0.048611in}}%
\pgfusepath{stroke,fill}%
}%
\begin{pgfscope}%
\pgfsys@transformshift{1.259887in}{0.402778in}%
\pgfsys@useobject{currentmarker}{}%
\end{pgfscope}%
\end{pgfscope}%
\begin{pgfscope}%
\pgftext[x=1.259887in,y=0.305556in,,top]{\rmfamily\fontsize{9.000000}{10.800000}\selectfont \(\displaystyle 4\)}%
\end{pgfscope}%
\begin{pgfscope}%
\pgfsetbuttcap%
\pgfsetroundjoin%
\definecolor{currentfill}{rgb}{0.000000,0.000000,0.000000}%
\pgfsetfillcolor{currentfill}%
\pgfsetlinewidth{0.803000pt}%
\definecolor{currentstroke}{rgb}{0.000000,0.000000,0.000000}%
\pgfsetstrokecolor{currentstroke}%
\pgfsetdash{}{0pt}%
\pgfsys@defobject{currentmarker}{\pgfqpoint{0.000000in}{-0.048611in}}{\pgfqpoint{0.000000in}{0.000000in}}{%
\pgfpathmoveto{\pgfqpoint{0.000000in}{0.000000in}}%
\pgfpathlineto{\pgfqpoint{0.000000in}{-0.048611in}}%
\pgfusepath{stroke,fill}%
}%
\begin{pgfscope}%
\pgfsys@transformshift{2.079382in}{0.402778in}%
\pgfsys@useobject{currentmarker}{}%
\end{pgfscope}%
\end{pgfscope}%
\begin{pgfscope}%
\pgftext[x=2.079382in,y=0.305556in,,top]{\rmfamily\fontsize{9.000000}{10.800000}\selectfont \(\displaystyle 8\)}%
\end{pgfscope}%
\begin{pgfscope}%
\pgfsetbuttcap%
\pgfsetroundjoin%
\definecolor{currentfill}{rgb}{0.000000,0.000000,0.000000}%
\pgfsetfillcolor{currentfill}%
\pgfsetlinewidth{0.803000pt}%
\definecolor{currentstroke}{rgb}{0.000000,0.000000,0.000000}%
\pgfsetstrokecolor{currentstroke}%
\pgfsetdash{}{0pt}%
\pgfsys@defobject{currentmarker}{\pgfqpoint{0.000000in}{-0.048611in}}{\pgfqpoint{0.000000in}{0.000000in}}{%
\pgfpathmoveto{\pgfqpoint{0.000000in}{0.000000in}}%
\pgfpathlineto{\pgfqpoint{0.000000in}{-0.048611in}}%
\pgfusepath{stroke,fill}%
}%
\begin{pgfscope}%
\pgfsys@transformshift{2.898877in}{0.402778in}%
\pgfsys@useobject{currentmarker}{}%
\end{pgfscope}%
\end{pgfscope}%
\begin{pgfscope}%
\pgftext[x=2.898877in,y=0.305556in,,top]{\rmfamily\fontsize{9.000000}{10.800000}\selectfont \(\displaystyle 12\)}%
\end{pgfscope}%
\begin{pgfscope}%
\pgftext[x=1.874508in,y=0.138889in,,top]{\rmfamily\fontsize{10.000000}{12.000000}\selectfont Time (years)}%
\end{pgfscope}%
\begin{pgfscope}%
\pgfsetbuttcap%
\pgfsetroundjoin%
\definecolor{currentfill}{rgb}{0.000000,0.000000,0.000000}%
\pgfsetfillcolor{currentfill}%
\pgfsetlinewidth{0.803000pt}%
\definecolor{currentstroke}{rgb}{0.000000,0.000000,0.000000}%
\pgfsetstrokecolor{currentstroke}%
\pgfsetdash{}{0pt}%
\pgfsys@defobject{currentmarker}{\pgfqpoint{-0.048611in}{0.000000in}}{\pgfqpoint{0.000000in}{0.000000in}}{%
\pgfpathmoveto{\pgfqpoint{0.000000in}{0.000000in}}%
\pgfpathlineto{\pgfqpoint{-0.048611in}{0.000000in}}%
\pgfusepath{stroke,fill}%
}%
\begin{pgfscope}%
\pgfsys@transformshift{0.440392in}{0.543267in}%
\pgfsys@useobject{currentmarker}{}%
\end{pgfscope}%
\end{pgfscope}%
\begin{pgfscope}%
\pgftext[x=0.179012in,y=0.499864in,left,base]{\rmfamily\fontsize{9.000000}{10.800000}\selectfont \(\displaystyle 0.6\)}%
\end{pgfscope}%
\begin{pgfscope}%
\pgfsetbuttcap%
\pgfsetroundjoin%
\definecolor{currentfill}{rgb}{0.000000,0.000000,0.000000}%
\pgfsetfillcolor{currentfill}%
\pgfsetlinewidth{0.803000pt}%
\definecolor{currentstroke}{rgb}{0.000000,0.000000,0.000000}%
\pgfsetstrokecolor{currentstroke}%
\pgfsetdash{}{0pt}%
\pgfsys@defobject{currentmarker}{\pgfqpoint{-0.048611in}{0.000000in}}{\pgfqpoint{0.000000in}{0.000000in}}{%
\pgfpathmoveto{\pgfqpoint{0.000000in}{0.000000in}}%
\pgfpathlineto{\pgfqpoint{-0.048611in}{0.000000in}}%
\pgfusepath{stroke,fill}%
}%
\begin{pgfscope}%
\pgfsys@transformshift{0.440392in}{1.220332in}%
\pgfsys@useobject{currentmarker}{}%
\end{pgfscope}%
\end{pgfscope}%
\begin{pgfscope}%
\pgftext[x=0.179012in,y=1.176930in,left,base]{\rmfamily\fontsize{9.000000}{10.800000}\selectfont \(\displaystyle 0.8\)}%
\end{pgfscope}%
\begin{pgfscope}%
\pgfsetbuttcap%
\pgfsetroundjoin%
\definecolor{currentfill}{rgb}{0.000000,0.000000,0.000000}%
\pgfsetfillcolor{currentfill}%
\pgfsetlinewidth{0.803000pt}%
\definecolor{currentstroke}{rgb}{0.000000,0.000000,0.000000}%
\pgfsetstrokecolor{currentstroke}%
\pgfsetdash{}{0pt}%
\pgfsys@defobject{currentmarker}{\pgfqpoint{-0.048611in}{0.000000in}}{\pgfqpoint{0.000000in}{0.000000in}}{%
\pgfpathmoveto{\pgfqpoint{0.000000in}{0.000000in}}%
\pgfpathlineto{\pgfqpoint{-0.048611in}{0.000000in}}%
\pgfusepath{stroke,fill}%
}%
\begin{pgfscope}%
\pgfsys@transformshift{0.440392in}{1.897398in}%
\pgfsys@useobject{currentmarker}{}%
\end{pgfscope}%
\end{pgfscope}%
\begin{pgfscope}%
\pgftext[x=0.179012in,y=1.853995in,left,base]{\rmfamily\fontsize{9.000000}{10.800000}\selectfont \(\displaystyle 1.0\)}%
\end{pgfscope}%
\begin{pgfscope}%
\pgftext[x=0.123457in,y=1.182502in,,bottom,rotate=90.000000]{\rmfamily\fontsize{10.000000}{12.000000}\selectfont Normalized IPC}%
\end{pgfscope}%
\begin{pgfscope}%
\pgfpathrectangle{\pgfqpoint{0.440392in}{0.402778in}}{\pgfqpoint{2.868232in}{1.559448in}}%
\pgfusepath{clip}%
\pgfsetrectcap%
\pgfsetroundjoin%
\pgfsetlinewidth{0.501875pt}%
\definecolor{currentstroke}{rgb}{0.800000,0.800000,0.800000}%
\pgfsetstrokecolor{currentstroke}%
\pgfsetdash{}{0pt}%
\pgfpathmoveto{\pgfqpoint{0.645266in}{0.402778in}}%
\pgfpathlineto{\pgfqpoint{0.645266in}{1.962226in}}%
\pgfusepath{stroke}%
\end{pgfscope}%
\begin{pgfscope}%
\pgfpathrectangle{\pgfqpoint{0.440392in}{0.402778in}}{\pgfqpoint{2.868232in}{1.559448in}}%
\pgfusepath{clip}%
\pgfsetrectcap%
\pgfsetroundjoin%
\pgfsetlinewidth{0.501875pt}%
\definecolor{currentstroke}{rgb}{0.800000,0.800000,0.800000}%
\pgfsetstrokecolor{currentstroke}%
\pgfsetdash{}{0pt}%
\pgfpathmoveto{\pgfqpoint{0.850140in}{0.402778in}}%
\pgfpathlineto{\pgfqpoint{0.850140in}{1.962226in}}%
\pgfusepath{stroke}%
\end{pgfscope}%
\begin{pgfscope}%
\pgfpathrectangle{\pgfqpoint{0.440392in}{0.402778in}}{\pgfqpoint{2.868232in}{1.559448in}}%
\pgfusepath{clip}%
\pgfsetrectcap%
\pgfsetroundjoin%
\pgfsetlinewidth{0.501875pt}%
\definecolor{currentstroke}{rgb}{0.800000,0.800000,0.800000}%
\pgfsetstrokecolor{currentstroke}%
\pgfsetdash{}{0pt}%
\pgfpathmoveto{\pgfqpoint{1.055014in}{0.402778in}}%
\pgfpathlineto{\pgfqpoint{1.055014in}{1.962226in}}%
\pgfusepath{stroke}%
\end{pgfscope}%
\begin{pgfscope}%
\pgfpathrectangle{\pgfqpoint{0.440392in}{0.402778in}}{\pgfqpoint{2.868232in}{1.559448in}}%
\pgfusepath{clip}%
\pgfsetrectcap%
\pgfsetroundjoin%
\pgfsetlinewidth{0.501875pt}%
\definecolor{currentstroke}{rgb}{0.800000,0.800000,0.800000}%
\pgfsetstrokecolor{currentstroke}%
\pgfsetdash{}{0pt}%
\pgfpathmoveto{\pgfqpoint{1.259887in}{0.402778in}}%
\pgfpathlineto{\pgfqpoint{1.259887in}{1.962226in}}%
\pgfusepath{stroke}%
\end{pgfscope}%
\begin{pgfscope}%
\pgfpathrectangle{\pgfqpoint{0.440392in}{0.402778in}}{\pgfqpoint{2.868232in}{1.559448in}}%
\pgfusepath{clip}%
\pgfsetrectcap%
\pgfsetroundjoin%
\pgfsetlinewidth{0.501875pt}%
\definecolor{currentstroke}{rgb}{0.800000,0.800000,0.800000}%
\pgfsetstrokecolor{currentstroke}%
\pgfsetdash{}{0pt}%
\pgfpathmoveto{\pgfqpoint{1.464761in}{0.402778in}}%
\pgfpathlineto{\pgfqpoint{1.464761in}{1.962226in}}%
\pgfusepath{stroke}%
\end{pgfscope}%
\begin{pgfscope}%
\pgfpathrectangle{\pgfqpoint{0.440392in}{0.402778in}}{\pgfqpoint{2.868232in}{1.559448in}}%
\pgfusepath{clip}%
\pgfsetrectcap%
\pgfsetroundjoin%
\pgfsetlinewidth{0.501875pt}%
\definecolor{currentstroke}{rgb}{0.800000,0.800000,0.800000}%
\pgfsetstrokecolor{currentstroke}%
\pgfsetdash{}{0pt}%
\pgfpathmoveto{\pgfqpoint{1.669635in}{0.402778in}}%
\pgfpathlineto{\pgfqpoint{1.669635in}{1.962226in}}%
\pgfusepath{stroke}%
\end{pgfscope}%
\begin{pgfscope}%
\pgfpathrectangle{\pgfqpoint{0.440392in}{0.402778in}}{\pgfqpoint{2.868232in}{1.559448in}}%
\pgfusepath{clip}%
\pgfsetrectcap%
\pgfsetroundjoin%
\pgfsetlinewidth{0.501875pt}%
\definecolor{currentstroke}{rgb}{0.800000,0.800000,0.800000}%
\pgfsetstrokecolor{currentstroke}%
\pgfsetdash{}{0pt}%
\pgfpathmoveto{\pgfqpoint{1.874508in}{0.402778in}}%
\pgfpathlineto{\pgfqpoint{1.874508in}{1.962226in}}%
\pgfusepath{stroke}%
\end{pgfscope}%
\begin{pgfscope}%
\pgfpathrectangle{\pgfqpoint{0.440392in}{0.402778in}}{\pgfqpoint{2.868232in}{1.559448in}}%
\pgfusepath{clip}%
\pgfsetrectcap%
\pgfsetroundjoin%
\pgfsetlinewidth{0.501875pt}%
\definecolor{currentstroke}{rgb}{0.800000,0.800000,0.800000}%
\pgfsetstrokecolor{currentstroke}%
\pgfsetdash{}{0pt}%
\pgfpathmoveto{\pgfqpoint{2.079382in}{0.402778in}}%
\pgfpathlineto{\pgfqpoint{2.079382in}{1.962226in}}%
\pgfusepath{stroke}%
\end{pgfscope}%
\begin{pgfscope}%
\pgfpathrectangle{\pgfqpoint{0.440392in}{0.402778in}}{\pgfqpoint{2.868232in}{1.559448in}}%
\pgfusepath{clip}%
\pgfsetrectcap%
\pgfsetroundjoin%
\pgfsetlinewidth{0.501875pt}%
\definecolor{currentstroke}{rgb}{0.800000,0.800000,0.800000}%
\pgfsetstrokecolor{currentstroke}%
\pgfsetdash{}{0pt}%
\pgfpathmoveto{\pgfqpoint{2.284256in}{0.402778in}}%
\pgfpathlineto{\pgfqpoint{2.284256in}{1.962226in}}%
\pgfusepath{stroke}%
\end{pgfscope}%
\begin{pgfscope}%
\pgfpathrectangle{\pgfqpoint{0.440392in}{0.402778in}}{\pgfqpoint{2.868232in}{1.559448in}}%
\pgfusepath{clip}%
\pgfsetrectcap%
\pgfsetroundjoin%
\pgfsetlinewidth{0.501875pt}%
\definecolor{currentstroke}{rgb}{0.800000,0.800000,0.800000}%
\pgfsetstrokecolor{currentstroke}%
\pgfsetdash{}{0pt}%
\pgfpathmoveto{\pgfqpoint{2.489130in}{0.402778in}}%
\pgfpathlineto{\pgfqpoint{2.489130in}{1.962226in}}%
\pgfusepath{stroke}%
\end{pgfscope}%
\begin{pgfscope}%
\pgfpathrectangle{\pgfqpoint{0.440392in}{0.402778in}}{\pgfqpoint{2.868232in}{1.559448in}}%
\pgfusepath{clip}%
\pgfsetrectcap%
\pgfsetroundjoin%
\pgfsetlinewidth{0.501875pt}%
\definecolor{currentstroke}{rgb}{0.800000,0.800000,0.800000}%
\pgfsetstrokecolor{currentstroke}%
\pgfsetdash{}{0pt}%
\pgfpathmoveto{\pgfqpoint{2.694003in}{0.402778in}}%
\pgfpathlineto{\pgfqpoint{2.694003in}{1.962226in}}%
\pgfusepath{stroke}%
\end{pgfscope}%
\begin{pgfscope}%
\pgfpathrectangle{\pgfqpoint{0.440392in}{0.402778in}}{\pgfqpoint{2.868232in}{1.559448in}}%
\pgfusepath{clip}%
\pgfsetrectcap%
\pgfsetroundjoin%
\pgfsetlinewidth{0.501875pt}%
\definecolor{currentstroke}{rgb}{0.800000,0.800000,0.800000}%
\pgfsetstrokecolor{currentstroke}%
\pgfsetdash{}{0pt}%
\pgfpathmoveto{\pgfqpoint{2.898877in}{0.402778in}}%
\pgfpathlineto{\pgfqpoint{2.898877in}{1.962226in}}%
\pgfusepath{stroke}%
\end{pgfscope}%
\begin{pgfscope}%
\pgfpathrectangle{\pgfqpoint{0.440392in}{0.402778in}}{\pgfqpoint{2.868232in}{1.559448in}}%
\pgfusepath{clip}%
\pgfsetrectcap%
\pgfsetroundjoin%
\pgfsetlinewidth{0.501875pt}%
\definecolor{currentstroke}{rgb}{0.800000,0.800000,0.800000}%
\pgfsetstrokecolor{currentstroke}%
\pgfsetdash{}{0pt}%
\pgfpathmoveto{\pgfqpoint{3.103751in}{0.402778in}}%
\pgfpathlineto{\pgfqpoint{3.103751in}{1.962226in}}%
\pgfusepath{stroke}%
\end{pgfscope}%
\begin{pgfscope}%
\pgfpathrectangle{\pgfqpoint{0.440392in}{0.402778in}}{\pgfqpoint{2.868232in}{1.559448in}}%
\pgfusepath{clip}%
\pgfsetrectcap%
\pgfsetroundjoin%
\pgfsetlinewidth{1.104125pt}%
\definecolor{currentstroke}{rgb}{0.145098,0.145098,0.145098}%
\pgfsetstrokecolor{currentstroke}%
\pgfsetdash{}{0pt}%
\pgfpathmoveto{\pgfqpoint{0.440392in}{1.891342in}}%
\pgfpathlineto{\pgfqpoint{1.309174in}{1.880856in}}%
\pgfpathlineto{\pgfqpoint{1.404683in}{1.616447in}}%
\pgfpathlineto{\pgfqpoint{1.547190in}{1.604606in}}%
\pgfpathlineto{\pgfqpoint{1.634045in}{1.600538in}}%
\pgfpathlineto{\pgfqpoint{1.697227in}{1.532560in}}%
\pgfpathlineto{\pgfqpoint{1.762007in}{1.468289in}}%
\pgfpathlineto{\pgfqpoint{1.824725in}{1.462865in}}%
\pgfpathlineto{\pgfqpoint{1.881612in}{1.223044in}}%
\pgfpathlineto{\pgfqpoint{2.007850in}{1.085462in}}%
\pgfpathlineto{\pgfqpoint{2.154824in}{1.085733in}}%
\pgfpathlineto{\pgfqpoint{2.290274in}{1.084467in}}%
\pgfpathlineto{\pgfqpoint{2.412994in}{1.059156in}}%
\pgfpathlineto{\pgfqpoint{2.546372in}{0.983405in}}%
\pgfpathlineto{\pgfqpoint{2.696879in}{0.975721in}}%
\pgfpathlineto{\pgfqpoint{2.837162in}{0.958003in}}%
\pgfpathlineto{\pgfqpoint{2.983534in}{0.790500in}}%
\pgfpathlineto{\pgfqpoint{3.242659in}{0.721437in}}%
\pgfusepath{stroke}%
\end{pgfscope}%
\begin{pgfscope}%
\pgfpathrectangle{\pgfqpoint{0.440392in}{0.402778in}}{\pgfqpoint{2.868232in}{1.559448in}}%
\pgfusepath{clip}%
\pgfsetbuttcap%
\pgfsetroundjoin%
\pgfsetlinewidth{1.104125pt}%
\definecolor{currentstroke}{rgb}{0.145098,0.145098,0.145098}%
\pgfsetstrokecolor{currentstroke}%
\pgfsetdash{{5.500000pt}{1.100000pt}}{0.000000pt}%
\pgfpathmoveto{\pgfqpoint{0.440392in}{1.891342in}}%
\pgfpathlineto{\pgfqpoint{1.189642in}{1.880042in}}%
\pgfpathlineto{\pgfqpoint{1.269898in}{1.633984in}}%
\pgfpathlineto{\pgfqpoint{1.389373in}{1.604425in}}%
\pgfpathlineto{\pgfqpoint{1.467390in}{1.601351in}}%
\pgfpathlineto{\pgfqpoint{1.527445in}{1.527136in}}%
\pgfpathlineto{\pgfqpoint{1.590092in}{1.467023in}}%
\pgfpathlineto{\pgfqpoint{1.652132in}{1.465034in}}%
\pgfpathlineto{\pgfqpoint{1.707663in}{1.259022in}}%
\pgfpathlineto{\pgfqpoint{1.813307in}{1.086908in}}%
\pgfpathlineto{\pgfqpoint{1.946384in}{1.084919in}}%
\pgfpathlineto{\pgfqpoint{2.074467in}{1.084829in}}%
\pgfpathlineto{\pgfqpoint{2.195404in}{1.068738in}}%
\pgfpathlineto{\pgfqpoint{2.326417in}{0.984580in}}%
\pgfpathlineto{\pgfqpoint{2.481078in}{0.976083in}}%
\pgfpathlineto{\pgfqpoint{2.635845in}{0.972105in}}%
\pgfpathlineto{\pgfqpoint{2.791329in}{0.817618in}}%
\pgfpathlineto{\pgfqpoint{3.030016in}{0.721618in}}%
\pgfusepath{stroke}%
\end{pgfscope}%
\begin{pgfscope}%
\pgfpathrectangle{\pgfqpoint{0.440392in}{0.402778in}}{\pgfqpoint{2.868232in}{1.559448in}}%
\pgfusepath{clip}%
\pgfsetbuttcap%
\pgfsetroundjoin%
\pgfsetlinewidth{1.104125pt}%
\definecolor{currentstroke}{rgb}{1.000000,0.000000,0.000000}%
\pgfsetstrokecolor{currentstroke}%
\pgfsetdash{{1.100000pt}{1.815000pt}}{0.000000pt}%
\pgfpathmoveto{\pgfqpoint{0.440392in}{0.473662in}}%
\pgfpathlineto{\pgfqpoint{3.308624in}{0.473662in}}%
\pgfusepath{stroke}%
\end{pgfscope}%
\begin{pgfscope}%
\pgfsetrectcap%
\pgfsetmiterjoin%
\pgfsetlinewidth{0.803000pt}%
\definecolor{currentstroke}{rgb}{0.000000,0.000000,0.000000}%
\pgfsetstrokecolor{currentstroke}%
\pgfsetdash{}{0pt}%
\pgfpathmoveto{\pgfqpoint{0.440392in}{0.402778in}}%
\pgfpathlineto{\pgfqpoint{0.440392in}{1.962226in}}%
\pgfusepath{stroke}%
\end{pgfscope}%
\begin{pgfscope}%
\pgfsetrectcap%
\pgfsetmiterjoin%
\pgfsetlinewidth{0.803000pt}%
\definecolor{currentstroke}{rgb}{0.000000,0.000000,0.000000}%
\pgfsetstrokecolor{currentstroke}%
\pgfsetdash{}{0pt}%
\pgfpathmoveto{\pgfqpoint{3.308624in}{0.402778in}}%
\pgfpathlineto{\pgfqpoint{3.308624in}{1.962226in}}%
\pgfusepath{stroke}%
\end{pgfscope}%
\begin{pgfscope}%
\pgfsetrectcap%
\pgfsetmiterjoin%
\pgfsetlinewidth{0.803000pt}%
\definecolor{currentstroke}{rgb}{0.000000,0.000000,0.000000}%
\pgfsetstrokecolor{currentstroke}%
\pgfsetdash{}{0pt}%
\pgfpathmoveto{\pgfqpoint{0.440392in}{0.402778in}}%
\pgfpathlineto{\pgfqpoint{3.308624in}{0.402778in}}%
\pgfusepath{stroke}%
\end{pgfscope}%
\begin{pgfscope}%
\pgfsetrectcap%
\pgfsetmiterjoin%
\pgfsetlinewidth{0.803000pt}%
\definecolor{currentstroke}{rgb}{0.000000,0.000000,0.000000}%
\pgfsetstrokecolor{currentstroke}%
\pgfsetdash{}{0pt}%
\pgfpathmoveto{\pgfqpoint{0.440392in}{1.962226in}}%
\pgfpathlineto{\pgfqpoint{3.308624in}{1.962226in}}%
\pgfusepath{stroke}%
\end{pgfscope}%
\begin{pgfscope}%
\pgfsetrectcap%
\pgfsetroundjoin%
\pgfsetlinewidth{1.104125pt}%
\definecolor{currentstroke}{rgb}{0.145098,0.145098,0.145098}%
\pgfsetstrokecolor{currentstroke}%
\pgfsetdash{}{0pt}%
\pgfpathmoveto{\pgfqpoint{0.566050in}{2.213199in}}%
\pgfpathlineto{\pgfqpoint{0.788273in}{2.213199in}}%
\pgfusepath{stroke}%
\end{pgfscope}%
\begin{pgfscope}%
\pgftext[x=0.877161in,y=2.174310in,left,base]{\rmfamily\fontsize{8.000000}{9.600000}\selectfont L2C2}%
\end{pgfscope}%
\begin{pgfscope}%
\pgfsetbuttcap%
\pgfsetroundjoin%
\pgfsetlinewidth{1.104125pt}%
\definecolor{currentstroke}{rgb}{0.145098,0.145098,0.145098}%
\pgfsetstrokecolor{currentstroke}%
\pgfsetdash{{5.500000pt}{1.100000pt}}{0.000000pt}%
\pgfpathmoveto{\pgfqpoint{1.376432in}{2.213199in}}%
\pgfpathlineto{\pgfqpoint{1.598655in}{2.213199in}}%
\pgfusepath{stroke}%
\end{pgfscope}%
\begin{pgfscope}%
\pgftext[x=1.687544in,y=2.174310in,left,base]{\rmfamily\fontsize{8.000000}{9.600000}\selectfont L2C2-NWL}%
\end{pgfscope}%
\begin{pgfscope}%
\pgfsetbuttcap%
\pgfsetroundjoin%
\pgfsetlinewidth{1.104125pt}%
\definecolor{currentstroke}{rgb}{1.000000,0.000000,0.000000}%
\pgfsetstrokecolor{currentstroke}%
\pgfsetdash{{1.100000pt}{1.815000pt}}{0.000000pt}%
\pgfpathmoveto{\pgfqpoint{2.509543in}{2.213199in}}%
\pgfpathlineto{\pgfqpoint{2.731765in}{2.213199in}}%
\pgfusepath{stroke}%
\end{pgfscope}%
\begin{pgfscope}%
\pgftext[x=2.820654in,y=2.174310in,left,base]{\rmfamily\fontsize{8.000000}{9.600000}\selectfont 0\% EC}%
\end{pgfscope}%
\end{pgfpicture}%
\makeatother%
\endgroup%

%% file: Figuras/ipc_best_fit.pgf
\begingroup%
\makeatletter%
\begin{pgfpicture}%
\pgfpathrectangle{\pgfpointorigin}{\pgfqpoint{3.343624in}{2.296532in}}%
\pgfusepath{use as bounding box, clip}%
\begin{pgfscope}%
\pgfsetbuttcap%
\pgfsetmiterjoin%
\definecolor{currentfill}{rgb}{1.000000,1.000000,1.000000}%
\pgfsetfillcolor{currentfill}%
\pgfsetlinewidth{0.000000pt}%
\definecolor{currentstroke}{rgb}{1.000000,1.000000,1.000000}%
\pgfsetstrokecolor{currentstroke}%
\pgfsetdash{}{0pt}%
\pgfpathmoveto{\pgfqpoint{0.000000in}{0.000000in}}%
\pgfpathlineto{\pgfqpoint{3.343624in}{0.000000in}}%
\pgfpathlineto{\pgfqpoint{3.343624in}{2.296532in}}%
\pgfpathlineto{\pgfqpoint{0.000000in}{2.296532in}}%
\pgfpathclose%
\pgfusepath{fill}%
\end{pgfscope}%
\begin{pgfscope}%
\pgfsetbuttcap%
\pgfsetmiterjoin%
\definecolor{currentfill}{rgb}{1.000000,1.000000,1.000000}%
\pgfsetfillcolor{currentfill}%
\pgfsetlinewidth{0.000000pt}%
\definecolor{currentstroke}{rgb}{0.000000,0.000000,0.000000}%
\pgfsetstrokecolor{currentstroke}%
\pgfsetstrokeopacity{0.000000}%
\pgfsetdash{}{0pt}%
\pgfpathmoveto{\pgfqpoint{0.440392in}{0.402778in}}%
\pgfpathlineto{\pgfqpoint{3.308624in}{0.402778in}}%
\pgfpathlineto{\pgfqpoint{3.308624in}{1.962226in}}%
\pgfpathlineto{\pgfqpoint{0.440392in}{1.962226in}}%
\pgfpathclose%
\pgfusepath{fill}%
\end{pgfscope}%
\begin{pgfscope}%
\pgfsetbuttcap%
\pgfsetroundjoin%
\definecolor{currentfill}{rgb}{0.000000,0.000000,0.000000}%
\pgfsetfillcolor{currentfill}%
\pgfsetlinewidth{0.803000pt}%
\definecolor{currentstroke}{rgb}{0.000000,0.000000,0.000000}%
\pgfsetstrokecolor{currentstroke}%
\pgfsetdash{}{0pt}%
\pgfsys@defobject{currentmarker}{\pgfqpoint{0.000000in}{-0.048611in}}{\pgfqpoint{0.000000in}{0.000000in}}{%
\pgfpathmoveto{\pgfqpoint{0.000000in}{0.000000in}}%
\pgfpathlineto{\pgfqpoint{0.000000in}{-0.048611in}}%
\pgfusepath{stroke,fill}%
}%
\begin{pgfscope}%
\pgfsys@transformshift{0.440392in}{0.402778in}%
\pgfsys@useobject{currentmarker}{}%
\end{pgfscope}%
\end{pgfscope}%
\begin{pgfscope}%
\pgftext[x=0.440392in,y=0.305556in,,top]{\rmfamily\fontsize{9.000000}{10.800000}\selectfont \(\displaystyle 0\)}%
\end{pgfscope}%
\begin{pgfscope}%
\pgfsetbuttcap%
\pgfsetroundjoin%
\definecolor{currentfill}{rgb}{0.000000,0.000000,0.000000}%
\pgfsetfillcolor{currentfill}%
\pgfsetlinewidth{0.803000pt}%
\definecolor{currentstroke}{rgb}{0.000000,0.000000,0.000000}%
\pgfsetstrokecolor{currentstroke}%
\pgfsetdash{}{0pt}%
\pgfsys@defobject{currentmarker}{\pgfqpoint{0.000000in}{-0.048611in}}{\pgfqpoint{0.000000in}{0.000000in}}{%
\pgfpathmoveto{\pgfqpoint{0.000000in}{0.000000in}}%
\pgfpathlineto{\pgfqpoint{0.000000in}{-0.048611in}}%
\pgfusepath{stroke,fill}%
}%
\begin{pgfscope}%
\pgfsys@transformshift{1.259887in}{0.402778in}%
\pgfsys@useobject{currentmarker}{}%
\end{pgfscope}%
\end{pgfscope}%
\begin{pgfscope}%
\pgftext[x=1.259887in,y=0.305556in,,top]{\rmfamily\fontsize{9.000000}{10.800000}\selectfont \(\displaystyle 4\)}%
\end{pgfscope}%
\begin{pgfscope}%
\pgfsetbuttcap%
\pgfsetroundjoin%
\definecolor{currentfill}{rgb}{0.000000,0.000000,0.000000}%
\pgfsetfillcolor{currentfill}%
\pgfsetlinewidth{0.803000pt}%
\definecolor{currentstroke}{rgb}{0.000000,0.000000,0.000000}%
\pgfsetstrokecolor{currentstroke}%
\pgfsetdash{}{0pt}%
\pgfsys@defobject{currentmarker}{\pgfqpoint{0.000000in}{-0.048611in}}{\pgfqpoint{0.000000in}{0.000000in}}{%
\pgfpathmoveto{\pgfqpoint{0.000000in}{0.000000in}}%
\pgfpathlineto{\pgfqpoint{0.000000in}{-0.048611in}}%
\pgfusepath{stroke,fill}%
}%
\begin{pgfscope}%
\pgfsys@transformshift{2.079382in}{0.402778in}%
\pgfsys@useobject{currentmarker}{}%
\end{pgfscope}%
\end{pgfscope}%
\begin{pgfscope}%
\pgftext[x=2.079382in,y=0.305556in,,top]{\rmfamily\fontsize{9.000000}{10.800000}\selectfont \(\displaystyle 8\)}%
\end{pgfscope}%
\begin{pgfscope}%
\pgfsetbuttcap%
\pgfsetroundjoin%
\definecolor{currentfill}{rgb}{0.000000,0.000000,0.000000}%
\pgfsetfillcolor{currentfill}%
\pgfsetlinewidth{0.803000pt}%
\definecolor{currentstroke}{rgb}{0.000000,0.000000,0.000000}%
\pgfsetstrokecolor{currentstroke}%
\pgfsetdash{}{0pt}%
\pgfsys@defobject{currentmarker}{\pgfqpoint{0.000000in}{-0.048611in}}{\pgfqpoint{0.000000in}{0.000000in}}{%
\pgfpathmoveto{\pgfqpoint{0.000000in}{0.000000in}}%
\pgfpathlineto{\pgfqpoint{0.000000in}{-0.048611in}}%
\pgfusepath{stroke,fill}%
}%
\begin{pgfscope}%
\pgfsys@transformshift{2.898877in}{0.402778in}%
\pgfsys@useobject{currentmarker}{}%
\end{pgfscope}%
\end{pgfscope}%
\begin{pgfscope}%
\pgftext[x=2.898877in,y=0.305556in,,top]{\rmfamily\fontsize{9.000000}{10.800000}\selectfont \(\displaystyle 12\)}%
\end{pgfscope}%
\begin{pgfscope}%
\pgftext[x=1.874508in,y=0.138889in,,top]{\rmfamily\fontsize{10.000000}{12.000000}\selectfont Time (years)}%
\end{pgfscope}%
\begin{pgfscope}%
\pgfsetbuttcap%
\pgfsetroundjoin%
\definecolor{currentfill}{rgb}{0.000000,0.000000,0.000000}%
\pgfsetfillcolor{currentfill}%
\pgfsetlinewidth{0.803000pt}%
\definecolor{currentstroke}{rgb}{0.000000,0.000000,0.000000}%
\pgfsetstrokecolor{currentstroke}%
\pgfsetdash{}{0pt}%
\pgfsys@defobject{currentmarker}{\pgfqpoint{-0.048611in}{0.000000in}}{\pgfqpoint{0.000000in}{0.000000in}}{%
\pgfpathmoveto{\pgfqpoint{0.000000in}{0.000000in}}%
\pgfpathlineto{\pgfqpoint{-0.048611in}{0.000000in}}%
\pgfusepath{stroke,fill}%
}%
\begin{pgfscope}%
\pgfsys@transformshift{0.440392in}{0.542852in}%
\pgfsys@useobject{currentmarker}{}%
\end{pgfscope}%
\end{pgfscope}%
\begin{pgfscope}%
\pgftext[x=0.179012in,y=0.499449in,left,base]{\rmfamily\fontsize{9.000000}{10.800000}\selectfont \(\displaystyle 0.6\)}%
\end{pgfscope}%
\begin{pgfscope}%
\pgfsetbuttcap%
\pgfsetroundjoin%
\definecolor{currentfill}{rgb}{0.000000,0.000000,0.000000}%
\pgfsetfillcolor{currentfill}%
\pgfsetlinewidth{0.803000pt}%
\definecolor{currentstroke}{rgb}{0.000000,0.000000,0.000000}%
\pgfsetstrokecolor{currentstroke}%
\pgfsetdash{}{0pt}%
\pgfsys@defobject{currentmarker}{\pgfqpoint{-0.048611in}{0.000000in}}{\pgfqpoint{0.000000in}{0.000000in}}{%
\pgfpathmoveto{\pgfqpoint{0.000000in}{0.000000in}}%
\pgfpathlineto{\pgfqpoint{-0.048611in}{0.000000in}}%
\pgfusepath{stroke,fill}%
}%
\begin{pgfscope}%
\pgfsys@transformshift{0.440392in}{1.215884in}%
\pgfsys@useobject{currentmarker}{}%
\end{pgfscope}%
\end{pgfscope}%
\begin{pgfscope}%
\pgftext[x=0.179012in,y=1.172481in,left,base]{\rmfamily\fontsize{9.000000}{10.800000}\selectfont \(\displaystyle 0.8\)}%
\end{pgfscope}%
\begin{pgfscope}%
\pgfsetbuttcap%
\pgfsetroundjoin%
\definecolor{currentfill}{rgb}{0.000000,0.000000,0.000000}%
\pgfsetfillcolor{currentfill}%
\pgfsetlinewidth{0.803000pt}%
\definecolor{currentstroke}{rgb}{0.000000,0.000000,0.000000}%
\pgfsetstrokecolor{currentstroke}%
\pgfsetdash{}{0pt}%
\pgfsys@defobject{currentmarker}{\pgfqpoint{-0.048611in}{0.000000in}}{\pgfqpoint{0.000000in}{0.000000in}}{%
\pgfpathmoveto{\pgfqpoint{0.000000in}{0.000000in}}%
\pgfpathlineto{\pgfqpoint{-0.048611in}{0.000000in}}%
\pgfusepath{stroke,fill}%
}%
\begin{pgfscope}%
\pgfsys@transformshift{0.440392in}{1.888915in}%
\pgfsys@useobject{currentmarker}{}%
\end{pgfscope}%
\end{pgfscope}%
\begin{pgfscope}%
\pgftext[x=0.179012in,y=1.845513in,left,base]{\rmfamily\fontsize{9.000000}{10.800000}\selectfont \(\displaystyle 1.0\)}%
\end{pgfscope}%
\begin{pgfscope}%
\pgftext[x=0.123457in,y=1.182502in,,bottom,rotate=90.000000]{\rmfamily\fontsize{10.000000}{12.000000}\selectfont Normalized IPC}%
\end{pgfscope}%
\begin{pgfscope}%
\pgfpathrectangle{\pgfqpoint{0.440392in}{0.402778in}}{\pgfqpoint{2.868232in}{1.559448in}}%
\pgfusepath{clip}%
\pgfsetrectcap%
\pgfsetroundjoin%
\pgfsetlinewidth{0.501875pt}%
\definecolor{currentstroke}{rgb}{0.800000,0.800000,0.800000}%
\pgfsetstrokecolor{currentstroke}%
\pgfsetdash{}{0pt}%
\pgfpathmoveto{\pgfqpoint{0.645266in}{0.402778in}}%
\pgfpathlineto{\pgfqpoint{0.645266in}{1.962226in}}%
\pgfusepath{stroke}%
\end{pgfscope}%
\begin{pgfscope}%
\pgfpathrectangle{\pgfqpoint{0.440392in}{0.402778in}}{\pgfqpoint{2.868232in}{1.559448in}}%
\pgfusepath{clip}%
\pgfsetrectcap%
\pgfsetroundjoin%
\pgfsetlinewidth{0.501875pt}%
\definecolor{currentstroke}{rgb}{0.800000,0.800000,0.800000}%
\pgfsetstrokecolor{currentstroke}%
\pgfsetdash{}{0pt}%
\pgfpathmoveto{\pgfqpoint{0.850140in}{0.402778in}}%
\pgfpathlineto{\pgfqpoint{0.850140in}{1.962226in}}%
\pgfusepath{stroke}%
\end{pgfscope}%
\begin{pgfscope}%
\pgfpathrectangle{\pgfqpoint{0.440392in}{0.402778in}}{\pgfqpoint{2.868232in}{1.559448in}}%
\pgfusepath{clip}%
\pgfsetrectcap%
\pgfsetroundjoin%
\pgfsetlinewidth{0.501875pt}%
\definecolor{currentstroke}{rgb}{0.800000,0.800000,0.800000}%
\pgfsetstrokecolor{currentstroke}%
\pgfsetdash{}{0pt}%
\pgfpathmoveto{\pgfqpoint{1.055014in}{0.402778in}}%
\pgfpathlineto{\pgfqpoint{1.055014in}{1.962226in}}%
\pgfusepath{stroke}%
\end{pgfscope}%
\begin{pgfscope}%
\pgfpathrectangle{\pgfqpoint{0.440392in}{0.402778in}}{\pgfqpoint{2.868232in}{1.559448in}}%
\pgfusepath{clip}%
\pgfsetrectcap%
\pgfsetroundjoin%
\pgfsetlinewidth{0.501875pt}%
\definecolor{currentstroke}{rgb}{0.800000,0.800000,0.800000}%
\pgfsetstrokecolor{currentstroke}%
\pgfsetdash{}{0pt}%
\pgfpathmoveto{\pgfqpoint{1.259887in}{0.402778in}}%
\pgfpathlineto{\pgfqpoint{1.259887in}{1.962226in}}%
\pgfusepath{stroke}%
\end{pgfscope}%
\begin{pgfscope}%
\pgfpathrectangle{\pgfqpoint{0.440392in}{0.402778in}}{\pgfqpoint{2.868232in}{1.559448in}}%
\pgfusepath{clip}%
\pgfsetrectcap%
\pgfsetroundjoin%
\pgfsetlinewidth{0.501875pt}%
\definecolor{currentstroke}{rgb}{0.800000,0.800000,0.800000}%
\pgfsetstrokecolor{currentstroke}%
\pgfsetdash{}{0pt}%
\pgfpathmoveto{\pgfqpoint{1.464761in}{0.402778in}}%
\pgfpathlineto{\pgfqpoint{1.464761in}{1.962226in}}%
\pgfusepath{stroke}%
\end{pgfscope}%
\begin{pgfscope}%
\pgfpathrectangle{\pgfqpoint{0.440392in}{0.402778in}}{\pgfqpoint{2.868232in}{1.559448in}}%
\pgfusepath{clip}%
\pgfsetrectcap%
\pgfsetroundjoin%
\pgfsetlinewidth{0.501875pt}%
\definecolor{currentstroke}{rgb}{0.800000,0.800000,0.800000}%
\pgfsetstrokecolor{currentstroke}%
\pgfsetdash{}{0pt}%
\pgfpathmoveto{\pgfqpoint{1.669635in}{0.402778in}}%
\pgfpathlineto{\pgfqpoint{1.669635in}{1.962226in}}%
\pgfusepath{stroke}%
\end{pgfscope}%
\begin{pgfscope}%
\pgfpathrectangle{\pgfqpoint{0.440392in}{0.402778in}}{\pgfqpoint{2.868232in}{1.559448in}}%
\pgfusepath{clip}%
\pgfsetrectcap%
\pgfsetroundjoin%
\pgfsetlinewidth{0.501875pt}%
\definecolor{currentstroke}{rgb}{0.800000,0.800000,0.800000}%
\pgfsetstrokecolor{currentstroke}%
\pgfsetdash{}{0pt}%
\pgfpathmoveto{\pgfqpoint{1.874508in}{0.402778in}}%
\pgfpathlineto{\pgfqpoint{1.874508in}{1.962226in}}%
\pgfusepath{stroke}%
\end{pgfscope}%
\begin{pgfscope}%
\pgfpathrectangle{\pgfqpoint{0.440392in}{0.402778in}}{\pgfqpoint{2.868232in}{1.559448in}}%
\pgfusepath{clip}%
\pgfsetrectcap%
\pgfsetroundjoin%
\pgfsetlinewidth{0.501875pt}%
\definecolor{currentstroke}{rgb}{0.800000,0.800000,0.800000}%
\pgfsetstrokecolor{currentstroke}%
\pgfsetdash{}{0pt}%
\pgfpathmoveto{\pgfqpoint{2.079382in}{0.402778in}}%
\pgfpathlineto{\pgfqpoint{2.079382in}{1.962226in}}%
\pgfusepath{stroke}%
\end{pgfscope}%
\begin{pgfscope}%
\pgfpathrectangle{\pgfqpoint{0.440392in}{0.402778in}}{\pgfqpoint{2.868232in}{1.559448in}}%
\pgfusepath{clip}%
\pgfsetrectcap%
\pgfsetroundjoin%
\pgfsetlinewidth{0.501875pt}%
\definecolor{currentstroke}{rgb}{0.800000,0.800000,0.800000}%
\pgfsetstrokecolor{currentstroke}%
\pgfsetdash{}{0pt}%
\pgfpathmoveto{\pgfqpoint{2.284256in}{0.402778in}}%
\pgfpathlineto{\pgfqpoint{2.284256in}{1.962226in}}%
\pgfusepath{stroke}%
\end{pgfscope}%
\begin{pgfscope}%
\pgfpathrectangle{\pgfqpoint{0.440392in}{0.402778in}}{\pgfqpoint{2.868232in}{1.559448in}}%
\pgfusepath{clip}%
\pgfsetrectcap%
\pgfsetroundjoin%
\pgfsetlinewidth{0.501875pt}%
\definecolor{currentstroke}{rgb}{0.800000,0.800000,0.800000}%
\pgfsetstrokecolor{currentstroke}%
\pgfsetdash{}{0pt}%
\pgfpathmoveto{\pgfqpoint{2.489130in}{0.402778in}}%
\pgfpathlineto{\pgfqpoint{2.489130in}{1.962226in}}%
\pgfusepath{stroke}%
\end{pgfscope}%
\begin{pgfscope}%
\pgfpathrectangle{\pgfqpoint{0.440392in}{0.402778in}}{\pgfqpoint{2.868232in}{1.559448in}}%
\pgfusepath{clip}%
\pgfsetrectcap%
\pgfsetroundjoin%
\pgfsetlinewidth{0.501875pt}%
\definecolor{currentstroke}{rgb}{0.800000,0.800000,0.800000}%
\pgfsetstrokecolor{currentstroke}%
\pgfsetdash{}{0pt}%
\pgfpathmoveto{\pgfqpoint{2.694003in}{0.402778in}}%
\pgfpathlineto{\pgfqpoint{2.694003in}{1.962226in}}%
\pgfusepath{stroke}%
\end{pgfscope}%
\begin{pgfscope}%
\pgfpathrectangle{\pgfqpoint{0.440392in}{0.402778in}}{\pgfqpoint{2.868232in}{1.559448in}}%
\pgfusepath{clip}%
\pgfsetrectcap%
\pgfsetroundjoin%
\pgfsetlinewidth{0.501875pt}%
\definecolor{currentstroke}{rgb}{0.800000,0.800000,0.800000}%
\pgfsetstrokecolor{currentstroke}%
\pgfsetdash{}{0pt}%
\pgfpathmoveto{\pgfqpoint{2.898877in}{0.402778in}}%
\pgfpathlineto{\pgfqpoint{2.898877in}{1.962226in}}%
\pgfusepath{stroke}%
\end{pgfscope}%
\begin{pgfscope}%
\pgfpathrectangle{\pgfqpoint{0.440392in}{0.402778in}}{\pgfqpoint{2.868232in}{1.559448in}}%
\pgfusepath{clip}%
\pgfsetrectcap%
\pgfsetroundjoin%
\pgfsetlinewidth{0.501875pt}%
\definecolor{currentstroke}{rgb}{0.800000,0.800000,0.800000}%
\pgfsetstrokecolor{currentstroke}%
\pgfsetdash{}{0pt}%
\pgfpathmoveto{\pgfqpoint{3.103751in}{0.402778in}}%
\pgfpathlineto{\pgfqpoint{3.103751in}{1.962226in}}%
\pgfusepath{stroke}%
\end{pgfscope}%
\begin{pgfscope}%
\pgfpathrectangle{\pgfqpoint{0.440392in}{0.402778in}}{\pgfqpoint{2.868232in}{1.559448in}}%
\pgfusepath{clip}%
\pgfsetrectcap%
\pgfsetroundjoin%
\pgfsetlinewidth{1.104125pt}%
\definecolor{currentstroke}{rgb}{0.631373,0.850980,0.607843}%
\pgfsetstrokecolor{currentstroke}%
\pgfsetdash{}{0pt}%
\pgfpathmoveto{\pgfqpoint{0.440392in}{1.889365in}}%
\pgfpathlineto{\pgfqpoint{0.723100in}{1.876875in}}%
\pgfpathlineto{\pgfqpoint{0.764265in}{1.867889in}}%
\pgfpathlineto{\pgfqpoint{0.789068in}{1.856297in}}%
\pgfpathlineto{\pgfqpoint{0.806517in}{1.845784in}}%
\pgfpathlineto{\pgfqpoint{0.820244in}{1.832395in}}%
\pgfpathlineto{\pgfqpoint{0.831566in}{1.813615in}}%
\pgfpathlineto{\pgfqpoint{0.841157in}{1.794116in}}%
\pgfpathlineto{\pgfqpoint{0.849450in}{1.765901in}}%
\pgfpathlineto{\pgfqpoint{0.856672in}{1.722859in}}%
\pgfpathlineto{\pgfqpoint{0.863078in}{1.670832in}}%
\pgfpathlineto{\pgfqpoint{0.868956in}{1.612784in}}%
\pgfpathlineto{\pgfqpoint{0.874465in}{1.546289in}}%
\pgfpathlineto{\pgfqpoint{0.879638in}{1.488601in}}%
\pgfpathlineto{\pgfqpoint{0.884562in}{1.422286in}}%
\pgfpathlineto{\pgfqpoint{0.889353in}{1.361543in}}%
\pgfpathlineto{\pgfqpoint{0.893883in}{1.296126in}}%
\pgfusepath{stroke}%
\end{pgfscope}%
\begin{pgfscope}%
\pgfpathrectangle{\pgfqpoint{0.440392in}{0.402778in}}{\pgfqpoint{2.868232in}{1.559448in}}%
\pgfusepath{clip}%
\pgfsetbuttcap%
\pgfsetroundjoin%
\pgfsetlinewidth{1.104125pt}%
\definecolor{currentstroke}{rgb}{0.631373,0.850980,0.607843}%
\pgfsetstrokecolor{currentstroke}%
\pgfsetdash{{5.500000pt}{1.100000pt}}{0.000000pt}%
\pgfpathmoveto{\pgfqpoint{0.440392in}{1.891342in}}%
\pgfpathlineto{\pgfqpoint{1.046969in}{1.879660in}}%
\pgfpathlineto{\pgfqpoint{1.060246in}{1.869147in}}%
\pgfpathlineto{\pgfqpoint{1.068564in}{1.857016in}}%
\pgfpathlineto{\pgfqpoint{1.074652in}{1.847581in}}%
\pgfpathlineto{\pgfqpoint{1.079517in}{1.832036in}}%
\pgfpathlineto{\pgfqpoint{1.083575in}{1.815143in}}%
\pgfpathlineto{\pgfqpoint{1.087121in}{1.795374in}}%
\pgfpathlineto{\pgfqpoint{1.090170in}{1.771382in}}%
\pgfpathlineto{\pgfqpoint{1.092878in}{1.740741in}}%
\pgfpathlineto{\pgfqpoint{1.095379in}{1.694554in}}%
\pgfpathlineto{\pgfqpoint{1.097723in}{1.642616in}}%
\pgfpathlineto{\pgfqpoint{1.099916in}{1.580794in}}%
\pgfpathlineto{\pgfqpoint{1.101984in}{1.523465in}}%
\pgfpathlineto{\pgfqpoint{1.103941in}{1.458678in}}%
\pgfpathlineto{\pgfqpoint{1.105807in}{1.398204in}}%
\pgfpathlineto{\pgfqpoint{1.107688in}{1.332698in}}%
\pgfusepath{stroke}%
\end{pgfscope}%
\begin{pgfscope}%
\pgfpathrectangle{\pgfqpoint{0.440392in}{0.402778in}}{\pgfqpoint{2.868232in}{1.559448in}}%
\pgfusepath{clip}%
\pgfsetrectcap%
\pgfsetroundjoin%
\pgfsetlinewidth{1.104125pt}%
\definecolor{currentstroke}{rgb}{0.145098,0.145098,0.145098}%
\pgfsetstrokecolor{currentstroke}%
\pgfsetdash{}{0pt}%
\pgfpathmoveto{\pgfqpoint{0.440392in}{1.882895in}}%
\pgfpathlineto{\pgfqpoint{1.309174in}{1.872472in}}%
\pgfpathlineto{\pgfqpoint{1.404683in}{1.609639in}}%
\pgfpathlineto{\pgfqpoint{1.547190in}{1.597867in}}%
\pgfpathlineto{\pgfqpoint{1.634045in}{1.593824in}}%
\pgfpathlineto{\pgfqpoint{1.697227in}{1.526251in}}%
\pgfpathlineto{\pgfqpoint{1.762007in}{1.462362in}}%
\pgfpathlineto{\pgfqpoint{1.824725in}{1.456971in}}%
\pgfpathlineto{\pgfqpoint{1.881612in}{1.218579in}}%
\pgfpathlineto{\pgfqpoint{2.007850in}{1.081816in}}%
\pgfpathlineto{\pgfqpoint{2.154824in}{1.082086in}}%
\pgfpathlineto{\pgfqpoint{2.290274in}{1.080828in}}%
\pgfpathlineto{\pgfqpoint{2.412994in}{1.055668in}}%
\pgfpathlineto{\pgfqpoint{2.546372in}{0.980368in}}%
\pgfpathlineto{\pgfqpoint{2.696879in}{0.972730in}}%
\pgfpathlineto{\pgfqpoint{2.837162in}{0.955118in}}%
\pgfpathlineto{\pgfqpoint{2.983534in}{0.788612in}}%
\pgfpathlineto{\pgfqpoint{3.242659in}{0.719961in}}%
\pgfusepath{stroke}%
\end{pgfscope}%
\begin{pgfscope}%
\pgfpathrectangle{\pgfqpoint{0.440392in}{0.402778in}}{\pgfqpoint{2.868232in}{1.559448in}}%
\pgfusepath{clip}%
\pgfsetbuttcap%
\pgfsetroundjoin%
\pgfsetlinewidth{1.104125pt}%
\definecolor{currentstroke}{rgb}{0.145098,0.145098,0.145098}%
\pgfsetstrokecolor{currentstroke}%
\pgfsetdash{{5.500000pt}{1.100000pt}}{0.000000pt}%
\pgfpathmoveto{\pgfqpoint{0.440392in}{1.879660in}}%
\pgfpathlineto{\pgfqpoint{0.869381in}{1.662565in}}%
\pgfpathlineto{\pgfqpoint{1.089287in}{1.407190in}}%
\pgfpathlineto{\pgfqpoint{1.203583in}{1.368911in}}%
\pgfpathlineto{\pgfqpoint{1.320327in}{1.353725in}}%
\pgfpathlineto{\pgfqpoint{1.413880in}{1.356510in}}%
\pgfpathlineto{\pgfqpoint{1.496865in}{1.357319in}}%
\pgfpathlineto{\pgfqpoint{1.573058in}{1.343212in}}%
\pgfpathlineto{\pgfqpoint{1.645425in}{1.288219in}}%
\pgfpathlineto{\pgfqpoint{1.718153in}{1.209684in}}%
\pgfpathlineto{\pgfqpoint{1.791140in}{1.161700in}}%
\pgfpathlineto{\pgfqpoint{1.860968in}{1.143549in}}%
\pgfpathlineto{\pgfqpoint{1.926193in}{1.132586in}}%
\pgfpathlineto{\pgfqpoint{1.987831in}{1.120365in}}%
\pgfpathlineto{\pgfqpoint{2.048635in}{1.090263in}}%
\pgfpathlineto{\pgfqpoint{2.114898in}{1.054051in}}%
\pgfpathlineto{\pgfqpoint{2.186580in}{1.026824in}}%
\pgfpathlineto{\pgfqpoint{2.266547in}{0.970843in}}%
\pgfusepath{stroke}%
\end{pgfscope}%
\begin{pgfscope}%
\pgfpathrectangle{\pgfqpoint{0.440392in}{0.402778in}}{\pgfqpoint{2.868232in}{1.559448in}}%
\pgfusepath{clip}%
\pgfsetbuttcap%
\pgfsetroundjoin%
\pgfsetlinewidth{1.104125pt}%
\definecolor{currentstroke}{rgb}{1.000000,0.000000,0.000000}%
\pgfsetstrokecolor{currentstroke}%
\pgfsetdash{{1.100000pt}{1.815000pt}}{0.000000pt}%
\pgfpathmoveto{\pgfqpoint{0.440392in}{0.473662in}}%
\pgfpathlineto{\pgfqpoint{3.308624in}{0.473662in}}%
\pgfusepath{stroke}%
\end{pgfscope}%
\begin{pgfscope}%
\pgfsetrectcap%
\pgfsetmiterjoin%
\pgfsetlinewidth{0.803000pt}%
\definecolor{currentstroke}{rgb}{0.000000,0.000000,0.000000}%
\pgfsetstrokecolor{currentstroke}%
\pgfsetdash{}{0pt}%
\pgfpathmoveto{\pgfqpoint{0.440392in}{0.402778in}}%
\pgfpathlineto{\pgfqpoint{0.440392in}{1.962226in}}%
\pgfusepath{stroke}%
\end{pgfscope}%
\begin{pgfscope}%
\pgfsetrectcap%
\pgfsetmiterjoin%
\pgfsetlinewidth{0.803000pt}%
\definecolor{currentstroke}{rgb}{0.000000,0.000000,0.000000}%
\pgfsetstrokecolor{currentstroke}%
\pgfsetdash{}{0pt}%
\pgfpathmoveto{\pgfqpoint{3.308624in}{0.402778in}}%
\pgfpathlineto{\pgfqpoint{3.308624in}{1.962226in}}%
\pgfusepath{stroke}%
\end{pgfscope}%
\begin{pgfscope}%
\pgfsetrectcap%
\pgfsetmiterjoin%
\pgfsetlinewidth{0.803000pt}%
\definecolor{currentstroke}{rgb}{0.000000,0.000000,0.000000}%
\pgfsetstrokecolor{currentstroke}%
\pgfsetdash{}{0pt}%
\pgfpathmoveto{\pgfqpoint{0.440392in}{0.402778in}}%
\pgfpathlineto{\pgfqpoint{3.308624in}{0.402778in}}%
\pgfusepath{stroke}%
\end{pgfscope}%
\begin{pgfscope}%
\pgfsetrectcap%
\pgfsetmiterjoin%
\pgfsetlinewidth{0.803000pt}%
\definecolor{currentstroke}{rgb}{0.000000,0.000000,0.000000}%
\pgfsetstrokecolor{currentstroke}%
\pgfsetdash{}{0pt}%
\pgfpathmoveto{\pgfqpoint{0.440392in}{1.962226in}}%
\pgfpathlineto{\pgfqpoint{3.308624in}{1.962226in}}%
\pgfusepath{stroke}%
\end{pgfscope}%
\begin{pgfscope}%
\pgfsetrectcap%
\pgfsetroundjoin%
\pgfsetlinewidth{1.104125pt}%
\definecolor{currentstroke}{rgb}{0.631373,0.850980,0.607843}%
\pgfsetstrokecolor{currentstroke}%
\pgfsetdash{}{0pt}%
\pgfpathmoveto{\pgfqpoint{0.606985in}{2.213199in}}%
\pgfpathlineto{\pgfqpoint{0.829207in}{2.213199in}}%
\pgfusepath{stroke}%
\end{pgfscope}%
\begin{pgfscope}%
\pgftext[x=0.918096in,y=2.174310in,left,base]{\rmfamily\fontsize{8.000000}{9.600000}\selectfont FD}%
\end{pgfscope}%
\begin{pgfscope}%
\pgfsetbuttcap%
\pgfsetroundjoin%
\pgfsetlinewidth{1.104125pt}%
\definecolor{currentstroke}{rgb}{0.631373,0.850980,0.607843}%
\pgfsetstrokecolor{currentstroke}%
\pgfsetdash{{5.500000pt}{1.100000pt}}{0.000000pt}%
\pgfpathmoveto{\pgfqpoint{0.606985in}{2.058260in}}%
\pgfpathlineto{\pgfqpoint{0.829207in}{2.058260in}}%
\pgfusepath{stroke}%
\end{pgfscope}%
\begin{pgfscope}%
\pgftext[x=0.918096in,y=2.019371in,left,base]{\rmfamily\fontsize{8.000000}{9.600000}\selectfont FD+6}%
\end{pgfscope}%
\begin{pgfscope}%
\pgfsetrectcap%
\pgfsetroundjoin%
\pgfsetlinewidth{1.104125pt}%
\definecolor{currentstroke}{rgb}{0.145098,0.145098,0.145098}%
\pgfsetstrokecolor{currentstroke}%
\pgfsetdash{}{0pt}%
\pgfpathmoveto{\pgfqpoint{1.458301in}{2.213199in}}%
\pgfpathlineto{\pgfqpoint{1.680523in}{2.213199in}}%
\pgfusepath{stroke}%
\end{pgfscope}%
\begin{pgfscope}%
\pgftext[x=1.769412in,y=2.174310in,left,base]{\rmfamily\fontsize{8.000000}{9.600000}\selectfont L2C2}%
\end{pgfscope}%
\begin{pgfscope}%
\pgfsetbuttcap%
\pgfsetroundjoin%
\pgfsetlinewidth{1.104125pt}%
\definecolor{currentstroke}{rgb}{0.145098,0.145098,0.145098}%
\pgfsetstrokecolor{currentstroke}%
\pgfsetdash{{5.500000pt}{1.100000pt}}{0.000000pt}%
\pgfpathmoveto{\pgfqpoint{1.458301in}{2.058260in}}%
\pgfpathlineto{\pgfqpoint{1.680523in}{2.058260in}}%
\pgfusepath{stroke}%
\end{pgfscope}%
\begin{pgfscope}%
\pgftext[x=1.769412in,y=2.019371in,left,base]{\rmfamily\fontsize{8.000000}{9.600000}\selectfont L2C2-BF}%
\end{pgfscope}%
\begin{pgfscope}%
\pgfsetbuttcap%
\pgfsetroundjoin%
\pgfsetlinewidth{1.104125pt}%
\definecolor{currentstroke}{rgb}{1.000000,0.000000,0.000000}%
\pgfsetstrokecolor{currentstroke}%
\pgfsetdash{{1.100000pt}{1.815000pt}}{0.000000pt}%
\pgfpathmoveto{\pgfqpoint{2.468609in}{2.213199in}}%
\pgfpathlineto{\pgfqpoint{2.690831in}{2.213199in}}%
\pgfusepath{stroke}%
\end{pgfscope}%
\begin{pgfscope}%
\pgftext[x=2.779720in,y=2.174310in,left,base]{\rmfamily\fontsize{8.000000}{9.600000}\selectfont 0\% EC}%
\end{pgfscope}%
\end{pgfpicture}%
\makeatother%
\endgroup%

%% file: Figuras/ipc_32MB.pgf
\begingroup%
\makeatletter%
\begin{pgfpicture}%
\pgfpathrectangle{\pgfpointorigin}{\pgfqpoint{2.222509in}{1.591882in}}%
\pgfusepath{use as bounding box, clip}%
\begin{pgfscope}%
\pgfsetbuttcap%
\pgfsetmiterjoin%
\definecolor{currentfill}{rgb}{1.000000,1.000000,1.000000}%
\pgfsetfillcolor{currentfill}%
\pgfsetlinewidth{0.000000pt}%
\definecolor{currentstroke}{rgb}{1.000000,1.000000,1.000000}%
\pgfsetstrokecolor{currentstroke}%
\pgfsetdash{}{0pt}%
\pgfpathmoveto{\pgfqpoint{0.000000in}{0.000000in}}%
\pgfpathlineto{\pgfqpoint{2.222509in}{0.000000in}}%
\pgfpathlineto{\pgfqpoint{2.222509in}{1.591882in}}%
\pgfpathlineto{\pgfqpoint{0.000000in}{1.591882in}}%
\pgfpathclose%
\pgfusepath{fill}%
\end{pgfscope}%
\begin{pgfscope}%
\pgfsetbuttcap%
\pgfsetmiterjoin%
\definecolor{currentfill}{rgb}{1.000000,1.000000,1.000000}%
\pgfsetfillcolor{currentfill}%
\pgfsetlinewidth{0.000000pt}%
\definecolor{currentstroke}{rgb}{0.000000,0.000000,0.000000}%
\pgfsetstrokecolor{currentstroke}%
\pgfsetstrokeopacity{0.000000}%
\pgfsetdash{}{0pt}%
\pgfpathmoveto{\pgfqpoint{0.365549in}{0.311728in}}%
\pgfpathlineto{\pgfqpoint{2.175248in}{0.311728in}}%
\pgfpathlineto{\pgfqpoint{2.175248in}{1.556882in}}%
\pgfpathlineto{\pgfqpoint{0.365549in}{1.556882in}}%
\pgfpathclose%
\pgfusepath{fill}%
\end{pgfscope}%
\begin{pgfscope}%
\pgfsetbuttcap%
\pgfsetroundjoin%
\definecolor{currentfill}{rgb}{0.000000,0.000000,0.000000}%
\pgfsetfillcolor{currentfill}%
\pgfsetlinewidth{0.803000pt}%
\definecolor{currentstroke}{rgb}{0.000000,0.000000,0.000000}%
\pgfsetstrokecolor{currentstroke}%
\pgfsetdash{}{0pt}%
\pgfsys@defobject{currentmarker}{\pgfqpoint{0.000000in}{-0.048611in}}{\pgfqpoint{0.000000in}{0.000000in}}{%
\pgfpathmoveto{\pgfqpoint{0.000000in}{0.000000in}}%
\pgfpathlineto{\pgfqpoint{0.000000in}{-0.048611in}}%
\pgfusepath{stroke,fill}%
}%
\begin{pgfscope}%
\pgfsys@transformshift{0.365549in}{0.311728in}%
\pgfsys@useobject{currentmarker}{}%
\end{pgfscope}%
\end{pgfscope}%
\begin{pgfscope}%
\pgftext[x=0.365549in,y=0.214506in,,top]{\rmfamily\fontsize{5.000000}{6.000000}\selectfont \(\displaystyle 0\)}%
\end{pgfscope}%
\begin{pgfscope}%
\pgfsetbuttcap%
\pgfsetroundjoin%
\definecolor{currentfill}{rgb}{0.000000,0.000000,0.000000}%
\pgfsetfillcolor{currentfill}%
\pgfsetlinewidth{0.803000pt}%
\definecolor{currentstroke}{rgb}{0.000000,0.000000,0.000000}%
\pgfsetstrokecolor{currentstroke}%
\pgfsetdash{}{0pt}%
\pgfsys@defobject{currentmarker}{\pgfqpoint{0.000000in}{-0.048611in}}{\pgfqpoint{0.000000in}{0.000000in}}{%
\pgfpathmoveto{\pgfqpoint{0.000000in}{0.000000in}}%
\pgfpathlineto{\pgfqpoint{0.000000in}{-0.048611in}}%
\pgfusepath{stroke,fill}%
}%
\begin{pgfscope}%
\pgfsys@transformshift{0.624078in}{0.311728in}%
\pgfsys@useobject{currentmarker}{}%
\end{pgfscope}%
\end{pgfscope}%
\begin{pgfscope}%
\pgftext[x=0.624078in,y=0.214506in,,top]{\rmfamily\fontsize{5.000000}{6.000000}\selectfont \(\displaystyle 4\)}%
\end{pgfscope}%
\begin{pgfscope}%
\pgfsetbuttcap%
\pgfsetroundjoin%
\definecolor{currentfill}{rgb}{0.000000,0.000000,0.000000}%
\pgfsetfillcolor{currentfill}%
\pgfsetlinewidth{0.803000pt}%
\definecolor{currentstroke}{rgb}{0.000000,0.000000,0.000000}%
\pgfsetstrokecolor{currentstroke}%
\pgfsetdash{}{0pt}%
\pgfsys@defobject{currentmarker}{\pgfqpoint{0.000000in}{-0.048611in}}{\pgfqpoint{0.000000in}{0.000000in}}{%
\pgfpathmoveto{\pgfqpoint{0.000000in}{0.000000in}}%
\pgfpathlineto{\pgfqpoint{0.000000in}{-0.048611in}}%
\pgfusepath{stroke,fill}%
}%
\begin{pgfscope}%
\pgfsys@transformshift{0.882606in}{0.311728in}%
\pgfsys@useobject{currentmarker}{}%
\end{pgfscope}%
\end{pgfscope}%
\begin{pgfscope}%
\pgftext[x=0.882606in,y=0.214506in,,top]{\rmfamily\fontsize{5.000000}{6.000000}\selectfont \(\displaystyle 8\)}%
\end{pgfscope}%
\begin{pgfscope}%
\pgfsetbuttcap%
\pgfsetroundjoin%
\definecolor{currentfill}{rgb}{0.000000,0.000000,0.000000}%
\pgfsetfillcolor{currentfill}%
\pgfsetlinewidth{0.803000pt}%
\definecolor{currentstroke}{rgb}{0.000000,0.000000,0.000000}%
\pgfsetstrokecolor{currentstroke}%
\pgfsetdash{}{0pt}%
\pgfsys@defobject{currentmarker}{\pgfqpoint{0.000000in}{-0.048611in}}{\pgfqpoint{0.000000in}{0.000000in}}{%
\pgfpathmoveto{\pgfqpoint{0.000000in}{0.000000in}}%
\pgfpathlineto{\pgfqpoint{0.000000in}{-0.048611in}}%
\pgfusepath{stroke,fill}%
}%
\begin{pgfscope}%
\pgfsys@transformshift{1.141134in}{0.311728in}%
\pgfsys@useobject{currentmarker}{}%
\end{pgfscope}%
\end{pgfscope}%
\begin{pgfscope}%
\pgftext[x=1.141134in,y=0.214506in,,top]{\rmfamily\fontsize{5.000000}{6.000000}\selectfont \(\displaystyle 12\)}%
\end{pgfscope}%
\begin{pgfscope}%
\pgfsetbuttcap%
\pgfsetroundjoin%
\definecolor{currentfill}{rgb}{0.000000,0.000000,0.000000}%
\pgfsetfillcolor{currentfill}%
\pgfsetlinewidth{0.803000pt}%
\definecolor{currentstroke}{rgb}{0.000000,0.000000,0.000000}%
\pgfsetstrokecolor{currentstroke}%
\pgfsetdash{}{0pt}%
\pgfsys@defobject{currentmarker}{\pgfqpoint{0.000000in}{-0.048611in}}{\pgfqpoint{0.000000in}{0.000000in}}{%
\pgfpathmoveto{\pgfqpoint{0.000000in}{0.000000in}}%
\pgfpathlineto{\pgfqpoint{0.000000in}{-0.048611in}}%
\pgfusepath{stroke,fill}%
}%
\begin{pgfscope}%
\pgfsys@transformshift{1.399663in}{0.311728in}%
\pgfsys@useobject{currentmarker}{}%
\end{pgfscope}%
\end{pgfscope}%
\begin{pgfscope}%
\pgftext[x=1.399663in,y=0.214506in,,top]{\rmfamily\fontsize{5.000000}{6.000000}\selectfont \(\displaystyle 16\)}%
\end{pgfscope}%
\begin{pgfscope}%
\pgfsetbuttcap%
\pgfsetroundjoin%
\definecolor{currentfill}{rgb}{0.000000,0.000000,0.000000}%
\pgfsetfillcolor{currentfill}%
\pgfsetlinewidth{0.803000pt}%
\definecolor{currentstroke}{rgb}{0.000000,0.000000,0.000000}%
\pgfsetstrokecolor{currentstroke}%
\pgfsetdash{}{0pt}%
\pgfsys@defobject{currentmarker}{\pgfqpoint{0.000000in}{-0.048611in}}{\pgfqpoint{0.000000in}{0.000000in}}{%
\pgfpathmoveto{\pgfqpoint{0.000000in}{0.000000in}}%
\pgfpathlineto{\pgfqpoint{0.000000in}{-0.048611in}}%
\pgfusepath{stroke,fill}%
}%
\begin{pgfscope}%
\pgfsys@transformshift{1.658191in}{0.311728in}%
\pgfsys@useobject{currentmarker}{}%
\end{pgfscope}%
\end{pgfscope}%
\begin{pgfscope}%
\pgftext[x=1.658191in,y=0.214506in,,top]{\rmfamily\fontsize{5.000000}{6.000000}\selectfont \(\displaystyle 20\)}%
\end{pgfscope}%
\begin{pgfscope}%
\pgfsetbuttcap%
\pgfsetroundjoin%
\definecolor{currentfill}{rgb}{0.000000,0.000000,0.000000}%
\pgfsetfillcolor{currentfill}%
\pgfsetlinewidth{0.803000pt}%
\definecolor{currentstroke}{rgb}{0.000000,0.000000,0.000000}%
\pgfsetstrokecolor{currentstroke}%
\pgfsetdash{}{0pt}%
\pgfsys@defobject{currentmarker}{\pgfqpoint{0.000000in}{-0.048611in}}{\pgfqpoint{0.000000in}{0.000000in}}{%
\pgfpathmoveto{\pgfqpoint{0.000000in}{0.000000in}}%
\pgfpathlineto{\pgfqpoint{0.000000in}{-0.048611in}}%
\pgfusepath{stroke,fill}%
}%
\begin{pgfscope}%
\pgfsys@transformshift{1.916719in}{0.311728in}%
\pgfsys@useobject{currentmarker}{}%
\end{pgfscope}%
\end{pgfscope}%
\begin{pgfscope}%
\pgftext[x=1.916719in,y=0.214506in,,top]{\rmfamily\fontsize{5.000000}{6.000000}\selectfont \(\displaystyle 24\)}%
\end{pgfscope}%
\begin{pgfscope}%
\pgfsetbuttcap%
\pgfsetroundjoin%
\definecolor{currentfill}{rgb}{0.000000,0.000000,0.000000}%
\pgfsetfillcolor{currentfill}%
\pgfsetlinewidth{0.803000pt}%
\definecolor{currentstroke}{rgb}{0.000000,0.000000,0.000000}%
\pgfsetstrokecolor{currentstroke}%
\pgfsetdash{}{0pt}%
\pgfsys@defobject{currentmarker}{\pgfqpoint{0.000000in}{-0.048611in}}{\pgfqpoint{0.000000in}{0.000000in}}{%
\pgfpathmoveto{\pgfqpoint{0.000000in}{0.000000in}}%
\pgfpathlineto{\pgfqpoint{0.000000in}{-0.048611in}}%
\pgfusepath{stroke,fill}%
}%
\begin{pgfscope}%
\pgfsys@transformshift{2.175248in}{0.311728in}%
\pgfsys@useobject{currentmarker}{}%
\end{pgfscope}%
\end{pgfscope}%
\begin{pgfscope}%
\pgftext[x=2.175248in,y=0.214506in,,top]{\rmfamily\fontsize{5.000000}{6.000000}\selectfont \(\displaystyle 28\)}%
\end{pgfscope}%
\begin{pgfscope}%
\pgftext[x=1.270398in,y=0.097222in,,top]{\rmfamily\fontsize{7.000000}{8.400000}\selectfont Time (years)}%
\end{pgfscope}%
\begin{pgfscope}%
\pgfsetbuttcap%
\pgfsetroundjoin%
\definecolor{currentfill}{rgb}{0.000000,0.000000,0.000000}%
\pgfsetfillcolor{currentfill}%
\pgfsetlinewidth{0.803000pt}%
\definecolor{currentstroke}{rgb}{0.000000,0.000000,0.000000}%
\pgfsetstrokecolor{currentstroke}%
\pgfsetdash{}{0pt}%
\pgfsys@defobject{currentmarker}{\pgfqpoint{-0.048611in}{0.000000in}}{\pgfqpoint{0.000000in}{0.000000in}}{%
\pgfpathmoveto{\pgfqpoint{0.000000in}{0.000000in}}%
\pgfpathlineto{\pgfqpoint{-0.048611in}{0.000000in}}%
\pgfusepath{stroke,fill}%
}%
\begin{pgfscope}%
\pgfsys@transformshift{0.365549in}{0.477516in}%
\pgfsys@useobject{currentmarker}{}%
\end{pgfscope}%
\end{pgfscope}%
\begin{pgfscope}%
\pgftext[x=0.141975in,y=0.453404in,left,base]{\rmfamily\fontsize{5.000000}{6.000000}\selectfont \(\displaystyle 0.6\)}%
\end{pgfscope}%
\begin{pgfscope}%
\pgfsetbuttcap%
\pgfsetroundjoin%
\definecolor{currentfill}{rgb}{0.000000,0.000000,0.000000}%
\pgfsetfillcolor{currentfill}%
\pgfsetlinewidth{0.803000pt}%
\definecolor{currentstroke}{rgb}{0.000000,0.000000,0.000000}%
\pgfsetstrokecolor{currentstroke}%
\pgfsetdash{}{0pt}%
\pgfsys@defobject{currentmarker}{\pgfqpoint{-0.048611in}{0.000000in}}{\pgfqpoint{0.000000in}{0.000000in}}{%
\pgfpathmoveto{\pgfqpoint{0.000000in}{0.000000in}}%
\pgfpathlineto{\pgfqpoint{-0.048611in}{0.000000in}}%
\pgfusepath{stroke,fill}%
}%
\begin{pgfscope}%
\pgfsys@transformshift{0.365549in}{0.988900in}%
\pgfsys@useobject{currentmarker}{}%
\end{pgfscope}%
\end{pgfscope}%
\begin{pgfscope}%
\pgftext[x=0.141975in,y=0.964787in,left,base]{\rmfamily\fontsize{5.000000}{6.000000}\selectfont \(\displaystyle 0.8\)}%
\end{pgfscope}%
\begin{pgfscope}%
\pgfsetbuttcap%
\pgfsetroundjoin%
\definecolor{currentfill}{rgb}{0.000000,0.000000,0.000000}%
\pgfsetfillcolor{currentfill}%
\pgfsetlinewidth{0.803000pt}%
\definecolor{currentstroke}{rgb}{0.000000,0.000000,0.000000}%
\pgfsetstrokecolor{currentstroke}%
\pgfsetdash{}{0pt}%
\pgfsys@defobject{currentmarker}{\pgfqpoint{-0.048611in}{0.000000in}}{\pgfqpoint{0.000000in}{0.000000in}}{%
\pgfpathmoveto{\pgfqpoint{0.000000in}{0.000000in}}%
\pgfpathlineto{\pgfqpoint{-0.048611in}{0.000000in}}%
\pgfusepath{stroke,fill}%
}%
\begin{pgfscope}%
\pgfsys@transformshift{0.365549in}{1.500284in}%
\pgfsys@useobject{currentmarker}{}%
\end{pgfscope}%
\end{pgfscope}%
\begin{pgfscope}%
\pgftext[x=0.141975in,y=1.476171in,left,base]{\rmfamily\fontsize{5.000000}{6.000000}\selectfont \(\displaystyle 1.0\)}%
\end{pgfscope}%
\begin{pgfscope}%
\pgftext[x=0.086419in,y=0.934305in,,bottom,rotate=90.000000]{\rmfamily\fontsize{7.000000}{8.400000}\selectfont Normalized IPC}%
\end{pgfscope}%
\begin{pgfscope}%
\pgfpathrectangle{\pgfqpoint{0.365549in}{0.311728in}}{\pgfqpoint{1.809698in}{1.245154in}}%
\pgfusepath{clip}%
\pgfsetrectcap%
\pgfsetroundjoin%
\pgfsetlinewidth{0.501875pt}%
\definecolor{currentstroke}{rgb}{0.800000,0.800000,0.800000}%
\pgfsetstrokecolor{currentstroke}%
\pgfsetdash{}{0pt}%
\pgfpathmoveto{\pgfqpoint{0.430181in}{0.311728in}}%
\pgfpathlineto{\pgfqpoint{0.430181in}{1.556882in}}%
\pgfusepath{stroke}%
\end{pgfscope}%
\begin{pgfscope}%
\pgfpathrectangle{\pgfqpoint{0.365549in}{0.311728in}}{\pgfqpoint{1.809698in}{1.245154in}}%
\pgfusepath{clip}%
\pgfsetrectcap%
\pgfsetroundjoin%
\pgfsetlinewidth{0.501875pt}%
\definecolor{currentstroke}{rgb}{0.800000,0.800000,0.800000}%
\pgfsetstrokecolor{currentstroke}%
\pgfsetdash{}{0pt}%
\pgfpathmoveto{\pgfqpoint{0.494813in}{0.311728in}}%
\pgfpathlineto{\pgfqpoint{0.494813in}{1.556882in}}%
\pgfusepath{stroke}%
\end{pgfscope}%
\begin{pgfscope}%
\pgfpathrectangle{\pgfqpoint{0.365549in}{0.311728in}}{\pgfqpoint{1.809698in}{1.245154in}}%
\pgfusepath{clip}%
\pgfsetrectcap%
\pgfsetroundjoin%
\pgfsetlinewidth{0.501875pt}%
\definecolor{currentstroke}{rgb}{0.800000,0.800000,0.800000}%
\pgfsetstrokecolor{currentstroke}%
\pgfsetdash{}{0pt}%
\pgfpathmoveto{\pgfqpoint{0.559446in}{0.311728in}}%
\pgfpathlineto{\pgfqpoint{0.559446in}{1.556882in}}%
\pgfusepath{stroke}%
\end{pgfscope}%
\begin{pgfscope}%
\pgfpathrectangle{\pgfqpoint{0.365549in}{0.311728in}}{\pgfqpoint{1.809698in}{1.245154in}}%
\pgfusepath{clip}%
\pgfsetrectcap%
\pgfsetroundjoin%
\pgfsetlinewidth{0.501875pt}%
\definecolor{currentstroke}{rgb}{0.800000,0.800000,0.800000}%
\pgfsetstrokecolor{currentstroke}%
\pgfsetdash{}{0pt}%
\pgfpathmoveto{\pgfqpoint{0.624078in}{0.311728in}}%
\pgfpathlineto{\pgfqpoint{0.624078in}{1.556882in}}%
\pgfusepath{stroke}%
\end{pgfscope}%
\begin{pgfscope}%
\pgfpathrectangle{\pgfqpoint{0.365549in}{0.311728in}}{\pgfqpoint{1.809698in}{1.245154in}}%
\pgfusepath{clip}%
\pgfsetrectcap%
\pgfsetroundjoin%
\pgfsetlinewidth{0.501875pt}%
\definecolor{currentstroke}{rgb}{0.800000,0.800000,0.800000}%
\pgfsetstrokecolor{currentstroke}%
\pgfsetdash{}{0pt}%
\pgfpathmoveto{\pgfqpoint{0.688710in}{0.311728in}}%
\pgfpathlineto{\pgfqpoint{0.688710in}{1.556882in}}%
\pgfusepath{stroke}%
\end{pgfscope}%
\begin{pgfscope}%
\pgfpathrectangle{\pgfqpoint{0.365549in}{0.311728in}}{\pgfqpoint{1.809698in}{1.245154in}}%
\pgfusepath{clip}%
\pgfsetrectcap%
\pgfsetroundjoin%
\pgfsetlinewidth{0.501875pt}%
\definecolor{currentstroke}{rgb}{0.800000,0.800000,0.800000}%
\pgfsetstrokecolor{currentstroke}%
\pgfsetdash{}{0pt}%
\pgfpathmoveto{\pgfqpoint{0.753342in}{0.311728in}}%
\pgfpathlineto{\pgfqpoint{0.753342in}{1.556882in}}%
\pgfusepath{stroke}%
\end{pgfscope}%
\begin{pgfscope}%
\pgfpathrectangle{\pgfqpoint{0.365549in}{0.311728in}}{\pgfqpoint{1.809698in}{1.245154in}}%
\pgfusepath{clip}%
\pgfsetrectcap%
\pgfsetroundjoin%
\pgfsetlinewidth{0.501875pt}%
\definecolor{currentstroke}{rgb}{0.800000,0.800000,0.800000}%
\pgfsetstrokecolor{currentstroke}%
\pgfsetdash{}{0pt}%
\pgfpathmoveto{\pgfqpoint{0.817974in}{0.311728in}}%
\pgfpathlineto{\pgfqpoint{0.817974in}{1.556882in}}%
\pgfusepath{stroke}%
\end{pgfscope}%
\begin{pgfscope}%
\pgfpathrectangle{\pgfqpoint{0.365549in}{0.311728in}}{\pgfqpoint{1.809698in}{1.245154in}}%
\pgfusepath{clip}%
\pgfsetrectcap%
\pgfsetroundjoin%
\pgfsetlinewidth{0.501875pt}%
\definecolor{currentstroke}{rgb}{0.800000,0.800000,0.800000}%
\pgfsetstrokecolor{currentstroke}%
\pgfsetdash{}{0pt}%
\pgfpathmoveto{\pgfqpoint{0.882606in}{0.311728in}}%
\pgfpathlineto{\pgfqpoint{0.882606in}{1.556882in}}%
\pgfusepath{stroke}%
\end{pgfscope}%
\begin{pgfscope}%
\pgfpathrectangle{\pgfqpoint{0.365549in}{0.311728in}}{\pgfqpoint{1.809698in}{1.245154in}}%
\pgfusepath{clip}%
\pgfsetrectcap%
\pgfsetroundjoin%
\pgfsetlinewidth{0.501875pt}%
\definecolor{currentstroke}{rgb}{0.800000,0.800000,0.800000}%
\pgfsetstrokecolor{currentstroke}%
\pgfsetdash{}{0pt}%
\pgfpathmoveto{\pgfqpoint{0.947238in}{0.311728in}}%
\pgfpathlineto{\pgfqpoint{0.947238in}{1.556882in}}%
\pgfusepath{stroke}%
\end{pgfscope}%
\begin{pgfscope}%
\pgfpathrectangle{\pgfqpoint{0.365549in}{0.311728in}}{\pgfqpoint{1.809698in}{1.245154in}}%
\pgfusepath{clip}%
\pgfsetrectcap%
\pgfsetroundjoin%
\pgfsetlinewidth{0.501875pt}%
\definecolor{currentstroke}{rgb}{0.800000,0.800000,0.800000}%
\pgfsetstrokecolor{currentstroke}%
\pgfsetdash{}{0pt}%
\pgfpathmoveto{\pgfqpoint{1.011870in}{0.311728in}}%
\pgfpathlineto{\pgfqpoint{1.011870in}{1.556882in}}%
\pgfusepath{stroke}%
\end{pgfscope}%
\begin{pgfscope}%
\pgfpathrectangle{\pgfqpoint{0.365549in}{0.311728in}}{\pgfqpoint{1.809698in}{1.245154in}}%
\pgfusepath{clip}%
\pgfsetrectcap%
\pgfsetroundjoin%
\pgfsetlinewidth{0.501875pt}%
\definecolor{currentstroke}{rgb}{0.800000,0.800000,0.800000}%
\pgfsetstrokecolor{currentstroke}%
\pgfsetdash{}{0pt}%
\pgfpathmoveto{\pgfqpoint{1.076502in}{0.311728in}}%
\pgfpathlineto{\pgfqpoint{1.076502in}{1.556882in}}%
\pgfusepath{stroke}%
\end{pgfscope}%
\begin{pgfscope}%
\pgfpathrectangle{\pgfqpoint{0.365549in}{0.311728in}}{\pgfqpoint{1.809698in}{1.245154in}}%
\pgfusepath{clip}%
\pgfsetrectcap%
\pgfsetroundjoin%
\pgfsetlinewidth{0.501875pt}%
\definecolor{currentstroke}{rgb}{0.800000,0.800000,0.800000}%
\pgfsetstrokecolor{currentstroke}%
\pgfsetdash{}{0pt}%
\pgfpathmoveto{\pgfqpoint{1.141134in}{0.311728in}}%
\pgfpathlineto{\pgfqpoint{1.141134in}{1.556882in}}%
\pgfusepath{stroke}%
\end{pgfscope}%
\begin{pgfscope}%
\pgfpathrectangle{\pgfqpoint{0.365549in}{0.311728in}}{\pgfqpoint{1.809698in}{1.245154in}}%
\pgfusepath{clip}%
\pgfsetrectcap%
\pgfsetroundjoin%
\pgfsetlinewidth{0.501875pt}%
\definecolor{currentstroke}{rgb}{0.800000,0.800000,0.800000}%
\pgfsetstrokecolor{currentstroke}%
\pgfsetdash{}{0pt}%
\pgfpathmoveto{\pgfqpoint{1.205766in}{0.311728in}}%
\pgfpathlineto{\pgfqpoint{1.205766in}{1.556882in}}%
\pgfusepath{stroke}%
\end{pgfscope}%
\begin{pgfscope}%
\pgfpathrectangle{\pgfqpoint{0.365549in}{0.311728in}}{\pgfqpoint{1.809698in}{1.245154in}}%
\pgfusepath{clip}%
\pgfsetrectcap%
\pgfsetroundjoin%
\pgfsetlinewidth{0.501875pt}%
\definecolor{currentstroke}{rgb}{0.800000,0.800000,0.800000}%
\pgfsetstrokecolor{currentstroke}%
\pgfsetdash{}{0pt}%
\pgfpathmoveto{\pgfqpoint{1.270398in}{0.311728in}}%
\pgfpathlineto{\pgfqpoint{1.270398in}{1.556882in}}%
\pgfusepath{stroke}%
\end{pgfscope}%
\begin{pgfscope}%
\pgfpathrectangle{\pgfqpoint{0.365549in}{0.311728in}}{\pgfqpoint{1.809698in}{1.245154in}}%
\pgfusepath{clip}%
\pgfsetrectcap%
\pgfsetroundjoin%
\pgfsetlinewidth{0.501875pt}%
\definecolor{currentstroke}{rgb}{0.800000,0.800000,0.800000}%
\pgfsetstrokecolor{currentstroke}%
\pgfsetdash{}{0pt}%
\pgfpathmoveto{\pgfqpoint{1.335031in}{0.311728in}}%
\pgfpathlineto{\pgfqpoint{1.335031in}{1.556882in}}%
\pgfusepath{stroke}%
\end{pgfscope}%
\begin{pgfscope}%
\pgfpathrectangle{\pgfqpoint{0.365549in}{0.311728in}}{\pgfqpoint{1.809698in}{1.245154in}}%
\pgfusepath{clip}%
\pgfsetrectcap%
\pgfsetroundjoin%
\pgfsetlinewidth{0.501875pt}%
\definecolor{currentstroke}{rgb}{0.800000,0.800000,0.800000}%
\pgfsetstrokecolor{currentstroke}%
\pgfsetdash{}{0pt}%
\pgfpathmoveto{\pgfqpoint{1.399663in}{0.311728in}}%
\pgfpathlineto{\pgfqpoint{1.399663in}{1.556882in}}%
\pgfusepath{stroke}%
\end{pgfscope}%
\begin{pgfscope}%
\pgfpathrectangle{\pgfqpoint{0.365549in}{0.311728in}}{\pgfqpoint{1.809698in}{1.245154in}}%
\pgfusepath{clip}%
\pgfsetrectcap%
\pgfsetroundjoin%
\pgfsetlinewidth{0.501875pt}%
\definecolor{currentstroke}{rgb}{0.800000,0.800000,0.800000}%
\pgfsetstrokecolor{currentstroke}%
\pgfsetdash{}{0pt}%
\pgfpathmoveto{\pgfqpoint{1.464295in}{0.311728in}}%
\pgfpathlineto{\pgfqpoint{1.464295in}{1.556882in}}%
\pgfusepath{stroke}%
\end{pgfscope}%
\begin{pgfscope}%
\pgfpathrectangle{\pgfqpoint{0.365549in}{0.311728in}}{\pgfqpoint{1.809698in}{1.245154in}}%
\pgfusepath{clip}%
\pgfsetrectcap%
\pgfsetroundjoin%
\pgfsetlinewidth{0.501875pt}%
\definecolor{currentstroke}{rgb}{0.800000,0.800000,0.800000}%
\pgfsetstrokecolor{currentstroke}%
\pgfsetdash{}{0pt}%
\pgfpathmoveto{\pgfqpoint{1.528927in}{0.311728in}}%
\pgfpathlineto{\pgfqpoint{1.528927in}{1.556882in}}%
\pgfusepath{stroke}%
\end{pgfscope}%
\begin{pgfscope}%
\pgfpathrectangle{\pgfqpoint{0.365549in}{0.311728in}}{\pgfqpoint{1.809698in}{1.245154in}}%
\pgfusepath{clip}%
\pgfsetrectcap%
\pgfsetroundjoin%
\pgfsetlinewidth{0.501875pt}%
\definecolor{currentstroke}{rgb}{0.800000,0.800000,0.800000}%
\pgfsetstrokecolor{currentstroke}%
\pgfsetdash{}{0pt}%
\pgfpathmoveto{\pgfqpoint{1.593559in}{0.311728in}}%
\pgfpathlineto{\pgfqpoint{1.593559in}{1.556882in}}%
\pgfusepath{stroke}%
\end{pgfscope}%
\begin{pgfscope}%
\pgfpathrectangle{\pgfqpoint{0.365549in}{0.311728in}}{\pgfqpoint{1.809698in}{1.245154in}}%
\pgfusepath{clip}%
\pgfsetrectcap%
\pgfsetroundjoin%
\pgfsetlinewidth{0.501875pt}%
\definecolor{currentstroke}{rgb}{0.800000,0.800000,0.800000}%
\pgfsetstrokecolor{currentstroke}%
\pgfsetdash{}{0pt}%
\pgfpathmoveto{\pgfqpoint{1.658191in}{0.311728in}}%
\pgfpathlineto{\pgfqpoint{1.658191in}{1.556882in}}%
\pgfusepath{stroke}%
\end{pgfscope}%
\begin{pgfscope}%
\pgfpathrectangle{\pgfqpoint{0.365549in}{0.311728in}}{\pgfqpoint{1.809698in}{1.245154in}}%
\pgfusepath{clip}%
\pgfsetrectcap%
\pgfsetroundjoin%
\pgfsetlinewidth{0.501875pt}%
\definecolor{currentstroke}{rgb}{0.800000,0.800000,0.800000}%
\pgfsetstrokecolor{currentstroke}%
\pgfsetdash{}{0pt}%
\pgfpathmoveto{\pgfqpoint{1.722823in}{0.311728in}}%
\pgfpathlineto{\pgfqpoint{1.722823in}{1.556882in}}%
\pgfusepath{stroke}%
\end{pgfscope}%
\begin{pgfscope}%
\pgfpathrectangle{\pgfqpoint{0.365549in}{0.311728in}}{\pgfqpoint{1.809698in}{1.245154in}}%
\pgfusepath{clip}%
\pgfsetrectcap%
\pgfsetroundjoin%
\pgfsetlinewidth{0.501875pt}%
\definecolor{currentstroke}{rgb}{0.800000,0.800000,0.800000}%
\pgfsetstrokecolor{currentstroke}%
\pgfsetdash{}{0pt}%
\pgfpathmoveto{\pgfqpoint{1.787455in}{0.311728in}}%
\pgfpathlineto{\pgfqpoint{1.787455in}{1.556882in}}%
\pgfusepath{stroke}%
\end{pgfscope}%
\begin{pgfscope}%
\pgfpathrectangle{\pgfqpoint{0.365549in}{0.311728in}}{\pgfqpoint{1.809698in}{1.245154in}}%
\pgfusepath{clip}%
\pgfsetrectcap%
\pgfsetroundjoin%
\pgfsetlinewidth{0.501875pt}%
\definecolor{currentstroke}{rgb}{0.800000,0.800000,0.800000}%
\pgfsetstrokecolor{currentstroke}%
\pgfsetdash{}{0pt}%
\pgfpathmoveto{\pgfqpoint{1.852087in}{0.311728in}}%
\pgfpathlineto{\pgfqpoint{1.852087in}{1.556882in}}%
\pgfusepath{stroke}%
\end{pgfscope}%
\begin{pgfscope}%
\pgfpathrectangle{\pgfqpoint{0.365549in}{0.311728in}}{\pgfqpoint{1.809698in}{1.245154in}}%
\pgfusepath{clip}%
\pgfsetrectcap%
\pgfsetroundjoin%
\pgfsetlinewidth{0.501875pt}%
\definecolor{currentstroke}{rgb}{0.800000,0.800000,0.800000}%
\pgfsetstrokecolor{currentstroke}%
\pgfsetdash{}{0pt}%
\pgfpathmoveto{\pgfqpoint{1.916719in}{0.311728in}}%
\pgfpathlineto{\pgfqpoint{1.916719in}{1.556882in}}%
\pgfusepath{stroke}%
\end{pgfscope}%
\begin{pgfscope}%
\pgfpathrectangle{\pgfqpoint{0.365549in}{0.311728in}}{\pgfqpoint{1.809698in}{1.245154in}}%
\pgfusepath{clip}%
\pgfsetrectcap%
\pgfsetroundjoin%
\pgfsetlinewidth{0.501875pt}%
\definecolor{currentstroke}{rgb}{0.800000,0.800000,0.800000}%
\pgfsetstrokecolor{currentstroke}%
\pgfsetdash{}{0pt}%
\pgfpathmoveto{\pgfqpoint{1.981351in}{0.311728in}}%
\pgfpathlineto{\pgfqpoint{1.981351in}{1.556882in}}%
\pgfusepath{stroke}%
\end{pgfscope}%
\begin{pgfscope}%
\pgfpathrectangle{\pgfqpoint{0.365549in}{0.311728in}}{\pgfqpoint{1.809698in}{1.245154in}}%
\pgfusepath{clip}%
\pgfsetrectcap%
\pgfsetroundjoin%
\pgfsetlinewidth{0.501875pt}%
\definecolor{currentstroke}{rgb}{0.800000,0.800000,0.800000}%
\pgfsetstrokecolor{currentstroke}%
\pgfsetdash{}{0pt}%
\pgfpathmoveto{\pgfqpoint{2.045984in}{0.311728in}}%
\pgfpathlineto{\pgfqpoint{2.045984in}{1.556882in}}%
\pgfusepath{stroke}%
\end{pgfscope}%
\begin{pgfscope}%
\pgfpathrectangle{\pgfqpoint{0.365549in}{0.311728in}}{\pgfqpoint{1.809698in}{1.245154in}}%
\pgfusepath{clip}%
\pgfsetrectcap%
\pgfsetroundjoin%
\pgfsetlinewidth{0.501875pt}%
\definecolor{currentstroke}{rgb}{0.800000,0.800000,0.800000}%
\pgfsetstrokecolor{currentstroke}%
\pgfsetdash{}{0pt}%
\pgfpathmoveto{\pgfqpoint{2.110616in}{0.311728in}}%
\pgfpathlineto{\pgfqpoint{2.110616in}{1.556882in}}%
\pgfusepath{stroke}%
\end{pgfscope}%
\begin{pgfscope}%
\pgfpathrectangle{\pgfqpoint{0.365549in}{0.311728in}}{\pgfqpoint{1.809698in}{1.245154in}}%
\pgfusepath{clip}%
\pgfsetrectcap%
\pgfsetroundjoin%
\pgfsetlinewidth{1.104125pt}%
\definecolor{currentstroke}{rgb}{0.631373,0.850980,0.607843}%
\pgfsetstrokecolor{currentstroke}%
\pgfsetdash{}{0pt}%
\pgfpathmoveto{\pgfqpoint{0.365549in}{1.500087in}}%
\pgfpathlineto{\pgfqpoint{0.538781in}{1.491485in}}%
\pgfpathlineto{\pgfqpoint{0.563922in}{1.484196in}}%
\pgfpathlineto{\pgfqpoint{0.579012in}{1.477104in}}%
\pgfpathlineto{\pgfqpoint{0.589698in}{1.469355in}}%
\pgfpathlineto{\pgfqpoint{0.598062in}{1.463970in}}%
\pgfpathlineto{\pgfqpoint{0.604893in}{1.456550in}}%
\pgfpathlineto{\pgfqpoint{0.610585in}{1.446372in}}%
\pgfpathlineto{\pgfqpoint{0.615467in}{1.432648in}}%
\pgfpathlineto{\pgfqpoint{0.619769in}{1.414983in}}%
\pgfpathlineto{\pgfqpoint{0.623539in}{1.387206in}}%
\pgfpathlineto{\pgfqpoint{0.626959in}{1.350827in}}%
\pgfpathlineto{\pgfqpoint{0.630178in}{1.308604in}}%
\pgfpathlineto{\pgfqpoint{0.633152in}{1.251474in}}%
\pgfpathlineto{\pgfqpoint{0.635970in}{1.189879in}}%
\pgfpathlineto{\pgfqpoint{0.638770in}{1.134194in}}%
\pgfpathlineto{\pgfqpoint{0.641441in}{1.083630in}}%
\pgfusepath{stroke}%
\end{pgfscope}%
\begin{pgfscope}%
\pgfpathrectangle{\pgfqpoint{0.365549in}{0.311728in}}{\pgfqpoint{1.809698in}{1.245154in}}%
\pgfusepath{clip}%
\pgfsetbuttcap%
\pgfsetroundjoin%
\pgfsetlinewidth{1.104125pt}%
\definecolor{currentstroke}{rgb}{0.631373,0.850980,0.607843}%
\pgfsetstrokecolor{currentstroke}%
\pgfsetdash{{5.500000pt}{1.100000pt}}{0.000000pt}%
\pgfpathmoveto{\pgfqpoint{0.365549in}{1.500284in}}%
\pgfpathlineto{\pgfqpoint{0.738291in}{1.492207in}}%
\pgfpathlineto{\pgfqpoint{0.746447in}{1.482620in}}%
\pgfpathlineto{\pgfqpoint{0.751472in}{1.476710in}}%
\pgfpathlineto{\pgfqpoint{0.755187in}{1.470537in}}%
\pgfpathlineto{\pgfqpoint{0.758128in}{1.463445in}}%
\pgfpathlineto{\pgfqpoint{0.760591in}{1.459242in}}%
\pgfpathlineto{\pgfqpoint{0.762709in}{1.449458in}}%
\pgfpathlineto{\pgfqpoint{0.764525in}{1.438032in}}%
\pgfpathlineto{\pgfqpoint{0.766134in}{1.425884in}}%
\pgfpathlineto{\pgfqpoint{0.767597in}{1.409599in}}%
\pgfpathlineto{\pgfqpoint{0.768946in}{1.382084in}}%
\pgfpathlineto{\pgfqpoint{0.770187in}{1.347544in}}%
\pgfpathlineto{\pgfqpoint{0.771370in}{1.304401in}}%
\pgfpathlineto{\pgfqpoint{0.772491in}{1.250358in}}%
\pgfpathlineto{\pgfqpoint{0.773573in}{1.191323in}}%
\pgfpathlineto{\pgfqpoint{0.774639in}{1.133012in}}%
\pgfusepath{stroke}%
\end{pgfscope}%
\begin{pgfscope}%
\pgfpathrectangle{\pgfqpoint{0.365549in}{0.311728in}}{\pgfqpoint{1.809698in}{1.245154in}}%
\pgfusepath{clip}%
\pgfsetrectcap%
\pgfsetroundjoin%
\pgfsetlinewidth{1.104125pt}%
\definecolor{currentstroke}{rgb}{0.145098,0.145098,0.145098}%
\pgfsetstrokecolor{currentstroke}%
\pgfsetdash{}{0pt}%
\pgfpathmoveto{\pgfqpoint{0.365549in}{1.494571in}}%
\pgfpathlineto{\pgfqpoint{0.945357in}{1.454186in}}%
\pgfpathlineto{\pgfqpoint{1.021309in}{1.321868in}}%
\pgfpathlineto{\pgfqpoint{1.071902in}{1.306305in}}%
\pgfpathlineto{\pgfqpoint{1.110686in}{1.265526in}}%
\pgfpathlineto{\pgfqpoint{1.150942in}{1.212993in}}%
\pgfpathlineto{\pgfqpoint{1.189163in}{1.212271in}}%
\pgfpathlineto{\pgfqpoint{1.222315in}{1.035497in}}%
\pgfpathlineto{\pgfqpoint{1.306207in}{0.902391in}}%
\pgfpathlineto{\pgfqpoint{1.402824in}{0.902457in}}%
\pgfpathlineto{\pgfqpoint{1.491839in}{0.901143in}}%
\pgfpathlineto{\pgfqpoint{1.570926in}{0.878095in}}%
\pgfpathlineto{\pgfqpoint{1.650824in}{0.813544in}}%
\pgfpathlineto{\pgfqpoint{1.743259in}{0.801068in}}%
\pgfpathlineto{\pgfqpoint{1.830987in}{0.782813in}}%
\pgfpathlineto{\pgfqpoint{1.923984in}{0.644979in}}%
\pgfpathlineto{\pgfqpoint{2.091823in}{0.588637in}}%
\pgfusepath{stroke}%
\end{pgfscope}%
\begin{pgfscope}%
\pgfpathrectangle{\pgfqpoint{0.365549in}{0.311728in}}{\pgfqpoint{1.809698in}{1.245154in}}%
\pgfusepath{clip}%
\pgfsetbuttcap%
\pgfsetroundjoin%
\pgfsetlinewidth{1.104125pt}%
\definecolor{currentstroke}{rgb}{0.145098,0.145098,0.145098}%
\pgfsetstrokecolor{currentstroke}%
\pgfsetdash{{5.500000pt}{1.100000pt}}{0.000000pt}%
\pgfpathmoveto{\pgfqpoint{0.365549in}{1.495556in}}%
\pgfpathlineto{\pgfqpoint{1.102146in}{1.493323in}}%
\pgfpathlineto{\pgfqpoint{1.131600in}{1.360283in}}%
\pgfpathlineto{\pgfqpoint{1.157524in}{1.306765in}}%
\pgfpathlineto{\pgfqpoint{1.180692in}{1.283256in}}%
\pgfpathlineto{\pgfqpoint{1.205515in}{1.214766in}}%
\pgfpathlineto{\pgfqpoint{1.232242in}{1.212140in}}%
\pgfpathlineto{\pgfqpoint{1.257528in}{1.025122in}}%
\pgfpathlineto{\pgfqpoint{1.322679in}{0.902457in}}%
\pgfpathlineto{\pgfqpoint{1.401657in}{0.902457in}}%
\pgfpathlineto{\pgfqpoint{1.473942in}{0.900749in}}%
\pgfpathlineto{\pgfqpoint{1.540743in}{0.875731in}}%
\pgfpathlineto{\pgfqpoint{1.611593in}{0.809473in}}%
\pgfpathlineto{\pgfqpoint{1.692578in}{0.801593in}}%
\pgfpathlineto{\pgfqpoint{1.769724in}{0.783141in}}%
\pgfpathlineto{\pgfqpoint{1.851968in}{0.641498in}}%
\pgfpathlineto{\pgfqpoint{2.003822in}{0.588243in}}%
\pgfusepath{stroke}%
\end{pgfscope}%
\begin{pgfscope}%
\pgfpathrectangle{\pgfqpoint{0.365549in}{0.311728in}}{\pgfqpoint{1.809698in}{1.245154in}}%
\pgfusepath{clip}%
\pgfsetbuttcap%
\pgfsetroundjoin%
\pgfsetlinewidth{1.104125pt}%
\definecolor{currentstroke}{rgb}{1.000000,0.000000,0.000000}%
\pgfsetstrokecolor{currentstroke}%
\pgfsetdash{{1.100000pt}{1.815000pt}}{0.000000pt}%
\pgfpathmoveto{\pgfqpoint{0.365549in}{0.368326in}}%
\pgfpathlineto{\pgfqpoint{2.175248in}{0.368326in}}%
\pgfusepath{stroke}%
\end{pgfscope}%
\begin{pgfscope}%
\pgfsetrectcap%
\pgfsetmiterjoin%
\pgfsetlinewidth{0.803000pt}%
\definecolor{currentstroke}{rgb}{0.000000,0.000000,0.000000}%
\pgfsetstrokecolor{currentstroke}%
\pgfsetdash{}{0pt}%
\pgfpathmoveto{\pgfqpoint{0.365549in}{0.311728in}}%
\pgfpathlineto{\pgfqpoint{0.365549in}{1.556882in}}%
\pgfusepath{stroke}%
\end{pgfscope}%
\begin{pgfscope}%
\pgfsetrectcap%
\pgfsetmiterjoin%
\pgfsetlinewidth{0.803000pt}%
\definecolor{currentstroke}{rgb}{0.000000,0.000000,0.000000}%
\pgfsetstrokecolor{currentstroke}%
\pgfsetdash{}{0pt}%
\pgfpathmoveto{\pgfqpoint{2.175248in}{0.311728in}}%
\pgfpathlineto{\pgfqpoint{2.175248in}{1.556882in}}%
\pgfusepath{stroke}%
\end{pgfscope}%
\begin{pgfscope}%
\pgfsetrectcap%
\pgfsetmiterjoin%
\pgfsetlinewidth{0.803000pt}%
\definecolor{currentstroke}{rgb}{0.000000,0.000000,0.000000}%
\pgfsetstrokecolor{currentstroke}%
\pgfsetdash{}{0pt}%
\pgfpathmoveto{\pgfqpoint{0.365549in}{0.311728in}}%
\pgfpathlineto{\pgfqpoint{2.175248in}{0.311728in}}%
\pgfusepath{stroke}%
\end{pgfscope}%
\begin{pgfscope}%
\pgfsetrectcap%
\pgfsetmiterjoin%
\pgfsetlinewidth{0.803000pt}%
\definecolor{currentstroke}{rgb}{0.000000,0.000000,0.000000}%
\pgfsetstrokecolor{currentstroke}%
\pgfsetdash{}{0pt}%
\pgfpathmoveto{\pgfqpoint{0.365549in}{1.556882in}}%
\pgfpathlineto{\pgfqpoint{2.175248in}{1.556882in}}%
\pgfusepath{stroke}%
\end{pgfscope}%
\begin{pgfscope}%
\pgfsetbuttcap%
\pgfsetmiterjoin%
\definecolor{currentfill}{rgb}{1.000000,1.000000,1.000000}%
\pgfsetfillcolor{currentfill}%
\pgfsetfillopacity{0.800000}%
\pgfsetlinewidth{1.003750pt}%
\definecolor{currentstroke}{rgb}{0.800000,0.800000,0.800000}%
\pgfsetstrokecolor{currentstroke}%
\pgfsetstrokeopacity{0.800000}%
\pgfsetdash{}{0pt}%
\pgfpathmoveto{\pgfqpoint{1.484203in}{0.909196in}}%
\pgfpathlineto{\pgfqpoint{2.116914in}{0.909196in}}%
\pgfpathquadraticcurveto{\pgfqpoint{2.133581in}{0.909196in}}{\pgfqpoint{2.133581in}{0.925863in}}%
\pgfpathlineto{\pgfqpoint{2.133581in}{1.498549in}}%
\pgfpathquadraticcurveto{\pgfqpoint{2.133581in}{1.515215in}}{\pgfqpoint{2.116914in}{1.515215in}}%
\pgfpathlineto{\pgfqpoint{1.484203in}{1.515215in}}%
\pgfpathquadraticcurveto{\pgfqpoint{1.467536in}{1.515215in}}{\pgfqpoint{1.467536in}{1.498549in}}%
\pgfpathlineto{\pgfqpoint{1.467536in}{0.925863in}}%
\pgfpathquadraticcurveto{\pgfqpoint{1.467536in}{0.909196in}}{\pgfqpoint{1.484203in}{0.909196in}}%
\pgfpathclose%
\pgfusepath{stroke,fill}%
\end{pgfscope}%
\begin{pgfscope}%
\pgfsetrectcap%
\pgfsetroundjoin%
\pgfsetlinewidth{1.104125pt}%
\definecolor{currentstroke}{rgb}{0.631373,0.850980,0.607843}%
\pgfsetstrokecolor{currentstroke}%
\pgfsetdash{}{0pt}%
\pgfpathmoveto{\pgfqpoint{1.500870in}{1.452715in}}%
\pgfpathlineto{\pgfqpoint{1.667536in}{1.452715in}}%
\pgfusepath{stroke}%
\end{pgfscope}%
\begin{pgfscope}%
\pgftext[x=1.734203in,y=1.423549in,left,base]{\rmfamily\fontsize{6.000000}{7.200000}\selectfont FD}%
\end{pgfscope}%
\begin{pgfscope}%
\pgfsetbuttcap%
\pgfsetroundjoin%
\pgfsetlinewidth{1.104125pt}%
\definecolor{currentstroke}{rgb}{0.631373,0.850980,0.607843}%
\pgfsetstrokecolor{currentstroke}%
\pgfsetdash{{5.500000pt}{1.100000pt}}{0.000000pt}%
\pgfpathmoveto{\pgfqpoint{1.500870in}{1.336511in}}%
\pgfpathlineto{\pgfqpoint{1.667536in}{1.336511in}}%
\pgfusepath{stroke}%
\end{pgfscope}%
\begin{pgfscope}%
\pgftext[x=1.734203in,y=1.307345in,left,base]{\rmfamily\fontsize{6.000000}{7.200000}\selectfont FD+6}%
\end{pgfscope}%
\begin{pgfscope}%
\pgfsetrectcap%
\pgfsetroundjoin%
\pgfsetlinewidth{1.104125pt}%
\definecolor{currentstroke}{rgb}{0.145098,0.145098,0.145098}%
\pgfsetstrokecolor{currentstroke}%
\pgfsetdash{}{0pt}%
\pgfpathmoveto{\pgfqpoint{1.500870in}{1.220308in}}%
\pgfpathlineto{\pgfqpoint{1.667536in}{1.220308in}}%
\pgfusepath{stroke}%
\end{pgfscope}%
\begin{pgfscope}%
\pgftext[x=1.734203in,y=1.191141in,left,base]{\rmfamily\fontsize{6.000000}{7.200000}\selectfont L2C2}%
\end{pgfscope}%
\begin{pgfscope}%
\pgfsetbuttcap%
\pgfsetroundjoin%
\pgfsetlinewidth{1.104125pt}%
\definecolor{currentstroke}{rgb}{0.145098,0.145098,0.145098}%
\pgfsetstrokecolor{currentstroke}%
\pgfsetdash{{5.500000pt}{1.100000pt}}{0.000000pt}%
\pgfpathmoveto{\pgfqpoint{1.500870in}{1.104104in}}%
\pgfpathlineto{\pgfqpoint{1.667536in}{1.104104in}}%
\pgfusepath{stroke}%
\end{pgfscope}%
\begin{pgfscope}%
\pgftext[x=1.734203in,y=1.074937in,left,base]{\rmfamily\fontsize{6.000000}{7.200000}\selectfont L2C2+6}%
\end{pgfscope}%
\begin{pgfscope}%
\pgfsetbuttcap%
\pgfsetroundjoin%
\pgfsetlinewidth{1.104125pt}%
\definecolor{currentstroke}{rgb}{1.000000,0.000000,0.000000}%
\pgfsetstrokecolor{currentstroke}%
\pgfsetdash{{1.100000pt}{1.815000pt}}{0.000000pt}%
\pgfpathmoveto{\pgfqpoint{1.500870in}{0.987900in}}%
\pgfpathlineto{\pgfqpoint{1.667536in}{0.987900in}}%
\pgfusepath{stroke}%
\end{pgfscope}%
\begin{pgfscope}%
\pgftext[x=1.734203in,y=0.958734in,left,base]{\rmfamily\fontsize{6.000000}{7.200000}\selectfont 0\% EC}%
\end{pgfscope}%
\end{pgfpicture}%
\makeatother%
\endgroup%

%% file: Figuras/ipc_8c.pgf
\begingroup%
\makeatletter%
\begin{pgfpicture}%
\pgfpathrectangle{\pgfpointorigin}{\pgfqpoint{2.222509in}{1.591882in}}%
\pgfusepath{use as bounding box, clip}%
\begin{pgfscope}%
\pgfsetbuttcap%
\pgfsetmiterjoin%
\definecolor{currentfill}{rgb}{1.000000,1.000000,1.000000}%
\pgfsetfillcolor{currentfill}%
\pgfsetlinewidth{0.000000pt}%
\definecolor{currentstroke}{rgb}{1.000000,1.000000,1.000000}%
\pgfsetstrokecolor{currentstroke}%
\pgfsetdash{}{0pt}%
\pgfpathmoveto{\pgfqpoint{0.000000in}{0.000000in}}%
\pgfpathlineto{\pgfqpoint{2.222509in}{0.000000in}}%
\pgfpathlineto{\pgfqpoint{2.222509in}{1.591882in}}%
\pgfpathlineto{\pgfqpoint{0.000000in}{1.591882in}}%
\pgfpathclose%
\pgfusepath{fill}%
\end{pgfscope}%
\begin{pgfscope}%
\pgfsetbuttcap%
\pgfsetmiterjoin%
\definecolor{currentfill}{rgb}{1.000000,1.000000,1.000000}%
\pgfsetfillcolor{currentfill}%
\pgfsetlinewidth{0.000000pt}%
\definecolor{currentstroke}{rgb}{0.000000,0.000000,0.000000}%
\pgfsetstrokecolor{currentstroke}%
\pgfsetstrokeopacity{0.000000}%
\pgfsetdash{}{0pt}%
\pgfpathmoveto{\pgfqpoint{0.365549in}{0.311728in}}%
\pgfpathlineto{\pgfqpoint{2.175248in}{0.311728in}}%
\pgfpathlineto{\pgfqpoint{2.175248in}{1.556882in}}%
\pgfpathlineto{\pgfqpoint{0.365549in}{1.556882in}}%
\pgfpathclose%
\pgfusepath{fill}%
\end{pgfscope}%
\begin{pgfscope}%
\pgfsetbuttcap%
\pgfsetroundjoin%
\definecolor{currentfill}{rgb}{0.000000,0.000000,0.000000}%
\pgfsetfillcolor{currentfill}%
\pgfsetlinewidth{0.803000pt}%
\definecolor{currentstroke}{rgb}{0.000000,0.000000,0.000000}%
\pgfsetstrokecolor{currentstroke}%
\pgfsetdash{}{0pt}%
\pgfsys@defobject{currentmarker}{\pgfqpoint{0.000000in}{-0.048611in}}{\pgfqpoint{0.000000in}{0.000000in}}{%
\pgfpathmoveto{\pgfqpoint{0.000000in}{0.000000in}}%
\pgfpathlineto{\pgfqpoint{0.000000in}{-0.048611in}}%
\pgfusepath{stroke,fill}%
}%
\begin{pgfscope}%
\pgfsys@transformshift{0.365549in}{0.311728in}%
\pgfsys@useobject{currentmarker}{}%
\end{pgfscope}%
\end{pgfscope}%
\begin{pgfscope}%
\pgftext[x=0.365549in,y=0.214506in,,top]{\rmfamily\fontsize{5.000000}{6.000000}\selectfont \(\displaystyle 0\)}%
\end{pgfscope}%
\begin{pgfscope}%
\pgfsetbuttcap%
\pgfsetroundjoin%
\definecolor{currentfill}{rgb}{0.000000,0.000000,0.000000}%
\pgfsetfillcolor{currentfill}%
\pgfsetlinewidth{0.803000pt}%
\definecolor{currentstroke}{rgb}{0.000000,0.000000,0.000000}%
\pgfsetstrokecolor{currentstroke}%
\pgfsetdash{}{0pt}%
\pgfsys@defobject{currentmarker}{\pgfqpoint{0.000000in}{-0.048611in}}{\pgfqpoint{0.000000in}{0.000000in}}{%
\pgfpathmoveto{\pgfqpoint{0.000000in}{0.000000in}}%
\pgfpathlineto{\pgfqpoint{0.000000in}{-0.048611in}}%
\pgfusepath{stroke,fill}%
}%
\begin{pgfscope}%
\pgfsys@transformshift{0.817974in}{0.311728in}%
\pgfsys@useobject{currentmarker}{}%
\end{pgfscope}%
\end{pgfscope}%
\begin{pgfscope}%
\pgftext[x=0.817974in,y=0.214506in,,top]{\rmfamily\fontsize{5.000000}{6.000000}\selectfont \(\displaystyle 4\)}%
\end{pgfscope}%
\begin{pgfscope}%
\pgfsetbuttcap%
\pgfsetroundjoin%
\definecolor{currentfill}{rgb}{0.000000,0.000000,0.000000}%
\pgfsetfillcolor{currentfill}%
\pgfsetlinewidth{0.803000pt}%
\definecolor{currentstroke}{rgb}{0.000000,0.000000,0.000000}%
\pgfsetstrokecolor{currentstroke}%
\pgfsetdash{}{0pt}%
\pgfsys@defobject{currentmarker}{\pgfqpoint{0.000000in}{-0.048611in}}{\pgfqpoint{0.000000in}{0.000000in}}{%
\pgfpathmoveto{\pgfqpoint{0.000000in}{0.000000in}}%
\pgfpathlineto{\pgfqpoint{0.000000in}{-0.048611in}}%
\pgfusepath{stroke,fill}%
}%
\begin{pgfscope}%
\pgfsys@transformshift{1.270398in}{0.311728in}%
\pgfsys@useobject{currentmarker}{}%
\end{pgfscope}%
\end{pgfscope}%
\begin{pgfscope}%
\pgftext[x=1.270398in,y=0.214506in,,top]{\rmfamily\fontsize{5.000000}{6.000000}\selectfont \(\displaystyle 8\)}%
\end{pgfscope}%
\begin{pgfscope}%
\pgfsetbuttcap%
\pgfsetroundjoin%
\definecolor{currentfill}{rgb}{0.000000,0.000000,0.000000}%
\pgfsetfillcolor{currentfill}%
\pgfsetlinewidth{0.803000pt}%
\definecolor{currentstroke}{rgb}{0.000000,0.000000,0.000000}%
\pgfsetstrokecolor{currentstroke}%
\pgfsetdash{}{0pt}%
\pgfsys@defobject{currentmarker}{\pgfqpoint{0.000000in}{-0.048611in}}{\pgfqpoint{0.000000in}{0.000000in}}{%
\pgfpathmoveto{\pgfqpoint{0.000000in}{0.000000in}}%
\pgfpathlineto{\pgfqpoint{0.000000in}{-0.048611in}}%
\pgfusepath{stroke,fill}%
}%
\begin{pgfscope}%
\pgfsys@transformshift{1.722823in}{0.311728in}%
\pgfsys@useobject{currentmarker}{}%
\end{pgfscope}%
\end{pgfscope}%
\begin{pgfscope}%
\pgftext[x=1.722823in,y=0.214506in,,top]{\rmfamily\fontsize{5.000000}{6.000000}\selectfont \(\displaystyle 12\)}%
\end{pgfscope}%
\begin{pgfscope}%
\pgfsetbuttcap%
\pgfsetroundjoin%
\definecolor{currentfill}{rgb}{0.000000,0.000000,0.000000}%
\pgfsetfillcolor{currentfill}%
\pgfsetlinewidth{0.803000pt}%
\definecolor{currentstroke}{rgb}{0.000000,0.000000,0.000000}%
\pgfsetstrokecolor{currentstroke}%
\pgfsetdash{}{0pt}%
\pgfsys@defobject{currentmarker}{\pgfqpoint{0.000000in}{-0.048611in}}{\pgfqpoint{0.000000in}{0.000000in}}{%
\pgfpathmoveto{\pgfqpoint{0.000000in}{0.000000in}}%
\pgfpathlineto{\pgfqpoint{0.000000in}{-0.048611in}}%
\pgfusepath{stroke,fill}%
}%
\begin{pgfscope}%
\pgfsys@transformshift{2.175248in}{0.311728in}%
\pgfsys@useobject{currentmarker}{}%
\end{pgfscope}%
\end{pgfscope}%
\begin{pgfscope}%
\pgftext[x=2.175248in,y=0.214506in,,top]{\rmfamily\fontsize{5.000000}{6.000000}\selectfont \(\displaystyle 16\)}%
\end{pgfscope}%
\begin{pgfscope}%
\pgftext[x=1.270398in,y=0.097222in,,top]{\rmfamily\fontsize{7.000000}{8.400000}\selectfont Time (years)}%
\end{pgfscope}%
\begin{pgfscope}%
\pgfsetbuttcap%
\pgfsetroundjoin%
\definecolor{currentfill}{rgb}{0.000000,0.000000,0.000000}%
\pgfsetfillcolor{currentfill}%
\pgfsetlinewidth{0.803000pt}%
\definecolor{currentstroke}{rgb}{0.000000,0.000000,0.000000}%
\pgfsetstrokecolor{currentstroke}%
\pgfsetdash{}{0pt}%
\pgfsys@defobject{currentmarker}{\pgfqpoint{-0.048611in}{0.000000in}}{\pgfqpoint{0.000000in}{0.000000in}}{%
\pgfpathmoveto{\pgfqpoint{0.000000in}{0.000000in}}%
\pgfpathlineto{\pgfqpoint{-0.048611in}{0.000000in}}%
\pgfusepath{stroke,fill}%
}%
\begin{pgfscope}%
\pgfsys@transformshift{0.365549in}{0.354199in}%
\pgfsys@useobject{currentmarker}{}%
\end{pgfscope}%
\end{pgfscope}%
\begin{pgfscope}%
\pgftext[x=0.141975in,y=0.330086in,left,base]{\rmfamily\fontsize{5.000000}{6.000000}\selectfont \(\displaystyle 0.6\)}%
\end{pgfscope}%
\begin{pgfscope}%
\pgfsetbuttcap%
\pgfsetroundjoin%
\definecolor{currentfill}{rgb}{0.000000,0.000000,0.000000}%
\pgfsetfillcolor{currentfill}%
\pgfsetlinewidth{0.803000pt}%
\definecolor{currentstroke}{rgb}{0.000000,0.000000,0.000000}%
\pgfsetstrokecolor{currentstroke}%
\pgfsetdash{}{0pt}%
\pgfsys@defobject{currentmarker}{\pgfqpoint{-0.048611in}{0.000000in}}{\pgfqpoint{0.000000in}{0.000000in}}{%
\pgfpathmoveto{\pgfqpoint{0.000000in}{0.000000in}}%
\pgfpathlineto{\pgfqpoint{-0.048611in}{0.000000in}}%
\pgfusepath{stroke,fill}%
}%
\begin{pgfscope}%
\pgfsys@transformshift{0.365549in}{0.927241in}%
\pgfsys@useobject{currentmarker}{}%
\end{pgfscope}%
\end{pgfscope}%
\begin{pgfscope}%
\pgftext[x=0.141975in,y=0.903129in,left,base]{\rmfamily\fontsize{5.000000}{6.000000}\selectfont \(\displaystyle 0.8\)}%
\end{pgfscope}%
\begin{pgfscope}%
\pgfsetbuttcap%
\pgfsetroundjoin%
\definecolor{currentfill}{rgb}{0.000000,0.000000,0.000000}%
\pgfsetfillcolor{currentfill}%
\pgfsetlinewidth{0.803000pt}%
\definecolor{currentstroke}{rgb}{0.000000,0.000000,0.000000}%
\pgfsetstrokecolor{currentstroke}%
\pgfsetdash{}{0pt}%
\pgfsys@defobject{currentmarker}{\pgfqpoint{-0.048611in}{0.000000in}}{\pgfqpoint{0.000000in}{0.000000in}}{%
\pgfpathmoveto{\pgfqpoint{0.000000in}{0.000000in}}%
\pgfpathlineto{\pgfqpoint{-0.048611in}{0.000000in}}%
\pgfusepath{stroke,fill}%
}%
\begin{pgfscope}%
\pgfsys@transformshift{0.365549in}{1.500284in}%
\pgfsys@useobject{currentmarker}{}%
\end{pgfscope}%
\end{pgfscope}%
\begin{pgfscope}%
\pgftext[x=0.141975in,y=1.476171in,left,base]{\rmfamily\fontsize{5.000000}{6.000000}\selectfont \(\displaystyle 1.0\)}%
\end{pgfscope}%
\begin{pgfscope}%
\pgftext[x=0.086419in,y=0.934305in,,bottom,rotate=90.000000]{\rmfamily\fontsize{7.000000}{8.400000}\selectfont Normalized IPC}%
\end{pgfscope}%
\begin{pgfscope}%
\pgfpathrectangle{\pgfqpoint{0.365549in}{0.311728in}}{\pgfqpoint{1.809698in}{1.245154in}}%
\pgfusepath{clip}%
\pgfsetrectcap%
\pgfsetroundjoin%
\pgfsetlinewidth{0.501875pt}%
\definecolor{currentstroke}{rgb}{0.800000,0.800000,0.800000}%
\pgfsetstrokecolor{currentstroke}%
\pgfsetdash{}{0pt}%
\pgfpathmoveto{\pgfqpoint{0.478655in}{0.311728in}}%
\pgfpathlineto{\pgfqpoint{0.478655in}{1.556882in}}%
\pgfusepath{stroke}%
\end{pgfscope}%
\begin{pgfscope}%
\pgfpathrectangle{\pgfqpoint{0.365549in}{0.311728in}}{\pgfqpoint{1.809698in}{1.245154in}}%
\pgfusepath{clip}%
\pgfsetrectcap%
\pgfsetroundjoin%
\pgfsetlinewidth{0.501875pt}%
\definecolor{currentstroke}{rgb}{0.800000,0.800000,0.800000}%
\pgfsetstrokecolor{currentstroke}%
\pgfsetdash{}{0pt}%
\pgfpathmoveto{\pgfqpoint{0.591762in}{0.311728in}}%
\pgfpathlineto{\pgfqpoint{0.591762in}{1.556882in}}%
\pgfusepath{stroke}%
\end{pgfscope}%
\begin{pgfscope}%
\pgfpathrectangle{\pgfqpoint{0.365549in}{0.311728in}}{\pgfqpoint{1.809698in}{1.245154in}}%
\pgfusepath{clip}%
\pgfsetrectcap%
\pgfsetroundjoin%
\pgfsetlinewidth{0.501875pt}%
\definecolor{currentstroke}{rgb}{0.800000,0.800000,0.800000}%
\pgfsetstrokecolor{currentstroke}%
\pgfsetdash{}{0pt}%
\pgfpathmoveto{\pgfqpoint{0.704868in}{0.311728in}}%
\pgfpathlineto{\pgfqpoint{0.704868in}{1.556882in}}%
\pgfusepath{stroke}%
\end{pgfscope}%
\begin{pgfscope}%
\pgfpathrectangle{\pgfqpoint{0.365549in}{0.311728in}}{\pgfqpoint{1.809698in}{1.245154in}}%
\pgfusepath{clip}%
\pgfsetrectcap%
\pgfsetroundjoin%
\pgfsetlinewidth{0.501875pt}%
\definecolor{currentstroke}{rgb}{0.800000,0.800000,0.800000}%
\pgfsetstrokecolor{currentstroke}%
\pgfsetdash{}{0pt}%
\pgfpathmoveto{\pgfqpoint{0.817974in}{0.311728in}}%
\pgfpathlineto{\pgfqpoint{0.817974in}{1.556882in}}%
\pgfusepath{stroke}%
\end{pgfscope}%
\begin{pgfscope}%
\pgfpathrectangle{\pgfqpoint{0.365549in}{0.311728in}}{\pgfqpoint{1.809698in}{1.245154in}}%
\pgfusepath{clip}%
\pgfsetrectcap%
\pgfsetroundjoin%
\pgfsetlinewidth{0.501875pt}%
\definecolor{currentstroke}{rgb}{0.800000,0.800000,0.800000}%
\pgfsetstrokecolor{currentstroke}%
\pgfsetdash{}{0pt}%
\pgfpathmoveto{\pgfqpoint{0.931080in}{0.311728in}}%
\pgfpathlineto{\pgfqpoint{0.931080in}{1.556882in}}%
\pgfusepath{stroke}%
\end{pgfscope}%
\begin{pgfscope}%
\pgfpathrectangle{\pgfqpoint{0.365549in}{0.311728in}}{\pgfqpoint{1.809698in}{1.245154in}}%
\pgfusepath{clip}%
\pgfsetrectcap%
\pgfsetroundjoin%
\pgfsetlinewidth{0.501875pt}%
\definecolor{currentstroke}{rgb}{0.800000,0.800000,0.800000}%
\pgfsetstrokecolor{currentstroke}%
\pgfsetdash{}{0pt}%
\pgfpathmoveto{\pgfqpoint{1.044186in}{0.311728in}}%
\pgfpathlineto{\pgfqpoint{1.044186in}{1.556882in}}%
\pgfusepath{stroke}%
\end{pgfscope}%
\begin{pgfscope}%
\pgfpathrectangle{\pgfqpoint{0.365549in}{0.311728in}}{\pgfqpoint{1.809698in}{1.245154in}}%
\pgfusepath{clip}%
\pgfsetrectcap%
\pgfsetroundjoin%
\pgfsetlinewidth{0.501875pt}%
\definecolor{currentstroke}{rgb}{0.800000,0.800000,0.800000}%
\pgfsetstrokecolor{currentstroke}%
\pgfsetdash{}{0pt}%
\pgfpathmoveto{\pgfqpoint{1.157292in}{0.311728in}}%
\pgfpathlineto{\pgfqpoint{1.157292in}{1.556882in}}%
\pgfusepath{stroke}%
\end{pgfscope}%
\begin{pgfscope}%
\pgfpathrectangle{\pgfqpoint{0.365549in}{0.311728in}}{\pgfqpoint{1.809698in}{1.245154in}}%
\pgfusepath{clip}%
\pgfsetrectcap%
\pgfsetroundjoin%
\pgfsetlinewidth{0.501875pt}%
\definecolor{currentstroke}{rgb}{0.800000,0.800000,0.800000}%
\pgfsetstrokecolor{currentstroke}%
\pgfsetdash{}{0pt}%
\pgfpathmoveto{\pgfqpoint{1.270398in}{0.311728in}}%
\pgfpathlineto{\pgfqpoint{1.270398in}{1.556882in}}%
\pgfusepath{stroke}%
\end{pgfscope}%
\begin{pgfscope}%
\pgfpathrectangle{\pgfqpoint{0.365549in}{0.311728in}}{\pgfqpoint{1.809698in}{1.245154in}}%
\pgfusepath{clip}%
\pgfsetrectcap%
\pgfsetroundjoin%
\pgfsetlinewidth{0.501875pt}%
\definecolor{currentstroke}{rgb}{0.800000,0.800000,0.800000}%
\pgfsetstrokecolor{currentstroke}%
\pgfsetdash{}{0pt}%
\pgfpathmoveto{\pgfqpoint{1.383505in}{0.311728in}}%
\pgfpathlineto{\pgfqpoint{1.383505in}{1.556882in}}%
\pgfusepath{stroke}%
\end{pgfscope}%
\begin{pgfscope}%
\pgfpathrectangle{\pgfqpoint{0.365549in}{0.311728in}}{\pgfqpoint{1.809698in}{1.245154in}}%
\pgfusepath{clip}%
\pgfsetrectcap%
\pgfsetroundjoin%
\pgfsetlinewidth{0.501875pt}%
\definecolor{currentstroke}{rgb}{0.800000,0.800000,0.800000}%
\pgfsetstrokecolor{currentstroke}%
\pgfsetdash{}{0pt}%
\pgfpathmoveto{\pgfqpoint{1.496611in}{0.311728in}}%
\pgfpathlineto{\pgfqpoint{1.496611in}{1.556882in}}%
\pgfusepath{stroke}%
\end{pgfscope}%
\begin{pgfscope}%
\pgfpathrectangle{\pgfqpoint{0.365549in}{0.311728in}}{\pgfqpoint{1.809698in}{1.245154in}}%
\pgfusepath{clip}%
\pgfsetrectcap%
\pgfsetroundjoin%
\pgfsetlinewidth{0.501875pt}%
\definecolor{currentstroke}{rgb}{0.800000,0.800000,0.800000}%
\pgfsetstrokecolor{currentstroke}%
\pgfsetdash{}{0pt}%
\pgfpathmoveto{\pgfqpoint{1.609717in}{0.311728in}}%
\pgfpathlineto{\pgfqpoint{1.609717in}{1.556882in}}%
\pgfusepath{stroke}%
\end{pgfscope}%
\begin{pgfscope}%
\pgfpathrectangle{\pgfqpoint{0.365549in}{0.311728in}}{\pgfqpoint{1.809698in}{1.245154in}}%
\pgfusepath{clip}%
\pgfsetrectcap%
\pgfsetroundjoin%
\pgfsetlinewidth{0.501875pt}%
\definecolor{currentstroke}{rgb}{0.800000,0.800000,0.800000}%
\pgfsetstrokecolor{currentstroke}%
\pgfsetdash{}{0pt}%
\pgfpathmoveto{\pgfqpoint{1.722823in}{0.311728in}}%
\pgfpathlineto{\pgfqpoint{1.722823in}{1.556882in}}%
\pgfusepath{stroke}%
\end{pgfscope}%
\begin{pgfscope}%
\pgfpathrectangle{\pgfqpoint{0.365549in}{0.311728in}}{\pgfqpoint{1.809698in}{1.245154in}}%
\pgfusepath{clip}%
\pgfsetrectcap%
\pgfsetroundjoin%
\pgfsetlinewidth{0.501875pt}%
\definecolor{currentstroke}{rgb}{0.800000,0.800000,0.800000}%
\pgfsetstrokecolor{currentstroke}%
\pgfsetdash{}{0pt}%
\pgfpathmoveto{\pgfqpoint{1.835929in}{0.311728in}}%
\pgfpathlineto{\pgfqpoint{1.835929in}{1.556882in}}%
\pgfusepath{stroke}%
\end{pgfscope}%
\begin{pgfscope}%
\pgfpathrectangle{\pgfqpoint{0.365549in}{0.311728in}}{\pgfqpoint{1.809698in}{1.245154in}}%
\pgfusepath{clip}%
\pgfsetrectcap%
\pgfsetroundjoin%
\pgfsetlinewidth{0.501875pt}%
\definecolor{currentstroke}{rgb}{0.800000,0.800000,0.800000}%
\pgfsetstrokecolor{currentstroke}%
\pgfsetdash{}{0pt}%
\pgfpathmoveto{\pgfqpoint{1.949035in}{0.311728in}}%
\pgfpathlineto{\pgfqpoint{1.949035in}{1.556882in}}%
\pgfusepath{stroke}%
\end{pgfscope}%
\begin{pgfscope}%
\pgfpathrectangle{\pgfqpoint{0.365549in}{0.311728in}}{\pgfqpoint{1.809698in}{1.245154in}}%
\pgfusepath{clip}%
\pgfsetrectcap%
\pgfsetroundjoin%
\pgfsetlinewidth{0.501875pt}%
\definecolor{currentstroke}{rgb}{0.800000,0.800000,0.800000}%
\pgfsetstrokecolor{currentstroke}%
\pgfsetdash{}{0pt}%
\pgfpathmoveto{\pgfqpoint{2.062142in}{0.311728in}}%
\pgfpathlineto{\pgfqpoint{2.062142in}{1.556882in}}%
\pgfusepath{stroke}%
\end{pgfscope}%
\begin{pgfscope}%
\pgfpathrectangle{\pgfqpoint{0.365549in}{0.311728in}}{\pgfqpoint{1.809698in}{1.245154in}}%
\pgfusepath{clip}%
\pgfsetrectcap%
\pgfsetroundjoin%
\pgfsetlinewidth{1.104125pt}%
\definecolor{currentstroke}{rgb}{0.631373,0.850980,0.607843}%
\pgfsetstrokecolor{currentstroke}%
\pgfsetdash{}{0pt}%
\pgfpathmoveto{\pgfqpoint{0.365549in}{1.499305in}}%
\pgfpathlineto{\pgfqpoint{0.537332in}{1.491883in}}%
\pgfpathlineto{\pgfqpoint{0.562360in}{1.479199in}}%
\pgfpathlineto{\pgfqpoint{0.577477in}{1.470268in}}%
\pgfpathlineto{\pgfqpoint{0.588193in}{1.462886in}}%
\pgfpathlineto{\pgfqpoint{0.596631in}{1.451386in}}%
\pgfpathlineto{\pgfqpoint{0.603560in}{1.433074in}}%
\pgfpathlineto{\pgfqpoint{0.609349in}{1.413254in}}%
\pgfpathlineto{\pgfqpoint{0.614370in}{1.389519in}}%
\pgfpathlineto{\pgfqpoint{0.618838in}{1.352692in}}%
\pgfpathlineto{\pgfqpoint{0.622773in}{1.311013in}}%
\pgfpathlineto{\pgfqpoint{0.626382in}{1.262481in}}%
\pgfpathlineto{\pgfqpoint{0.629725in}{1.197352in}}%
\pgfpathlineto{\pgfqpoint{0.632832in}{1.143927in}}%
\pgfpathlineto{\pgfqpoint{0.635771in}{1.082019in}}%
\pgfpathlineto{\pgfqpoint{0.638684in}{1.026025in}}%
\pgfpathlineto{\pgfqpoint{0.641447in}{0.970234in}}%
\pgfusepath{stroke}%
\end{pgfscope}%
\begin{pgfscope}%
\pgfpathrectangle{\pgfqpoint{0.365549in}{0.311728in}}{\pgfqpoint{1.809698in}{1.245154in}}%
\pgfusepath{clip}%
\pgfsetbuttcap%
\pgfsetroundjoin%
\pgfsetlinewidth{1.104125pt}%
\definecolor{currentstroke}{rgb}{0.631373,0.850980,0.607843}%
\pgfsetstrokecolor{currentstroke}%
\pgfsetdash{{5.500000pt}{1.100000pt}}{0.000000pt}%
\pgfpathmoveto{\pgfqpoint{0.365549in}{1.500284in}}%
\pgfpathlineto{\pgfqpoint{0.735345in}{1.492821in}}%
\pgfpathlineto{\pgfqpoint{0.743465in}{1.478710in}}%
\pgfpathlineto{\pgfqpoint{0.748491in}{1.470798in}}%
\pgfpathlineto{\pgfqpoint{0.752218in}{1.462275in}}%
\pgfpathlineto{\pgfqpoint{0.755184in}{1.453180in}}%
\pgfpathlineto{\pgfqpoint{0.757678in}{1.433564in}}%
\pgfpathlineto{\pgfqpoint{0.759843in}{1.416191in}}%
\pgfpathlineto{\pgfqpoint{0.761711in}{1.399347in}}%
\pgfpathlineto{\pgfqpoint{0.763380in}{1.365987in}}%
\pgfpathlineto{\pgfqpoint{0.764909in}{1.336257in}}%
\pgfpathlineto{\pgfqpoint{0.766323in}{1.297554in}}%
\pgfpathlineto{\pgfqpoint{0.767626in}{1.234097in}}%
\pgfpathlineto{\pgfqpoint{0.768868in}{1.181365in}}%
\pgfpathlineto{\pgfqpoint{0.770039in}{1.126309in}}%
\pgfpathlineto{\pgfqpoint{0.771172in}{1.064360in}}%
\pgfpathlineto{\pgfqpoint{0.772283in}{1.009304in}}%
\pgfusepath{stroke}%
\end{pgfscope}%
\begin{pgfscope}%
\pgfpathrectangle{\pgfqpoint{0.365549in}{0.311728in}}{\pgfqpoint{1.809698in}{1.245154in}}%
\pgfusepath{clip}%
\pgfsetrectcap%
\pgfsetroundjoin%
\pgfsetlinewidth{1.104125pt}%
\definecolor{currentstroke}{rgb}{0.145098,0.145098,0.145098}%
\pgfsetstrokecolor{currentstroke}%
\pgfsetdash{}{0pt}%
\pgfpathmoveto{\pgfqpoint{0.365549in}{1.493840in}}%
\pgfpathlineto{\pgfqpoint{0.899883in}{1.491720in}}%
\pgfpathlineto{\pgfqpoint{0.958967in}{1.281486in}}%
\pgfpathlineto{\pgfqpoint{1.042594in}{1.278387in}}%
\pgfpathlineto{\pgfqpoint{1.094503in}{1.279161in}}%
\pgfpathlineto{\pgfqpoint{1.132338in}{1.262074in}}%
\pgfpathlineto{\pgfqpoint{1.167791in}{1.214970in}}%
\pgfpathlineto{\pgfqpoint{1.201897in}{1.210239in}}%
\pgfpathlineto{\pgfqpoint{1.232166in}{0.994296in}}%
\pgfpathlineto{\pgfqpoint{1.308138in}{0.904493in}}%
\pgfpathlineto{\pgfqpoint{1.395032in}{0.904330in}}%
\pgfpathlineto{\pgfqpoint{1.473178in}{0.903270in}}%
\pgfpathlineto{\pgfqpoint{1.542135in}{0.881981in}}%
\pgfpathlineto{\pgfqpoint{1.613441in}{0.821174in}}%
\pgfpathlineto{\pgfqpoint{1.694735in}{0.812162in}}%
\pgfpathlineto{\pgfqpoint{1.771859in}{0.794543in}}%
\pgfpathlineto{\pgfqpoint{1.853567in}{0.613469in}}%
\pgfpathlineto{\pgfqpoint{2.010807in}{0.561023in}}%
\pgfusepath{stroke}%
\end{pgfscope}%
\begin{pgfscope}%
\pgfpathrectangle{\pgfqpoint{0.365549in}{0.311728in}}{\pgfqpoint{1.809698in}{1.245154in}}%
\pgfusepath{clip}%
\pgfsetbuttcap%
\pgfsetroundjoin%
\pgfsetlinewidth{1.104125pt}%
\definecolor{currentstroke}{rgb}{0.145098,0.145098,0.145098}%
\pgfsetstrokecolor{currentstroke}%
\pgfsetdash{{5.500000pt}{1.100000pt}}{0.000000pt}%
\pgfpathmoveto{\pgfqpoint{0.365549in}{1.492902in}}%
\pgfpathlineto{\pgfqpoint{1.124107in}{1.490659in}}%
\pgfpathlineto{\pgfqpoint{1.154959in}{1.301306in}}%
\pgfpathlineto{\pgfqpoint{1.181740in}{1.279896in}}%
\pgfpathlineto{\pgfqpoint{1.204652in}{1.269904in}}%
\pgfpathlineto{\pgfqpoint{1.227068in}{1.217702in}}%
\pgfpathlineto{\pgfqpoint{1.250928in}{1.206079in}}%
\pgfpathlineto{\pgfqpoint{1.273638in}{0.993643in}}%
\pgfpathlineto{\pgfqpoint{1.331576in}{0.904575in}}%
\pgfpathlineto{\pgfqpoint{1.400356in}{0.904615in}}%
\pgfpathlineto{\pgfqpoint{1.463164in}{0.902535in}}%
\pgfpathlineto{\pgfqpoint{1.521605in}{0.875334in}}%
\pgfpathlineto{\pgfqpoint{1.583882in}{0.818524in}}%
\pgfpathlineto{\pgfqpoint{1.655375in}{0.812162in}}%
\pgfpathlineto{\pgfqpoint{1.723501in}{0.793483in}}%
\pgfpathlineto{\pgfqpoint{1.796843in}{0.606047in}}%
\pgfpathlineto{\pgfqpoint{1.941855in}{0.560656in}}%
\pgfusepath{stroke}%
\end{pgfscope}%
\begin{pgfscope}%
\pgfpathrectangle{\pgfqpoint{0.365549in}{0.311728in}}{\pgfqpoint{1.809698in}{1.245154in}}%
\pgfusepath{clip}%
\pgfsetbuttcap%
\pgfsetroundjoin%
\pgfsetlinewidth{1.104125pt}%
\definecolor{currentstroke}{rgb}{1.000000,0.000000,0.000000}%
\pgfsetstrokecolor{currentstroke}%
\pgfsetdash{{1.100000pt}{1.815000pt}}{0.000000pt}%
\pgfpathmoveto{\pgfqpoint{0.365549in}{0.368326in}}%
\pgfpathlineto{\pgfqpoint{2.175248in}{0.368326in}}%
\pgfusepath{stroke}%
\end{pgfscope}%
\begin{pgfscope}%
\pgfsetrectcap%
\pgfsetmiterjoin%
\pgfsetlinewidth{0.803000pt}%
\definecolor{currentstroke}{rgb}{0.000000,0.000000,0.000000}%
\pgfsetstrokecolor{currentstroke}%
\pgfsetdash{}{0pt}%
\pgfpathmoveto{\pgfqpoint{0.365549in}{0.311728in}}%
\pgfpathlineto{\pgfqpoint{0.365549in}{1.556882in}}%
\pgfusepath{stroke}%
\end{pgfscope}%
\begin{pgfscope}%
\pgfsetrectcap%
\pgfsetmiterjoin%
\pgfsetlinewidth{0.803000pt}%
\definecolor{currentstroke}{rgb}{0.000000,0.000000,0.000000}%
\pgfsetstrokecolor{currentstroke}%
\pgfsetdash{}{0pt}%
\pgfpathmoveto{\pgfqpoint{2.175248in}{0.311728in}}%
\pgfpathlineto{\pgfqpoint{2.175248in}{1.556882in}}%
\pgfusepath{stroke}%
\end{pgfscope}%
\begin{pgfscope}%
\pgfsetrectcap%
\pgfsetmiterjoin%
\pgfsetlinewidth{0.803000pt}%
\definecolor{currentstroke}{rgb}{0.000000,0.000000,0.000000}%
\pgfsetstrokecolor{currentstroke}%
\pgfsetdash{}{0pt}%
\pgfpathmoveto{\pgfqpoint{0.365549in}{0.311728in}}%
\pgfpathlineto{\pgfqpoint{2.175248in}{0.311728in}}%
\pgfusepath{stroke}%
\end{pgfscope}%
\begin{pgfscope}%
\pgfsetrectcap%
\pgfsetmiterjoin%
\pgfsetlinewidth{0.803000pt}%
\definecolor{currentstroke}{rgb}{0.000000,0.000000,0.000000}%
\pgfsetstrokecolor{currentstroke}%
\pgfsetdash{}{0pt}%
\pgfpathmoveto{\pgfqpoint{0.365549in}{1.556882in}}%
\pgfpathlineto{\pgfqpoint{2.175248in}{1.556882in}}%
\pgfusepath{stroke}%
\end{pgfscope}%
\begin{pgfscope}%
\pgfsetbuttcap%
\pgfsetmiterjoin%
\definecolor{currentfill}{rgb}{1.000000,1.000000,1.000000}%
\pgfsetfillcolor{currentfill}%
\pgfsetfillopacity{0.800000}%
\pgfsetlinewidth{1.003750pt}%
\definecolor{currentstroke}{rgb}{0.800000,0.800000,0.800000}%
\pgfsetstrokecolor{currentstroke}%
\pgfsetstrokeopacity{0.800000}%
\pgfsetdash{}{0pt}%
\pgfpathmoveto{\pgfqpoint{1.484203in}{0.909196in}}%
\pgfpathlineto{\pgfqpoint{2.116914in}{0.909196in}}%
\pgfpathquadraticcurveto{\pgfqpoint{2.133581in}{0.909196in}}{\pgfqpoint{2.133581in}{0.925863in}}%
\pgfpathlineto{\pgfqpoint{2.133581in}{1.498549in}}%
\pgfpathquadraticcurveto{\pgfqpoint{2.133581in}{1.515215in}}{\pgfqpoint{2.116914in}{1.515215in}}%
\pgfpathlineto{\pgfqpoint{1.484203in}{1.515215in}}%
\pgfpathquadraticcurveto{\pgfqpoint{1.467536in}{1.515215in}}{\pgfqpoint{1.467536in}{1.498549in}}%
\pgfpathlineto{\pgfqpoint{1.467536in}{0.925863in}}%
\pgfpathquadraticcurveto{\pgfqpoint{1.467536in}{0.909196in}}{\pgfqpoint{1.484203in}{0.909196in}}%
\pgfpathclose%
\pgfusepath{stroke,fill}%
\end{pgfscope}%
\begin{pgfscope}%
\pgfsetrectcap%
\pgfsetroundjoin%
\pgfsetlinewidth{1.104125pt}%
\definecolor{currentstroke}{rgb}{0.631373,0.850980,0.607843}%
\pgfsetstrokecolor{currentstroke}%
\pgfsetdash{}{0pt}%
\pgfpathmoveto{\pgfqpoint{1.500870in}{1.452715in}}%
\pgfpathlineto{\pgfqpoint{1.667536in}{1.452715in}}%
\pgfusepath{stroke}%
\end{pgfscope}%
\begin{pgfscope}%
\pgftext[x=1.734203in,y=1.423549in,left,base]{\rmfamily\fontsize{6.000000}{7.200000}\selectfont FD}%
\end{pgfscope}%
\begin{pgfscope}%
\pgfsetbuttcap%
\pgfsetroundjoin%
\pgfsetlinewidth{1.104125pt}%
\definecolor{currentstroke}{rgb}{0.631373,0.850980,0.607843}%
\pgfsetstrokecolor{currentstroke}%
\pgfsetdash{{5.500000pt}{1.100000pt}}{0.000000pt}%
\pgfpathmoveto{\pgfqpoint{1.500870in}{1.336511in}}%
\pgfpathlineto{\pgfqpoint{1.667536in}{1.336511in}}%
\pgfusepath{stroke}%
\end{pgfscope}%
\begin{pgfscope}%
\pgftext[x=1.734203in,y=1.307345in,left,base]{\rmfamily\fontsize{6.000000}{7.200000}\selectfont FD+6}%
\end{pgfscope}%
\begin{pgfscope}%
\pgfsetrectcap%
\pgfsetroundjoin%
\pgfsetlinewidth{1.104125pt}%
\definecolor{currentstroke}{rgb}{0.145098,0.145098,0.145098}%
\pgfsetstrokecolor{currentstroke}%
\pgfsetdash{}{0pt}%
\pgfpathmoveto{\pgfqpoint{1.500870in}{1.220308in}}%
\pgfpathlineto{\pgfqpoint{1.667536in}{1.220308in}}%
\pgfusepath{stroke}%
\end{pgfscope}%
\begin{pgfscope}%
\pgftext[x=1.734203in,y=1.191141in,left,base]{\rmfamily\fontsize{6.000000}{7.200000}\selectfont L2C2}%
\end{pgfscope}%
\begin{pgfscope}%
\pgfsetbuttcap%
\pgfsetroundjoin%
\pgfsetlinewidth{1.104125pt}%
\definecolor{currentstroke}{rgb}{0.145098,0.145098,0.145098}%
\pgfsetstrokecolor{currentstroke}%
\pgfsetdash{{5.500000pt}{1.100000pt}}{0.000000pt}%
\pgfpathmoveto{\pgfqpoint{1.500870in}{1.104104in}}%
\pgfpathlineto{\pgfqpoint{1.667536in}{1.104104in}}%
\pgfusepath{stroke}%
\end{pgfscope}%
\begin{pgfscope}%
\pgftext[x=1.734203in,y=1.074937in,left,base]{\rmfamily\fontsize{6.000000}{7.200000}\selectfont L2C2+6}%
\end{pgfscope}%
\begin{pgfscope}%
\pgfsetbuttcap%
\pgfsetroundjoin%
\pgfsetlinewidth{1.104125pt}%
\definecolor{currentstroke}{rgb}{1.000000,0.000000,0.000000}%
\pgfsetstrokecolor{currentstroke}%
\pgfsetdash{{1.100000pt}{1.815000pt}}{0.000000pt}%
\pgfpathmoveto{\pgfqpoint{1.500870in}{0.987900in}}%
\pgfpathlineto{\pgfqpoint{1.667536in}{0.987900in}}%
\pgfusepath{stroke}%
\end{pgfscope}%
\begin{pgfscope}%
\pgftext[x=1.734203in,y=0.958734in,left,base]{\rmfamily\fontsize{6.000000}{7.200000}\selectfont 0\% EC}%
\end{pgfscope}%
\end{pgfpicture}%
\makeatother%
\endgroup%

%% file: Figuras/ipc_MI.pgf
\begingroup%
\makeatletter%
\begin{pgfpicture}%
\pgfpathrectangle{\pgfpointorigin}{\pgfqpoint{2.222509in}{1.591882in}}%
\pgfusepath{use as bounding box, clip}%
\begin{pgfscope}%
\pgfsetbuttcap%
\pgfsetmiterjoin%
\definecolor{currentfill}{rgb}{1.000000,1.000000,1.000000}%
\pgfsetfillcolor{currentfill}%
\pgfsetlinewidth{0.000000pt}%
\definecolor{currentstroke}{rgb}{1.000000,1.000000,1.000000}%
\pgfsetstrokecolor{currentstroke}%
\pgfsetdash{}{0pt}%
\pgfpathmoveto{\pgfqpoint{0.000000in}{0.000000in}}%
\pgfpathlineto{\pgfqpoint{2.222509in}{0.000000in}}%
\pgfpathlineto{\pgfqpoint{2.222509in}{1.591882in}}%
\pgfpathlineto{\pgfqpoint{0.000000in}{1.591882in}}%
\pgfpathclose%
\pgfusepath{fill}%
\end{pgfscope}%
\begin{pgfscope}%
\pgfsetbuttcap%
\pgfsetmiterjoin%
\definecolor{currentfill}{rgb}{1.000000,1.000000,1.000000}%
\pgfsetfillcolor{currentfill}%
\pgfsetlinewidth{0.000000pt}%
\definecolor{currentstroke}{rgb}{0.000000,0.000000,0.000000}%
\pgfsetstrokecolor{currentstroke}%
\pgfsetstrokeopacity{0.000000}%
\pgfsetdash{}{0pt}%
\pgfpathmoveto{\pgfqpoint{0.365549in}{0.311728in}}%
\pgfpathlineto{\pgfqpoint{2.175248in}{0.311728in}}%
\pgfpathlineto{\pgfqpoint{2.175248in}{1.556882in}}%
\pgfpathlineto{\pgfqpoint{0.365549in}{1.556882in}}%
\pgfpathclose%
\pgfusepath{fill}%
\end{pgfscope}%
\begin{pgfscope}%
\pgfsetbuttcap%
\pgfsetroundjoin%
\definecolor{currentfill}{rgb}{0.000000,0.000000,0.000000}%
\pgfsetfillcolor{currentfill}%
\pgfsetlinewidth{0.803000pt}%
\definecolor{currentstroke}{rgb}{0.000000,0.000000,0.000000}%
\pgfsetstrokecolor{currentstroke}%
\pgfsetdash{}{0pt}%
\pgfsys@defobject{currentmarker}{\pgfqpoint{0.000000in}{-0.048611in}}{\pgfqpoint{0.000000in}{0.000000in}}{%
\pgfpathmoveto{\pgfqpoint{0.000000in}{0.000000in}}%
\pgfpathlineto{\pgfqpoint{0.000000in}{-0.048611in}}%
\pgfusepath{stroke,fill}%
}%
\begin{pgfscope}%
\pgfsys@transformshift{0.365549in}{0.311728in}%
\pgfsys@useobject{currentmarker}{}%
\end{pgfscope}%
\end{pgfscope}%
\begin{pgfscope}%
\pgftext[x=0.365549in,y=0.214506in,,top]{\rmfamily\fontsize{5.000000}{6.000000}\selectfont \(\displaystyle 0\)}%
\end{pgfscope}%
\begin{pgfscope}%
\pgfsetbuttcap%
\pgfsetroundjoin%
\definecolor{currentfill}{rgb}{0.000000,0.000000,0.000000}%
\pgfsetfillcolor{currentfill}%
\pgfsetlinewidth{0.803000pt}%
\definecolor{currentstroke}{rgb}{0.000000,0.000000,0.000000}%
\pgfsetstrokecolor{currentstroke}%
\pgfsetdash{}{0pt}%
\pgfsys@defobject{currentmarker}{\pgfqpoint{0.000000in}{-0.048611in}}{\pgfqpoint{0.000000in}{0.000000in}}{%
\pgfpathmoveto{\pgfqpoint{0.000000in}{0.000000in}}%
\pgfpathlineto{\pgfqpoint{0.000000in}{-0.048611in}}%
\pgfusepath{stroke,fill}%
}%
\begin{pgfscope}%
\pgfsys@transformshift{0.817974in}{0.311728in}%
\pgfsys@useobject{currentmarker}{}%
\end{pgfscope}%
\end{pgfscope}%
\begin{pgfscope}%
\pgftext[x=0.817974in,y=0.214506in,,top]{\rmfamily\fontsize{5.000000}{6.000000}\selectfont \(\displaystyle 4\)}%
\end{pgfscope}%
\begin{pgfscope}%
\pgfsetbuttcap%
\pgfsetroundjoin%
\definecolor{currentfill}{rgb}{0.000000,0.000000,0.000000}%
\pgfsetfillcolor{currentfill}%
\pgfsetlinewidth{0.803000pt}%
\definecolor{currentstroke}{rgb}{0.000000,0.000000,0.000000}%
\pgfsetstrokecolor{currentstroke}%
\pgfsetdash{}{0pt}%
\pgfsys@defobject{currentmarker}{\pgfqpoint{0.000000in}{-0.048611in}}{\pgfqpoint{0.000000in}{0.000000in}}{%
\pgfpathmoveto{\pgfqpoint{0.000000in}{0.000000in}}%
\pgfpathlineto{\pgfqpoint{0.000000in}{-0.048611in}}%
\pgfusepath{stroke,fill}%
}%
\begin{pgfscope}%
\pgfsys@transformshift{1.270398in}{0.311728in}%
\pgfsys@useobject{currentmarker}{}%
\end{pgfscope}%
\end{pgfscope}%
\begin{pgfscope}%
\pgftext[x=1.270398in,y=0.214506in,,top]{\rmfamily\fontsize{5.000000}{6.000000}\selectfont \(\displaystyle 8\)}%
\end{pgfscope}%
\begin{pgfscope}%
\pgfsetbuttcap%
\pgfsetroundjoin%
\definecolor{currentfill}{rgb}{0.000000,0.000000,0.000000}%
\pgfsetfillcolor{currentfill}%
\pgfsetlinewidth{0.803000pt}%
\definecolor{currentstroke}{rgb}{0.000000,0.000000,0.000000}%
\pgfsetstrokecolor{currentstroke}%
\pgfsetdash{}{0pt}%
\pgfsys@defobject{currentmarker}{\pgfqpoint{0.000000in}{-0.048611in}}{\pgfqpoint{0.000000in}{0.000000in}}{%
\pgfpathmoveto{\pgfqpoint{0.000000in}{0.000000in}}%
\pgfpathlineto{\pgfqpoint{0.000000in}{-0.048611in}}%
\pgfusepath{stroke,fill}%
}%
\begin{pgfscope}%
\pgfsys@transformshift{1.722823in}{0.311728in}%
\pgfsys@useobject{currentmarker}{}%
\end{pgfscope}%
\end{pgfscope}%
\begin{pgfscope}%
\pgftext[x=1.722823in,y=0.214506in,,top]{\rmfamily\fontsize{5.000000}{6.000000}\selectfont \(\displaystyle 12\)}%
\end{pgfscope}%
\begin{pgfscope}%
\pgfsetbuttcap%
\pgfsetroundjoin%
\definecolor{currentfill}{rgb}{0.000000,0.000000,0.000000}%
\pgfsetfillcolor{currentfill}%
\pgfsetlinewidth{0.803000pt}%
\definecolor{currentstroke}{rgb}{0.000000,0.000000,0.000000}%
\pgfsetstrokecolor{currentstroke}%
\pgfsetdash{}{0pt}%
\pgfsys@defobject{currentmarker}{\pgfqpoint{0.000000in}{-0.048611in}}{\pgfqpoint{0.000000in}{0.000000in}}{%
\pgfpathmoveto{\pgfqpoint{0.000000in}{0.000000in}}%
\pgfpathlineto{\pgfqpoint{0.000000in}{-0.048611in}}%
\pgfusepath{stroke,fill}%
}%
\begin{pgfscope}%
\pgfsys@transformshift{2.175248in}{0.311728in}%
\pgfsys@useobject{currentmarker}{}%
\end{pgfscope}%
\end{pgfscope}%
\begin{pgfscope}%
\pgftext[x=2.175248in,y=0.214506in,,top]{\rmfamily\fontsize{5.000000}{6.000000}\selectfont \(\displaystyle 16\)}%
\end{pgfscope}%
\begin{pgfscope}%
\pgftext[x=1.270398in,y=0.097222in,,top]{\rmfamily\fontsize{7.000000}{8.400000}\selectfont Time (years)}%
\end{pgfscope}%
\begin{pgfscope}%
\pgfsetbuttcap%
\pgfsetroundjoin%
\definecolor{currentfill}{rgb}{0.000000,0.000000,0.000000}%
\pgfsetfillcolor{currentfill}%
\pgfsetlinewidth{0.803000pt}%
\definecolor{currentstroke}{rgb}{0.000000,0.000000,0.000000}%
\pgfsetstrokecolor{currentstroke}%
\pgfsetdash{}{0pt}%
\pgfsys@defobject{currentmarker}{\pgfqpoint{-0.048611in}{0.000000in}}{\pgfqpoint{0.000000in}{0.000000in}}{%
\pgfpathmoveto{\pgfqpoint{0.000000in}{0.000000in}}%
\pgfpathlineto{\pgfqpoint{-0.048611in}{0.000000in}}%
\pgfusepath{stroke,fill}%
}%
\begin{pgfscope}%
\pgfsys@transformshift{0.365549in}{0.566562in}%
\pgfsys@useobject{currentmarker}{}%
\end{pgfscope}%
\end{pgfscope}%
\begin{pgfscope}%
\pgftext[x=0.141975in,y=0.542450in,left,base]{\rmfamily\fontsize{5.000000}{6.000000}\selectfont \(\displaystyle 0.6\)}%
\end{pgfscope}%
\begin{pgfscope}%
\pgfsetbuttcap%
\pgfsetroundjoin%
\definecolor{currentfill}{rgb}{0.000000,0.000000,0.000000}%
\pgfsetfillcolor{currentfill}%
\pgfsetlinewidth{0.803000pt}%
\definecolor{currentstroke}{rgb}{0.000000,0.000000,0.000000}%
\pgfsetstrokecolor{currentstroke}%
\pgfsetdash{}{0pt}%
\pgfsys@defobject{currentmarker}{\pgfqpoint{-0.048611in}{0.000000in}}{\pgfqpoint{0.000000in}{0.000000in}}{%
\pgfpathmoveto{\pgfqpoint{0.000000in}{0.000000in}}%
\pgfpathlineto{\pgfqpoint{-0.048611in}{0.000000in}}%
\pgfusepath{stroke,fill}%
}%
\begin{pgfscope}%
\pgfsys@transformshift{0.365549in}{1.032784in}%
\pgfsys@useobject{currentmarker}{}%
\end{pgfscope}%
\end{pgfscope}%
\begin{pgfscope}%
\pgftext[x=0.141975in,y=1.008672in,left,base]{\rmfamily\fontsize{5.000000}{6.000000}\selectfont \(\displaystyle 0.8\)}%
\end{pgfscope}%
\begin{pgfscope}%
\pgfsetbuttcap%
\pgfsetroundjoin%
\definecolor{currentfill}{rgb}{0.000000,0.000000,0.000000}%
\pgfsetfillcolor{currentfill}%
\pgfsetlinewidth{0.803000pt}%
\definecolor{currentstroke}{rgb}{0.000000,0.000000,0.000000}%
\pgfsetstrokecolor{currentstroke}%
\pgfsetdash{}{0pt}%
\pgfsys@defobject{currentmarker}{\pgfqpoint{-0.048611in}{0.000000in}}{\pgfqpoint{0.000000in}{0.000000in}}{%
\pgfpathmoveto{\pgfqpoint{0.000000in}{0.000000in}}%
\pgfpathlineto{\pgfqpoint{-0.048611in}{0.000000in}}%
\pgfusepath{stroke,fill}%
}%
\begin{pgfscope}%
\pgfsys@transformshift{0.365549in}{1.499007in}%
\pgfsys@useobject{currentmarker}{}%
\end{pgfscope}%
\end{pgfscope}%
\begin{pgfscope}%
\pgftext[x=0.141975in,y=1.474894in,left,base]{\rmfamily\fontsize{5.000000}{6.000000}\selectfont \(\displaystyle 1.0\)}%
\end{pgfscope}%
\begin{pgfscope}%
\pgftext[x=0.086419in,y=0.934305in,,bottom,rotate=90.000000]{\rmfamily\fontsize{7.000000}{8.400000}\selectfont Normalized IPC}%
\end{pgfscope}%
\begin{pgfscope}%
\pgfpathrectangle{\pgfqpoint{0.365549in}{0.311728in}}{\pgfqpoint{1.809698in}{1.245154in}}%
\pgfusepath{clip}%
\pgfsetrectcap%
\pgfsetroundjoin%
\pgfsetlinewidth{0.501875pt}%
\definecolor{currentstroke}{rgb}{0.800000,0.800000,0.800000}%
\pgfsetstrokecolor{currentstroke}%
\pgfsetdash{}{0pt}%
\pgfpathmoveto{\pgfqpoint{0.478655in}{0.311728in}}%
\pgfpathlineto{\pgfqpoint{0.478655in}{1.556882in}}%
\pgfusepath{stroke}%
\end{pgfscope}%
\begin{pgfscope}%
\pgfpathrectangle{\pgfqpoint{0.365549in}{0.311728in}}{\pgfqpoint{1.809698in}{1.245154in}}%
\pgfusepath{clip}%
\pgfsetrectcap%
\pgfsetroundjoin%
\pgfsetlinewidth{0.501875pt}%
\definecolor{currentstroke}{rgb}{0.800000,0.800000,0.800000}%
\pgfsetstrokecolor{currentstroke}%
\pgfsetdash{}{0pt}%
\pgfpathmoveto{\pgfqpoint{0.591762in}{0.311728in}}%
\pgfpathlineto{\pgfqpoint{0.591762in}{1.556882in}}%
\pgfusepath{stroke}%
\end{pgfscope}%
\begin{pgfscope}%
\pgfpathrectangle{\pgfqpoint{0.365549in}{0.311728in}}{\pgfqpoint{1.809698in}{1.245154in}}%
\pgfusepath{clip}%
\pgfsetrectcap%
\pgfsetroundjoin%
\pgfsetlinewidth{0.501875pt}%
\definecolor{currentstroke}{rgb}{0.800000,0.800000,0.800000}%
\pgfsetstrokecolor{currentstroke}%
\pgfsetdash{}{0pt}%
\pgfpathmoveto{\pgfqpoint{0.704868in}{0.311728in}}%
\pgfpathlineto{\pgfqpoint{0.704868in}{1.556882in}}%
\pgfusepath{stroke}%
\end{pgfscope}%
\begin{pgfscope}%
\pgfpathrectangle{\pgfqpoint{0.365549in}{0.311728in}}{\pgfqpoint{1.809698in}{1.245154in}}%
\pgfusepath{clip}%
\pgfsetrectcap%
\pgfsetroundjoin%
\pgfsetlinewidth{0.501875pt}%
\definecolor{currentstroke}{rgb}{0.800000,0.800000,0.800000}%
\pgfsetstrokecolor{currentstroke}%
\pgfsetdash{}{0pt}%
\pgfpathmoveto{\pgfqpoint{0.817974in}{0.311728in}}%
\pgfpathlineto{\pgfqpoint{0.817974in}{1.556882in}}%
\pgfusepath{stroke}%
\end{pgfscope}%
\begin{pgfscope}%
\pgfpathrectangle{\pgfqpoint{0.365549in}{0.311728in}}{\pgfqpoint{1.809698in}{1.245154in}}%
\pgfusepath{clip}%
\pgfsetrectcap%
\pgfsetroundjoin%
\pgfsetlinewidth{0.501875pt}%
\definecolor{currentstroke}{rgb}{0.800000,0.800000,0.800000}%
\pgfsetstrokecolor{currentstroke}%
\pgfsetdash{}{0pt}%
\pgfpathmoveto{\pgfqpoint{0.931080in}{0.311728in}}%
\pgfpathlineto{\pgfqpoint{0.931080in}{1.556882in}}%
\pgfusepath{stroke}%
\end{pgfscope}%
\begin{pgfscope}%
\pgfpathrectangle{\pgfqpoint{0.365549in}{0.311728in}}{\pgfqpoint{1.809698in}{1.245154in}}%
\pgfusepath{clip}%
\pgfsetrectcap%
\pgfsetroundjoin%
\pgfsetlinewidth{0.501875pt}%
\definecolor{currentstroke}{rgb}{0.800000,0.800000,0.800000}%
\pgfsetstrokecolor{currentstroke}%
\pgfsetdash{}{0pt}%
\pgfpathmoveto{\pgfqpoint{1.044186in}{0.311728in}}%
\pgfpathlineto{\pgfqpoint{1.044186in}{1.556882in}}%
\pgfusepath{stroke}%
\end{pgfscope}%
\begin{pgfscope}%
\pgfpathrectangle{\pgfqpoint{0.365549in}{0.311728in}}{\pgfqpoint{1.809698in}{1.245154in}}%
\pgfusepath{clip}%
\pgfsetrectcap%
\pgfsetroundjoin%
\pgfsetlinewidth{0.501875pt}%
\definecolor{currentstroke}{rgb}{0.800000,0.800000,0.800000}%
\pgfsetstrokecolor{currentstroke}%
\pgfsetdash{}{0pt}%
\pgfpathmoveto{\pgfqpoint{1.157292in}{0.311728in}}%
\pgfpathlineto{\pgfqpoint{1.157292in}{1.556882in}}%
\pgfusepath{stroke}%
\end{pgfscope}%
\begin{pgfscope}%
\pgfpathrectangle{\pgfqpoint{0.365549in}{0.311728in}}{\pgfqpoint{1.809698in}{1.245154in}}%
\pgfusepath{clip}%
\pgfsetrectcap%
\pgfsetroundjoin%
\pgfsetlinewidth{0.501875pt}%
\definecolor{currentstroke}{rgb}{0.800000,0.800000,0.800000}%
\pgfsetstrokecolor{currentstroke}%
\pgfsetdash{}{0pt}%
\pgfpathmoveto{\pgfqpoint{1.270398in}{0.311728in}}%
\pgfpathlineto{\pgfqpoint{1.270398in}{1.556882in}}%
\pgfusepath{stroke}%
\end{pgfscope}%
\begin{pgfscope}%
\pgfpathrectangle{\pgfqpoint{0.365549in}{0.311728in}}{\pgfqpoint{1.809698in}{1.245154in}}%
\pgfusepath{clip}%
\pgfsetrectcap%
\pgfsetroundjoin%
\pgfsetlinewidth{0.501875pt}%
\definecolor{currentstroke}{rgb}{0.800000,0.800000,0.800000}%
\pgfsetstrokecolor{currentstroke}%
\pgfsetdash{}{0pt}%
\pgfpathmoveto{\pgfqpoint{1.383505in}{0.311728in}}%
\pgfpathlineto{\pgfqpoint{1.383505in}{1.556882in}}%
\pgfusepath{stroke}%
\end{pgfscope}%
\begin{pgfscope}%
\pgfpathrectangle{\pgfqpoint{0.365549in}{0.311728in}}{\pgfqpoint{1.809698in}{1.245154in}}%
\pgfusepath{clip}%
\pgfsetrectcap%
\pgfsetroundjoin%
\pgfsetlinewidth{0.501875pt}%
\definecolor{currentstroke}{rgb}{0.800000,0.800000,0.800000}%
\pgfsetstrokecolor{currentstroke}%
\pgfsetdash{}{0pt}%
\pgfpathmoveto{\pgfqpoint{1.496611in}{0.311728in}}%
\pgfpathlineto{\pgfqpoint{1.496611in}{1.556882in}}%
\pgfusepath{stroke}%
\end{pgfscope}%
\begin{pgfscope}%
\pgfpathrectangle{\pgfqpoint{0.365549in}{0.311728in}}{\pgfqpoint{1.809698in}{1.245154in}}%
\pgfusepath{clip}%
\pgfsetrectcap%
\pgfsetroundjoin%
\pgfsetlinewidth{0.501875pt}%
\definecolor{currentstroke}{rgb}{0.800000,0.800000,0.800000}%
\pgfsetstrokecolor{currentstroke}%
\pgfsetdash{}{0pt}%
\pgfpathmoveto{\pgfqpoint{1.609717in}{0.311728in}}%
\pgfpathlineto{\pgfqpoint{1.609717in}{1.556882in}}%
\pgfusepath{stroke}%
\end{pgfscope}%
\begin{pgfscope}%
\pgfpathrectangle{\pgfqpoint{0.365549in}{0.311728in}}{\pgfqpoint{1.809698in}{1.245154in}}%
\pgfusepath{clip}%
\pgfsetrectcap%
\pgfsetroundjoin%
\pgfsetlinewidth{0.501875pt}%
\definecolor{currentstroke}{rgb}{0.800000,0.800000,0.800000}%
\pgfsetstrokecolor{currentstroke}%
\pgfsetdash{}{0pt}%
\pgfpathmoveto{\pgfqpoint{1.722823in}{0.311728in}}%
\pgfpathlineto{\pgfqpoint{1.722823in}{1.556882in}}%
\pgfusepath{stroke}%
\end{pgfscope}%
\begin{pgfscope}%
\pgfpathrectangle{\pgfqpoint{0.365549in}{0.311728in}}{\pgfqpoint{1.809698in}{1.245154in}}%
\pgfusepath{clip}%
\pgfsetrectcap%
\pgfsetroundjoin%
\pgfsetlinewidth{0.501875pt}%
\definecolor{currentstroke}{rgb}{0.800000,0.800000,0.800000}%
\pgfsetstrokecolor{currentstroke}%
\pgfsetdash{}{0pt}%
\pgfpathmoveto{\pgfqpoint{1.835929in}{0.311728in}}%
\pgfpathlineto{\pgfqpoint{1.835929in}{1.556882in}}%
\pgfusepath{stroke}%
\end{pgfscope}%
\begin{pgfscope}%
\pgfpathrectangle{\pgfqpoint{0.365549in}{0.311728in}}{\pgfqpoint{1.809698in}{1.245154in}}%
\pgfusepath{clip}%
\pgfsetrectcap%
\pgfsetroundjoin%
\pgfsetlinewidth{0.501875pt}%
\definecolor{currentstroke}{rgb}{0.800000,0.800000,0.800000}%
\pgfsetstrokecolor{currentstroke}%
\pgfsetdash{}{0pt}%
\pgfpathmoveto{\pgfqpoint{1.949035in}{0.311728in}}%
\pgfpathlineto{\pgfqpoint{1.949035in}{1.556882in}}%
\pgfusepath{stroke}%
\end{pgfscope}%
\begin{pgfscope}%
\pgfpathrectangle{\pgfqpoint{0.365549in}{0.311728in}}{\pgfqpoint{1.809698in}{1.245154in}}%
\pgfusepath{clip}%
\pgfsetrectcap%
\pgfsetroundjoin%
\pgfsetlinewidth{0.501875pt}%
\definecolor{currentstroke}{rgb}{0.800000,0.800000,0.800000}%
\pgfsetstrokecolor{currentstroke}%
\pgfsetdash{}{0pt}%
\pgfpathmoveto{\pgfqpoint{2.062142in}{0.311728in}}%
\pgfpathlineto{\pgfqpoint{2.062142in}{1.556882in}}%
\pgfusepath{stroke}%
\end{pgfscope}%
\begin{pgfscope}%
\pgfpathrectangle{\pgfqpoint{0.365549in}{0.311728in}}{\pgfqpoint{1.809698in}{1.245154in}}%
\pgfusepath{clip}%
\pgfsetrectcap%
\pgfsetroundjoin%
\pgfsetlinewidth{1.104125pt}%
\definecolor{currentstroke}{rgb}{0.631373,0.850980,0.607843}%
\pgfsetstrokecolor{currentstroke}%
\pgfsetdash{}{0pt}%
\pgfpathmoveto{\pgfqpoint{0.365549in}{1.500284in}}%
\pgfpathlineto{\pgfqpoint{0.490783in}{1.477290in}}%
\pgfpathlineto{\pgfqpoint{0.509055in}{1.453657in}}%
\pgfpathlineto{\pgfqpoint{0.520207in}{1.434112in}}%
\pgfpathlineto{\pgfqpoint{0.528094in}{1.413289in}}%
\pgfpathlineto{\pgfqpoint{0.534352in}{1.389145in}}%
\pgfpathlineto{\pgfqpoint{0.539571in}{1.362957in}}%
\pgfpathlineto{\pgfqpoint{0.543995in}{1.326422in}}%
\pgfpathlineto{\pgfqpoint{0.547761in}{1.279794in}}%
\pgfpathlineto{\pgfqpoint{0.551121in}{1.227418in}}%
\pgfpathlineto{\pgfqpoint{0.554062in}{1.161629in}}%
\pgfpathlineto{\pgfqpoint{0.556786in}{1.098395in}}%
\pgfpathlineto{\pgfqpoint{0.559310in}{1.033755in}}%
\pgfpathlineto{\pgfqpoint{0.561685in}{0.972054in}}%
\pgfpathlineto{\pgfqpoint{0.563916in}{0.911247in}}%
\pgfpathlineto{\pgfqpoint{0.566132in}{0.858615in}}%
\pgfpathlineto{\pgfqpoint{0.568193in}{0.806878in}}%
\pgfusepath{stroke}%
\end{pgfscope}%
\begin{pgfscope}%
\pgfpathrectangle{\pgfqpoint{0.365549in}{0.311728in}}{\pgfqpoint{1.809698in}{1.245154in}}%
\pgfusepath{clip}%
\pgfsetbuttcap%
\pgfsetroundjoin%
\pgfsetlinewidth{1.104125pt}%
\definecolor{currentstroke}{rgb}{0.631373,0.850980,0.607843}%
\pgfsetstrokecolor{currentstroke}%
\pgfsetdash{{5.500000pt}{1.100000pt}}{0.000000pt}%
\pgfpathmoveto{\pgfqpoint{0.365549in}{1.499007in}}%
\pgfpathlineto{\pgfqpoint{0.635291in}{1.476268in}}%
\pgfpathlineto{\pgfqpoint{0.641224in}{1.454423in}}%
\pgfpathlineto{\pgfqpoint{0.644953in}{1.434750in}}%
\pgfpathlineto{\pgfqpoint{0.647713in}{1.414311in}}%
\pgfpathlineto{\pgfqpoint{0.649925in}{1.390678in}}%
\pgfpathlineto{\pgfqpoint{0.651789in}{1.364107in}}%
\pgfpathlineto{\pgfqpoint{0.653427in}{1.331531in}}%
\pgfpathlineto{\pgfqpoint{0.654836in}{1.292058in}}%
\pgfpathlineto{\pgfqpoint{0.656081in}{1.247602in}}%
\pgfpathlineto{\pgfqpoint{0.657232in}{1.193438in}}%
\pgfpathlineto{\pgfqpoint{0.658299in}{1.131353in}}%
\pgfpathlineto{\pgfqpoint{0.659312in}{1.067097in}}%
\pgfpathlineto{\pgfqpoint{0.660256in}{1.004501in}}%
\pgfpathlineto{\pgfqpoint{0.661143in}{0.945611in}}%
\pgfpathlineto{\pgfqpoint{0.661993in}{0.889913in}}%
\pgfpathlineto{\pgfqpoint{0.662837in}{0.837410in}}%
\pgfusepath{stroke}%
\end{pgfscope}%
\begin{pgfscope}%
\pgfpathrectangle{\pgfqpoint{0.365549in}{0.311728in}}{\pgfqpoint{1.809698in}{1.245154in}}%
\pgfusepath{clip}%
\pgfsetrectcap%
\pgfsetroundjoin%
\pgfsetlinewidth{1.104125pt}%
\definecolor{currentstroke}{rgb}{0.145098,0.145098,0.145098}%
\pgfsetstrokecolor{currentstroke}%
\pgfsetdash{}{0pt}%
\pgfpathmoveto{\pgfqpoint{0.365549in}{1.486743in}}%
\pgfpathlineto{\pgfqpoint{0.782098in}{1.483549in}}%
\pgfpathlineto{\pgfqpoint{0.838177in}{1.475374in}}%
\pgfpathlineto{\pgfqpoint{0.874740in}{1.459022in}}%
\pgfpathlineto{\pgfqpoint{0.901922in}{1.347116in}}%
\pgfpathlineto{\pgfqpoint{0.930564in}{1.280944in}}%
\pgfpathlineto{\pgfqpoint{0.957619in}{1.267658in}}%
\pgfpathlineto{\pgfqpoint{0.981940in}{0.816587in}}%
\pgfpathlineto{\pgfqpoint{1.061931in}{0.664186in}}%
\pgfpathlineto{\pgfqpoint{1.153352in}{0.664186in}}%
\pgfpathlineto{\pgfqpoint{1.236223in}{0.663930in}}%
\pgfpathlineto{\pgfqpoint{1.315244in}{0.588305in}}%
\pgfpathlineto{\pgfqpoint{1.455219in}{0.555729in}}%
\pgfpathlineto{\pgfqpoint{1.601223in}{0.555218in}}%
\pgfpathlineto{\pgfqpoint{1.736298in}{0.553430in}}%
\pgfpathlineto{\pgfqpoint{1.870676in}{0.519705in}}%
\pgfpathlineto{\pgfqpoint{2.036501in}{0.507569in}}%
\pgfusepath{stroke}%
\end{pgfscope}%
\begin{pgfscope}%
\pgfpathrectangle{\pgfqpoint{0.365549in}{0.311728in}}{\pgfqpoint{1.809698in}{1.245154in}}%
\pgfusepath{clip}%
\pgfsetbuttcap%
\pgfsetroundjoin%
\pgfsetlinewidth{1.104125pt}%
\definecolor{currentstroke}{rgb}{0.145098,0.145098,0.145098}%
\pgfsetstrokecolor{currentstroke}%
\pgfsetdash{{5.500000pt}{1.100000pt}}{0.000000pt}%
\pgfpathmoveto{\pgfqpoint{0.365549in}{1.486998in}}%
\pgfpathlineto{\pgfqpoint{0.899875in}{1.499901in}}%
\pgfpathlineto{\pgfqpoint{0.921684in}{1.474735in}}%
\pgfpathlineto{\pgfqpoint{0.940385in}{1.458639in}}%
\pgfpathlineto{\pgfqpoint{0.956743in}{1.371644in}}%
\pgfpathlineto{\pgfqpoint{0.974644in}{1.281455in}}%
\pgfpathlineto{\pgfqpoint{0.993639in}{1.258716in}}%
\pgfpathlineto{\pgfqpoint{1.011973in}{0.807900in}}%
\pgfpathlineto{\pgfqpoint{1.073562in}{0.663930in}}%
\pgfpathlineto{\pgfqpoint{1.148087in}{0.664569in}}%
\pgfpathlineto{\pgfqpoint{1.213415in}{0.661886in}}%
\pgfpathlineto{\pgfqpoint{1.283467in}{0.582812in}}%
\pgfpathlineto{\pgfqpoint{1.406735in}{0.556240in}}%
\pgfpathlineto{\pgfqpoint{1.532676in}{0.555218in}}%
\pgfpathlineto{\pgfqpoint{1.651183in}{0.553558in}}%
\pgfpathlineto{\pgfqpoint{1.769399in}{0.519450in}}%
\pgfpathlineto{\pgfqpoint{1.919360in}{0.507569in}}%
\pgfusepath{stroke}%
\end{pgfscope}%
\begin{pgfscope}%
\pgfpathrectangle{\pgfqpoint{0.365549in}{0.311728in}}{\pgfqpoint{1.809698in}{1.245154in}}%
\pgfusepath{clip}%
\pgfsetbuttcap%
\pgfsetroundjoin%
\pgfsetlinewidth{1.104125pt}%
\definecolor{currentstroke}{rgb}{1.000000,0.000000,0.000000}%
\pgfsetstrokecolor{currentstroke}%
\pgfsetdash{{1.100000pt}{1.815000pt}}{0.000000pt}%
\pgfpathmoveto{\pgfqpoint{0.365549in}{0.368326in}}%
\pgfpathlineto{\pgfqpoint{2.175248in}{0.368326in}}%
\pgfusepath{stroke}%
\end{pgfscope}%
\begin{pgfscope}%
\pgfsetrectcap%
\pgfsetmiterjoin%
\pgfsetlinewidth{0.803000pt}%
\definecolor{currentstroke}{rgb}{0.000000,0.000000,0.000000}%
\pgfsetstrokecolor{currentstroke}%
\pgfsetdash{}{0pt}%
\pgfpathmoveto{\pgfqpoint{0.365549in}{0.311728in}}%
\pgfpathlineto{\pgfqpoint{0.365549in}{1.556882in}}%
\pgfusepath{stroke}%
\end{pgfscope}%
\begin{pgfscope}%
\pgfsetrectcap%
\pgfsetmiterjoin%
\pgfsetlinewidth{0.803000pt}%
\definecolor{currentstroke}{rgb}{0.000000,0.000000,0.000000}%
\pgfsetstrokecolor{currentstroke}%
\pgfsetdash{}{0pt}%
\pgfpathmoveto{\pgfqpoint{2.175248in}{0.311728in}}%
\pgfpathlineto{\pgfqpoint{2.175248in}{1.556882in}}%
\pgfusepath{stroke}%
\end{pgfscope}%
\begin{pgfscope}%
\pgfsetrectcap%
\pgfsetmiterjoin%
\pgfsetlinewidth{0.803000pt}%
\definecolor{currentstroke}{rgb}{0.000000,0.000000,0.000000}%
\pgfsetstrokecolor{currentstroke}%
\pgfsetdash{}{0pt}%
\pgfpathmoveto{\pgfqpoint{0.365549in}{0.311728in}}%
\pgfpathlineto{\pgfqpoint{2.175248in}{0.311728in}}%
\pgfusepath{stroke}%
\end{pgfscope}%
\begin{pgfscope}%
\pgfsetrectcap%
\pgfsetmiterjoin%
\pgfsetlinewidth{0.803000pt}%
\definecolor{currentstroke}{rgb}{0.000000,0.000000,0.000000}%
\pgfsetstrokecolor{currentstroke}%
\pgfsetdash{}{0pt}%
\pgfpathmoveto{\pgfqpoint{0.365549in}{1.556882in}}%
\pgfpathlineto{\pgfqpoint{2.175248in}{1.556882in}}%
\pgfusepath{stroke}%
\end{pgfscope}%
\begin{pgfscope}%
\pgfsetbuttcap%
\pgfsetmiterjoin%
\definecolor{currentfill}{rgb}{1.000000,1.000000,1.000000}%
\pgfsetfillcolor{currentfill}%
\pgfsetfillopacity{0.800000}%
\pgfsetlinewidth{1.003750pt}%
\definecolor{currentstroke}{rgb}{0.800000,0.800000,0.800000}%
\pgfsetstrokecolor{currentstroke}%
\pgfsetstrokeopacity{0.800000}%
\pgfsetdash{}{0pt}%
\pgfpathmoveto{\pgfqpoint{1.484203in}{0.909196in}}%
\pgfpathlineto{\pgfqpoint{2.116914in}{0.909196in}}%
\pgfpathquadraticcurveto{\pgfqpoint{2.133581in}{0.909196in}}{\pgfqpoint{2.133581in}{0.925863in}}%
\pgfpathlineto{\pgfqpoint{2.133581in}{1.498549in}}%
\pgfpathquadraticcurveto{\pgfqpoint{2.133581in}{1.515215in}}{\pgfqpoint{2.116914in}{1.515215in}}%
\pgfpathlineto{\pgfqpoint{1.484203in}{1.515215in}}%
\pgfpathquadraticcurveto{\pgfqpoint{1.467536in}{1.515215in}}{\pgfqpoint{1.467536in}{1.498549in}}%
\pgfpathlineto{\pgfqpoint{1.467536in}{0.925863in}}%
\pgfpathquadraticcurveto{\pgfqpoint{1.467536in}{0.909196in}}{\pgfqpoint{1.484203in}{0.909196in}}%
\pgfpathclose%
\pgfusepath{stroke,fill}%
\end{pgfscope}%
\begin{pgfscope}%
\pgfsetrectcap%
\pgfsetroundjoin%
\pgfsetlinewidth{1.104125pt}%
\definecolor{currentstroke}{rgb}{0.631373,0.850980,0.607843}%
\pgfsetstrokecolor{currentstroke}%
\pgfsetdash{}{0pt}%
\pgfpathmoveto{\pgfqpoint{1.500870in}{1.452715in}}%
\pgfpathlineto{\pgfqpoint{1.667536in}{1.452715in}}%
\pgfusepath{stroke}%
\end{pgfscope}%
\begin{pgfscope}%
\pgftext[x=1.734203in,y=1.423549in,left,base]{\rmfamily\fontsize{6.000000}{7.200000}\selectfont FD}%
\end{pgfscope}%
\begin{pgfscope}%
\pgfsetbuttcap%
\pgfsetroundjoin%
\pgfsetlinewidth{1.104125pt}%
\definecolor{currentstroke}{rgb}{0.631373,0.850980,0.607843}%
\pgfsetstrokecolor{currentstroke}%
\pgfsetdash{{5.500000pt}{1.100000pt}}{0.000000pt}%
\pgfpathmoveto{\pgfqpoint{1.500870in}{1.336511in}}%
\pgfpathlineto{\pgfqpoint{1.667536in}{1.336511in}}%
\pgfusepath{stroke}%
\end{pgfscope}%
\begin{pgfscope}%
\pgftext[x=1.734203in,y=1.307345in,left,base]{\rmfamily\fontsize{6.000000}{7.200000}\selectfont FD+6}%
\end{pgfscope}%
\begin{pgfscope}%
\pgfsetrectcap%
\pgfsetroundjoin%
\pgfsetlinewidth{1.104125pt}%
\definecolor{currentstroke}{rgb}{0.145098,0.145098,0.145098}%
\pgfsetstrokecolor{currentstroke}%
\pgfsetdash{}{0pt}%
\pgfpathmoveto{\pgfqpoint{1.500870in}{1.220308in}}%
\pgfpathlineto{\pgfqpoint{1.667536in}{1.220308in}}%
\pgfusepath{stroke}%
\end{pgfscope}%
\begin{pgfscope}%
\pgftext[x=1.734203in,y=1.191141in,left,base]{\rmfamily\fontsize{6.000000}{7.200000}\selectfont L2C2}%
\end{pgfscope}%
\begin{pgfscope}%
\pgfsetbuttcap%
\pgfsetroundjoin%
\pgfsetlinewidth{1.104125pt}%
\definecolor{currentstroke}{rgb}{0.145098,0.145098,0.145098}%
\pgfsetstrokecolor{currentstroke}%
\pgfsetdash{{5.500000pt}{1.100000pt}}{0.000000pt}%
\pgfpathmoveto{\pgfqpoint{1.500870in}{1.104104in}}%
\pgfpathlineto{\pgfqpoint{1.667536in}{1.104104in}}%
\pgfusepath{stroke}%
\end{pgfscope}%
\begin{pgfscope}%
\pgftext[x=1.734203in,y=1.074937in,left,base]{\rmfamily\fontsize{6.000000}{7.200000}\selectfont L2C2+6}%
\end{pgfscope}%
\begin{pgfscope}%
\pgfsetbuttcap%
\pgfsetroundjoin%
\pgfsetlinewidth{1.104125pt}%
\definecolor{currentstroke}{rgb}{1.000000,0.000000,0.000000}%
\pgfsetstrokecolor{currentstroke}%
\pgfsetdash{{1.100000pt}{1.815000pt}}{0.000000pt}%
\pgfpathmoveto{\pgfqpoint{1.500870in}{0.987900in}}%
\pgfpathlineto{\pgfqpoint{1.667536in}{0.987900in}}%
\pgfusepath{stroke}%
\end{pgfscope}%
\begin{pgfscope}%
\pgftext[x=1.734203in,y=0.958734in,left,base]{\rmfamily\fontsize{6.000000}{7.200000}\selectfont 0\% EC}%
\end{pgfscope}%
\end{pgfpicture}%
\makeatother%
\endgroup%

%% file: VIII-Conclusions.tex
\section{Conclusion}
\label{sec:conclusions}

We have introduced L2C2, a new fault-tolerant NV-LLC organization that achieves per-byte write rate reduction without performance loss and allows compressed blocks to be placed in degraded frames. L2C2 evenly distributes the write wear within each frame, uses an appropriate replacement policy, and inherently allows adding redundant capacity in each cache frame, further extending the time in which the cache remains without performance degradation. Compression and decompression circuits have been synthesized, considering intra-frame wear-leveling, concluding that their inclusion seems very feasible in terms of area, power and latency.

On the other hand, we have developed a procedure that allows to forecast in detail the temporal evolution of such NV-LLCs. 
To the best of our knowledge, the proposed forecast procedure is the first in its class. It couples simulation phases in which statistics are gathered from the system with prediction phases in which the bitcells that become faulty are predicted. This methodology has allowed us to compare several NV-LLC organizations in terms of lifetime and performance. It has also allowed us to measure the influence of manufacturing process variability on these results. 

Our evaluation shows that, with an affordable hardware overhead, L2C2 achieves a large lifetime improvement compared to a reference NV-LLC provided with frame disabling. The lifetime is multiplied by a factor from 6 to 37 times depending on the variability in the manufacturing process.
Increasing redundancy significantly increases the time to loss of performance by one to two years in all configurations, regardless of the variability in the manufacturing process. However, it does not increase the lifetime of the L2C2.

Knowledge of how performance evolves through time could be essential for manufacturers to be able to incorporate NVM technologies with the confidence that they can guarantee certain performance for a reasonably appealing time period.

Finally, the new forecast procedure leaves the door open to detailed evaluation of different cache organizations, varying, for example, content management policy between cache levels, replacement policy, or wear-leveling.

%% file: Appendix.tex
\section*{Appendix A. Time scaling of forecasted indexes when considering bitcells with more endurance.}
\label{appendix}

Let
\begin{equation} 
\label{eq1}
    N(w_{b};\ \mu,\ \sigma) = \ \frac{1}{\sigma\sqrt{2\pi}}e^{\frac{{- (w}_{b} - \mu)}{2\sigma^{2}}^{2}}
\end{equation}
be the normal probability distribution function that estimates the number of writes $w_{b}$ causing failure in a \emph{baseline} bitcell. Assuming a constant write rate WR (writes/s) on the baseline bitcell, the probability distribution function of the failure time $t_{b}$ can be obtained from Eq. \ref{eq1} by the linear transformation $t_{b} = \ \frac{w_{b}}{\text{WR}}$:
\begin{equation}
\label{eq2}
    N(t_{b};\ \frac{\mu}{\text{WR}},\ \frac{\sigma}{\text{WR}})
\end{equation}

We can characterize an \emph{improved} bitcell, with $k$ times higher endurance, by applying to Eq. \ref{eq1} and \ref{eq2} the linear transformation
$w_{i} = w_{b} \cdot k$. Thus, the probability distribution functions of the number of writes and failure time for the improved bitcell are, respectively:
\begin{equation}
\label{eq3}
    N(w_{i};\ \mu \cdot k,\ \ \sigma \cdot k)\text{\ \ \ \ }\text{and}\text{\ \ \ \ }N(t_{i};\ \frac{\mu \cdot k}{\text{WR}},\ \frac{\sigma \cdot k}{\text{WR}})    
\end{equation}

On the other hand, the probability of failure of the baseline bitcell, $P_{b}$, at a time $t_{b} \leq t$ is:
\begin{equation}
\label{eq4}
    P_{b}(t_{b} \leq t) = \ \int_{0}^{t}{N(t_{b};\ \frac{\mu}{\text{WR}},\ \frac{\sigma}{\text{WR}})dt_{b}}
\end{equation}

To know the probability of failure of the improved bitcell, $P_{i}$, at a time $t_{i} \leq t$ from $P_{b}$, it is necessary to apply another linear transformation: $t_{b} = \frac{t_{i}}{k}$. Thus:
\begin{equation}
\label{eq5}   
    P_{i}(t_{i} \leq t) = P_{b}(t_{b} \leq {\text{lin}\_\text{trans}}_{i \rightarrow b}(t)) = P_{b}(t_{b} \leq \frac{t}{k}))
\end{equation}

Rewriting the two probabilities as a function of $t$, we have:
\begin{equation}
\label{eq6}
    P_{i}(t) = P_{b}(\frac{t}{k})    
\end{equation}

To conclude, let us consider a cache with $c$ baseline bitcells, each with an endurance approximated by the probability distribution of Eq. \ref{eq1} and subjected to a constant per-cell write rate WR (writes/s). Assuming bit granularity the decrease of its effective capacity with time, $\text{Ceff}_{b}(t)$ is:
\begin{equation}
\label{eq7}
    \text{Ceff}_{b}(t) = C \cdot (1 - P_{b}(t))    
\end{equation}

And for a cache of the same size made with improved bitcells:
\begin{equation}
\label{eq8}
    \text{Ceff}_{i}(t) = C \cdot (1 - P_{b}(\frac{t}{k}))
\end{equation}

In this case, with byte granularity and different write rates in each frame, it can be reasoned in the same way. That is, any forecasted index with enhanced cells at time t matches the same index forecasted with base cells but at time $\frac{t}{k}$.